\documentclass[fleqn]{modacmtrans2e}

\usepackage{amsmath}
\usepackage{amssymb}
\usepackage{latexsym}
\usepackage{stmaryrd}
\usepackage{url}
\usepackage{verbatim}
\usepackage{ifthen}
\usepackage{pstricks}
\usepackage{graphicx}
\usepackage{rotating}
\usepackage{enumerate}

\allowdisplaybreaks

\newdimen\proofrulebreadth \proofrulebreadth=.05em
\newdimen\proofdotseparation \proofdotseparation=1.25ex
\newdimen\proofrulebaseline \proofrulebaseline=2ex
\newcount\proofdotnumber \proofdotnumber=3
\let\then\relax
\def\hfi{\hskip0pt plus.0001fil}
\mathchardef\squigto="3A3B
\newif\ifinsideprooftree\insideprooftreefalse
\newif\ifonleftofproofrule\onleftofproofrulefalse
\newif\ifproofdots\proofdotsfalse
\newif\ifdoubleproof\doubleprooffalse
\let\wereinproofbit\relax
\newdimen\shortenproofleft
\newdimen\shortenproofright
\newdimen\proofbelowshift
\newbox\proofabove
\newbox\proofbelow
\newbox\proofrulename
\def\shiftproofbelow{\let\next\relax\afterassignment\setshiftproofbelow\dimen0 }
\def\shiftproofbelowneg{\def\next{\multiply\dimen0 by-1 }%
\afterassignment\setshiftproofbelow\dimen0 }
\def\setshiftproofbelow{\next\proofbelowshift=\dimen0 }
\def\setproofrulebreadth{\proofrulebreadth}

\def\prooftree{% NESTED ZERO (\ifonleftofproofrule)
\ifnum  \lastpenalty=1
\then   \unpenalty
\else   \onleftofproofrulefalse
\fi
\ifonleftofproofrule
\else   \ifinsideprooftree
        \then   \hskip.5em plus1fil
        \fi
\fi
\bgroup% NESTED ONE (\proofbelow, \proofrulename, \proofabove,
\setbox\proofbelow=\hbox{}\setbox\proofrulename=\hbox{}%
\let\justifies\proofover\let\leadsto\proofoverdots\let\Justifies\proofoverdbl
\let\using\proofusing\let\[\prooftree
\ifinsideprooftree\let\]\endprooftree\fi
\proofdotsfalse\doubleprooffalse
\let\thickness\setproofrulebreadth
\let\shiftright\shiftproofbelow \let\shift\shiftproofbelow
\let\shiftleft\shiftproofbelowneg
\let\ifwasinsideprooftree\ifinsideprooftree
\insideprooftreetrue
\setbox\proofabove=\hbox\bgroup$\displaystyle % NESTED TWO
\let\wereinproofbit\prooftree
\shortenproofleft=0pt \shortenproofright=0pt \proofbelowshift=0pt
\onleftofproofruletrue\penalty1
}

\def\eproofbit{% NESTED TWO
\ifx    \wereinproofbit\prooftree
\then   \ifcase \lastpenalty
        \then   \shortenproofright=0pt  % 0: some other object, no indentation
        \or     \unpenalty\hfil         % 1: empty hypotheses, just glue
        \or     \unpenalty\unskip       % 2: just had a tree, remove glue
        \else   \shortenproofright=0pt  % eh?
        \fi
\fi
\global\dimen0=\shortenproofleft
\global\dimen1=\shortenproofright
\global\dimen2=\proofrulebreadth
\global\dimen3=\proofbelowshift
\global\dimen4=\proofdotseparation
\global\count255=\proofdotnumber
$\egroup  % NESTED ONE
\shortenproofleft=\dimen0
\shortenproofright=\dimen1
\proofrulebreadth=\dimen2
\proofbelowshift=\dimen3
\proofdotseparation=\dimen4
\proofdotnumber=\count255
}

\def\proofover{% NESTED TWO
\eproofbit % NESTED ONE
\setbox\proofbelow=\hbox\bgroup % NESTED TWO
\let\wereinproofbit\proofover
$\displaystyle
}%
\def\proofoverdbl{% NESTED TWO
\eproofbit % NESTED ONE
\doubleprooftrue
\setbox\proofbelow=\hbox\bgroup % NESTED TWO
\let\wereinproofbit\proofoverdbl
$\displaystyle
}%
\def\proofoverdots{% NESTED TWO
\eproofbit % NESTED ONE
\proofdotstrue
\setbox\proofbelow=\hbox\bgroup % NESTED TWO
\let\wereinproofbit\proofoverdots
$\displaystyle
}%
\def\proofusing{% NESTED TWO
\eproofbit % NESTED ONE
\setbox\proofrulename=\hbox\bgroup % NESTED TWO
\let\wereinproofbit\proofusing
\kern0.3em$
}

\def\endprooftree{% NESTED TWO
\eproofbit % NESTED ONE
  \dimen5 =0pt% spread of hypotheses
\dimen0=\wd\proofabove \advance\dimen0-\shortenproofleft
\advance\dimen0-\shortenproofright
\dimen1=.5\dimen0 \advance\dimen1-.5\wd\proofbelow
\dimen4=\dimen1
\advance\dimen1\proofbelowshift \advance\dimen4-\proofbelowshift
\ifdim  \dimen1<0pt
\then   \advance\shortenproofleft\dimen1
        \advance\dimen0-\dimen1
        \dimen1=0pt
        \ifdim  \shortenproofleft<0pt
        \then   \setbox\proofabove=\hbox{%
                        \kern-\shortenproofleft\unhbox\proofabove}%
                \shortenproofleft=0pt
        \fi
\fi
\ifdim  \dimen4<0pt
\then   \advance\shortenproofright\dimen4
        \advance\dimen0-\dimen4
        \dimen4=0pt
\fi
\ifdim  \shortenproofright<\wd\proofrulename
\then   \shortenproofright=\wd\proofrulename
\fi
\dimen2=\shortenproofleft \advance\dimen2 by\dimen1
\dimen3=\shortenproofright\advance\dimen3 by\dimen4
\ifproofdots
\then
        \dimen6=\shortenproofleft \advance\dimen6 .5\dimen0
        \setbox1=\vbox to\proofdotseparation{\vss\hbox{$\cdot$}\vss}%
        \setbox0=\hbox{%
                \advance\dimen6-.5\wd1
                \kern\dimen6
                $\vcenter to\proofdotnumber\proofdotseparation
                        {\leaders\box1\vfill}$%
                \unhbox\proofrulename}%
\else   \dimen6=\fontdimen22\the\textfont2 % height of maths axis
        \dimen7=\dimen6
        \advance\dimen6by.5\proofrulebreadth
        \advance\dimen7by-.5\proofrulebreadth
        \setbox0=\hbox{%
                \kern\shortenproofleft
                \ifdoubleproof
                \then   \hbox to\dimen0{%
                        $\mathsurround0pt\mathord=\mkern-6mu%
                        \cleaders\hbox{$\mkern-2mu=\mkern-2mu$}\hfill
                        \mkern-6mu\mathord=$}%
                \else   \vrule height\dimen6 depth-\dimen7 width\dimen0
                \fi
                \unhbox\proofrulename}%
        \ht0=\dimen6 \dp0=-\dimen7
\fi
\let\doll\relax
\ifwasinsideprooftree
\then   \let\VBOX\vbox
\else   \ifmmode\else$\let\doll=$\fi
        \let\VBOX\vcenter
\fi
\VBOX   {\baselineskip\proofrulebaseline \lineskip.2ex
        \expandafter\lineskiplimit\ifproofdots0ex\else-0.6ex\fi
        \hbox   spread\dimen5   {\hfi\unhbox\proofabove\hfi}%
        \hbox{\box0}%
        \hbox   {\kern\dimen2 \box\proofbelow}}\doll%
\global\dimen2=\dimen2
\global\dimen3=\dimen3
\egroup % NESTED ZERO
\ifonleftofproofrule
\then   \shortenproofleft=\dimen2
\fi
\shortenproofright=\dimen3
\onleftofproofrulefalse
\ifinsideprooftree
\then   \hskip.5em plus 1fil \penalty2
\fi
}

\newcommand*{\nohyp}{\phantom{x}}

\newcommand*{\Cplusplus}{{C\nolinebreak[4]\hspace{-.05em}\raisebox{.4ex}
{\tiny\bf ++}}}

\newcommand*{\vbar}{\mathrel{\mid}}

\newcommand*{\Type}{\mathrm{Type}}
\newcommand*{\dType}{\mathrm{dType}}
\newcommand*{\dT}{\mathrm{dT}}
\newcommand*{\sType}{\mathrm{sType}}
\newcommand*{\sT}{\mathrm{sT}}
\newcommand*{\cType}{\mathrm{cType}}
\newcommand*{\cT}{\mathrm{cT}}
\newcommand*{\Integer}{\mathrm{Integer}}
\newcommand*{\Bool}{\mathrm{Bool}}
\newcommand*{\Id}{\mathrm{Id}}
\newcommand*{\id}{\mathrm{id}}
\newcommand*{\rId}{\mathrm{rId}}
\newcommand*{\idx}{\mathrm{x}}
\newcommand*{\ridx}{\underline{\mathrm{x}}}
\newcommand*{\Exp}{\mathrm{Exp}}
\newcommand*{\Exps}{\mathrm{Exps}}
\newcommand*{\Decl}{\mathrm{Decl}}
\newcommand*{\exceptDecl}{\mathrm{exceptDecl}}
\newcommand*{\Catch}{\mathrm{Catch}}
\newcommand*{\Stmt}{\mathrm{Stmt}}
\newcommand*{\Label}{\mathrm{Label}}
\newcommand*{\Con}{\mathrm{Con}}
\newcommand*{\con}{\mathrm{con}}
\newcommand*{\fps}{\mathrm{fps}}
\newcommand*{\funBody}{\mathrm{Body}}
\newcommand*{\funbody}{\mathrm{body}}
\newcommand*{\main}{\mathrm{main}}
\newcommand*{\es}{\mathrm{es}}
\newcommand*{\formParams}{\mathrm{formParams}}
\newcommand*{\emptysequence}{\boxempty}
\newcommand*{\Glob}{\mathrm{Glob}}

\newcommand*{\NTe}{\Gamma_\mathrm{e}}
\newcommand*{\NTb}{\Gamma_\mathrm{b}}
\newcommand*{\NTd}{\Gamma_\mathrm{d}}
\newcommand*{\NTg}{\Gamma_\mathrm{g}}
\newcommand*{\NTs}{\Gamma_\mathrm{s}}
\newcommand*{\NTk}{\Gamma_\mathrm{k}}
\newcommand*{\Te}{T_\mathrm{e}}
\newcommand*{\Tb}{T_\mathrm{b}}
\newcommand*{\Td}{T_\mathrm{d}}
\newcommand*{\Tg}{T_\mathrm{g}}
\newcommand*{\Ts}{T_\mathrm{s}}
\newcommand*{\Tk}{T_\mathrm{k}}

\newcommand*{\lambdaop}{\mathop{\lambda}\nolimits}

\newcommand*{\NT}[1]{\Gamma_{#1}}
\newcommand*{\NTq}{\Gamma_q}
\newcommand*{\Tq}{T_q}

\newcommand*{\dVal}{\mathrm{dVal}}
\newcommand*{\sVal}{\mathrm{sVal}}
\newcommand*{\sval}{\mathrm{sval}}

\newcommand*{\CtrlMode}{\mathord{\mathrm{CtrlMode}}}
\newcommand*{\cm}{\mathrm{cm}}
\newcommand*{\GotoMode}{\mathord{\mathrm{GotoMode}}}
\newcommand*{\SwitchMode}{\mathord{\mathrm{SwitchMode}}}
\newcommand*{\cmgoto}{\mathop{\mathrm{goto}}\nolimits}
\newcommand*{\cmswitch}{\mathop{\mathrm{switch}}\nolimits}
\newcommand*{\cmbreak}{\mathop{\mathrm{break}}\nolimits}
\newcommand*{\cmcontinue}{\mathop{\mathrm{continue}}\nolimits}
\newcommand*{\cmreturn}{\mathop{\mathrm{return}}\nolimits}
\newcommand*{\cmexec}{\mathrm{exec}}
\newcommand*{\ValMode}{\mathord{\mathrm{ValMode}}}
\newcommand*{\cmvalue}{\mathop{\mathrm{value}}\nolimits}
\newcommand*{\EnvMode}{\mathord{\mathrm{EnvMode}}}
\newcommand*{\cmenv}{\mathrm{env}}
\newcommand*{\ExceptMode}{\mathord{\mathrm{ExceptMode}}}
\newcommand*{\cmexcept}{\mathrm{except}}

\newcommand*{\CtrlState}{\mathord{\mathrm{CtrlState}}}
\newcommand*{\cs}{\mathord{\mathrm{cs}}}
\newcommand*{\ValState}{\mathord{\mathrm{ValState}}}
\newcommand*{\valstate}{\upsilon}
\newcommand*{\ExceptState}{\mathord{\mathrm{ExceptState}}}
\newcommand*{\exceptstate}{\varepsilon}

\newcommand*{\kw}[1]{\mathop{\textup{\textbf{#1}}}}

\newcommand{\defrel}[1]{\mathrel{\buildrel \mathrm{def} \over {#1}}}
\newcommand{\defeq}{\defrel{=}}

\newcommand{\diverges}{\,\mathord{\buildrel \infty \over \longrightarrow}}

\providecommand*{\Nset}{\mathbb{N}}             % Naturals
\providecommand*{\Zset}{\mathbb{Z}}             % Integers
\providecommand*{\Rset}{\mathbb{R}}             % Reals

\newcommand*{\calF}{\ensuremath{\mathcal{F}}}

\newcommand*{\calR}{\ensuremath{\mathcal{R}}}
\newcommand*{\calS}{\ensuremath{\mathcal{S}}}

\newcommand*{\calU}{\ensuremath{\mathcal{U}}}

\newcommand*{\reld}[3]{\mathord{#1}\subseteq#2\times#3}
\newcommand*{\fund}[3]{\mathord{#1}\colon#2\to#3}
\newcommand*{\pard}[3]{\mathord{#1}\colon#2\rightarrowtail#3}

\newcommand{\st}{\mathrel{.}}
\newcommand{\itc}{\mathrel{:}}

\newcommand*{\dom}{\mathop{\mathrm{dom}}\nolimits}
\newcommand*{\restrict}[1]{\mathop{\mid}\nolimits_{#1}}

\newcommand*{\type}{\mathop{\mathrm{type}}\nolimits}

\newcommand*{\lub}{\mathop{\mathrm{lub}}\nolimits}
\newcommand*{\lfp}{\mathop{\mathrm{lfp}}\nolimits}
\newcommand*{\gfp}{\mathop{\mathrm{gfp}}\nolimits}

\newcommand*{\widen}{\mathbin{\nabla}}

\renewcommand{\emptyset}{\varnothing}

\newcommand*{\sseq}{\subseteq}
\newcommand*{\sseqf}{\mathrel{\subseteq_\mathrm{f}}}

\newcommand*{\union}{\cup}
\newcommand*{\bigunion}{\bigcup}
\newcommand*{\inters}{\cap}
\newcommand*{\setdiff}{\setminus}

\newcommand{\sset}[2]{{\renewcommand{\arraystretch}{1.2}
                      \left\{\,#1 \,\left|\,
                               \begin{array}{@{}l@{}}#2\end{array}
                      \right.   \,\right\}}}

\newcommand*{\ttv}{\mathrm{tt}}
\newcommand*{\ffv}{\mathrm{ff}}
\newcommand*{\divop}{\mathbin{/}}
\newcommand*{\modop}{\mathbin{\%}}
\newcommand*{\andop}{\mathbin{\textbf{\textup{and}}}}
\newcommand*{\orop}{\mathbin{\textbf{\textup{or}}}}
\newcommand*{\notop}{\mathop{\textbf{\textup{not}}}}

\newcommand*{\FI}{\mathop{\mathrm{FI}}\nolimits}
\newcommand*{\DI}{\mathop{\mathrm{DI}}\nolimits}
\newcommand*{\SL}{\mathop{\mathrm{SL}}\nolimits}

\newcommand*{\Env}{\mathord{\mathrm{Env}}}
\newcommand*{\emptystring}{\mathord{\epsilon}}

\newcommand*{\RTSExcept}{\mathord{\mathrm{RTSExcept}}}
\newcommand*{\rtsexcept}{\chi}
\newcommand*{\Except}{\mathord{\mathrm{Except}}}
\newcommand*{\except}{\xi}
\newcommand*{\none}{\mathtt{none}}
\newcommand*{\divbyzero}{\mathtt{divbyzero}}
\newcommand*{\stkovflw}{\mathtt{stkovflw}}
\newcommand*{\datovflw}{\mathtt{datovflw}}
\newcommand*{\memerror}{\mathtt{memerror}}
\newcommand*{\caught}{\mathtt{caught}}
\newcommand*{\uncaught}{\mathtt{uncaught}}

\newcommand*{\TEnv}{\mathord{\mathrm{TEnv}}}
\newcommand*{\tinteger}{\mathrm{integer}}
\newcommand*{\tboolean}{\mathrm{boolean}}
\newcommand*{\trtsexcept}{\mathrm{rts\_exception}}

\newcommand*{\Loc}{\mathord{\mathrm{Loc}}}
\newcommand*{\Ind}{\mathrm{Ind}}
\newcommand*{\Addr}{\mathrm{Addr}}
\newcommand*{\Map}{\mathrm{Map}}
\newcommand*{\Stack}{\mathord{\mathrm{Stack}}}
\newcommand*{\Mem}{\mathord{\mathrm{Mem}}}
\newcommand*{\stknew}{\mathop{\mathrm{new}_\mathrm{s}}\nolimits}
\newcommand*{\datnew}{\mathop{\mathrm{new}_\mathrm{d}}\nolimits}
\newcommand*{\txtnew}{\mathop{\mathrm{new}_\mathrm{t}}\nolimits}
\newcommand*{\heapnew}{\mathop{\mathrm{new}_\mathrm{h}}\nolimits}
\newcommand*{\heapdel}{\mathop{\mathrm{delete}_\mathrm{h}}\nolimits}
\newcommand*{\datcleanup}{\mathop{\mathrm{cleanup}_\mathrm{d}}\nolimits}
\newcommand*{\smark}{\mathop{\mathrm{mark}_\mathrm{s}}\nolimits}
\newcommand*{\sunmark}{\mathop{\mathrm{unmark}_\mathrm{s}}\nolimits}
\newcommand*{\slink}{\mathop{\mathrm{link}_\mathrm{s}}\nolimits}
\newcommand*{\sunlink}{\mathop{\mathrm{unlink}_\mathrm{s}}\nolimits}
\newcommand*{\asmark}{\mathop{\mathrm{mark}_\mathrm{s}^\sharp}\nolimits}
\newcommand*{\asunmark}{\mathop{\mathrm{unmark}_\mathrm{s}^\sharp}\nolimits}
\newcommand*{\aslink}{\mathop{\mathrm{link}_\mathrm{s}^\sharp}\nolimits}
\newcommand*{\asunlink}{\mathop{\mathrm{unlink}_\mathrm{s}^\sharp}\nolimits}
\newcommand*{\sm}{\dag}
\newcommand*{\fm}{\ddag}
\newcommand*{\topmost}{\mathop{\mathrm{tf}}\nolimits}
\newcommand*{\datcleanupshort}{\mathop{\mathrm{cu}_\mathrm{d}}\nolimits}
\newcommand*{\sunmarkshort}{\mathop{\mathrm{um}_\mathrm{s}}\nolimits}
\newcommand*{\sunlinkshort}{\mathop{\mathrm{ul}_\mathrm{s}}\nolimits}

\newcommand*{\location}[1]{\mathord{#1 \; \mathrm{loc}}}
\newcommand*{\asupported}{\mathop{\mathrm{supported}^\sharp}\nolimits}
\newcommand*{\aeval}{\mathop{\mathrm{eval}^\sharp}\nolimits}
\newcommand*{\ceval}[1]{\mathop{\mathrm{eval}_{#1}}\nolimits}

\newcommand*{\Abstract}{\mathord{\mathrm{Abstract}}}
\newcommand*{\abs}{\mathord{\mathrm{abs}}}

\newcommand{\intp}{\mathop{\mathrm{int}}\nolimits}

\newcommand*{\bneg}{\mathop{\neg}\nolimits}

\newcommand*{\absuminus}{\mathop{\ominus}\nolimits}
\newcommand*{\absadd}{\mathbin{\oplus}}
\newcommand*{\abssub}{\mathbin{\ominus}}
\newcommand*{\absmul}{\mathbin{\odot}}
\newcommand*{\absdiv}{\mathbin{\oslash}}
\newcommand*{\absmod}{\mathbin{\obar}}
\newcommand*{\abseq}{\mathrel{\triangleq}}
\newcommand*{\absneq}{\mathrel{\not\triangleq}}
\newcommand*{\absleq}{\mathrel{\trianglelefteq}}
\newcommand*{\abslt}{\mathrel{\vartriangleleft}}
\newcommand*{\absgeq}{\mathrel{\trianglerighteq}}
\newcommand*{\absgt}{\mathrel{\vartriangleright}}
\newcommand*{\absneg}{\mathrel{\circleddash}}
\newcommand*{\absor}{\mathrel{\ovee}}
\newcommand*{\absand}{\mathrel{\owedge}}

\newcommand{\summary}[1]{\textrm{\textbf{\textup{#1}}}}

\newcommand*{\sel}{\mathop{\mathrm{sel}}\nolimits}
\newcommand*{\mem}{\mathop{\mathrm{mem}}\nolimits}

\newcommand*{\stimes}{\otimes}
\newcommand*{\spair}[2]{{#1} \otimes {#2}}
\newcommand*{\iT}{\mathrm{iT}}
\newcommand*{\iType}{\mathrm{iType}}
\newcommand*{\tschar}{\mathrm{signed\_char}}
\newcommand*{\tuchar}{\mathrm{unsigned\_char}}
\newcommand*{\flcon}{\mathrm{fl}}
\newcommand*{\Float}{\mathrm{Float}}
\newcommand*{\sccon}{\mathrm{sc}}
\newcommand*{\sChar}{\mathrm{sChar}}
\newcommand*{\uccon}{\mathrm{uc}}
\newcommand*{\uChar}{\mathrm{uChar}}

\newcommand*{\tfloat}{\mathrm{float}}
\newcommand*{\nType}{\mathrm{nType}}
\newcommand*{\nT}{\mathrm{nT}}

\newcommand*{\eType}{\mathrm{eType}}
\newcommand*{\eT}{\mathrm{eT}}
\newcommand*{\aType}{\mathrm{aType}}
\newcommand*{\aT}{\mathrm{aT}}
\newcommand*{\rType}{\mathrm{rType}}
\newcommand*{\rT}{\mathrm{rT}}
\newcommand*{\oType}{\mathrm{oType}}
\newcommand*{\oT}{\mathrm{oT}}
\newcommand*{\fType}{\mathrm{fType}}
\newcommand*{\fT}{\mathrm{fT}}
\newcommand*{\mType}{\mathrm{mType}}
\newcommand*{\mT}{\mathrm{mT}}
\newcommand*{\pType}{\mathrm{pType}}
\newcommand*{\pT}{\mathrm{pT}}
\newcommand*{\Offset}{\mathrm{Offset}}
\newcommand*{\nooffset}{\boxempty}
\newcommand*{\indexoffset}[1]{\mathopen{\boldsymbol{[}}{#1}\mathclose{\boldsymbol{]}}}
\newcommand*{\fieldoffset}[1]{\mathop{\boldsymbol{.}}{#1}}
\newcommand*{\lValue}{\mathrm{LValue}}
\newcommand*{\lvalue}{\mathrm{lval}}
\newcommand*{\pointer}[1]{{#1}\boldsymbol{\ast}}
\newcommand*{\maddress}[1]{\mathop{\&}{#1}}
\newcommand*{\indirection}[1]{\mathop{\boldsymbol{\ast}}{#1}}
\newcommand*{\locnull}{\mathord{l_\mathrm{null}}}
\newcommand*{\ptrmove}{{\mathop{\mathrm{ptrmove}}\nolimits}}
\newcommand*{\ptrdiff}{{\mathop{\mathrm{ptrdiff}}\nolimits}}
\newcommand*{\ptrcmp}{{\mathop{\mathrm{ptrcmp}}\nolimits}}
\newcommand*{\arraysyntax}[3]{\kw{#1} {#2} \kw{of}\,{#3}}
\newcommand*{\arraytype}[2]{\arraysyntax{array}{#1}{#2}}
\newcommand*{\firstof}{{\mathop{\mathrm{firstof}}\nolimits}}
\newcommand*{\arrayindex}{\mathop{\mathrm{index}}\nolimits}
\newcommand*{\locindex}{\mathop{\mathrm{locindex}}\nolimits}
\newcommand*{\recordsyntax}[3]{\kw{#1} {#2} \kw{of}\,{#3}}
\newcommand*{\recordtype}[2]{\recordsyntax{record}{#1}{#2}}
\newcommand*{\field}{\mathop{\mathrm{field}}\nolimits}
\newcommand*{\locfield}{\mathop{\mathrm{locfield}}\nolimits}
\newcommand*{\NTo}{\Gamma_\mathrm{o}}
\newcommand*{\To}{T_\mathrm{o}}
\newcommand*{\NTl}{\Gamma_\mathrm{l}}
\newcommand*{\Tl}{T_\mathrm{l}}
\newcommand*{\arraydatnew}{\mathop{\mathrm{newarray}_\mathrm{d}}\nolimits}
\newcommand*{\arraystknew}{\mathop{\mathrm{newarray}_\mathrm{s}}\nolimits}

\newboolean{TOPLAS}
\setboolean{TOPLAS}{true}
\newtheorem{theorem}{Theorem}[section]
\newtheorem{definition}[theorem]{Definition}

\newtheorem{proposition}[theorem]{Proposition}

\ifthenelse{\boolean{TOPLAS}}{

}{

\newcommand*{\citeNN}[1]{\cite{#1}}

\newenvironment{proof}
               {
                 \begin{pf}
               }
               {
                 \end{pf}
               }

}

\begin{document}

\ifthenelse{\boolean{TOPLAS}}{
}{
\begin{frontmatter}
}

\title{%
  On the Design of Generic Static Analyzers \\
  for Modern Imperative Languages
}

\begin{abstract}
The design and implementation of precise static analyzers for
significant fragments of modern imperative languages like C,
\Cplusplus{}, Java and Python is a challenging problem.
In this paper, we consider a core imperative language
that has several features found in mainstream languages such as those
including recursive functions,
run-time system and user-defined exceptions, and a realistic data and
memory model.
For this language we provide a concrete semantics
---characterizing both finite and infinite computations---
and a generic abstract semantics that we prove sound with
respect to the concrete one.
We say the abstract semantics is generic since it is designed to be
completely parametric on the analysis domains:
in particular, it provides support for
\emph{relational} domains (i.e., abstract domains that can capture the
relationships between different data objects).
We also sketch how the proposed methodology can be extended to accommodate
a larger language that includes pointers, compound data objects and
non-structured control flow mechanisms.
The approach, which is based on structured, big-step
$\mathrm{G}^\infty\mathrm{SOS}$ operational semantics and on abstract
interpretation, is modular in that the overall static analyzer is
naturally partitioned into components with clearly identified
responsibilities and interfaces, something that greatly simplifies
both the proof of correctness and the implementation.
\end{abstract}

\ifthenelse{\boolean{TOPLAS}}{

\author{%
ROBERTO BAGNARA \\
    Department of Mathematics,
    University of Parma,
    Italy
\and
PATRICIA M. HILL \\
    School of Computing,
    University of Leeds,
    UK
\and
ANDREA PESCETTI, and
ENEA ZAFFANELLA \\
    Department of Mathematics,
    University of Parma,
    Italy
}

\markboth{%
R.~Bagnara,
P.M.~Hill,
A.~Pescetti,
and E.~Zaffanella}%
{On the Design of Generic Static Analyzers for Modern Imperative Languages}

\category{F3.1}{Logics and Meanings of Programs}{Specifying and Verifying and Reasoning about Programs.}
\terms{Languages, Verification.}

\begin{bottomstuff}
This work has been partly supported by MIUR project
``AIDA --- Abstract Interpretation: Design and Applications''
and by a Royal Society (UK) International Joint Project (ESEP) award.
\end{bottomstuff}

}{
\author[Parma]{Roberto Bagnara},
\ead{bagnara@cs.unipr.it}
\author[Leeds]{Patricia M. Hill},
\ead{hill@comp.leeds.ac.uk}
\author[Parma]{Andrea Pescetti}
\ead{pescetti@cs.unipr.it}
\author[Parma]{Enea Zaffanella}
\ead{zaffanella@cs.unipr.it}

\address[Parma]{Department of Mathematics, University of Parma, Italy}
\address[Leeds]{School of Computing, University of Leeds, UK}

\end{frontmatter}
}
\maketitle

%\tableofcontents

\section{Introduction}

The last few years have witnessed significant progress toward
achieving the ideal of the program verification grand challenge
\cite{Hoare03}. Still, the distance separating us from that ideal can
be measured by the substantial lack of available tools that are able
to verify the absence of relevant classes of run-time errors in code
written in (reasonably rich fragments of) mainstream imperative
languages like C, \Cplusplus{}, Java and Python.  True: there is a
handful of commercial products that target generic applications
written in C, but little is known about them.  In contrast, several
papers explain the essence of the techniques employed by the
\emph{ASTR\'EE} analyzer to formally and automatically verify the
absence of run-time errors in large safety-critical embedded
control/command codes \cite{BlanchetCCFMMMR02,BlanchetCCFMMMR03};
however, ASTR\'EE is specially targeted at a particular class of
programs and program properties, so that widening its scope of
application is likely to require significant effort \cite{Cousot05b}.
It is interesting to observe that, among the dozens of software
development tools that are freely available, there are hardly any
that, by analyzing the program semantics, are able to certify the
absence of important classes of run-time hazards such as, say, the
widely known \emph{buffer overflows} in C code.

The reason for the current, extreme scarcity of the resource ``precise
analyzers for mainstream programming languages'' is that the design
and implementation of such analyzers is a very challenging problem.
The theory of abstract interpretation \cite{CousotC77,CousotC92fr} is
crucial to the management of the complexity of this problem and, in
fact, both ASTR\'EE and the existing commercial analyzers are
(as far as we know) based on it.
Static analysis via abstract interpretation is conducted by mimicking
the execution of the analyzed programs on an \emph{abstract domain}.
This is a set of computable representations of program properties
equipped with all the operations required to mirror, in an approximate
though correct way, the real, \emph{concrete} executions of the
program.
Over the last decade, research and development on the abstract domains
has led to the availability of several implementations of a wide range
of abstract domains: from the most efficient though imprecise, to the
most precise though inefficient.  Simplification and acceleration
techniques have also been developed to mitigate the effects of this
complexity/precision trade-off.  So the lack of semantics-based static
analyzers is not ascribable to a shortage of abstract domains and
their implementations.
The point is that there is more to a working analyzer than a collection
of abstract domains:
\begin{enumerate}[(i)]
\item
A \emph{concrete semantics} must be selected for the analyzed language
that models all the aspects of executions that are relevant to the
properties of interest.  This semantics must be recognizable as a
sound characterization of the language at the intended level of
abstraction.
\item
An \emph{abstract semantics} must be selected and correlated to the
concrete semantics.  This requires a proof of correctness that, while
greatly simplified by abstract interpretation theory,
can be a time-consuming task by highly qualified individuals.
\item
An algorithm to finitely and efficiently compute (approximations of)
the abstract semantics must be selected.
\item
For good results, the abstract domain needs to be an object that is
both complex and easily adaptable.  So, instead of designing a new
domain from scratch, it is often better if one can be obtained by
combining simpler, existing, abstract domains.  Even though the theory
of abstract interpretation provides important conceptual instruments
for the design of such a combination, a significant effort is still
needed to achieve, as far as possible, the desired precision and
efficiency levels.  Note that this point can have an impact on
points~(ii) and~(iii): a \emph{generic} abstract semantics has the
advantage of not requiring an entirely new proof and a new algorithm
each time the abstract domain changes.
\end{enumerate}
This paper, which is the first product of a long-term research plan
that is meant to deal with all of the points above, specifically
addresses points~(i) and~(ii) and refers to a slight generalization
of existing techniques for point~(iii).

\subsection{Contribution}

We build on ideas that have been around for quite some time but, as
far as we know, have never been sufficiently elaborated to be applied to
the description and analysis of realistic imperative languages.  In
extreme synthesis, the contribution consists in filling a good portion
of the gaps that have impeded the application of these ideas to
complex imperative programming languages such as C.%
\footnote{It is worth noticing that we improperly refer to the C language
to actually mean some more constrained language
---like CIL, the \emph{C Intermediate Language}
described in \cite{NeculaMRW02}---
where all ambiguities have been removed, in addition to an
ABI (\emph{Application Binary Interface}) that further
defines its semantics.  Similarly, by `Python' we
mean a tractable subset of the language, such as the \emph{RPython}
subset being developed by the \emph{PyPy} project (\url{http://pypy.org/}).}

More precisely, here we define the concrete and generic abstract
semantics constructions for a language
---called CPM---
that incorporates all the features of mainstream, single-threaded
imperative programming languages that can be somehow problematic
from the point of view of static analysis.
Most notably, the CPM language features:
a non-toy memory model;
exceptions;
run-time errors modeled via exceptions
(for instance, an exception is raised whenever a division by zero is
attempted, when a stack allocation request causes a stack overflow
or when other memory errors occur);
array types;
pointer types to both data objects and functions;
short-circuit evaluation of Boolean operators;
user-defined (possibly recursive) functions;
and
non-structured control flow mechanisms.

For the description of the concrete dynamic semantics of the language
we have used a structured operational semantics (SOS) approach extended to
deal with infinite computations, mainly building on the work of Kahn,
Plotkin and Cousot.  With respect to what can be found in the
literature, we have added the treatment of all non-structured control
flow mechanisms of the C language.  Of course, as the ultimate goal of
this research is to end up with practical analysis tools, the concrete dynamic
semantics has been defined in order to facilitate as much as possible
the subsequent abstraction phase.  Still, our dynamic semantics
retains all the traditional good features: in particular, the concrete
rule schemata are plainly readable (assuming the reader becomes
sufficiently familiar with the unavoidable notational conventions) and
fairly concise.

For the abstract semantics, we build on the work of Schmidt by providing
the concrete dynamic semantics rules with abstract counterparts.
As far as we know, this is the first time that Schmidt's proposal
is applied to the analysis of a realistic programming language
[D.~Schmidt, personal communication, 2004].
A remarkable feature of our abstract semantics is that it is truly
generic in that it fully supports relational abstract domains: the key
step in this direction is the identification and specification of a
suitable set of operators on (concrete and abstract) memory
structures, that allow for domain-independent approximations but
without inherent limitations on the obtainable precision.

Schmidt's proposal about the abstract interpretation of natural semantics
has, in our opinion, two important advantages: concrete and abstract rules
can be made executable and are easily correlated.
We review these two aspects in turn.

Even though here we do not provide details in this respect,
a prototype system ---called ECLAIR\footnote{The
`Extended CLAIR' system targets the analysis of mainstream programming
languages by building upon CLAIR, the `Combined Language and Abstract
Interpretation Resource', which was initially developed and used in a
teaching context (see \url{http://www.cs.unipr.it/clair/}).}---
has been developed in parallel with the writing of the present paper.
The Prolog implementation exploits nice features
of a semantics construction based on SOS approach:
the concrete semantics rule schemata can be directly translated into
Prolog clauses; and the resulting interpreter, with the help of a \Cplusplus{}
implementation of memory structures, is efficient enough to run non-trivial
programs.
Similar considerations apply to the modules implementing the abstract
semantics: the abstract semantics rules are almost directly translated
to generic Prolog code that is interfaced with specialized libraries
implementing several abstract domains, including accurate ones
such as the ones provided by the Parma Polyhedra Library
\cite{BagnaraHRZ05SCP,BagnaraHZ05FAC,BagnaraHZ06TR}.
So, following this approach, the distance between the expression
of the concrete semantics and its executable realization is,
as is well known, very little;  but the same can be said about the
distance between the specification of the abstract semantics and
the static analyzer that results from its implementation.
This prototype system therefore gives us confidence that
both the concrete and abstract semantics are correctly modeled and
that, in this paper, no real difficulties have been overlooked.

For space reasons, only a subset of CPM is treated in full depth in
the main body of the paper (the extension of the design to the full
language is only briefly described even though all the important points
are covered).  For this subset, we give a complete proof of
correctness that relates the abstract semantics to the concrete
semantics.  The proofs are not complicated and suggest (also because
of the way we present them) the possibility of their automatization.
To summarize, at this stage of the research work it does not seem
unreasonable that we may end up with: readable and executable
representations of the concrete semantics of mainstream programming
languages; readable and executable representations of program analyzers;
correctness of the analyzers established by automatic specialized
theorem provers; and, at last, availability of sophisticated program
analyzers for such languages.

A final word is due to address the following concern:
if the target languages
are ``real'' imperative programming languages, why choose CPM, an
unreal one?  The reason is indeed quite simple:
Java and Python miss some of the ``hard'' features of C;
C~misses exceptions; \Cplusplus{} is too hard, for the time being.
So, choosing any one of these real languages
would have been unlikely to provide us with the answer we were looking for,
which was about the adequacy of Schmidt's approach with respect to
the above goals.  Moreover, in its ECLAIR realization, the CPM language
is being extended so as to become a superset of C (i.e., with all the
floating-point and integer types, cast and bitwise operators and so forth).
Once that code has stabilized, a C and a Java subsystem will be forked.

\subsection{Related Work}

The literature on abstract interpretation proposes several
\emph{frameworks} for static analysis, where the more general
approaches put forward in foundational papers are partially specialized
according to a given criterion.
For a few examples of specializations based on the programming paradigm,
one can mention
the frameworks in~\cite{Bruynooghe91} and \cite{GiacobazziDL92}
for the analysis of (constraint) logic programs;
the approach in~\cite{CousotC94} for the analysis of functional programs;
and the so called ``Marktoberdorf'98 generic static analyzer''
specified in~\cite{Cousot98mark} for the analysis of imperative programs.

All of these frameworks are ``generic'' in that,
while fixing some of the parameters of the considered problem,
they are still characterized by several degrees of freedom.
It is therefore natural to reason on the similarities and differences
between these approaches.
However, independently from the programming paradigm under analysis,
direct comparisons between frameworks are extremely difficult in that
each proposal typically focuses on the solution of a subset of the
relevant issues, while partially disregarding other important problems.
For instance, both \cite{Bruynooghe91} and \cite{GiacobazziDL92}
study the generic algebraic properties that allow for a clean and safe
separation between the abstract domains and the abstract interpreter;
in contrast, \cite{Cousot98mark} provides full details
for a specific instance of the proposed framework, ranging from
the parsing of literal constants to the explicit implementation
of the abstract operators for the abstract domain of intervals.
On the other hand, the frameworks mentioned above differ from
the one presented in this paper in that they allow for significant
simplifications of the language analyzed.
Here we briefly discuss the main differences between
the language considered in our proposal and the one in~\cite{Cousot98mark}.

At the syntactic level, as already mentioned,
the language CPM is much richer than the simple imperative language
adopted in~\cite{Cousot98mark}, which has no support for functions,
nesting of block statements, exceptions, non-structured control flows
and it allows for a single data type (in particular, no pointers and arrays).
These syntactic differences are clearly mirrored at the semantics level.
In particular, even though the detection of initialization and arithmetic
errors is considered by the semantics in~\cite{Cousot98mark},
the actual process of error propagation is not modeled.
In contrast, the semantics construction we propose can easily accommodate
the sophisticated exception propagation and handling mechanisms
that can be found in modern languages such as \Cplusplus{}, Java and Python.
Note that this choice has a non-trivial impact on the specification
of the other components of the semantic construction. For example,
the short-circuit evaluation of Boolean expressions cannot be
normalized as proposed in~\cite{Cousot98mark}, because such a normalization
process, by influencing the order of evaluation of subexpressions,
is unable to preserve the concrete semantics as far as exceptional
computation paths are concerned.
A minor difference is in the modeling of integer variables and values:
while \cite{Cousot98mark} considers the case of possibly uninitialized
variables taking values in a finite set of machine-representable integers,
for ease of presentation we have opted for definitely initialized
variables storing arbitrary (i.e., unbounded) integer values.
Since the CPM language supports an extensible set of RTS exceptions,
the specification of a semantics modeling (the generation, propagation
and handling of) uninitialization errors is rather straightforward.
An extension of the semantics to the case of several sets of bounded
and unbounded numerical types, with suitable type conversion functions,
is under development.
Another difference is in the generality of the abstract semantics
construction: following the approach described here, an analyzer can
take full advantage of the more accurate information provided by a
relational domain such as that of polyhedra.  In contrast, the work in
\cite{Cousot98mark} only considers the simpler case of non-relational
abstract domains.
As mentioned above, the semantics we propose also models the case of
possibly recursive functions
(with a call-by-value parameter passing mechanism),
which are not supported by the language syntax
considered in~\cite{Cousot98mark}.
While both this paper and~\cite{Cousot98mark} consider
the specification of a \emph{forward} static analysis framework,
\cite{Cousot98mark} also provides a backward analysis for arithmetic
expressions, to be used in reductive iterations so as to improve precision
losses that are usually incurred by non-relational approximations.

\subsection{Plan of the Paper}

The paper is organized as follows.
Section~\ref{sec:prelims} introduces the notation and terminology
used throughout the paper;
Section~\ref{sec:syntax} defines the syntax of a subset of the
imperative language CPM, whereas Section~\ref{sec:static-semantics}
defines its static semantics;
the concrete dynamic semantics of this fragment is presented in
Section~\ref{sec:concrete-dynamic-semantics}, whereas its abstract
counterpart is defined in  Section~\ref{sec:abstract-dynamic-semantics}.
The proof of correctness of the abstract semantics is the subject
of Section~\ref{sec:abstract-semantics-correctness},
while the computation of further approximations is treated
in Section~\ref{sec:computable-approximations}.
The integration of the full CPM language in the analysis framework
presented in this paper is discussed in Section~\ref{sec:extensions}.
Section~\ref{sec:conclusion} concludes.

\section{Preliminaries}
\label{sec:prelims}

Let $S$ and $T$ be sets.
The notation $S \sseqf T$ means that $S$ is a \emph{finite} subset of $T$.
We write $S \uplus T$ to denote the union $S \union T$, yet
emphasizing the fact that $S \inters T = \emptyset$.
The set of total (resp., partial) functions from $S$ to $T$
is denoted by $S \to T$ (resp., $S \rightarrowtail T$).
We denote by $\dom(f)$ the \emph{domain} of a function
$\fund{f}{S}{T}$ (resp., $\pard{f}{S}{T}$),
where $\dom(f) = S$ (resp., $\dom(f) \sseq S$).
Let $(S, \preceq)$ be a partial order and $\fund{f}{S}{S}$ be a function.
An element $x \in S$ such that $x = f(x)$ (resp., $x \preceq f(x)$)
is called a \emph{fixpoint} (resp., \emph{post-fixpoint}) of $f$.
The notation $\lfp_{\mathord{\preceq}}(f)$
(resp., $\gfp_{\mathord{\preceq}}(f)$)
stands, if it exists, for the least (resp., greatest) fixpoint of $f$.
A complete lattice is a partial order $(S, \preceq)$ such that
$\lub T$ exists for each $T \sseq S$.  If $\fund{f}{S}{S}$ is monotonic
over the complete lattice $S$,
the Knaster-Tarski theorem ensures that the set of post-fixpoints of $f$
is itself a complete lattice.
The \emph{fixpoint coinduction} proof principle follows:
if $f$ is monotonic over the complete lattice $S$ then,
in order to prove that $x \preceq \gfp_{\mathord{\preceq}}(f)$,
it is sufficient to prove that $x \preceq f(x)$.

Let $S = \{s_1, \ldots, s_n \}$ be a finite set of cardinality $n \geq 0$.
Then, the notation
$\{ s_1 \mapsto t_1, \ldots, s_n \mapsto t_n \}$,
where $\{t_1, \ldots, t_n \} \sseq T$, stands for the function
$\fund{f}{S}{T}$ such that $f(s_i) = t_i$, for each $i = 1$, \ldots, $n$.
Note that, assuming that the codomain $T$ is clear from context,
the empty set $\emptyset$ denotes the (nowhere defined) function
$\fund{f}{\emptyset}{T}$.

When denoting the application of a function
$\fund{f}{(S_1 \times\cdots\times S_n)}{T}$
we omit, as customary, the outer parentheses and write
$f(s_1, \ldots, s_n)$ to mean $f\bigl((s_1, \ldots, s_n)\bigr)$.

Let $\pard{f_0}{S_0}{T_0}$ and $\pard{f_1}{S_1}{T_1}$ be partial functions.
Then the function
$\pard{f_0[f_1]}{(S_0 \union S_1)}{(T_0 \union T_1)}$
is defined, for each $x \in \dom(f_0) \union \dom(f_1)$, by
\[
  \bigl(f_0[f_1]\bigr)(x)
    \defeq
      \begin{cases}
        f_1(x), &\text{if $x \in \dom(f_1)$;} \\
        f_0(x), &\text{if $x \in \dom(f_0) \setminus \dom(f_1)$.}
      \end{cases}
\]
(Note that, if $f_0$ and $f_1$ are total functions,
then $f_0[f_1]$ is total too.)

For a partial function $\pard{f}{S}{T}$ and a set $S' \subseteq S$,
$f\restrict{S'}$ denotes the restriction of $f$ to $S'$,
i.e., the function $\pard{f\restrict{S'}}{S'}{T}$ defined,
for each $x \in S' \inters \dom(f)$, by
$f\restrict{S'}(x) = f(x)$.
(Note that, if $f$ is a total function, then $f\restrict{S'}$ is total too.)
With a minor abuse of notation, we will sometimes write
$f \setminus S''$ to denote $f \restrict{S \setminus S''}$.

$S^\star$ denotes the set of all finite, possibly empty strings
of symbols taken from $S$.
The empty string is denoted by $\emptystring$.
If $w, z \in S \union S^\star$, the concatenation of $w$ and $z$
is an element of $S^\star$ denoted by $wz$ or, to avoid ambiguities,
by $w \cdot z$. The length of a string $z$ is denoted by $|z|$.

The \emph{integer part} function $\fund{\intp}{\Rset}{\Zset}$
is given, for each $x \in \Rset$, by
$\intp(x) \defeq \lfloor x \rfloor$, if $x \geq 0$,
and
$\intp(x) \defeq \lceil  x \rceil$, if $x < 0$.
The \emph{integer division} and the \emph{modulo} operations
\(
  \fund{\mathord{\div}, \mathord{\bmod}}%
       {\bigl(\Zset\times\Zset\setdiff\{ 0 \}\bigr)}%
       {\Zset}
\)
are defined, for each $x, y \in \Zset$ with $y \neq 0$,
respectively by
$x \div y  \defeq \intp(x/y)$
and
$x \bmod y \defeq x - (x \div y) \cdot y$.

We assume familiarity with the field of program analysis and verification
via abstract interpretation.  The reader is referred to the literature
for the theory
(e.g., \cite{Cousot81,CousotC76,CousotC77,CousotC79,CousotC92fr,CousotC92})
and examples of applications
\cite{DorRS01,Halbwachs93,ShahamKS00}.

\section{The Language Syntax}
\label{sec:syntax}

The run-time support of CPM uses exceptions to communicate
run-time errors.  The set of RTS exceptions is left open so
that it can be extended if and when needed.
That said, the basic syntactic sets of the CPM language are:
\begin{description}
\item[Identifiers]
$\id \in \Id = \{ \main, \idx, \idx_0, \idx_1, \ldots \} \uplus \rId$,
  where $\rId \defeq \{ \ridx, \ridx_0, \ridx_1, \ldots \}$;
\item[Basic types]
$T \in \Type \defeq \{ \tinteger, \tboolean \}$;
\item[Integers]
$m \in \Integer \defeq \Zset$;
\item[Booleans]
$t \in \Bool \defeq \{ \ttv, \ffv \}$;
\item[RTS exceptions]
\(
  \rtsexcept
    \in \RTSExcept
          \defeq \{ \divbyzero, \stkovflw, \memerror, \ldots \}
\).
\end{description}
The identifiers in $\rId$ are ``reserved'' for the specification of the
concrete semantics.

From the basic sets, a number of syntactic categories are defined,
along with their syntactic meta-variables,
by means of the BNF rules:
\begin{description}
\item[Expressions]
\begin{align*}
  \Exp \ni
  e &::= m \vbar -e \vbar e_0 + e_1 \vbar e_0 - e_1
     \vbar e_0 * e_1 \vbar e_0 \divop e_1 \vbar e_0 \modop e_1 \\
    &\vbar t \vbar e_0 = e_1 \vbar e_0 \neq e_1 \vbar e_0 < e_1
     \vbar e_0 \leq e_1 \vbar e_0 \geq e_1 \vbar e_0 > e_1 \\
    &\vbar \notop e \vbar e_0 \andop e_1 \vbar e_0 \orop e_1
     \vbar \id
\end{align*}
\item[Sequences of expressions]
\begin{align*}
  \Exps \ni
  \es ::= \emptysequence \vbar e, \es
\end{align*}
\item[Storable types]
\[
  \sType \ni
  \sT ::= T
\]
\item[Formal parameters]
\[
  \formParams \ni
  \fps ::= \emptysequence \vbar \id : \sT, \fps
\]
\item[Function bodies]
\[
  \funBody \ni
  \funbody ::= \kw{let} d \,\kw{in} s \kw{result} e
    \vbar \kw{extern} : \sT
\]
\item[Global declarations]
\[
  \Glob \ni
  g ::= \kw{gvar} \id : \sT = e
      \vbar \kw{function} \; \id(\fps) = \funbody
      \vbar \kw{rec} g \vbar g_0 ; g_1
\]
\item[Local declarations]
\[
  \Decl \ni
  d ::= \kw{nil}
     \vbar \kw{lvar} \id : \sT = e
     \vbar d_0 ; d_1
\]
\item[Catchable types]
\[
  \cType \ni
  \cT ::= \trtsexcept \vbar \sT
\]
\item[Exception declarations]
\[
  \exceptDecl \ni
  p ::= \rtsexcept \vbar \cT \vbar \id : \sT \vbar \kw{any}
\]
\item[Catch clauses]
\[
  \Catch \ni
  k ::= (p) \, s \vbar k_0 ; k_1
\]
\item[Statements]
\begin{align*}
  \Stmt \ni
  s &::=  \kw{nop}
     \vbar \id := e
     \vbar \id_0 := \id(\es)
     \vbar s_0 ; s_1
     \vbar d ; s \\
    &\vbar \kw{if} e \kw{then} s_0 \kw{else} s_1
     \vbar \kw{while} e \kw{do} s \\
    &\vbar \kw{throw} \rtsexcept \vbar \kw{throw} e
     \vbar \kw{try} s \kw{catch} k
     \vbar \kw{try} s_0 \kw{finally} s_1
\end{align*}
\end{description}
Observe that there is no need of a separate syntactic category for programs:
as we will see, a CPM program is just a global declaration
defining the special function `$\main$', like in C and \Cplusplus{}.

It should be noted that some apparent limitations of the abstract syntax
of CPM are not real limitations.  For instance: the use of function
calls as expressions can be avoided by introducing temporary variables;
procedures can be rendered by functions that return a dummy value;
and so forth.  More generally, a slight elaboration of the abstract syntax
presented here and extended in Section~\ref{sec:extensions} is used
in the ECLAIR prototype to encode the C language almost in its entirety,
plus the basic exception handling mechanisms of \Cplusplus{} and Java.

For notational convenience,
we also define the syntactic categories of constants, storable values%%
\footnote{The reason for a distinction between the roles of constants
and storable values (as well as basic types and storable types)
will become clear when discussing language extensions
in Section~\ref{sec:extensions}.}
and exceptions:
\begin{description}
\item[Constants]
\[
  \Con \ni
  \con ::= m \vbar t
\]
\item[Storable values]
\[
  \sVal \ni
  \sval ::= \con
\]
\item[Exceptions]
\[
  \Except \ni
  \except ::= \rtsexcept \vbar \sval
\]
\end{description}

The (partial) function $\pard{\type}{\sVal}{\sType}$, mapping
a storable value to its type name `$\tinteger$' or `$\tboolean$',
is defined by:
\begin{align*}
  \type(\sval)
    &\defeq
      \begin{cases}
        \tinteger,        &\text{if $\sval = m \in \Integer$;} \\
        \tboolean,       &\text{if $\sval = t \in \Bool$.}
      \end{cases} \\
\intertext{%
For ease of notation, we also define the overloadings
$\pard{\type}{\Except}{\cType}$
and $\pard{\type}{\exceptDecl}{\cType}$ defined by
}
  \type(\except)
    &\defeq
      \begin{cases}
        \trtsexcept,  &\text{if $\except = \rtsexcept \in \RTSExcept$;} \\
        \type(\sval), &\text{if $\except = \sval \in \sVal$;}
      \end{cases} \\
  \type(p)
    &\defeq
      \begin{cases}
        \trtsexcept,  &\text{if $p = \rtsexcept \in \RTSExcept$;} \\
        \cT,          &\text{if $p = \cT \in \cType$;} \\
        \sT,          &\text{if $p = \id : \sT$ and $\sT \in \sType$.}
      \end{cases}
\end{align*}
Note that such an overloading is consistent and the resulting function
is not defined on value $\kw{any} \in \exceptDecl$.

The helper function
$\fund{\dom}{\cType}{\{\Integer, \Bool, \RTSExcept\}}$, which associates
a catchable type name to the corresponding domain, is defined by
\[
  \dom(\cT)
    \defeq
      \begin{cases}
        \Integer,        &\text{if $\cT = \tinteger$;} \\
        \Bool,       &\text{if $\cT = \tboolean$;} \\
        \RTSExcept,  &\text{if $\cT = \trtsexcept$.}
      \end{cases}
\]

\section{Static Semantics}
\label{sec:static-semantics}

The static semantics of the CPM language establishes the conditions under
which a program is well typed.
Only well-typed programs are given a dynamic semantics.

\subsection{Defined and Free Identifiers}

The set of identifiers defined by sequences of formal parameters,
(global or local) declarations or exception declarations
is defined as follows:
\begin{gather*}
  \DI(\emptysequence)
    \defeq \DI(\kw{nil})
    \defeq \DI(\funbody)
    \defeq \DI(\rtsexcept)
    \defeq \DI(\cT)
    \defeq \DI(\kw{any})
    \defeq \emptyset; \\
  \DI(\id : \sT)
    \defeq \DI(\kw{gvar} \id : \sT = e)
    \defeq \DI(\kw{lvar} \id : \sT = e) \\
  \quad
    \defeq \DI(\kw{function} \; \id(\fps) = \funbody)
    \defeq \{ \id \}; \\
  \DI(\id : \sT, \fps)
    \defeq \DI(\id : \sT) \union \DI(\fps);\\
  \DI(\kw{rec} g)
    \defeq \DI(g); \\
  \DI(g_0 ; g_1)
    \defeq \DI(g_0) \union \DI(g_1); \\
  \DI(d_0 ; d_1)
    \defeq \DI(d_0) \union \DI(d_1).
\end{gather*}
The set of identifiers that occur freely in (sequences of) expressions,
(exception) declarations, statements and catch clauses is defined by:
\begin{gather*}
  \FI(m)
    \defeq \FI(t)
    \defeq \FI(\kw{nop})
    \defeq \FI(\emptysequence)
    \defeq \FI(\id : \sT)
    \defeq \FI(\kw{nil}) \\
  \quad
    \defeq \FI(\rtsexcept)
    \defeq \FI(\cT)
    \defeq \FI(\kw{any})
    \defeq \FI(\kw{throw} \rtsexcept)
    \defeq \FI(\kw{extern} : \sT)
    \defeq \emptyset; \\
  \FI(-e)
    \defeq \FI(\notop e)
    \defeq \FI(\kw{lvar} \id : \sT = e) \\
  \quad
    \defeq \FI(\kw{gvar} \id : \sT = e)
    \defeq \FI(\kw{throw} e)
    \defeq \FI(e); \\
  \FI(e_0 \mathbin{\mathrm{op}} e_1)
    \defeq \FI(e_0) \union \FI(e_1), \text{ for
\(
  \mathrm{op}
    \in \{
           \mathord{+}, \ldots, \modop,
           \mathord{=}, \ldots, \mathord{>},
           \mathord{\andop}, \mathord{\orop}
        \}
\);} \\
  \FI(\id)
    \defeq \{ \id \}; \\
  \FI(\kw{let} d \,\kw{in} s \kw{result} e)
    \defeq
      \FI(d)
        \union
      \bigl(\FI(s) \setdiff \DI(d)\bigr)
        \union
      \bigl(\FI(e) \setdiff \DI(d)\bigr); \\
  \FI(\kw{function} \; \id(\fps) = \funbody)
    \defeq
      \FI(\funbody) \setdiff \DI(\fps); \\
  \FI(\kw{rec} g) \defeq \FI(g) \setdiff \DI(g); \\
  \FI(g_0 ; g_1)
    \defeq \FI(g_0) \union \bigl(\FI(g_1) \setminus \DI(g_0)\bigr); \\
  \FI(d_0 ; d_1) \defeq \FI(d_0) \union
    \bigl(\FI(d_1) \setminus \DI(d_0)\bigr); \\
  \FI(\id := e) \defeq \{ \id \} \union \FI(e); \\
  \FI(e, \es) \defeq \FI(e) \union \FI(\es); \\
  \FI\bigl(\id_0 := \id(\es)\bigr)
    \defeq \{ \id, \id_0 \} \union \FI(\es); \\
  \FI(d ; s) \defeq \FI(d) \union \bigl(\FI(s) \setdiff \DI(d)\bigr); \\
  \FI\bigl((p) \, s\bigr) \defeq \FI(s) \setminus \DI(p); \\
  \FI(k_0 ; k_1) \defeq \FI(k_0) \union \FI(k_1); \\
  \FI(s_0 ; s_1) \defeq \FI(\kw{try} s_0 \kw{finally} s_1)
    \defeq \FI(s_0) \union \FI(s_1); \\
  \FI(\kw{if} e \kw{then} s_0 \kw{else} s_1)
    \defeq \FI(e) \union \FI(s_0) \union \FI(s_1); \\
  \FI(\kw{while} e \kw{do} s) \defeq \FI(e) \union \FI(s); \\
  \FI(\kw{try} s \kw{catch} k) \defeq \FI(s) \union \FI(k).
\end{gather*}

\subsection{Type Environments}

We start by defining the convenience syntactic category of
\begin{description}
\item[Denotable types]
\[
  \dType \ni
  \dT ::= \location{\sT}
          \vbar \fps \rightarrow \sT
\]
\end{description}
A type environment associates a denotable type to each identifier of a
given, finite set of identifiers.

\begin{definition} \summary{($\TEnv_I$, $\TEnv$.)}
For each $I \sseqf \Id$, the set of \emph{type environments over $I$}
is $\TEnv_I \defeq  I \to \dType$; the set of all type environments
is given by $\TEnv \defeq \biguplus_{I \sseqf \Id} \TEnv_I$.
Type environments are denoted by $\beta$, $\beta_0$, $\beta_1$
and so forth.
The notation $\beta : I$ is a shorthand for $\beta \in \TEnv_I$.
\end{definition}

\subsection{Static Semantics Predicates}

Let $I \sseqf \Id$ and  $\beta \in \TEnv_I$.
The well-typedness of program constructs whose free identifiers
are contained in $I$ is encoded by the following
predicates, here listed along with their informal meaning:
\begin{align*}
\beta &\vdash_I e : \sT,
  &\text{$e$ is well-formed and has type $\sT$ in $\beta$;} \\
\beta &\vdash_I \funbody : \sT,
  &\text{$\funbody$ is well-formed and has type $\sT$ in $\beta$;} \\
\beta, \fps & \vdash_I \es,
  &\text{$\es$ is compatible with $\fps$ and well formed in $\beta$;}\\
  & \fps : \delta,
  &\text{$\fps$ is well formed and yields the type environment $\delta$;}\\
\beta & \vdash_I g : \delta,
  &\text{$g$ is well formed
         and yields the type environment $\delta$ in $\beta$;} \\
\beta &\vdash_I d : \delta,
  &\text{$d$ is well-formed
         and yields the type environment $\delta$ in $\beta$;}\\
  &\vdash_I p : \delta,
  &\text{$p$ is well-formed and yields the type environment $\delta$;} \\
\beta &\vdash_I k,
  &\text{$k$ is well-formed in $\beta$;}\\
\beta &\vdash_I s,
  &\text{$s$ is well-formed in $\beta$.} \\
\end{align*}
These predicates are defined inductively on the abstract syntax
by means of the following rules.
\begin{description}
\item[Expressions]
\begin{gather*}
\begin{aligned}
&
\prooftree
  \nohyp
\justifies
  \beta \vdash_I m : \tinteger
\endprooftree
&\quad&
\prooftree
  \nohyp
\justifies
  \beta \vdash_I t : \tboolean
\endprooftree \\[1ex]
&
\prooftree
 \beta \vdash_I e : \tinteger
\justifies
 \beta \vdash_I -e : \tinteger
\endprooftree
&\quad&
\prooftree
 \beta \vdash_I e : \tboolean
\justifies
 \beta \vdash_I \notop\ e : \tboolean
\endprooftree
\end{aligned} \\[1ex]
\begin{aligned}
&
\prooftree
 \beta \vdash_I e_0 : \tinteger
\quad
 \beta \vdash_I e_1 : \tinteger
\justifies
  \beta \vdash_I e_0 \boxcircle e_1 : \tinteger
\using\quad\text{if
    \(
      \mathord{\boxcircle}
        \in
          \{ \mathord{+}, \mathord{-}, \mathord{*}, \divop, \modop \}
    \)
  }
\endprooftree \\[1ex]
&
\prooftree
 \beta \vdash_I e_0 : \tinteger
\quad
 \beta \vdash_I e_1 : \tinteger
\justifies
  \beta \vdash_I e_0 \boxast e_1 : \tboolean
\using\quad\text{if
    \(
      \mathord{\boxast}
        \in
          \{
             \mathord{=}, \mathord{\neq}, \mathord{<},
             \mathord{\leq}, \mathord{\geq}, \mathord{>}
          \}
    \)
  }
\endprooftree \\[1ex]
&
\prooftree
 \beta \vdash_I e_0 : \tboolean
\quad
 \beta \vdash_I e_1 : \tboolean
\justifies
  \beta \vdash_I e_0 \diamond e_1 : \tboolean
\using\quad
  \text{if $\mathord{\diamond} \in \{ \mathord{\andop}, \mathord{\orop} \}$}
\endprooftree
\end{aligned} \\[1ex]
\begin{aligned}
&
\prooftree
  \nohyp
\justifies
  \beta \vdash_I \id : \sT
\using\quad\text{if $\beta(\id) = \location{\sT}$}
\endprooftree
\end{aligned}
\end{gather*}
\item[Sequences of expressions]
\begin{align*}
&
\prooftree
  \nohyp
\justifies
  \beta, \emptysequence \vdash_I \emptysequence
\endprooftree
&\quad&
\prooftree
  \beta \vdash_I e : \sT
\quad
  \beta, \fps \vdash_I \es
\justifies
    \beta, (\id : \sT, \fps) \vdash_I (e, \es)
\endprooftree
\end{align*}
\item[Sequences of formal parameters]
\begin{align*}
&
\prooftree
  \nohyp
\justifies
  \emptysequence : \emptyset
\endprooftree
&\quad&
\prooftree
    \fps : \delta
\using\quad\text{if $\id \notin \DI(\fps)$}
\justifies
    (\id : \sT, \fps) : \{ \id \mapsto \location{\sT} \} \union \delta
\endprooftree
\end{align*}
\item[Function bodies]
\begin{align*}
\prooftree
  \beta \vdash_I d : \beta_0
\qquad
  \beta[\beta_0] \vdash_{I \union \DI(d)} s
\qquad
  \beta[\beta_0] \vdash_{I \union \DI(d)} e : \sT
\justifies
    \beta
      \vdash_I
        (\kw{let} d \,\kw{in} s \kw{result} e) : \sT
\endprooftree
\end{align*}
\begin{align*}
\prooftree
  \nohyp
\justifies
    \beta
      \vdash_I
        (\kw{extern} : \sT) : \sT
\endprooftree
\end{align*}
\item[Declarations]
\begin{gather}
\notag
\begin{aligned}
&
\prooftree
  \nohyp
\justifies
  \beta \vdash_I \kw{nil} : \emptyset
\endprooftree
&\quad&
\prooftree
  \beta \vdash_I e : \sT
\justifies
  \beta \vdash_I \kw{gvar} \id : \sT = e : \{ \id \mapsto \location{\sT} \}
\endprooftree \\[1ex]
&
&\quad&
\prooftree
  \beta \vdash_I e : \sT
\justifies
  \beta \vdash_I \kw{lvar} \id : \sT = e : \{ \id \mapsto \location{\sT} \}
\endprooftree
\end{aligned} \\[1ex]
\notag
\begin{aligned}
\prooftree
  \fps : \delta
\qquad
  \beta[\delta] \vdash_{I \union \DI(\fps)} \funbody : \sT
\justifies
    \beta
      \vdash_I
        \bigl(
          \kw{function} \; \id(\fps) = \funbody
        \bigr)
          : \bigl\{ \id \mapsto (\fps \rightarrow \sT) \bigr\}
\endprooftree
\end{aligned} \\[1ex]
\label{rule:well-formed-recursive-declaration}
\begin{aligned}
\prooftree
  \beta[\delta \restrict{J}] \vdash_{I \union J} g : \delta
\justifies
    \beta \vdash_I (\kw{rec} g) : \delta
\using\quad\text{if
\(
  J = \FI(g) \inters \DI(g)
\)
and
\(
  \forall \id, \sT : (\id \mapsto \location{\sT}) \notin \delta
\)
}
\endprooftree
\end{aligned} \\[1ex]
\notag
\begin{aligned}
&
\prooftree
  \beta \vdash_I g_0 : \beta_0
\quad
  \beta[\beta_0] \vdash_{I \cup \DI(g_0)} g_1 : \beta_1
\justifies
  \beta \vdash_I g_0;g_1 : \beta_0[\beta_1]
\endprooftree
&\quad&
\prooftree
 \beta \vdash_I d_0 : \beta_0
\quad
 \beta[\beta_0] \vdash_{I \union \DI(d_0)} d_1 : \beta_1
\justifies
  \beta \vdash_I d_0 ; d_1 : \beta_0[\beta_1]
\endprooftree
\end{aligned}
\end{gather}
Note that rule~\eqref{rule:well-formed-recursive-declaration} seems
to suggest that $\delta$ must be guessed.  Indeed, this is not the case,
as it can be proved that the environment generated by a declaration $g$
only depends on $g$ and not on the environment used to establish whether
$g$ is well formed.  While the right thing to do is to define two static
semantics predicates for declarations ---one for the generated environments
and the other for well-formedness \cite{Plotkin04b}---
we opted for a more concise presentation.
Also notice that the side condition
in rule~\eqref{rule:well-formed-recursive-declaration}
explicitly forbids recursive declarations of variables.%
\footnote{Namely, a recursive declaration such as
$\kw{rec} \; \kw{gvar} \id : \sT = e$ is not well-typed.}
\item[Exception declarations]
\begin{align*}
&
\prooftree
  \nohyp
\justifies
  \vdash_I \rtsexcept : \emptyset
\endprooftree
&\quad&
\prooftree
  \nohyp
\justifies
  \vdash_I \cT : \emptyset
\endprooftree \\[1ex]
&
\prooftree
  \nohyp
\justifies
  \vdash_I \id : \sT : \{ \id \mapsto \location{\sT} \}
\endprooftree
&\quad&
\prooftree
  \nohyp
\justifies
  \vdash_I \kw{any} : \emptyset
\endprooftree
\end{align*}
\item[Catch clauses]
\begin{align*}
&
\prooftree
  \vdash_I p : \delta
  \quad \beta[\delta] \vdash_{I \union \DI(p)} s
\justifies
  \beta \vdash_I (p) \, s
\endprooftree
&\quad&
\prooftree
  \beta \vdash_I k_0
  \quad \beta \vdash_I k_1
\justifies
  \beta \vdash_I k_0 ; k_1
\endprooftree
\end{align*}
\item[Statements]
\begin{align*}
&
\prooftree
  \nohyp
\justifies
  \beta \vdash_I \kw{nop}
\endprooftree
&\quad&
\prooftree
  \beta \vdash_I e : \sT
\justifies
  \beta \vdash_I \id := e
\using\quad\text{if $\beta(\id) = \location{\sT}$}
\endprooftree
\end{align*}
\begin{align*}
&
\prooftree
  \beta, \fps \vdash_I \es
\using\quad\text{if $\beta(\id_0) = \location{\sT}$ and
                 $\beta(\id) = \fps \rightarrow \sT$}
\justifies
  \beta \vdash_I \id_0 := \id(\es)
\endprooftree
\end{align*}
\begin{gather*}
\begin{aligned}
&
\prooftree
  \beta \vdash_I s_0
\quad
  \beta \vdash_I s_1
\justifies
  \beta \vdash_I s_0 ; s_1
\endprooftree
&\quad&
\prooftree
  \beta \vdash_I d : \beta_0
\quad
  \beta[\beta_0] \vdash_{I \union \DI(d)} s
\justifies
  \beta \vdash_I d ; s
\endprooftree \\[1ex]
&
\prooftree
  \beta \vdash_I e : \tboolean
\quad
  \beta \vdash_I s_0
\quad
  \beta \vdash_I s_1
\justifies
  \beta \vdash_I \kw{if} e \kw{then} s_0 \kw{else} s_1
\endprooftree
&\quad&
\prooftree
  \beta \vdash_I e : \tboolean
\quad
  \beta \vdash_I s
\justifies
  \beta \vdash_I \kw{while} e \kw{do} s
\endprooftree \\[1ex]
&
\prooftree
  \nohyp
\justifies
  \beta \vdash_I \kw{throw} \rtsexcept
\endprooftree
&\quad&
\prooftree
  \beta \vdash_I e : \sT
\justifies
  \beta \vdash_I \kw{throw} e
\endprooftree \\[1ex]
&
\prooftree
  \beta \vdash_I s
  \quad \beta \vdash_I k
\justifies
  \beta \vdash_I \kw{try} s \kw{catch} k
\endprooftree
&\quad&
\prooftree
  \beta \vdash_I s_0
  \quad \beta \vdash_I s_1
\justifies
  \beta \vdash_I \kw{try} s_0 \kw{finally} s_1
\endprooftree
\end{aligned}
\end{gather*}
\end{description}

A program $g$ is said to be \emph{valid} if and only if
it does not contain any occurrence of a reserved identifier $\id \in \rId$,
$\emptyset \vdash_\emptyset g : \beta$
and $\beta(\main) = \emptysequence \rightarrow \tinteger$.

\section{Concrete Dynamic Semantics}
\label{sec:concrete-dynamic-semantics}

For the specification of the concrete dynamic semantics for CPM, we
adopt the $\mathrm{G}^\infty\mathrm{SOS}$ approach of Cousot and Cousot
\cite{CousotC92}.
This generalizes with infinite computations
the \emph{natural} semantics approach by Kahn \cite{Kahn87}, which,
in turn, is a ``big-step'' operational semantics defined by structural
induction on program structures in the style of
Plotkin \cite{Plotkin04b}.

\subsection{Absolute Locations and Indirect Locators}

An \emph{absolute location} (or, simply, \emph{location}) is a unique
identifier for a memory area of unspecified size.  The (possibly
infinite) set of all locations is denoted by $\Loc$, while individual
locations are denoted by $l$, $l_0$, $l_1$ and so forth.
We also postulate the existence of a set $\Ind \defeq \Nset$ of
\emph{indirect (stack) locators} such that $\Loc \inters \Ind = \emptyset$.
Indirect locators are denoted by $i$, $i_0$, $i_1$ and so forth.
For notational convenience, we define the set of \emph{addresses} as
$\Addr \defeq \Loc \uplus \Ind$.
Addresses are denoted by $a$, $a_0$, $a_1$ and so forth.

\subsection{Concrete Execution Environments}

The concrete dynamic aspect of declarations is captured by
concrete execution environments.
These map a finite set of identifiers to concrete denotable values.
In the sequel we will simply write `environment' to refer to
execution environments.

\begin{definition}\summary{($\Abstract$, $\dVal$, $\Env_I$.)}
We define
\begin{equation*}
  \Abstract
    \defeq
      \{\,
        \lambdaop \fps \st \funbody
      \mid
        \fps \in \formParams, \funbody \in \funBody
      \,\}.
\end{equation*}
The set of \emph{concrete denotable values} is
\[
  \dVal
    \defeq
      (\Addr \times \sType) \uplus \Abstract.
\]

For $I \sseqf \Id$, $\Env_I \defeq I \to \dVal$ is
the set of \emph{concrete environments over $I$}.
The set of all environments is given by
$\Env \defeq \biguplus_{I \sseqf \Id} \Env_I$.
Environments in $\Env_I$ are denoted by
$\rho$, $\rho_0$, $\rho_1$ and so forth.
We write $\rho : I$ as a shorthand for $\rho \in \Env_I$.
For $\rho : I$ and $\beta : I$, we write $\rho : \beta$ to signify
that
\begin{multline*}
  \forall \id \in I
    \itc
      \bigl(
        \exists (a, \sT) \in \Addr \times \sType
          \st \beta(\id) = \location{\sT} \land \rho(\id) = (a, \sT)
      \bigr) \\
    \lor
      \bigl(
        \exists \abs = (\lambdaop \fps \st \funbody) \in \Abstract
	  \st
            \beta(\id) = \fps \rightarrow \sT
              \land
            \beta \vdash_I \funbody : \sT\\
              \land
	    \rho(\id) = \abs
      \bigr).
\end{multline*}
\end{definition}

\subsection{Memory Structures, Value States and Exception States}

A \emph{memory structure} uses a stack and suitable operators
to allocate/deallocate, organize, read and update
the locations of an absolute memory map, which is a partial function
mapping a location and a storable type to a storable value.
Memory structures model all the memory areas that are used in the
most common implementations of imperative programming languages:
the \emph{data segment} (for global variables) and the \emph{stack segment}
(for local variables) are of interest for the language fragment we are
considering; the \emph{text segment} (where pointers to function point to)
and the \emph{heap segment} (for dynamically allocated memory) are required
to deal with the extensions of Section~\ref{sec:extensions}.
As it will be clear from the following definition, our notion of
memory structure is underspecified: while we define it and its
operations so that the semantics of programs is the expected one, we
allow for many possible implementations by leaving out many details
that are inessential to the achievement of that objective.
It is for this same reason that we treat locations as unique identifiers
neglecting the mathematical structure they may or may not have.
More generally, what we call ``concrete semantics'' is indeed
an abstraction of an infinite number of machines
and compilation schemes that could be used to execute our programs.
Furthermore, since the considered fragment of CPM
does not support pointers, arrays, type casts and unions, we can
here make the simplifying assumption that there is no overlap between the
storage cells associated to different locations.
In Section~\ref{sec:extensions} we will hint at how these assumptions
must be modified in order to accommodate the full language.

Memory structures will be used to describe the outcome of computations
whose only observable behavior is given by their side effects.
Computations yielding a proper value will be described by
a \emph{value state}, which pairs the value computed with a memory
structure recording the side effects of the execution.
Exceptional behavior must, of course, be taken into proper account:
thus, the result of an exceptional computation path will be described
by pairing the memory structure with an exception,
yielding what we call an \emph{exception state}.

\begin{definition}
\summary{($\Map$, $\Stack$, $\Mem$, $\ValState$, $\ExceptState$.)}
\label{def:concrete-memory-structure}
The set of all absolute maps is the set of partial functions
\[
  \Map \defeq (\Loc \times \sType) \rightarrowtail \sVal.
\]
Absolute maps are denoted by $\mu$, $\mu_0$, $\mu_1$ and so forth.
The absolute map update partial function
\[
  \pard{\cdot[\cdot := \cdot]}%
       {\bigl(\Map \times (\Loc \times \sType) \times \sVal\bigr)}%
       {\Map}
\]
is defined,
for each $\mu \in \Map$,
$(l, \sT) \in \Loc \times \sType$ such that $(l, \sT) \in \dom(\mu)$
and $\sval \in \sVal$ such that $\sT = \type(\sval)$, by
\[
  \mu\bigl[(l, \sT) := \sval\bigr] \defeq \mu',
\]
where $\mu' \in \Map$ is any absolute map satisfying the following
conditions:
\begin{enumerate}[(i)]
\item
$\dom(\mu') = \dom(\mu)$;
\item
$\mu'(l, \sT) = \sval$;
\item
$\mu'(l', \sT') = \mu(l', \sT')$,
for each $(l', \sT') \in \dom(\mu)$ such that $l' \neq l$.
\end{enumerate}

Let $W \defeq \bigl( \Loc \union \{\sm, \fm\} \bigr)^{\star}$.
An element $w \in W$ is a \emph{stack} if and only if no location
occurs more than once in it.
The set of all stacks is denoted by $\Stack$.
`$\,\sm$' is called \emph{stack marker} and
`$\,\fm$' is called \emph{frame marker}.
The \emph{top-most frame} of $w \in \Stack$,
denoted by $\topmost(w)$,
is the longest suffix of $w$ containing no frame marker;
formally,
$\topmost(w) \in \bigl(\Loc \union \{\sm\}\bigr)^{\star}$
satisfies either $w = \topmost(w)$ or $w = w' \fm \topmost(w)$.
The partial infix operator
$\pard{\mathord{@}}{\Stack \times \Ind}{\Loc}$
maps, when defined, a stack $w$ and an indirect locator $i$
into an absolute location to be found in the top-most frame;
formally, if $i < n = |\topmost(w)|$,
$\topmost(w) = z_0 \cdots z_{n-1}$ and $z_i = l$,
then $w \mathbin{@} i \defeq l$.

A \emph{memory structure} is an element of
$\Mem \defeq \Map \times \Stack$.
Memory structures are denoted by
$\sigma$, $\sigma_0$, $\sigma_1$ and so forth.

A \emph{value state} is an element of
$\ValState \defeq \sVal \times \Mem$.
Value states are denoted by
$\valstate$, $\valstate_0$, $\valstate_1$ and so forth.

An \emph{exception state} is an element of
$\ExceptState \defeq \Mem \times \Except$.
Exception states are denoted by
$\exceptstate$, $\exceptstate_0$, $\exceptstate_1$ and so forth.

The overloading
$\pard{\mathord{@}}{\Mem \times \Addr}{\Loc}$
of the partial infix operator $\mathord{@}$ is defined,
for each $\sigma = (\mu, w)$ and $a \in \Addr$, as follows
and under the following conditions:
\[
  \sigma \mathbin{@} a
    \defeq
      \begin{cases}
        a, &\text{if $a \in \Loc$;} \\
        l, &\text{if $a \in \Ind$ and $l = w \mathbin{@} a$ is defined.}
      \end{cases}
\]
The memory structure read and update operators
\begin{align*}
  \fund{\cdot[\cdot, \cdot]}%
       {&\bigl(
          \Mem \times \Addr \times \sType
        \bigr)}%
       {(\ValState \uplus \ExceptState)}, \\
  \fund{\cdot[\cdot := \cdot]}%
       {&\bigl(
          \Mem \times (\Addr \times \sType) \times \sVal
        \bigr)}%
       {(\Mem \uplus \ExceptState)}
\end{align*}
are respectively defined,
for each $\sigma = (\mu, w) \in \Mem$,
$a \in \Addr$, $\sT \in \sType$ and $\sval \in \sVal$, as follows:
let $d = (\sigma \mathbin{@} a, \sT)$; then
\begin{align*}
  \sigma[a, \sT]
    &\defeq
      \begin{cases}
        \bigl( \mu(d), \sigma \bigr),
          &\text{if $d \in \dom(\mu)$;} \\
        (\sigma, \memerror),
          &\text{otherwise;} \\
      \end{cases} \\
  \sigma\bigl[ (a, \sT) := \sval \bigr]
    &\defeq
      \begin{cases}
        \bigl(\mu[d := \sval], w\bigr),
          &\text{if $d \in \dom(\mu)$ and $\sT = \type(\sval)$;} \\
        (\sigma, \memerror),
          &\text{otherwise.}
      \end{cases}
\end{align*}

The data and stack memory allocation functions
\begin{align*}
  \fund{\datnew}%
       {&\ValState}%
       {\bigl( (\Mem \times \Loc) \uplus \ExceptState \bigr)}, \\
  \fund{\stknew}%
       {&\ValState}%
       {\bigl( (\Mem \times \Ind) \uplus \ExceptState \bigr)}
\end{align*}
are defined, for each $\valstate = (\sval, \sigma) \in \ValState$,
where $\sigma = (\mu, w)$, by
\begin{align*}
  \datnew(\valstate)
    &\defeq
      \begin{cases}
        ((\mu', w), l),
          &\text{if the data segment of $\sigma$ can be extended;} \\
        (\sigma, \datovflw),
          &\text{otherwise;}
      \end{cases} \\
  \stknew(\valstate)
    &\defeq
      \begin{cases}
        ((\mu', w'), i),
          &\text{if the stack segment of $\sigma$ can be extended;} \\
        (\sigma, \stkovflw),
          &\text{otherwise;}
      \end{cases}
\end{align*}
where, in the case of $\stknew$,
$w' \in \Stack$ and $i \in \Ind$ are such that:
\begin{enumerate}[(i)]
\item
$w' = w \cdot l$;
\item
$i = |\topmost(w)|$;
\end{enumerate}
and, for both $\datnew$ and $\stknew$, $\mu' \in \Map$
and $l \in \Loc$ are such that:
\begin{enumerate}[(i)]
\setcounter{enumi}{2}
\item
for each $\sT \in \sType$, $(l, \sT) \notin \dom(\mu)$;
\item
for each $(l', \sT') \in \dom(\mu)$, $\mu'(l', \sT') = \mu(l', \sT')$;
\item
$\mu'\bigl(l, \type(\sval)\bigr) = \sval$.
\end{enumerate}
The memory structure data cleanup function
\(
  \fund{\datcleanup}{\ExceptState}{\ExceptState}
\)
is given, for each $\exceptstate = (\sigma, \except) \in \ExceptState$, by
\[
  \datcleanup(\exceptstate)
    \defeq
      \bigl( (\emptyset, \emptystring), \except \bigr).
\]
The stack mark function
\(
  \fund{\smark}{\Mem}{\Mem}
\)
is given, for each $\sigma \in \Mem$, by
\[
  \smark(\sigma)
    \defeq
      (\mu, w \sm),
        \qquad \text{where $\sigma = (\mu, w)$.}
\]
The stack unmark partial function
\(
  \pard{\sunmark}{\Mem}{\Mem}
\)
is given,
for each $\sigma \in \Mem$ such that
$\sigma = (\mu, w' \sm w'')$ and $w'' \in \Loc^\star$, by
\begin{equation*}
  \sunmark (\mu, w' \sm w'')
    \defeq
      (\mu', w'),
\end{equation*}
where the absolute map $\mu' \in \Map$ satisfies:
\begin{enumerate}[(i)]
\item
\(
  \dom(\mu')
    = \bigl\{\,
        (l, \sT) \in \dom(\mu)
      \bigm|
        \text{$l$ does not occur in $w''$}
      \,\bigr\}
\);
\item
$\mu' = \mu \restrict{\dom(\mu')}$.
\end{enumerate}
The frame link partial function
\(
  \pard{\slink}{\Mem}{\Mem}
\)
is given,
for each $\sigma \in \Mem$ such that
$\sigma = (\mu, w' \sm w'')$ and
$w'' \in \Loc^{\star}$, by
\begin{align*}
  \slink(\mu, w' \sm  w'')
    &\defeq
      (\mu, w' \fm w'').
\intertext{%
The frame unlink partial function
\(
  \pard{\sunlink}{\Mem}{\Mem}
\)
is given,
for each $\sigma \in \Mem$ such that
$\sigma = (\mu, w' \fm w'')$ and $w'' \in \Loc^\star$, by
}
  \sunlink(\mu, w' \fm  w'')
    &\defeq
      (\mu, w' \sm w'').
\end{align*}
\end{definition}
For ease of notation, the stack unmark and the frame unlink
partial functions are lifted to also work on exception states.
Namely, for each $\exceptstate = (\sigma, \except) \in \ExceptState$,
\begin{align*}
  \sunmark(\sigma, \except)
    &\defeq
        \bigl( \sunmark(\sigma), \except\bigr); \\[1ex]
  \sunlink(\sigma, \except)
    &\defeq
        \bigl( \sunlink(\sigma), \except\bigr).
\end{align*}

Intuitively,
global variables are allocated in the data segment using $\datnew$
and are accessed through absolute locations; function $\datcleanup$
models their deallocation due to an RTS exception thrown during the
program start-up phase.
The functions $\smark$ and $\sunmark$ use the stack marker `$\sm$'
to implement the automatic allocation (through $\stknew$) and
deallocation of stack slots for storing local variables,
return values and actual arguments of function calls.
The functions $\slink$ and $\sunlink$ use the frame marker `$\fm$'
to partition the stack into activation frames, each frame corresponding
to a function call.
All accesses to the top-most frame can be expressed in terms of indirect
locators (i.e., offsets from the top-most frame marker), because at each
program point the layout of the current top-most frame is statically known.
As it will be clearer when considering the concrete rules for function calls,
the frame marker is used to move the return value and the actual arguments,
which are allocated by the caller, from the activation frame of the caller
to the activation frame of the callee, and vice versa.

The memory structures and operations satisfy the following property:
for each pair of memory structures $\sigma_0$ and $\sigma_1$ such that
$\sigma_1$ has been obtained from $\sigma_0$
by any sequence of operations where each $\mathord{\slink}$
is matched by a corresponding $\mathord{\sunlink}$,
for each indirect locator $i \in \Ind$,
if $\sigma_0 \mathbin{@} i$ and $\sigma_1 \mathbin{@} i$ are both defined,
then $\sigma_0 \mathbin{@} i = \sigma_1 \mathbin{@} i$.

As anticipated, we profit from the lack of aliasing in the fragment of
CPM considered here, i.e., we assume there is no overlap between the storage
cells associated to $(l_0, \sT_0)$ and the ones associated to $(l_1,
\sT_1)$, unless $l_0 = l_1$.
Moreover, we need not specify the relationship between $\mu(l, \sT_0)$
and $\mu(l, \sT_1)$ for the case where $\sT_0 \neq \sT_1$.
This also implies that the absolute map update operator is underspecified,
resulting in a nondeterministic operator.
Of course, any real implementation will be characterized by a complete
specification: for instance, a precise definition of the memory overflow
conditions will take the place of the informal conditions
``if the data (resp., stack) segment of $\sigma$ can be extended''
in the definitions of $\datnew$ and $\stknew$.
As is clear from the definition above, where memory is writable
if and only if it is readable, we do not attempt to model read-only memory.
It is also worth observing that, in the sequel, the ``meaning'' of
variable identifiers will depend on unrestricted
elements of $\Env \times \Mem$.
As a consequence we can have \emph{dangling references}, that is,
a pair $(\rho, \sigma) \in \Env \times \Mem$ with $\rho : I$
can be such that there exists an identifier $\id \in I$ for which
$\rho(\id) = (a, \sT)$ and $\sigma[a, \sT] = \memerror$.

\subsection{Configurations}

The dynamic semantics of CPM is expressed by means of an
\emph{evaluation (or reduction) relation}, which specifies how a
\emph{non-terminal configuration} is reduced to a
\emph{terminal configuration}.
The sets of non-terminal configurations are parametric with respect
to a type environment associating every identifier to its type.

\begin{definition} \summary{(Non-terminal configurations.)}
\label{def:nonterminal-concrete-configuration}
The sets of \emph{non-terminal configurations} for expressions,
local and global declarations, statements, function bodies and catch
clauses are given, respectively and for each $\beta \in \TEnv_I$, by
\begin{align*}
  \NTe^\beta
    &\defeq
      \bigl\{\,
        \langle e, \sigma \rangle
          \in \Exp \times \Mem
      \bigm|
        \exists \sT \in \sType \st \beta \vdash_I e : \sT
      \,\bigr\}, \\
  \NTd^\beta
    &\defeq
      \bigl\{\,
        \langle d, \sigma \rangle
          \in \Decl \times \Mem
      \bigm|
         \exists \delta \in \TEnv \st \beta \vdash_I d : \delta
      \,\bigr\}, \\
  \NTg^\beta
    &\defeq
      \bigl\{\,
        \langle g, \sigma \rangle
          \in \Glob \times \Mem
      \bigm|
         \exists \delta \in \TEnv \st \beta \vdash_I g : \delta
      \,\bigr\}, \\
  \NTs^\beta
    &\defeq
      \bigl\{\,
        \langle s, \sigma \rangle
          \in \Stmt \times \Mem
      \bigm|
        \beta \vdash_I s
      \,\bigr\}, \\
  \NTb^\beta
    &\defeq
      \bigl\{\,
        \langle \funbody, \sigma \rangle
          \in \funBody \times \Mem
      \bigm|
        \exists \sT \in \sType \st \beta \vdash_I \funbody : \sT
      \,\bigr\}, \\
  \NTk^\beta
    &\defeq
      \bigl\{\,
        \langle k, \exceptstate \rangle
          \in \Catch \times \ExceptState
      \bigm|
        \beta \vdash_I k
      \,\bigr\}.
\end{align*}
\end{definition}

Each kind of terminal configuration has to allow for the possibility
of both a non-exceptional and an exceptional computation path.

\begin{definition} \summary{(Terminal configurations.)}
\label{def:terminal-concrete-configuration}
The sets of \emph{terminal configurations} for expressions,
local and global declarations, statements, function bodies and catch clauses
are given, respectively, by
\begin{align*}
  \Te
    &\defeq
      \ValState \uplus \ExceptState, \\
  \Td
    &\defeq
  \Tg \defeq
      (\Env \times \Mem)
        \uplus \ExceptState, \\
  \Ts
    &\defeq
  \Tb
    \defeq
      \Mem \uplus \ExceptState, \\
  \Tk
    &\defeq
      \bigl(
        \{ \caught \}
          \times
        \Ts
      \bigr)
    \uplus
      \bigl(
        \{ \uncaught \}
          \times
        \ExceptState
      \bigr).
\end{align*}
\end{definition}
Note that $\Te$ is defined as $\ValState \uplus \ExceptState$;
as it will be apparent from the concrete semantics, expressions
never modify the memory structure, so $\Te$ could have been
defined as $\sVal \uplus \Except$; but defining it as
$\ValState \uplus \ExceptState$ simplifies the approximation
relations in Section~\ref{sec:abstract-dynamic-semantics}.

In the following, we write $N$ and $\eta$ to denote
a non-terminal and a terminal concrete configuration, respectively.
For clarity of notation, we often use angle brackets to highlight
that a tuple is indeed representing a configuration. Angle brackets
are not normally used for configurations made of a single element.
Therefore, when $\exceptstate = (\sigma, \except) \in \ExceptState$,
we indifferently write
$\exceptstate \in \Ts$ or
$\langle \sigma, \except \rangle \in \Ts$,
as well as
$\langle \caught, \exceptstate \rangle \in \Tk$ or
$\bigl\langle \caught, (\sigma, \except) \bigr\rangle \in \Tk$.

A few explanatory words are needed for $\Tk$.
When the evaluation of a non-terminal configuration
for catch clauses $\langle k, \exceptstate \rangle \in \NTk^\beta$
yields the terminal configuration $\langle \caught, \eta \rangle \in \Tk$,
then the exception $\except$ in $\exceptstate = (\sigma, \except)$
was caught inside $k$ and $\eta \in \Ts$
is the result of evaluating the corresponding exception handler statement;
note that $\eta \in \Ts$ may itself be another exception state,
meaning that another exception was thrown during the evaluation of the
exception handler statement.
In contrast, when the resulting terminal configuration is
$\langle \uncaught, \exceptstate \rangle \in \Tk$,
then the exception in $\exceptstate$ was not caught inside $k$ and
will be propagated to the outer context.%
\footnote{Note that the names of the labels $\caught$ and $\uncaught$
have been chosen as such for clarity, but provide no special meaning:
they are only needed for a correct application of the disjoint union
construction, since we have $\Ts \inters \ExceptState \neq \emptyset$.}

\subsection{Concrete Evaluation Relations}
\label{sec:concrete-evaluation-relations}

For convenience, in order to represent function closures, we extend the
syntactic category of local declarations with (recursive) execution
environments.  These syntactic constructs are meant to be only
available in the dynamic semantics (in non-terminal configurations):
they cannot occur in the program text.
Thus we have
\[
  \Decl \ni
  d ::= \ldots
    \vbar \rho
    \vbar \kw{rec} \rho
\]
Consequently, if $\rho : I$ we define
\(
  \DI(\rho)
    \defeq
  \DI(\kw{rec} \rho)
    \defeq
      I
\),
\(
  \FI(\rho)
    \defeq
      \bigunion_{\id \in I} \FI\bigl(\rho(\id)\bigr)
\)
and
\(
  \FI(\kw{rec} \rho)
    \defeq
      \FI(\rho) \setminus I,
\)
where the function $\FI$ is defined on elements of $\dVal$ by
\(
  \FI(l, \sT)
    \defeq \FI(i, \sT)
    \defeq \emptyset
\)
and
\(
  \FI(\lambdaop \fps \st \funbody)
    \defeq \FI(\funbody) \setminus \DI(\fps)
\).
The static semantics is extended by adding the rules
\[
\prooftree
  \rho : \delta
\justifies
  \beta \vdash_I \rho : \delta
\endprooftree
\quad
\prooftree
  \beta[\delta \restrict{J}] \vdash_{I \union J} \rho : \delta
\justifies
  \beta \vdash_I \kw{rec} \rho : \delta
\using\quad\text{if $J = \FI(\rho) \inters \DI(\rho)$ and
\(
  \forall \id : (\id \mapsto \location{\sT}) \notin \delta.
\)
}
\endprooftree
\]

The concrete evaluation relations that complete the definition of the
concrete semantics for CPM are defined, as usual, by structural
induction from a set of rule schemata.
The evaluation relations are of the form
\(
  \rho \vdash_\beta N \rightarrow \eta,
\)
where
$\beta \in \TEnv_I$, $\rho \in \Env_J$, $\rho : \beta \restrict{J}$ and,
for some
\(
  q
    \in
      \{
        \mathrm{e}, \mathrm{d}, \mathrm{g}, \mathrm{s}, \mathrm{b}, \mathrm{k}
      \}
\),
$N \in  \NTq^\beta$ and $\eta \in \Tq$.

%% CHECKME: why this page break?
%%\pagebreak
\subsubsection{Expressions}
\label{subsubsec:conc_expressions}

\begin{description}
\item[Constant]
\begin{gather}
\label{rule:conc_constant}
\prooftree
  \nohyp
\justifies
  \rho \vdash_\beta \langle \con, \sigma \rangle
    \rightarrow
      \langle \con, \sigma \rangle
\endprooftree
\end{gather}
\item[Identifier]
\begin{gather}
\label{rule:conc_identifier}
\prooftree
  \nohyp
\justifies
  \rho \vdash_\beta \langle \id, \sigma \rangle
    \rightarrow
      \sigma\bigl[\rho(\id)\bigr]
\endprooftree
\end{gather}
\item[Unary minus]
\begin{gather}
\label{rule:conc_uminus_error}
\prooftree
  \rho \vdash_\beta \langle e, \sigma \rangle
    \rightarrow
      \exceptstate
\justifies
  \rho \vdash_\beta \langle -e, \sigma \rangle
    \rightarrow
      \exceptstate
\endprooftree \\[1ex]
\label{rule:conc_uminus_ok}
\prooftree
  \rho \vdash_\beta \langle e, \sigma \rangle
    \rightarrow
      \langle m, \sigma_0 \rangle
\justifies
  \rho \vdash_\beta \langle -e, \sigma \rangle
    \rightarrow
      \langle -m, \sigma_0 \rangle
\endprooftree
\end{gather}
\item[Binary arithmetic operations]
Letting $\mathord{\boxcircle}$ denote any abstract syntax operator in
\(
   \{ \mathord{+}, \mathord{-}, \mathord{*}, \divop, \modop \}
\)
and
\(
  \mathord{\circ}
  \in
    \{
      \mathord{+}, \mathord{-}, \mathord{\cdot},
      \mathord{\div}, \mathord{\bmod}
    \}
\)
the corresponding arithmetic operation.
Then the rules for addition, subtraction, multiplication,
division and remainder are given by the following schemata:
\begin{gather}
\label{rule:conc_arith_bop_0}
\prooftree
  \rho \vdash_\beta \langle e_0, \sigma \rangle
    \rightarrow
      \exceptstate
\justifies
  \rho \vdash_\beta \langle e_0 \boxcircle e_1, \sigma \rangle
    \rightarrow
      \exceptstate
\endprooftree \\[1ex]
\label{rule:conc_arith_bop_1}
\prooftree
  \rho \vdash_\beta \langle e_0, \sigma \rangle
    \rightarrow
      \langle m_0, \sigma_0 \rangle
\quad
  \rho \vdash_\beta \langle e_1, \sigma_0 \rangle
    \rightarrow
      \exceptstate
\justifies
  \rho \vdash_\beta \langle e_0 \boxcircle e_1, \sigma \rangle
    \rightarrow
      \exceptstate
\endprooftree \\[1ex]
\label{rule:conc_arith_bop_2}
\prooftree
  \rho \vdash_\beta \langle e_0, \sigma \rangle
    \rightarrow
      \langle m_0, \sigma_0 \rangle
\quad
  \rho \vdash_\beta \langle e_1, \sigma_0 \rangle
    \rightarrow
      \langle m_1, \sigma_1 \rangle
\justifies
  \rho \vdash_\beta \langle e_0 \boxcircle e_1, \sigma \rangle
    \rightarrow
      \langle m_0 \circ m_1, \sigma_1 \rangle
\using\quad\text{if $\mathord{\boxcircle} \notin \{\divop, \modop \}$
                 or $m_1 \neq 0$}
\endprooftree \\[1ex]
\label{rule:conc_arith_bop_exc_0}
\prooftree
  \rho \vdash_\beta \langle e_0, \sigma \rangle
    \rightarrow
      \langle m_0, \sigma_0 \rangle
\quad
  \rho \vdash_\beta \langle e_1, \sigma_0 \rangle
    \rightarrow
      \langle 0, \sigma_1 \rangle
\justifies
  \rho \vdash_\beta \langle e_0 \boxcircle e_1, \sigma \rangle
    \rightarrow
      \langle \sigma_1, \divbyzero \rangle
\using\quad\text{if $\mathord{\boxcircle} \in \{\divop, \modop \}$}
\endprooftree
\end{gather}
\item[Arithmetic tests]
Let
\(
  \mathord{\boxast}
  \in
  \{ \mathord{=}, \mathord{\neq}, \mathord{<},
     \mathord{\leq}, \mathord{\geq}, \mathord{>} \}
\)
be an abstract syntax operator and denote with
`$\mathord{\lessgtr}$' the corresponding test operation
in $\Zset \times \Zset \rightarrow \Bool$.
The rules for the arithmetic tests are then given by the following
schemata:
\begin{gather}
\label{rule:conc_arith_test_error_0}
\prooftree
  \rho \vdash_\beta \langle e_0, \sigma \rangle
    \rightarrow
      \exceptstate
\justifies
  \rho \vdash_\beta \langle e_0 \boxast e_1, \sigma \rangle
    \rightarrow
      \exceptstate
\endprooftree \\[1ex]
\label{rule:conc_arith_test_error_1}
\prooftree
  \rho \vdash_\beta \langle e_0, \sigma \rangle
    \rightarrow
      \langle m_0, \sigma_0 \rangle
\quad
  \rho \vdash_\beta \langle e_1, \sigma_0 \rangle
    \rightarrow
      \exceptstate
\justifies
  \rho \vdash_\beta \langle e_0 \boxast e_1, \sigma \rangle
    \rightarrow
      \exceptstate
\endprooftree \\[1ex]
\label{rule:conc_arith_test_ok}
\prooftree
  \rho \vdash_\beta \langle e_0, \sigma \rangle
    \rightarrow
      \langle m_0, \sigma_0 \rangle
\quad
  \rho \vdash_\beta \langle e_1, \sigma_0 \rangle
    \rightarrow
      \langle m_1, \sigma_1 \rangle
\justifies
  \rho \vdash_\beta \langle e_0 \boxast e_1, \sigma \rangle
    \rightarrow
      \langle m_0 \lessgtr m_1, \sigma_1 \rangle
\endprooftree
\end{gather}
\item[Negation]
\begin{gather}
\label{rule:conc_negation_error}
\prooftree
  \rho \vdash_\beta \langle b, \sigma \rangle
    \rightarrow
      \exceptstate
\justifies
  \rho \vdash_\beta \langle \notop\ b, \sigma \rangle
    \rightarrow
      \exceptstate
\endprooftree \\[1ex]
\label{rule:conc_negation_ok}
\prooftree
  \rho \vdash_\beta \langle b, \sigma \rangle
    \rightarrow
      \langle t, \sigma_0 \rangle
\justifies
  \rho \vdash_\beta \langle \notop\ b, \sigma \rangle
    \rightarrow
    \langle \bneg t, \sigma_0 \rangle
\endprooftree
\end{gather}
\item[Conjunction]
\begin{gather}
\label{rule:conc_conjunction_0}
\prooftree
  \rho \vdash_\beta \langle b_0, \sigma \rangle
    \rightarrow
      \exceptstate
\justifies
  \rho \vdash_\beta \langle b_0 \andop b_1, \sigma \rangle
    \rightarrow
      \exceptstate
\endprooftree \\[1ex]
\label{rule:conc_conjunction_1}
\prooftree
  \rho \vdash_\beta \langle b_0, \sigma \rangle
    \rightarrow
      \langle \ffv, \sigma_0 \rangle
\justifies
  \rho \vdash_\beta \langle b_0 \andop b_1, \sigma \rangle
    \rightarrow
      \langle \ffv, \sigma_0 \rangle
\endprooftree \\[1ex]
\label{rule:conc_conjunction_2}
\prooftree
  \rho \vdash_\beta \langle b_0, \sigma \rangle
    \rightarrow
      \langle \ttv, \sigma_0 \rangle
\quad
  \rho \vdash_\beta \langle b_1, \sigma_0 \rangle
    \rightarrow
      \eta
\justifies
  \rho \vdash_\beta \langle b_0 \andop b_1, \sigma \rangle
    \rightarrow
      \eta
\endprooftree
\end{gather}
\item[Disjunction]
\begin{gather}
\label{rule:conc_disjunction_0}
\prooftree
  \rho \vdash_\beta \langle b_0, \sigma \rangle
    \rightarrow
      \exceptstate
\justifies
  \rho \vdash_\beta \langle b_0 \orop b_1, \sigma \rangle
    \rightarrow
      \exceptstate
\endprooftree \\[1ex]
\label{rule:conc_disjunction_1}
\prooftree
  \rho \vdash_\beta \langle b_0, \sigma \rangle
    \rightarrow
      \langle \ttv, \sigma_0 \rangle
\justifies
  \rho \vdash_\beta \langle b_0 \orop b_1, \sigma \rangle
    \rightarrow
      \langle \ttv, \sigma_0 \rangle
\endprooftree \\[1ex]
\label{rule:conc_disjunction_2}
\prooftree
  \rho \vdash_\beta \langle b_0, \sigma \rangle
    \rightarrow
      \langle \ffv, \sigma_0 \rangle
\quad
  \rho \vdash_\beta \langle b_1, \sigma_0 \rangle
    \rightarrow
      \eta
\justifies
  \rho \vdash_\beta \langle b_0 \orop b_1, \sigma \rangle
    \rightarrow
      \eta
\endprooftree
\end{gather}
\end{description}

\subsubsection{Declarations}

\begin{description}
\item[Nil]
\begin{gather}
\label{rule:conc_decl_nil}
\prooftree
   \nohyp
\justifies
 \rho \vdash_\beta \langle \kw{nil}, \sigma \rangle
   \rightarrow
  \langle \emptyset, \sigma \rangle
\endprooftree
\end{gather}
\item[Environment]
\begin{gather}
\label{rule:conc_decl_environment}
\prooftree
  \nohyp
\justifies
  \rho
    \vdash_\beta
      \langle \rho_0, \sigma \rangle
        \rightarrow
          \langle \rho_0, \sigma \rangle
\endprooftree
\end{gather}
\item[Recursive environment]
\begin{gather}
\label{rule:conc_decl_rec_environment}
\prooftree
  \nohyp
\justifies
  \rho
    \vdash_\beta
      \langle \kw{rec} \rho_0, \sigma \rangle
        \rightarrow
          \langle \rho_1, \sigma \rangle
\endprooftree
\end{gather}
\begin{align*}
&\text{if }
  \rho_1
    = \bigl\{\,
         \id \mapsto \rho_0(\id)
       \bigm|
         \rho_0(\id) = \lambda \fps \st \kw{extern} : \sT
       \,\bigr\} \\
& \qquad
    \union
       \sset{
         \id \mapsto \abs_1
       }{
         \forall i \in \{0,1\}
           \itc
             \abs_i = \lambda \fps
                        \st \kw{let} d_i \,\kw{in} s \kw{result} e, \\
         \rho_0(\id) = \abs_0,
         d_1 = \kw{rec} \bigl( \rho_0 \setminus \DI(\fps) \bigr); d_0
       }.
\end{align*}
\item[Global variable declaration]
\begin{gather}
\label{rule:conc_gvar_decl_expr_err}
\prooftree
  \rho \vdash_\beta \langle e, \sigma \rangle
    \rightarrow
      \exceptstate
\justifies
  \rho \vdash_\beta \langle \kw{gvar} \id : \sT = e, \sigma \rangle
    \rightarrow
      \datcleanup(\exceptstate)
\endprooftree \\[1ex]
\label{rule:conc_gvar_decl_dat_err}
\prooftree
  \rho \vdash_\beta \langle e, \sigma \rangle
    \rightarrow
      \valstate
\justifies
  \rho \vdash_\beta \langle \kw{gvar} \id : \sT = e, \sigma \rangle
    \rightarrow
      \datcleanup(\exceptstate)
\using\quad\text{if $\datnew(\valstate) = \exceptstate$}
\endprooftree \\[1ex]
\label{rule:conc_gvar_decl_ok}
\prooftree
  \rho \vdash_\beta \langle e, \sigma \rangle
    \rightarrow
      \valstate
\justifies
  \rho \vdash_\beta \langle \kw{gvar} \id : \sT = e, \sigma \rangle
    \rightarrow
      \langle
        \rho_1, \sigma_1
      \rangle
\endprooftree
\end{gather}
if $\datnew(\valstate) = (\sigma_1, l)$ and
$\rho_1 = \bigl\{ \id \mapsto (l, \sT) \bigr\}$.
\item[Local variable declaration]
\begin{gather}
\label{rule:conc_lvar_decl_expr_err}
\prooftree
  \rho \vdash_\beta \langle e, \sigma \rangle
    \rightarrow
      \exceptstate
\justifies
  \rho \vdash_\beta \langle \kw{lvar} \id : \sT = e, \sigma \rangle
    \rightarrow
      \sunmark(\exceptstate)
\endprooftree \\[1ex]
\label{rule:conc_lvar_decl_stk_err}
\prooftree
  \rho \vdash_\beta \langle e, \sigma \rangle
    \rightarrow
      \valstate
\justifies
  \rho \vdash_\beta \langle \kw{lvar} \id : \sT = e, \sigma \rangle
    \rightarrow
      \sunmark(\exceptstate)
\using\quad\text{if $\stknew(\valstate) = \exceptstate$}
\endprooftree \\[1ex]
\label{rule:conc_lvar_decl_ok}
\prooftree
  \rho \vdash_\beta \langle e, \sigma \rangle
    \rightarrow
      \valstate
\justifies
  \rho \vdash_\beta \langle \kw{lvar} \id : \sT = e, \sigma \rangle
    \rightarrow
      \langle
        \rho_1, \sigma_1
      \rangle
\endprooftree
\end{gather}
if $\stknew(\valstate) = (\sigma_1, i)$ and
$\rho_1 = \bigl\{ \id \mapsto (i, \sT) \bigr\}$.
\item[Function declaration]
\begin{gather}
\label{rule:conc_decl_function}
\prooftree
  \nohyp
\justifies
  \rho
    \vdash_\beta
      \bigl\langle
        \kw{function} \id(\fps) = \funbody_0,
        \sigma
      \bigr\rangle
  \rightarrow
    \langle
      \rho_0, \sigma
    \rangle
\endprooftree
\end{gather}
if $\rho_0 = \{ \id \mapsto \lambdaop \fps \st \funbody_1 \}$ and
either $\funbody_0 = \funbody_1 = \kw{extern} : \sT$
or, for each $i \in \{0, 1\}$,
$\funbody_i = \kw{let} d_i \,\kw{in} s \kw{result} e$,
$I = \FI(\funbody_0) \setminus \DI(\fps)$ and
$d_1 = \rho \restrict{I}; d_0$.
\item[Recursive declaration]
\begin{gather}
\label{rule:conc_decl_rec}
\prooftree
  (\rho \setminus J) \vdash_{\beta[\beta_1]} \langle g, \sigma \rangle
    \rightarrow
      \langle \rho_0, \sigma_0 \rangle
\quad
  \rho \vdash_\beta \langle \kw{rec} \rho_0, \sigma_0 \rangle
    \rightarrow
      \eta
\justifies
  \rho \vdash_\beta \langle \kw{rec} g, \sigma \rangle
    \rightarrow
      \eta
\endprooftree
\end{gather}
if $J = \FI(g) \inters \DI(g)$,
$\beta \vdash_{\FI(g)} g : \beta_0$ and
$\beta_1 = \beta_0 \restrict{J}$.
\item[Global sequential composition]
\begin{gather}
\label{rule:conc_glob_seq_err_0}
\prooftree
  \rho \vdash_\beta \langle g_0, \sigma \rangle
    \rightarrow
      \exceptstate
\justifies
  \rho \vdash_\beta \langle g_0;g_1, \sigma \rangle
    \rightarrow
      \exceptstate
\endprooftree \\[1ex]
\label{rule:conc_glob_seq_err_1}
\prooftree
  \rho \vdash_\beta \langle g_0, \sigma \rangle
    \rightarrow
      \langle \rho_0, \sigma_0 \rangle
\quad
  \rho[\rho_0] \vdash_{\beta[\beta_0]} \langle g_1, \sigma_0 \rangle
    \rightarrow
      \exceptstate
\justifies
  \rho \vdash_\beta \langle g_0;g_1, \sigma \rangle
    \rightarrow
      \exceptstate
\using\quad\text{if $\beta \vdash_{\FI(g_0)} g_0 : \beta_0$}
\endprooftree \\[1ex]
\label{rule:conc_glob_seq_ok}
\prooftree
  \rho \vdash_\beta \langle g_0, \sigma \rangle
    \rightarrow
      \langle \rho_0, \sigma_0 \rangle
\quad
  \rho[\rho_0] \vdash_{\beta[\beta_0]} \langle g_1, \sigma_0 \rangle
    \rightarrow
      \langle \rho_1, \sigma_1 \rangle
\justifies
  \rho \vdash_\beta \langle g_0;g_1, \sigma \rangle
    \rightarrow
      \bigl\langle \rho_0[\rho_1], \sigma_1 \bigr\rangle
\using\quad\text{if $\beta \vdash_{\FI(g_0)} g_0 : \beta_0$}
\endprooftree
\end{gather}
\item[Local sequential composition]
\begin{gather}
\label{rule:conc_decl_seq_err_0}
\prooftree
  \rho \vdash_\beta \langle d_0 , \sigma \rangle
    \rightarrow
      \exceptstate
\justifies
  \rho \vdash_\beta \langle d_0;d_1 , \sigma \rangle
    \rightarrow
      \exceptstate
\endprooftree \\[1ex]
\label{rule:conc_decl_seq_err_1}
\prooftree
  \rho \vdash_\beta \langle d_0 , \sigma \rangle
    \rightarrow
      \langle \rho_0, \sigma_0 \rangle
\quad
  \rho[\rho_0] \vdash_{\beta[\beta_0]} \langle d_1 , \sigma_0 \rangle
    \rightarrow
      \exceptstate
\justifies
  \rho \vdash_\beta \langle d_0;d_1 , \sigma \rangle
    \rightarrow
      \exceptstate
\using\quad\text{if $\beta \vdash_{\FI(d_0)} d_0 : \beta_0$}
\endprooftree \\[1ex]
\label{rule:conc_decl_seq_ok}
\prooftree
  \rho \vdash_\beta \langle d_0 , \sigma \rangle
    \rightarrow
      \langle \rho_0, \sigma_0 \rangle
\quad
  \rho[\rho_0] \vdash_{\beta[\beta_0]} \langle d_1 , \sigma_0 \rangle
    \rightarrow
      \langle \rho_1, \sigma_1 \rangle
\justifies
  \rho \vdash_\beta \langle d_0;d_1 , \sigma \rangle
    \rightarrow
      \bigl\langle \rho_0[\rho_1], \sigma_1 \bigr\rangle
\using\quad\text{if $\beta \vdash_{\FI(d_0)} d_0 : \beta_0$}
\endprooftree
\end{gather}
\end{description}

\subsubsection{Statements}

\begin{description}
\item[Nop]
\begin{gather}
\label{rule:conc_nop}
\prooftree
  \nohyp
\justifies
  \rho \vdash_\beta \langle \kw{nop}, \sigma \rangle
    \rightarrow
      \sigma
\endprooftree
\end{gather}
\item[Assignment]
\begin{gather}
\label{rule:conc_assignment_error}
\prooftree
  \rho \vdash_\beta \langle e, \sigma \rangle
    \rightarrow
      \exceptstate
\justifies
  \rho \vdash_\beta \langle \id := e, \sigma \rangle
    \rightarrow
      \exceptstate
\endprooftree \\[1ex]
\label{rule:conc_assignment_ok}
\prooftree
  \rho \vdash_\beta \langle e, \sigma \rangle
    \rightarrow
      \langle \sval, \sigma_0 \rangle
\justifies
  \rho \vdash_\beta \langle \id := e, \sigma \rangle
    \rightarrow
      \sigma_0\bigl[ \rho(\id) := \sval \bigr]
\endprooftree
\end{gather}

\item[Statement sequence]
\begin{gather}
\label{rule:conc_sequence_0}
\prooftree
  \rho \vdash_\beta \langle s_0, \sigma \rangle
    \rightarrow
      \exceptstate
\justifies
  \rho \vdash_\beta \langle s_0 ; s_1, \sigma \rangle
    \rightarrow
      \exceptstate
\endprooftree \\[1ex]
\label{rule:conc_sequence_1}
\prooftree
  \rho \vdash_\beta \langle s_0, \sigma \rangle
    \rightarrow
      \sigma_0
\quad
  \rho \vdash_\beta \langle s_1, \sigma_0 \rangle
    \rightarrow
      \eta
\justifies
  \rho \vdash_\beta \langle s_0 ; s_1, \sigma \rangle
    \rightarrow
      \eta
\endprooftree
\end{gather}
\item[Block]
\begin{gather}
\label{rule:conc_block_0}
\prooftree
  \rho \vdash_\beta \bigl\langle d, \smark(\sigma) \bigr\rangle
    \rightarrow
      \exceptstate
\justifies
  \rho \vdash_\beta \langle d ; s, \sigma \rangle
    \rightarrow
      \exceptstate
\endprooftree \\[1ex]
\label{rule:conc_block_1}
\prooftree
  \rho \vdash_\beta \bigl\langle d, \smark(\sigma) \bigr\rangle
    \rightarrow
      \langle \rho_0, \sigma_0 \rangle
\quad
  \rho[\rho_0] \vdash_{\beta[\beta_0]} \langle s, \sigma_0 \rangle
    \rightarrow
       \eta
\justifies
  \rho \vdash_\beta \langle d ; s, \sigma \rangle
    \rightarrow
       \sunmark(\eta)
\using\quad\text{if $\beta \vdash_{\FI(d)} d : \beta_0$}
\endprooftree
\end{gather}
\item[Conditional]
\begin{gather}
\label{rule:conc_conditional_error}
\prooftree
  \rho \vdash_\beta \langle e, \sigma \rangle
    \rightarrow
      \exceptstate
\justifies
  \rho \vdash_\beta
         \langle \kw{if} e \kw{then} s_0 \kw{else} s_1, \sigma \rangle
    \rightarrow
      \exceptstate
\endprooftree \\[1ex]
\label{rule:conc_conditional_true}
\prooftree
  \rho \vdash_\beta \langle e, \sigma \rangle
    \rightarrow
      \langle \ttv, \sigma_0 \rangle
\quad
  \rho \vdash_\beta \langle s_0, \sigma_0 \rangle
    \rightarrow
      \eta
\justifies
  \rho \vdash_\beta
         \langle \kw{if} e \kw{then} s_0 \kw{else} s_1, \sigma \rangle
    \rightarrow
      \eta
\endprooftree \\[1ex]
\label{rule:conc_conditional_false}
\prooftree
  \rho \vdash_\beta \langle e, \sigma \rangle
    \rightarrow
      \langle \ffv, \sigma_0 \rangle
\quad
  \rho \vdash_\beta \langle s_1, \sigma_0 \rangle
    \rightarrow
      \eta
\justifies
  \rho \vdash_\beta
         \langle \kw{if} e \kw{then} s_0 \kw{else} s_1, \sigma \rangle
    \rightarrow
      \eta
\endprooftree
\end{gather}
\item[While]
\begin{gather}
\label{rule:conc_while_guard_error}
\prooftree
  \rho \vdash_\beta \langle e, \sigma \rangle
    \rightarrow
      \exceptstate
\justifies
  \rho \vdash_\beta \langle \kw{while} e \kw{do} s, \sigma \rangle
    \rightarrow
      \exceptstate
\endprooftree \\[1ex]
\label{rule:conc_while_false_ok}
\prooftree
  \rho \vdash_\beta \langle e, \sigma \rangle
    \rightarrow
      \langle \ffv, \sigma_0 \rangle
\justifies
  \rho \vdash_\beta \langle \kw{while} e \kw{do} s, \sigma \rangle
    \rightarrow
      \sigma_0
\endprooftree \\[1ex]
\label{rule:conc_while_true_body_error}
\prooftree
  \rho \vdash_\beta \langle e, \sigma \rangle
    \rightarrow
      \langle \ttv, \sigma_0 \rangle
\quad
  \rho \vdash_\beta \langle s, \sigma_0 \rangle
    \rightarrow
      \exceptstate
\justifies
  \rho \vdash_\beta \langle \kw{while} e \kw{do} s, \sigma \rangle
    \rightarrow
      \exceptstate
\endprooftree \\[1ex]
\label{rule:conc_while_true_body_ok}
\prooftree
  \rho \vdash_\beta \langle e, \sigma \rangle
    \rightarrow
      \langle \ttv, \sigma_0 \rangle
\quad
  \rho \vdash_\beta \langle s, \sigma_0 \rangle
    \rightarrow
      \sigma_1
\quad
  \rho \vdash_\beta \langle \kw{while} e \kw{do} s, \sigma_1 \rangle
    \rightarrow
      \eta
\justifies
  \rho \vdash_\beta \langle \kw{while} e \kw{do} s, \sigma \rangle
    \rightarrow
      \eta
\endprooftree
\end{gather}
\item[Throw]
\begin{gather}
\label{rule:conc_throw_except}
\prooftree
  \nohyp
\justifies
  \rho \vdash_\beta \langle \kw{throw} \rtsexcept, \sigma \rangle
    \rightarrow
      \langle \sigma, \rtsexcept \rangle
\endprooftree \\[1ex]
\label{rule:conc_throw_expr_error}
\prooftree
  \rho \vdash_\beta \langle e, \sigma \rangle
    \rightarrow
      \exceptstate
\justifies
  \rho \vdash_\beta \langle \kw{throw} e, \sigma \rangle
    \rightarrow
      \exceptstate
\endprooftree \\[1ex]
\label{rule:conc_throw_expr_ok}
\prooftree
  \rho \vdash_\beta \langle e, \sigma \rangle
    \rightarrow
      \langle \sval, \sigma_0 \rangle
\justifies
  \rho \vdash_\beta \langle \kw{throw} e, \sigma \rangle
    \rightarrow
      \langle \sigma_0, \sval \rangle
\endprooftree
\end{gather}
\item[Try blocks]
\begin{gather}
\label{rule:conc_try_no_except}
\prooftree
  \rho \vdash_\beta \langle s, \sigma \rangle
    \rightarrow
      \sigma_0
\justifies
  \rho
    \vdash_\beta
      \langle
        \kw{try} s \kw{catch} k,
        \sigma
      \rangle
    \rightarrow
      \sigma_0
\endprooftree \\[1ex]
\label{rule:conc_try_except}
\prooftree
  \rho \vdash_\beta \langle s, \sigma \rangle
    \rightarrow
      \exceptstate_0
\quad
  \rho \vdash_\beta \langle k, \exceptstate_0 \rangle
    \rightarrow
      \langle u, \eta \rangle
\justifies
  \rho
    \vdash_\beta
      \langle
        \kw{try} s \kw{catch} k,
        \sigma
      \rangle
    \rightarrow
      \eta
\using\quad\text{if $u \in \{ \caught, \uncaught \}$}
\endprooftree
\end{gather}

\begin{gather}
\label{rule:conc_try_finally}
\prooftree
  \rho \vdash_\beta \langle s_0, \sigma \rangle
    \rightarrow
      \sigma_0
\quad
  \rho \vdash_\beta \langle s_1, \sigma_0 \rangle
    \rightarrow
      \eta
\justifies
  \rho
    \vdash_\beta
      \langle
        \kw{try} s_0 \kw{finally} s_1,
        \sigma
      \rangle
    \rightarrow
      \eta
\endprooftree \\[1ex]
\label{rule:conc_try_finally_exc_0}
\prooftree
  \rho \vdash_\beta \langle s_0, \sigma \rangle
    \rightarrow
      \langle \sigma_0, \except_0 \rangle
\quad
  \rho \vdash_\beta \langle s_1, \sigma_0 \rangle
    \rightarrow
      \sigma_1
\justifies
  \rho
    \vdash_\beta
      \langle
        \kw{try} s_0 \kw{finally} s_1,
        \sigma
      \rangle
    \rightarrow
      \langle \sigma_1, \except_0 \rangle
\endprooftree \\[1ex]
\label{rule:conc_try_finally_exc_1}
\prooftree
  \rho \vdash_\beta \langle s_0, \sigma \rangle
    \rightarrow
      \langle \sigma_0, \except_0 \rangle
\quad
  \rho \vdash_\beta \langle s_1, \sigma_0 \rangle
    \rightarrow
      \exceptstate
\justifies
  \rho
    \vdash_\beta
      \langle
        \kw{try} s_0 \kw{finally} s_1,
        \sigma
      \rangle
    \rightarrow
      \exceptstate
\endprooftree
\end{gather}
\item[Function call]
Consider the following conditions:
\begin{gather}
\label{eq:function-call-condition-beta-rho-d}
\left.
\begin{aligned}
  \beta(\id)
    &= (\fps \rightarrow \sT_0) \\
  \rho(\id)
    &= \lambdaop \id_1 : \sT_1, \ldots, \id_n : \sT_n \st \funbody \\
  d &= (\kw{lvar} \ridx_0 : \sT_0 = \id_0;
          \kw{lvar} \ridx_1 : \sT_1 = e_1;
            \ldots;
              \kw{lvar} \ridx_n : \sT_n = e_n)
\end{aligned}
\right\} \\
\label{eq:function-call-condition-rho0-rho1}
  \rho_1
    = \bigl\{
        \ridx_0 \mapsto (0, \sT_0)
      \bigr\}
        \union
      \bigl\{\,
        \id_j \mapsto (j, \sT_j)
      \bigm|
        j = 1, \ldots, n
      \,\bigr\},
    \;
  \rho_0 : \beta_0,
    \;
  \rho_1 : \beta_1.
\end{gather}
Then the rule schemata for function calls are the following:
\begin{gather}
\label{rule:conc_function_call_param_err}
\prooftree
  \rho
    \vdash_\beta
      \bigl\langle
        d, \smark(\sigma)
      \bigr\rangle
    \rightarrow
      \exceptstate
\justifies
  \rho
    \vdash_\beta
      \bigl\langle
        \id_0 := \id(e_1, \ldots, e_n),
        \sigma
      \bigr\rangle
    \rightarrow
      \exceptstate
\using\quad\text{if \eqref{eq:function-call-condition-beta-rho-d} holds}
\endprooftree \\[1ex]
\label{rule:conc_function_call_eval_err}
\prooftree
\begin{aligned}
& \rho
    \vdash_\beta
      \bigl\langle
        d, \smark(\sigma)
      \bigr\rangle
    \rightarrow
      \langle \rho_0, \sigma_0 \rangle \\
& \rho[\rho_1]
    \vdash_{\beta[\beta_1]}
      \bigl\langle \funbody, \slink(\sigma_0) \bigr\rangle
        \rightarrow
          \exceptstate
\end{aligned}
\justifies
  \rho
    \vdash_\beta
      \bigl\langle
        \id_0 := \id(e_1, \ldots, e_n),
        \sigma
      \bigr\rangle
    \rightarrow
      \sunmark\bigl(\sunlink(\exceptstate)\bigr)
\using\quad\text{if \eqref{eq:function-call-condition-beta-rho-d}
                 and \eqref{eq:function-call-condition-rho0-rho1} hold}
\endprooftree \\[1ex]
\label{rule:conc_function_call_eval_ok}
\prooftree
\begin{aligned}
& \rho
    \vdash_\beta
      \bigl\langle
        d, \smark(\sigma)
      \bigr\rangle
    \rightarrow
      \langle \rho_0, \sigma_0 \rangle \\
& \rho[\rho_1]
    \vdash_{\beta[\beta_1]}
      \bigl\langle \funbody, \slink(\sigma_0) \bigr\rangle
        \rightarrow
          \sigma_1 \\
& \rho[\rho_0]
    \vdash_{\beta[\beta_0]}
      \bigl\langle
        \id_0 := \ridx_0,
        \sunlink(\sigma_1)
      \bigr\rangle
        \rightarrow
          \eta_2
\end{aligned}
\justifies
  \rho
    \vdash_\beta
      \bigl\langle
        \id_0 := \id(e_1, \ldots, e_n),
        \sigma
      \bigr\rangle
    \rightarrow
      \sunmark(\eta_2)
\using\quad\text{if \eqref{eq:function-call-condition-beta-rho-d}
                 and \eqref{eq:function-call-condition-rho0-rho1} hold}
\endprooftree
\end{gather}
\end{description}
Note that parameter passing is implemented by using reserved identifiers
that reference the return value ($\ridx_0$) and the actual
arguments ($\ridx_1$, \ldots, $\ridx_n$).
When evaluating the function body
(i.e., after linking a new activation frame),
the callee can get access to the return value and the arguments' values
by using the indirect locators $0$ and $1$, \ldots, $n$, respectively;
to this end, the callee uses the environment $\rho_1$, where
the reserved identifier $\ridx_0$ is still mapped to the return value,
whereas the arguments are accessible using the formal parameters' names
$\id_1$, \ldots, $\id_n$.

\subsubsection{Function Bodies}

\begin{gather}
\label{rule:conc_function_body_0}
\prooftree
  \rho \vdash_\beta \bigl\langle d, \smark(\sigma) \bigr\rangle
    \rightarrow
      \exceptstate
\justifies
  \rho
    \vdash_\beta
      \langle \kw{let} d \,\kw{in} s \kw{result} e, \sigma \rangle
    \rightarrow
      \exceptstate
\endprooftree \\[1ex]
\label{rule:conc_function_body_1}
\prooftree
  \rho \vdash_\beta \bigl\langle d, \smark(\sigma) \bigr\rangle
    \rightarrow
      \langle \rho_0, \sigma_0 \rangle
\quad
  \rho[\rho_0] \vdash_{\beta[\beta_0]} \langle s, \sigma_0 \rangle
    \rightarrow
      \exceptstate
\justifies
  \rho
    \vdash_\beta
      \langle \kw{let} d \,\kw{in} s \kw{result} e, \sigma \rangle
    \rightarrow
      \sunmark(\exceptstate)
\using\quad\text{if $\beta \vdash_{\FI(d)} d : \beta_0$}
\endprooftree \\[1ex]
\label{rule:conc_function_body_2}
\prooftree
\begin{aligned}
& \rho \vdash_\beta \bigl\langle d, \smark(\sigma) \bigr\rangle
    \rightarrow
      \langle \rho_0, \sigma_0 \rangle \\
%%\quad
& \rho[\rho_0] \vdash_{\beta[\beta_0]} \langle s, \sigma_0 \rangle
    \rightarrow
      \sigma_1 \\
& \rho[\rho_0]
    \vdash_{\beta[\beta_0]}
      \langle \ridx_0 := e, \sigma_1 \rangle
    \rightarrow
      \eta_0
\end{aligned}
\justifies
  \rho
    \vdash_\beta
      \langle \kw{let} d \,\kw{in} s \kw{result} e, \sigma \rangle
    \rightarrow
      \sunmark(\eta_0)
\using\quad\text{if $\beta \vdash_{\FI(d)} d : \beta_0$}
\endprooftree
\end{gather}

\begin{gather}
\label{rule:conc_function_body_extern}
\prooftree
  \nohyp
\justifies
  \rho
    \vdash_\beta
      \bigl\langle
        \kw{extern} : \sT,
        (\mu, w)
      \bigr\rangle
    \rightarrow
      \eta
\endprooftree
\end{gather}
if
\(
  \exists \sigma_0 = (\mu_0, w) \in \Mem, \except \in \Except
    \st
      \eta = \sigma_0
        \lor
      \eta = \langle \sigma_0, \except \rangle
\).

\subsubsection{Catch Clauses}

\begin{description}
\item[Catch]
\begin{gather}
\label{rule:conc_catch_caught}
\prooftree
  \rho \vdash_\beta \langle s, \sigma \rangle
    \rightarrow
      \eta_0
\justifies
  \rho
    \vdash_\beta
      \bigl\langle
        (p) \, s,
        (\sigma, \except)
      \bigr\rangle
    \rightarrow
      \langle \caught, \eta_0 \rangle
\endprooftree \\[1ex]
\intertext{%
if $p = \except \in \RTSExcept$,
or $p = \type(\except)$,
or $p = \kw{any}$.
}
\label{rule:conc_catch_expr_caught_stkerr}
\prooftree
  \nohyp
\justifies
  \rho
    \vdash_\beta
      \bigl\langle
        (\id : \sT) \, s,
        (\sigma, \sval)
      \bigr\rangle
    \rightarrow
      \bigl\langle \caught, \sunmark(\exceptstate_0) \bigr\rangle
\endprooftree \\[1ex]
\intertext{%
if $\sT = \type(\sval)$ and
$\exceptstate_0 = \stknew\bigl( \sval, \smark(\sigma) \bigr)$.
}
\label{rule:conc_catch_expr_caught_noerr}
\prooftree
  \rho\bigl[\{\id \mapsto (i,\sT)\}\bigr]
    \vdash_{\beta[ \{\id \mapsto \location{\sT}\} ]}
      \langle s, \sigma_0 \rangle
    \rightarrow
      \eta_0
\justifies
  \rho
    \vdash_\beta
      \bigl\langle
        (\id : \sT) \, s,
        (\sigma, \sval)
      \bigr\rangle
    \rightarrow
      \bigl\langle \caught, \sunmark(\eta_0) \bigr\rangle
\endprooftree \\[1ex]
\intertext{%
if $\sT = \type(\sval)$ and
$(\sigma_0, i) = \stknew\bigl( \sval, \smark(\sigma) \bigr)$.
}
\label{rule:conc_catch_uncaught}
\prooftree
  \nohyp
\justifies
  \rho
    \vdash_\beta
      \bigl\langle
        (p) \, s,
        (\sigma, \except)
      \bigr\rangle
    \rightarrow
      \bigl\langle \uncaught, (\sigma, \except) \bigr\rangle
\endprooftree
\end{gather}
%% if $p \notin \bigl\{ \except, \cT, \kw{any} \bigr\}$
%% and $\forall \id \in \Id \itc p \neq \id : \cT$,
%% where $\cT = \type(\except)$.
if, letting $\cT = \type(\except)$, we have
$p \notin \bigl\{ \except, \cT, \kw{any} \bigr\}$
and $\forall \id \in \Id \itc p \neq \id : \cT$.
\item[Catch sequence]
\begin{gather}
\label{rule:conc_catch_seq_caught}
\prooftree
  \rho
    \vdash_\beta
      \langle k_0, \exceptstate \rangle
    \rightarrow
      \langle \caught, \eta_0 \rangle
\justifies
  \rho
    \vdash_\beta
      \langle
        k_0 ; k_1,
        \exceptstate
      \rangle
    \rightarrow
      \langle \caught, \eta_0 \rangle
\endprooftree \\[1ex]
\label{rule:conc_catch_seq_uncaught}
\prooftree
  \rho
    \vdash_\beta
      \langle k_0, \exceptstate \rangle
    \rightarrow
      \langle \uncaught, \exceptstate_0 \rangle
\quad
  \rho
    \vdash_\beta
      \langle k_1, \exceptstate_0 \rangle
    \rightarrow
      \eta
\justifies
  \rho
    \vdash_\beta
      \langle
        k_0 ; k_1,
        \exceptstate
      \rangle
    \rightarrow
      \eta
\endprooftree
\end{gather}
\end{description}

\subsection{Concrete Divergence Relation}
\label{sec:concrete-divergence-relation}

In order to capture divergent computations, we follow the approach
of Cousot and Cousot \cite{CousotC92}, also advocated by
Schmidt \cite{Schmidt98}
and Leroy \cite{Leroy06}.
This consists in introducing a \emph{divergence relation} by means
of sequents of the form
\(
  \rho \vdash_\beta N \diverges,
\)
where $N \in \NTq$ and
$q \in \{ \mathrm{s}, \mathrm{b}, \mathrm{k} \}$.
Intuitively, a divergence sequent of the form, say,
$\rho \vdash_\beta \langle s, \sigma \rangle \diverges$
means that, in the context given by $\rho$ and $\sigma$,
the execution of statement $s$ diverges.
We now give a set of rules that
(interpreted coinductively, as we will see later)
allow to characterize the behavior of divergent computations.
For instance, the following rule schemata characterize the divergence
behavior of statement sequences:
\[
  \prooftree
    \rho \vdash_\beta \langle s_0, \sigma \rangle
      \diverges
  \Justifies
    \rho \vdash_\beta \langle s_0 ; s_1, \sigma \rangle
      \diverges
  \endprooftree \\[1ex]
\quad
  \prooftree
    \rho \vdash_\beta \langle s_0, \sigma \rangle
      \rightarrow
        \sigma_0
  \quad
    \rho \vdash_\beta \langle s_1, \sigma_0 \rangle
      \diverges
  \Justifies
    \rho \vdash_\beta \langle s_0 ; s_1, \sigma \rangle
      \diverges
  \endprooftree
\]
Notice that, once the set of concrete rules characterizing finite computations
is known, the concrete rules modeling divergences can be specified
systematically (and thus implicitly).
Namely, for each concrete rule
\begin{equation}
\label{rule:conc_generic}
\prooftree
  P_0 \quad \cdots \quad P_{i-1}
    \quad
  \rho_i \vdash_{\beta_i} N_i \rightarrow \eta_i
    \quad
  P_{i+1} \quad \cdots \quad P_{h-1}
\justifies
  \rho \vdash_\beta N \rightarrow \eta
\using\quad\text{(side condition)}
\endprooftree
\end{equation}
such that $0 \leq i < h$ and,
for $q \in \{ \mathrm{s}, \mathrm{b}, \mathrm{k} \}$,
$N_i \in \biguplus \NTq^{\beta_i}$ and
$N \in \biguplus \NTq^\beta$,
there is the corresponding divergence rule where
the $i$-th premise is diverging, i.e.,
\begin{equation*}
\prooftree
  P_0 \quad \cdots \quad P_{i-1} \quad \rho_i \vdash_{\beta_i} N_i \diverges
\Justifies
  \rho \vdash_\beta N \diverges
\using\quad\text{(side condition)}
\endprooftree
\end{equation*}
Therefore, there are two rules above modeling the divergence of statement
sequences, which can be obtained from rule \eqref{rule:conc_sequence_1}.
It is worth noting that
a single divergence rule schema can be obtained from more than
one of the concrete rules in Section~\ref{sec:concrete-evaluation-relations}.

We will use the terms \emph{negative} and \emph{positive}
to distinguish the different kinds of rules constructed in
this and the previous section, respectively.
\begin{definition} \summary{(Concrete semantics rules.)}
\label{def:conc-rules}
The set $\calR_+$ (resp., $\calR_-$)
of \emph{positive} (resp., \emph{negative}) \emph{concrete semantics rules}
is the infinite set obtained by instantiating the rule schemata
of \textup{Section~\ref{sec:concrete-evaluation-relations}}
(resp., \textup{Section~\ref{sec:concrete-divergence-relation}})
in all possible ways (respecting, of course, the side conditions).
Moreover, $\calR \defeq \calR_+ \uplus \calR_-$.
\end{definition}

\subsection{Concrete Semantics Trees}
\label{sec:concrete-semantics-trees}

The concrete semantics of a program is a (possibly infinite)
set of finite or infinite trees.  Such trees are defined in
terms of the (infinite) set of instances of the rules defined
in the previous two sections.

Let $\calS$ be the (infinite) set of sequents occurring in the premises
and conclusions of the rules in $\calR$.
The \emph{concrete semantics universe}, denoted by $\calU$, is the
set of finitely branching trees of at most $\omega$-depth with labels
in $\calS$.
\begin{definition} \summary{(Concrete semantics universe.)}
\label{def:concrete-semantics-universe}
A set $P \sseq \Nset^\star$ is \emph{prefix-closed} if, for each
$z \in \Nset^\star$ and each $n \in \Nset$, $zn \in P$ implies $z \in P$.
A set $P \sseq \Nset^\star$ is \emph{canonical} if, for each
$z \in \Nset^\star$ there exists $h \in \Nset$ such that
\[
  \{\, n \in \Nset \mid zn \in P \,\} = \{ 0, \ldots, h-1 \}.
\]
An \emph{$\calS$-tree} is a partial function
$\pard{\theta}{\Nset^\star}{\calS}$
such that $\dom(\theta)$ is prefix-closed and canonical.
The \emph{concrete semantics universe} $\calU$ is the set of all $\calS$-trees.
\end{definition}
For each $p \in \dom(\theta)$, the tree $\theta_{[p]}$ defined,
for each $z \in \Nset^\star$, by $\theta_{[p]}(z) \defeq \theta(pz)$,
is called a \emph{subtree} of $\theta$;  it is called a \emph{proper subtree}
if $p \neq \emptystring$.
If $\dom(\theta) = \emptyset$, then $\theta$ is the empty tree.
If $\theta$ is not empty,
then $\theta(\emptystring)$ is the \emph{root} of $\theta$ and,
if $\{ 0, \ldots, h-1 \} \subseteq \dom(\theta)$
and $h \notin \dom(\theta)$,
then $\theta_{[0]}$, \dots,~$\theta_{[h-1]}$
are its \emph{immediate subtrees}
(note that $h \in \Nset$ may be zero);
in this case $\theta$ can be denoted by
$\frac{\theta_{[0]} \; \cdots \; \theta_{[h-1]}}{\theta(\emptystring)}$.

\begin{definition} \summary{(Concrete semantics trees.)}
\label{def:concrete-semantics-trees}
Let $\fund{\calF_+}{\wp(\calU)}{\wp(\calU)}$ be the continuous function
over the complete lattice $\bigl(\wp(\calU), \mathord{\sseq}\bigl)$ given,
for all $U \in \wp(\calU)$, by
\begin{align*}
  \calF_+(U)
    &\defeq
      \sset{%
        \prooftree
          \theta_0 \; \cdots \; \theta_{h-1}
        \justifies
          s
        \endprooftree
      }{%
        \theta_0, \ldots, \theta_{h-1} \in U, \\[1ex]
        \prooftree
          \theta_0(\emptystring) \; \cdots \; \theta_{h-1}(\emptystring)
        \justifies
          s
        \endprooftree
          \in \calR_+
      }. \\
\intertext{%
The set of \emph{positive concrete semantics trees} is
$\Theta_+ \defeq \lfp_{\mathord{\subseteq}}(\calF_+)$.
Consider now the co-continuous function
$\fund{\calF_-}{\wp(\calU)}{\wp(\calU)}$
given, for each $U \in \wp(\calU)$, by
}
  \calF_-(U)
    &\defeq
      \sset{%
        \prooftree
          \theta_0 \; \cdots \; \theta_{h-1}
        \justifies
          s
        \endprooftree
      }{%
        \theta_0, \ldots, \theta_{h-2} \in \Theta_+,
        \quad
        \theta_{h-1} \in U, \\[1ex]
        \prooftree
          \theta_0(\emptystring) \; \cdots \; \theta_{h-1}(\emptystring)
        \justifies
          s
        \endprooftree
          \in \calR_-
      }.
\end{align*}
The set of \emph{negative concrete semantics trees} is
$\Theta_- \defeq \gfp_{\mathord{\subseteq}}(\calF_-)$.
The set of all \emph{concrete semantics trees} is
$\Theta \defeq \Theta_+ \uplus \Theta_-$.
\end{definition}

We now show that, for every concrete non-terminal configuration,
there exists a concrete semantics tree with that in the root.
\begin{proposition}
\label{prop:concrete-tree-exists}
For each $\beta \in \TEnv$,
$\rho \in \Env$ such that $\rho : \beta$ and
$N \in \NTq^{\beta}$, where
\(
  q
    \in
      \{
        \mathrm{e}, \mathrm{d}, \mathrm{g},
        \mathrm{s}, \mathrm{b}, \mathrm{k}
      \}
\),
there exists $\theta \in \Theta$ such that
\[
  \theta(\emptystring)
  \in \bigl\{\,
        (\rho \vdash_\beta N \rightarrow \eta)
      \bigm|
        \eta \in \Tq
      \,\bigr\}
  \uplus
    \bigl\{\, (\rho \vdash_\beta N \diverges) \,\bigr\}.
\]
\end{proposition}
\begin{proof}
If $q = \mathrm{e}$ and $\eta \in \Te$,
we say that the sequent
$(\rho \vdash_\beta N \rightarrow \eta)$
is well-typed if
$N = \langle e, \sigma_0 \rangle$
and
$\eta = \langle \sval, \sigma_1 \rangle$
imply
$\beta \vdash e : \type(\sval)$.
For the proof, let
\begin{equation*}
  S_+(\rho, \beta, N)
    \defeq
      \bigl\{\,
        s
      \bigm|
        s = (\rho \vdash_\beta N \rightarrow \eta),
        \eta \in \Tq,
        (q = \mathrm{e} \implies \text{$s$ is well-typed})
      \,\bigr\}.
\end{equation*}

We now assume that $N \in \NTq^{\beta}$
is a fixed but arbitrary non-terminal configuration.
It suffices
to show there exists $\theta \in \Theta$ such that
\(
  \theta(\emptystring)
    \in
      S_+(\rho, \beta, N)
        \uplus
      \bigl\{\, (\rho \vdash_\beta N \diverges) \,\bigr\}
\).
Let $R_0$ be the set of all rules in  $\calR_+$
whose conclusions are in $S_+(\rho, \beta, N)$.
By inspecting the concrete evaluation rule schemata in
Section~\ref{sec:concrete-evaluation-relations},
$R_0 \neq \emptyset$.
Let $j \geq 0$ be the maximal value for which
there exist finite trees
$\theta_0, \ldots, \theta_{j-1} \in \Theta_+$ where
$P_0 = \theta_0(\emptystring), \ldots, P_{j-1} = \theta_{j-1}(\emptystring)$
are the first $j$ premises of a rule in $R_0$.
Let $R_j \sseq R_0$ be the set of all rules in $R_0$
with $P_0, \ldots, P_{j-1}$ as their first $j$ premises;
then $R_j \neq \emptyset$.
By inspecting the rule schemata
in Section~\ref{sec:concrete-evaluation-relations},
it can be seen that, if there exists
$\frac{P_0 \; \cdots \; P_{j-1} \; P'_j \; \cdots}{s'} \in R_j$
for some $P'_j \in S_+(\rho_j, \beta_j, N_j)$ and
$s' \in S_+(\rho, \beta, N)$,
then%
\footnote{
To help understand this property,
we illustrate it in the case that
$q = \mathrm{e}$
and the non-terminal configuration is
$N = \langle b_0 \andop b_1, \sigma \rangle$;
hence the concrete rule schemata
\eqref{rule:conc_conjunction_0}--\eqref{rule:conc_conjunction_2}
will apply.
In all the rule instances, the first premise is of the form
$P_0 = (\rho \vdash_\beta N_0 \rightarrow \eta_0)$,
where $N_0 = \langle b_0, \sigma \rangle$;
as a consequence, we have
\(
   S_+(\rho, \beta, N_0) =
     \bigl\{\,
       (\rho \vdash_\beta N_0 \rightarrow \eta_0)
     \bigm|
       \eta_0 \in B
     \,\bigr\}
\),
where
\(
  B \defeq
      \ExceptState
        \uplus
          \{\,
            \langle t, \sigma_0 \rangle \in \Te
          \mid
            t \in \Bool, \sigma_0 \in \Mem
          \,\}
\).
Thus, for each terminal configuration $\eta_0 \in B$,
there is a rule instance having $\eta_0$ in its first premise ---
that is we instantiate
rule~\eqref{rule:conc_conjunction_0}
when $\eta_0 = \exceptstate$,
rule~\eqref{rule:conc_conjunction_1}
when $\eta_0 = \langle \ffv, \sigma_0 \rangle$
and rule~\eqref{rule:conc_conjunction_2}
when $\eta_0 = \langle \ttv, \sigma_0 \rangle$.
Thus property~\eqref{prop:concrete-tree-exists:j_plus_1_premise}
holds for $j = 0$.
Moreover, although only
rule~\eqref{rule:conc_conjunction_2} applies when $j = 1$,
the terminal configuration for the second premise ($P_1$)
is just any terminal configuration in $\Te$.
Thus property~\eqref{prop:concrete-tree-exists:j_plus_1_premise}
also holds for $j = 1$.
}
\begin{equation}
\label{prop:concrete-tree-exists:j_plus_1_premise}
  \forall P_j \in S_+(\rho_j, \beta_j, N_j)
    \itc
      \exists s \in S_+(\rho, \beta, N)
        \st
          \frac{P_0 \; \cdots \; P_{j-1} \; P_j \; \cdots}{s} \in R_j.
\end{equation}

Suppose that $q \in \{ \mathrm{e}, \mathrm{d}, \mathrm{g} \}$
so that we can also assume $N = \langle u, \sigma \rangle$.
We show by structural induction on $u$ that
there exists $\theta \in \Theta_+$ such that
\(
  \theta(\emptystring) \in S_+(\rho, \beta, N)
\).
By inspecting the rule schemata in
Section~\ref{sec:concrete-evaluation-relations},
it can be seen that, if $u$ is atomic,
the rules in $R_0$ have no premises (so that $j = 0$) and hence,
letting $\theta \in \Theta_+$ be the singleton tree consisting of the
conclusion of a rule in $R_0$,
we obtain that $\theta(\emptystring) \in S_+(\rho, \beta, N)$.
Otherwise, $u$ is not atomic,
we show that each of the rules in $R_j$ has exactly $j$ premises;
to do this, we assume there exists a rule in $R_j$
with a $(j+1)$-th premise
$P_j$
and derive a contradiction.
Let
$N_j \in \NT{q_j}^{\beta_j}$ be the non-terminal configuration
in $P_j$.
By inspecting the rule schemata in
Section~\ref{sec:concrete-evaluation-relations}
in the case that $q \in \{ \mathrm{e}, \mathrm{d}, \mathrm{g} \}$,
it can be seen that:
\begin{enumerate}[(i)]
\item
\label{prop:concrete-tree-exists:q_j-in-edg}
$q_j \in \{ \mathrm{e}, \mathrm{d}, \mathrm{g}\}$
so that $N_j$ has the form $\langle u_j, \sigma_j \rangle$;
\item
\label{prop:concrete-tree-exists:u_j-in-u}
$u_j$ is a substructure of $u$
unless $R_j$ consists of instances of
the schematic rule~\eqref{rule:conc_decl_rec} and $j = 1$.
\end{enumerate}
If $u_j$ is a substructure of $u$,
by property~\eqref{prop:concrete-tree-exists:q_j-in-edg},
we can apply structural induction to obtain that
there exists a finite tree $\theta_j \in \Theta_+$ such that
$P_j = \theta_j(\emptystring) \in S_+(\rho_j, \beta_j, N_j)$;
hence,
by property~\eqref{prop:concrete-tree-exists:j_plus_1_premise},
there exists a rule in $R_j$ having
$P_j$ as its $(j+1)$-th premise;
contradicting the assumption that $j$ was maximal.
Otherwise, by property~\eqref{prop:concrete-tree-exists:u_j-in-u},
if $u_j$ is not a substructure of $u$,
the rules in $R_0$ must be instances of rule
schema~\eqref{rule:conc_decl_rec} and  $j = 1$;
in this case, rule schema~\eqref{rule:conc_decl_rec_environment},
which has no premises,
can be instantiated with the second premise of a rule in $R_j$
as its conclusion; and again we have a contradiction.
Thus, for any $u_j$, all rules in $R_j$ have exactly $j$ premises.
By Definition~\ref{def:concrete-semantics-trees},
\(
  \theta
    =
      \frac{\theta_0 \; \cdots \; \theta_{j-1}}%
           {s}
    \in
      \Theta_+
\)
for some $s \in S_+(\rho, \beta, N)$.
Therefore, since $\Theta_+ \sseq \Theta$, the thesis holds when
$q \in \{ \mathrm{e}, \mathrm{d}, \mathrm{g} \}$.

Suppose now that $q \in \{ \mathrm{s}, \mathrm{b}, \mathrm{k} \}$.
We prove that,
if there does not exist a tree $\theta \in \Theta_+$ such that
$\theta(\emptystring) \in S_+(\rho, \beta, N)$,
then, for all $n \geq 0$, there exists a tree $\theta$ such that
$\theta(\emptystring) = s_\infty \defeq (\rho \vdash_\beta N \diverges)$ and
$\theta \in \calF^n_-(\calU)$.
To this end, we reason by induction on $n \geq 0$.
By our assumption that there is no tree $\theta \in \Theta_+$
such that $\theta(\emptystring) \in S_+(\rho, \beta, N)$,
there must exist a rule
\begin{equation*}
  \frac{P_0 \; \cdots \; P_{j-1} \; P_j \; \cdots}%
         {s}
    \in
      R_j
\end{equation*}
for some $P_j \in S_+(\rho_j, \beta_j, N_j)$;
let $q_j$ be such that $N_j \in \NT{q_j}^{\beta_j}$.
By the maximality of $j$, there is no tree in $\Theta_+$ whose root is $P_j$.
We have already shown that,
if $q_j \in \{ \mathrm{e}, \mathrm{d}, \mathrm{g} \}$,
then there exists a tree
$\theta_j \in \Theta_+$ such that
$\theta_j(\emptystring) \in S_+(\rho_j, \beta_j, N_j)$;
thus, by property~\eqref{prop:concrete-tree-exists:j_plus_1_premise},
there must be a rule in $R_j$ whose $(j+1)$-th premise
is $\theta_j(\emptystring)$; contradicting the assumption that
$j \geq 0$ is maximal.
Hence $q_j \in \{ \mathrm{s}, \mathrm{b}, \mathrm{k} \}$.
By the definition of the negative concrete semantics rules
in Section~\ref{sec:concrete-divergence-relation},
there exists a corresponding negative rule
\begin{equation*}
  \frac{P_0 \; \cdots \; P_{j-1}  \; P_\infty}%
       {s_\infty}
    \in \calR_-
\end{equation*}
such that
$P_\infty = (\rho_j \vdash_{\beta_j} N_j \diverges)$.
Hence, by Definition~\ref{def:concrete-semantics-universe},
there exists a tree in $\calU = \calF^0_-(\calU)$
with root $s_\infty$,
so that the inductive hypothesis holds for $n = 0$.
Suppose now that $n > 0$.
By the inductive hypothesis,
there exists a tree $\theta_\infty \in \calF^{n-1}_-(\calU)$
such that $\theta_\infty(\emptystring) = P_\infty$.
Hence, by Definition~\ref{def:concrete-semantics-trees},
\(
 \frac{\theta_0 \; \cdots \; \theta_{j-1} \; \theta_\infty}%
           {s_\infty}
    \in
      \calF^n_-(\calU)
\).
Thus, for all $n \geq 0$, there exists a tree in $\calF^n_-(\calU)$
with root $s_\infty$ and hence,
by Definition~\ref{def:concrete-semantics-trees},
there exists a tree in $\Theta_-$ with root $s_\infty$.
Since $\Theta = \Theta_+ \uplus \Theta_-$,
the thesis holds when
$q \in \{ \mathrm{s}, \mathrm{b}, \mathrm{k} \}$.
\qed
\end{proof}

The concrete semantics of a valid program $g$ with respect to
the initial memory structure
$\sigma_\mathrm{i} \defeq (\emptyset, \emptystring) \in \Mem$
is a set of concrete semantics trees.
This set will always include a tree $\theta_0 \in \Theta$
(which, by Proposition~\ref{prop:concrete-tree-exists}, must exist)
such that
\[
  \theta_0(\emptystring)
    =
      \Bigl(
        \emptyset
          \vdash_\emptyset
            \bigl\langle
              (g; \textup{$\kw{gvar} \ridx : \tinteger = 0$}),
              \sigma_\mathrm{i}
            \bigr\rangle
              \rightarrow
                \eta_0
      \Bigr).
\]
If $\eta_0 = \exceptstate_0$,
i.e., an RTS exception is thrown during the evaluation of $g$,
then the concrete semantics is $\{\theta_0\}$.
If, instead, $\eta_0 =  \langle \rho_0, \sigma_0 \rangle$,
then the concrete semantics is
\[
  \{\theta_0 \} \union
  \bigl\{\,
    \theta \in \Theta
  \bigm|
    \theta(\emptystring)
      = (\rho_0 \vdash_\beta N \rightarrow \eta)
      \; \text{ or } \;
    \theta(\emptystring)
      = (\rho_0 \vdash_\beta N \diverges)
  \,\bigr\},
\]
where
\(
  N = \bigl\langle
        \bigl(
          \ridx := \main(\emptysequence)
        \bigr),
        \sigma_0
      \bigr\rangle
    \in \NTs^\beta
\)
and
\(
  \emptyset
    \vdash_\emptyset
      (g; \textup{$\kw{gvar} \ridx : \tinteger = 0$}) : \beta
\).

The concrete semantics for CPM we have just presented, extended
as indicated in Section~\ref{sec:extensions}, allows us to reason on a number
of interesting program safety properties (such as the absence of
division-by-zero and other run-time errors) as well as termination
and computational complexity.  In the next section, we will see how
the usually non-computable concrete semantics can be given an abstract
counterpart that is amenable to effective computation.

\section{Abstract Dynamic Semantics}
\label{sec:abstract-dynamic-semantics}

For the specification of the abstract semantics, we mainly
follow the approach outlined in the works
by Schmidt \citeNN{Schmidt95,Schmidt97,Schmidt98}.
The specification of the abstract semantics requires that appropriate
abstract domains are chosen to provide correct approximations for the
values that are involved in the concrete computation
\cite{CousotC77,CousotC79,CousotC92fr,CousotC92}.
For the sake of generality and extensibility, we will not target any
specific abstraction, but rather consider arbitrary abstract domains
that satisfy a limited set of properties that are sufficient to provide
the correctness of the overall analysis without compromising
its potential precision.

\subsection{Abstract Semantic Domains}
\label{sec:abstract-semantic-domains}

We adopt the framework proposed in~\cite[Section~7]{CousotC92fr},
where the correspondence between the concrete and the abstract domains
is induced from a concrete approximation relation and a concretization
function.
For the sole purpose of simplifying the presentation, we will consider
a particular instance of the framework by assuming a few additional
but non-essential domain properties.
The resulting construction is adequate for our purposes and still
allows for algebraically weak abstract domains, such as the domain of
convex polyhedra~\cite{CousotH78}.

A concrete domain is modeled as a complete lattice
$(C, \sqsubseteq, \bot, \top, \sqcap, \sqcup)$
of semantic properties;
as usual, the concrete approximation relation $c_1 \sqsubseteq c_2$ holds
if $c_1$ is a stronger property than $c_2$ (i.e., $c_2$ approximates $c_1$).
An abstract domain is modeled as a bounded join-semilattice
$(D^\sharp, \sqsubseteq^\sharp, \bot^\sharp, \sqcup^\sharp)$,
so that it has a bottom element $\bot^\sharp$
and the least upper bound
$d^\sharp_1 \sqcup^\sharp d^\sharp_2$ exists
for all $d^\sharp_1, d^\sharp_2 \in D^\sharp$.
When the abstract domain is also provided with a top element
$\top^\sharp \in D^\sharp$, we will write
$(D^\sharp, \sqsubseteq^\sharp, \bot^\sharp, \top^\sharp, \sqcup^\sharp)$.
The abstract domain $D^\sharp$ is related to $C$ by a monotonic
concretization function $\fund{\gamma}{D^\sharp}{C}$:
in words, $C$ is approximated by $D^\sharp$ through $\gamma$;
this approximation is said to be \emph{strict}
if $\gamma$ is a strict function.%
\footnote{Let $\fund{f}{D_1 \times \dots \times D_n}{D_0}$,
where $(D_i, \sqsubseteq_i, \bot_i, \sqcup_i)$
is a bounded join-semilattice, for each $i = 0$, \dots,~$n$.
Then, function $f$ is \emph{strict on the $i$-th argument}
if $d_i = \bot_i$ implies $f(d_1, \dots, d_n) = \bot_0$.}

In order to compute approximations for specific concrete objects,
we assume the existence of a partial abstraction function
$\pard{\alpha}{C}{D^\sharp}$ such that, for each $c \in C$,
if $\alpha(c)$ is defined then $c \sqsubseteq \gamma\bigl(\alpha(c)\bigr)$.
In particular, we assume that $\alpha(\bot) = \bot^\sharp$ is always defined;
if an abstract top element exists, then $\alpha(\top) = \top^\sharp$ is
also defined.
When needed or useful, we will require a few additional properties.

Most of the concrete domains used in the concrete semantics construction
are obtained as the powerset lattice
$\bigl( \wp(D), \sseq, \emptyset, D, \inters, \union \bigr)$
of some set of concrete objects $D$.
In such a situation, for each concrete object $d \in D$ and
abstract element $d^\sharp \in D^\sharp$ such that
the corresponding domains are related by the concretization function
$\fund{\gamma}{D^\sharp}{\wp(D)}$, we write
$d \propto d^\sharp$ and $d \not\propto d^\sharp$ to denote
the assertions $d \in \gamma(d^\sharp)$ and $d \notin \gamma(d^\sharp)$,
respectively.
For a lighter notation, we denote $\sqsubseteq^\sharp$,
$\bot^\sharp$, $\top^\sharp$ and $\sqcup^\sharp$ by $\sqsubseteq$, $\bot$,
$\top$ and $\sqcup$, respectively. We also overload the symbols
$\sqsubseteq$, $\bot$, $\top$, $\sqcup$, $\gamma$ and $\alpha$:
the context will always make clear which incarnation has to be
considered.

The approximations of composite concrete domains are
typically obtained by suitably combining the approximations already
available for their basic components.
For $i = 1$, $2$, let $D_i$ be a set of concrete objects and
consider the corresponding powerset lattice
$\bigl( \wp(D_i), \sseq, \emptyset, D_i, \inters, \union \bigr)$;
let also $D^\sharp_i$ be an abstract domain related to $\wp(D_i)$
by the concretization function
$\fund{\gamma_i}{D^\sharp_i}{\wp(D_i)}$.

\subsubsection{Approximation of Cartesian Products}

Values of the Cartesian product $D_1 \times D_2$
can be approximated by elements of the Cartesian product
$D^\sharp_1 \times D^\sharp_2$.
Namely, the component-wise ordered abstract domain
\(
  \bigl(D^\sharp_1 \times D^\sharp_2, \sqsubseteq, \bot, \sqcup \bigr)
\)
is related to the concrete powerset lattice
\(
  \bigl(
    \wp(D_1 \times D_2), \sseq, \emptyset, D_1 \times D_2, \inters, \union
  \bigr)
\)
by the concretization function
\(
  \fund{\gamma}%
       {(D^\sharp_1 \times D^\sharp_2)}%
       {\wp(D_1 \times D_2)}
\)
defined, for each
\(
  (d^\sharp_1, d^\sharp_2) \in D^\sharp_1 \times D^\sharp_2
\),
by
\begin{equation}
\label{eq:concretization-for-Cartesian-product}
  \gamma(d^\sharp_1, d^\sharp_2)
    \defeq
      \bigl\{\,
        (d_1, d_2) \in D_1 \times D_2
      \bigm|
        d_1 \in \gamma_1(d^\sharp_1), d_2 \in \gamma_2(d^\sharp_2)
      \,\bigr\}.
\end{equation}
Hence,
$(d_1, d_2) \propto (d^\sharp_1, d^\sharp_2)$ holds
if and only if
$d_1 \propto d^\sharp_1$ and $d_2 \propto d^\sharp_2$.

If the underlying approximations $D^\sharp_1$ and $D^\sharp_2$
are both strict, then a better approximation scheme can be obtained
by adopting the \emph{strict product} (also called \emph{smash product})
construction, which performs a simple form of reduction
by collapsing $(d^\sharp_1, d^\sharp_2)$ to the bottom element
whenever $d^\sharp_1 = \bot$ or $d^\sharp_2 = \bot$.
Namely,
\[
  D^\sharp_1 \stimes D^\sharp_2
    \defeq
      \bigl\{\,
        (d^\sharp_1, d^\sharp_2) \in D^\sharp_1 \times D^\sharp_2
      \bigm|
        \text{$d^\sharp_1 = \bot$ if and only if $d^\sharp_2 = \bot$}
      \,\bigr\}.
\]
The concretization function is defined exactly as
in~\eqref{eq:concretization-for-Cartesian-product}.
The constructor function
\(
  \fund{\spair{\cdot}{\cdot}}
       {(D^\sharp_1 \times D^\sharp_2)}
       {(D^\sharp_1 \stimes D^\sharp_2)}
\)
is defined by
\[
  \spair{d^\sharp_1}{d^\sharp_2}
    \defeq
      \begin{cases}
        (d^\sharp_1, d^\sharp_2),
          &\text{if $d^\sharp_1 \neq \bot$ and $d^\sharp_2 \neq \bot$}; \\
        \bot,
          &\text{otherwise.}
      \end{cases}
\]

\subsubsection{Approximation of Disjoint Unions}
\label{sec:approximation-disjoint-union}

In order to provide an abstract domain approximating
sets of concrete objects drawn from a disjoint union,
we use the following well-known construction several times.

Suppose that $D_1 \inters D_2 = \emptyset$.
Then, values of the disjoint union $D = D_1 \uplus D_2$
can be approximated by elements of the Cartesian product
$D^\sharp = D^\sharp_1 \times D^\sharp_2$.
In this case, the abstract domain $D^\sharp$
is related to the concrete powerset lattice
\(
  \bigl(
    \wp(D), \sseq, \emptyset, D, \inters, \union
  \bigr)
\)
by means of the concretization function
$\fund{\gamma}{(D^\sharp_1 \times D^\sharp_2)}{\wp(D_1 \uplus D_2)}$
defined, for each
\(
  (d^\sharp_1, d^\sharp_2) \in D^\sharp_1 \times D^\sharp_2
\),
by
\[
  \gamma(d^\sharp_1, d^\sharp_2)
    \defeq
      \gamma_1(d^\sharp_1) \uplus \gamma_2(d^\sharp_2).
\]
Therefore, the approximation provided by $D^\sharp$ is strict
if both $D^\sharp_1$ and $D^\sharp_2$ are so.
In order to simplify notation, if $d^\sharp_1 \in D^\sharp_1$
then we will sometimes write $d^\sharp_1$ to also denote
the abstract element $(d^\sharp_1, \bot) \in D^\sharp$;
similarly, $d^\sharp_2 \in D^\sharp_2$ also denotes
the abstract element $(\bot, d^\sharp_2) \in D^\sharp$.
As usual, for each $i = 1$, $2$ and $d_i \in D_i$, the notation
$d_i \propto (d^\sharp_1, d^\sharp_2)$
stands for the assertion
$d_i \in \gamma(d^\sharp_1, d^\sharp_2)$,
which is equivalent to $d_i \in \gamma_i(d^\sharp_i)$.
For the sake of clarity,
the abstract domain $D^\sharp$ as specified above
will be denoted by $D^\sharp_1 \uplus^\sharp D^\sharp_2$.
It is worth stressing that
\(
  D^\sharp_1 \uplus^\sharp D^\sharp_2
    \neq D^\sharp_1 \uplus D^\sharp_2
\).

\subsection{Approximation of Integers}
\label{sec:approximation-integers}

The concrete domain of integers
\(
  \bigl(\wp(\Integer), \sseq, \emptyset, \Integer, \inters, \union\bigr)
\)
is correctly approximated by an abstract domain
\(
  \bigl(\Integer^\sharp, \sqsubseteq, \bot, \top, \sqcup\bigr)
\),
where we assume that $\gamma$ is strict.
Elements of $\Integer^\sharp$ are denoted by
$m^\sharp$, $m^\sharp_0$, $m^\sharp_1$ and so forth.
We assume that the partial abstraction function
$\pard{\alpha}{\wp(\Integer)}{\Integer^\sharp}$ is defined
on all singletons $\{m\} \in \wp(\Integer)$.
We also assume that there are abstract binary operations
`$\absadd$', `$\abssub$', `$\absmul$', `$\absdiv$' and `$\absmod$'
on $\Integer^\sharp$ that are strict on each argument and
sound with respect to the corresponding operations on $\wp(\Integer)$
which, in turn, are the obvious pointwise extensions
of addition, subtraction, multiplication, division
and remainder over the integers.
More formally, we require
\(
  \gamma(m^\sharp_0 \absadd m^\sharp_1)
    \supseteq
      \bigl\{\,
        m_0 + m_1
      \bigm|
        m_0 \in \gamma(m^\sharp_0),
        m_1 \in \gamma(m^\sharp_1)
      \,\bigr\}
\)
for each $m^\sharp_0, m^\sharp_1 \in \Integer^\sharp$,
to ensure that `$\absadd$' is sound with respect to addition.
Likewise for `$\abssub$' and `$\absmul$' with respect
to subtraction and multiplication, respectively.
For the `$\absdiv$' operation we require soundness with respect to
integer division
i.e., that, for each $m^\sharp_0, m^\sharp_1 \in \Integer^\sharp$,
\(
  \gamma(m^\sharp_0 \absdiv m^\sharp_1)
    \supseteq
      \bigl\{\,
        m_0 \div m_1
      \bigm|
        m_0 \in \gamma(m^\sharp_0),
        m_1 \in \gamma(m^\sharp_1),
        m_1 \neq 0
      \,\bigr\}
\).
Likewise for `$\absmod$' with respect to the `$\mathord{\bmod}$' operation.
We also assume there is a unary abstract operation,
denoted by `$\abssub$', which is strict and sound with respect to
the unary minus concrete operation, that is,
\(
  \gamma(\abssub m^\sharp)
    \supseteq
      \bigl\{\,
        - m
      \bigm|
        m \in \gamma(m^\sharp)
      \,\bigr\}
\).

\subsection{Approximation of Booleans}
\label{sec:approximation-booleans}

We assume a complete lattice
\(
  \bigl(\Bool^\sharp, \sqsubseteq, \bot, \top, \sqcap, \sqcup\bigr)
\)
is given that is related to the concrete domain of Booleans
\(
  \bigl(\wp(\Bool), \sseq, \emptyset, \Bool, \inters, \union\bigr)
\)
by means of a Galois connection where $\gamma$ is strict.
Elements of $\Bool^\sharp$ are denoted by $t^\sharp$,
$t^\sharp_0$, $t^\sharp_1$ and so forth.
We assume that there are abstract operations
`$\absneg$', `$\absor$' and `$\absand$'
on $\Bool^\sharp$ that are strict on each argument and
sound with respect to the pointwise extensions
of Boolean negation, disjunction
and conjunction over $\wp(\Bool)$.
For instance, for the operation `$\absor$' to be sound with respect
to disjunction on $\wp(\Bool)$, it is required that,
\(
  \gamma(t^\sharp_0 \absor t^\sharp_1)
    \supseteq
      \bigl\{\,
        t_0 \lor t_1
      \bigm|
        t_0 \in \gamma(t^\sharp_0),
        t_1 \in \gamma(t^\sharp_1)
      \,\bigr\}
\)
for each $t^\sharp_0$ and $t^\sharp_1$ in $\Bool^\sharp$.
Likewise for `$\absand$'.
For operation `$\absneg$' to be sound with respect to negation on
$\wp(\Bool)$, we require that, for each $t^\sharp$ in $\Bool^\sharp$,
\(
  \gamma(\absneg t^\sharp)
    \supseteq
      \bigl\{\,
        \bneg t
      \bigm|
        t \in \gamma(t^\sharp)
      \,\bigr\}
\).

Furthermore, we assume that there are abstract operations
`$\abseq$', `$\absneq$', `$\abslt$', `$\absleq$', `$\absgeq$' and `$\absgt$'
on $\Integer^\sharp$ that are strict on each argument and
sound with respect to the pointwise extensions over $\wp(\Integer)$
of the corresponding relational operators
`$=$', `$\neq$', `$<$', `$\leq$', `$\geq$' and `$>$' over the integers,
considered as functions taking values in $\Bool$.
For instance, for the operation `$\abseq$' to be sound with respect
to equality on $\wp(\Integer)$, we require
that
\(
  \gamma(m^\sharp_0 \abseq m^\sharp_1)
    \supseteq
      \bigl\{\,
        m_0 = m_1
      \bigm|
        m_0 \in \gamma(m^\sharp_0),
        m_1 \in \gamma(m^\sharp_1)
      \,\bigr\}
\)
for each $m^\sharp_0, m^\sharp_1 \in \Integer^\sharp$.
Likewise for `$\absneq$', `$\abslt$', `$\absleq$', `$\absgeq$' and `$\absgt$'.

\subsection{Approximation of Storable Values}

The concrete domain of storable values
\(
  \bigl(\wp(\sVal), \sseq, \emptyset, \sVal, \inters, \union \bigr)
\),
including both integers and Booleans, is abstracted by the domain
$\sVal^\sharp \defeq \Integer^\sharp \uplus^\sharp \Bool^\sharp$.
The hypotheses on $\Integer^\sharp$ and $\Bool^\sharp$ imply that
the approximation is strict.

\subsection{Approximation of Exceptions}
\label{sec:approximation-exceptions}

For the approximation of RTS exceptions,
we assume that there is an abstract domain
\(
  \bigl(\RTSExcept^\sharp, \sqsubseteq, \bot, \top, \sqcup \bigr)
\),
which is related to the concrete powerset domain
\(
  \bigl(\wp(\RTSExcept), \sseq, \emptyset, \RTSExcept, \inters, \union\bigr)
\)
by a strict concretization function.
The partial abstraction function
$\pard{\alpha}{\wp(\RTSExcept)}{\RTSExcept^\sharp}$
is assumed to be defined on all singletons.
Elements of $\RTSExcept^\sharp$ are denoted by $\rtsexcept^\sharp$,
$\rtsexcept^\sharp_0$, $\rtsexcept^\sharp_1$ and so forth.

Generic exceptions, including both RTS exceptions and user-defined
exceptions, are approximated by elements of the domain
\(
  \Except^\sharp
    \defeq
      \RTSExcept^\sharp \uplus^\sharp \sVal^\sharp
\).
The hypotheses on its components imply that
the approximation is strict.
Elements of $\Except^\sharp$ are denoted by $\except^\sharp$,
$\except^\sharp_0$, $\except^\sharp_1$ and so forth.

\subsection{Approximation of Memory Structures, %
Value States and Exception States}

Here we differ from other published abstract semantics in that
we explicitly cater for \emph{relational} abstract domains as well as for
\emph{attribute-independent} ones \cite{CousotC79}.
While this complicates the presentation, it results in a truly
generic abstract semantics.  Moreover, the approach presented
here is ---all things considered--- quite simple and reflects into a modular,
clean design of the analyzer.

\begin{definition}
\summary{($\Mem^\sharp$, $\ValState^\sharp$, $\ExceptState^\sharp$.)}
\label{def:abstract-memory-structure}
\label{def:abstract-exception-state}
We assume there exists an abstract domain
\(
  \bigl(\Mem^\sharp, \sqsubseteq, \bot, \sqcup \bigr)
\)
that is related, by means of a strict concretization function,
to the concrete powerset domain
\(
  \bigl( \wp(\Mem), \sseq, \emptyset, \Mem, \inters, \union \bigr)
\).
Elements of $\Mem^\sharp$ are denoted by
$\sigma^\sharp$, $\sigma^\sharp_0$, $\sigma^\sharp_1$ and so forth.
We assume that, for each $\sigma \in \Mem$,
there exists $\sigma^\sharp \in \Mem^\sharp$
such that $\sigma \propto \sigma^\sharp$.

The abstract domain of value states is
\(
  \ValState^\sharp
    \defeq
      \sVal^\sharp \stimes \Mem^\sharp
\).
Elements of $\ValState^\sharp$ will be denoted by
$\valstate^\sharp$, $\valstate^\sharp_0$, $\valstate^\sharp_1$
and so forth.

The abstract domain of exception states is
\(
  \ExceptState^\sharp
    \defeq
      \Mem^\sharp \stimes \Except^\sharp
\).
Elements of $\ExceptState^\sharp$ will be denoted by
$\exceptstate^\sharp$, $\exceptstate^\sharp_0$, $\exceptstate^\sharp_1$
and so forth.
To improve readability,
$\none^\sharp$ will denote the bottom element
$\bot \in \ExceptState^\sharp$,
indicating that no exception is possible.

The abstract memory structure read and update operators
\begin{gather*}
  \fund{\cdot[\cdot, \cdot]}%
       {(\Mem^\sharp \times \Addr \times \sType)}%
       {(\ValState^\sharp \uplus^\sharp \ExceptState^\sharp)}, \\
  \fund{\cdot[\cdot :=^\sharp \cdot]}%
       {\bigl(
          \Mem^\sharp \times (\Addr \times \sType) \times \sVal^\sharp
        \bigr)}%
       {(\Mem^\sharp \uplus^\sharp \ExceptState^\sharp)}
\end{gather*}
are assumed to be such that,
for each $\sigma^\sharp \in \Mem^\sharp$,
$a \in \Addr$, $\sT \in \sType$ and $\sval^\sharp \in \sVal^\sharp$:
\begin{align*}
  \gamma\bigl( \sigma^\sharp[a,\sT]  \bigr)
    &\supseteq
      \bigl\{\,
        \sigma[a,\sT]
      \bigm|
        \sigma \in \gamma(\sigma^\sharp)
      \,\bigr\}, \\
  \gamma\bigl(
          \sigma^\sharp\bigl[(a,\sT) :=^\sharp \sval^\sharp\bigr]
        \bigr)
    &\supseteq
      \bigl\{\,
        \sigma\bigl[ (a,\sT) := \sval \bigr]
      \bigm|
        \sigma \in \gamma(\sigma^\sharp),
        \sval \in \gamma(\sval^\sharp)
      \,\bigr\}.
\end{align*}
The abstract data and stack memory allocation functions
\begin{gather*}
  \fund{\datnew^\sharp}
       {\ValState^\sharp}
       {\bigl(
          (\Mem^\sharp \times \Loc)
            \uplus^\sharp
          \ExceptState^\sharp
        \bigr)}, \\
  \fund{\stknew^\sharp}
       {\ValState^\sharp}
       {\bigl(
          (\Mem^\sharp \times \Ind)
            \uplus^\sharp
          \ExceptState^\sharp
        \bigr)}
\end{gather*}
are assumed to be such that,
for each $\valstate \in \ValState$ and
$\valstate^\sharp \in \ValState^\sharp$ such that
$\valstate \in \gamma(\valstate^\sharp)$,
and each $h \in \{ \mathrm{d}, \mathrm{s} \}$:
if $\mathop{\mathrm{new}_h}(\valstate) = (\sigma, a)$
(resp., $\mathop{\mathrm{new}_h}(\valstate) = \exceptstate$) and
\(
  \mathop{\mathrm{new}_h}^\sharp(\valstate^\sharp)
    = \bigl( (\sigma^\sharp, a'), \exceptstate^\sharp \bigr)
\),
then $\sigma \in \gamma(\sigma^\sharp)$ and $a = a'$
(resp., $\exceptstate \in \gamma(\exceptstate^\sharp)$).

The abstract memory structure data cleanup function
\[
  \fund{\datcleanup^\sharp}{\ExceptState^\sharp}{\ExceptState^\sharp}
\]
is such that, for each
$\exceptstate^\sharp \in \ExceptState^\sharp$,
we have
\[
  \gamma\bigl(\datcleanup^\sharp(\exceptstate^\sharp)\bigr)
    \supseteq
      \bigl\{\,
        \datcleanup(\exceptstate)
      \bigm|
        \exceptstate \in \gamma(\exceptstate^\sharp)
      \,\bigr\}.
\]

The abstract functions
\[
  \{ \asmark, \asunmark, \aslink, \asunlink \}
    \sseq
      \Mem^\sharp \to \Mem^\sharp
\]
are defined to be such that,
for each $\sigma^\sharp \in \Mem^\sharp$:
\begin{align*}
  \gamma\bigl( \asmark(\sigma^\sharp)  \bigr)
    &\supseteq
      \bigl\{\,
        \smark(\sigma)
      \bigm|
        \sigma \in \gamma(\sigma^\sharp)
      \,\bigr\}, \\
  \gamma\bigl( \asunmark(\sigma^\sharp) \bigr)
    &\supseteq
      \bigl\{\,
        \sunmark(\sigma)
      \bigm|
        \text{$\sigma \in \gamma(\sigma^\sharp)$
              and $\sunmark(\sigma)$ is defined}
      \,\bigr\}, \\
  \gamma\bigl( \aslink(\sigma^\sharp)  \bigr)
    &\supseteq
      \bigl\{\,
        \slink(\sigma)
      \bigm|
        \text{$\sigma \in \gamma(\sigma^\sharp)$
              and $\slink(\sigma)$ is defined}
      \,\bigr\}, \\
  \gamma\bigl( \asunlink(\sigma^\sharp) \bigr)
    &\supseteq
      \bigl\{\,
        \sunlink(\sigma)
      \bigm|
        \text{$\sigma \in \gamma(\sigma^\sharp)$
              and $\sunlink(\sigma)$ is defined}
      \,\bigr\}.
\end{align*}
It is assumed that all the abstract operators mentioned above
are strict on each of their arguments taken from an abstract domain.
\end{definition}
As done in the concrete,
the abstract stack unmark and the abstract frame unlink functions
are lifted to also work on abstract exception states.
Namely, for each
\(
  \exceptstate^\sharp
    = (\sigma^\sharp, \except^\sharp)
    \in \ExceptState^\sharp
\),
\begin{align*}
  \asunmark(\sigma^\sharp, \except^\sharp)
    &\defeq
      \bigl( \asunmark(\sigma^\sharp), \except^\sharp \bigr), \\
  \asunlink(\sigma^\sharp, \except^\sharp)
    &\defeq
      \bigl( \asunlink(\sigma^\sharp), \except^\sharp \bigr).
\end{align*}

Besides the abstract operators specified above,
which closely mimic the concrete operators related to
concrete memory structures and exception states,
other abstract operators will be used in the abstract semantics
construction so as to enhance its precision.

When dealing with Boolean guards during the abstract evaluation
of conditional and iteration statements,
it might be the case that no definite information is available.
In such a situation, the abstract execution can be made more precise if
the abstract memory structure is \emph{filtered} according to the condition
holding in the considered computation branch.

\begin{definition} \summary{(Memory structure filter.)}
\label{def:memstruct-filter}
An \emph{abstract memory structure filter} is any computable function
\(
  \fund{\phi}%
       {(\Env \times \Mem^\sharp \times \Exp)}%
       {\Mem^\sharp}
\)
such that, for each $e \in \Exp$, each $\beta : I$ with $\FI(e) \sseq I$
and $\beta \vdash_I e : \tboolean$, for each $\rho \in \Env$
with $\rho : \beta$ and each $\sigma^\sharp \in \Mem^\sharp$,
if $\phi(\rho, \sigma^\sharp, e) = \sigma^\sharp_\ttv$, then
\[
  \gamma(\sigma^\sharp_\ttv)
    \supseteq
      \bigl\{\,
        \sigma_\ttv \in \Mem
      \bigm|
        \sigma \in \gamma(\sigma^\sharp),
        \rho
          \vdash_\beta
            \langle e, \sigma \rangle
              \rightarrow
                \langle \ttv, \sigma_\ttv \rangle
      \,\bigr\}.
\]
\end{definition}

Similarly, abstract exception states can be filtered according to
whether or not they can be caught by the guard of a catch clause.

\begin{definition} \summary{(Exception state filters and selectors.)}
\label{def:exceptstate-filter}
The \emph{abstract exception state filters} are computable functions
\[
  \fund{\phi^+, \phi^-}%
    {(\exceptDecl \times \ExceptState^\sharp)}%
    {\ExceptState^\sharp}
\]
such that,
for each $p \in \exceptDecl$ and
each $\exceptstate^\sharp \in \ExceptState^\sharp$,
\begin{align*}
  \gamma\bigl( \phi^+(p, \exceptstate^\sharp) \bigr)
    &\supseteq
      \begin{cases}
        \gamma(\exceptstate^\sharp),
            &\text{if $p = \kw{\textup{any}}$;} \\
        \bigl\{\,
          (\sigma, \except) \in \gamma(\exceptstate^\sharp)
        \bigm|
          \except = p
        \,\bigr\},
            &\text{if $p \in \RTSExcept$;} \\
        \bigl\{\,
          (\sigma, \except) \in \gamma(\exceptstate^\sharp)
        \bigm|
          \except \in \dom\bigl(\type(p)\bigr)
        \,\bigr\},
            &\text{otherwise;}
      \end{cases} \\
  \gamma\bigl( \phi^-(p, \exceptstate^\sharp) \bigr)
    &\supseteq
      \begin{cases}
        \emptyset,
            &\text{if $p = \kw{\textup{any}}$;} \\
        \bigl\{\,
          (\sigma, \except) \in \gamma(\exceptstate^\sharp)
        \bigm|
          \except \neq p
        \,\bigr\},
            &\text{if $p \in \RTSExcept$;} \\
        \bigl\{\,
          (\sigma, \except) \in \gamma(\exceptstate^\sharp)
        \bigm|
          \except \notin \dom\bigl(\type(p)\bigr)
        \,\bigr\},
            &\text{otherwise.}
      \end{cases}
\end{align*}

The \emph{abstract memory structure and abstract exception selectors}
\begin{align*}
  \fund{\mem}{&\ExceptState^\sharp}{\Mem^\sharp}, \\
  \fund{\sel}{&(\cType \times \ExceptState^\sharp)}%
             {(\RTSExcept^\sharp \uplus \Integer^\sharp \uplus \Bool^\sharp)}
\end{align*}
are defined, for each
\(
  \exceptstate^\sharp
    = \bigl(
        \sigma^\sharp,
        \bigl(\rtsexcept^\sharp, (m^\sharp, t^\sharp)\bigr)
      \bigr)
    \in \ExceptState^\sharp
\)
and $\cT \in \cType$, by
\begin{align*}
  \mem(\exceptstate^\sharp)
    &\defeq
      \sigma^\sharp; \\
  \sel(\cT, \exceptstate^\sharp)
    &\defeq
      \begin{cases}
        \rtsexcept^\sharp,
            &\text{if $\cT = \trtsexcept$;} \\
        m^\sharp,
          &\text{if $\cT = \tinteger$;} \\
        t^\sharp,
          &\text{if $\cT = \tboolean$.}
      \end{cases}
\end{align*}
To simplify notation, we will write $\cT(\exceptstate^\sharp)$
to denote $\sel(\cT, \exceptstate^\sharp)$.
\end{definition}

The generic specification provided above for abstract memory structures
and the corresponding abstract operators plays a central role for the
modularity of the overall construction.
By exploiting this ``black box'' approach, we achieve orthogonality
not only from the specific abstract domains used to approximate
(sets of tuples of) storable values, but also from the critical
design decisions that have to be taken when approximating the
concrete stack, which may be unbounded in size due to recursive functions.
Hence, while still staying in the boundaries of the current framework,
we can flexibly explore, combine, and finely tune the sophisticated
proposals that have been put forward in the literature,
such as the work in~\cite{JeannetS03TR,JeannetS04},
which encompasses both the functional and the call string
approaches to interprocedural analysis \cite{CousotC77b,SharirP81}.

\subsection{Abstract Configurations}

Terminal and non-terminal configurations of the abstract transition
system are now defined.

\begin{definition} \summary{(Non-terminal abstract configurations.)}
\label{def:nonterminal-abstract-configuration}
The sets of \emph{non-terminal abstract configurations} for expressions,
local and global declarations, statements, function bodies and catch clauses
are given, for each $\beta \in \TEnv_I$ and respectively, by
\begin{align*}
  \NTe^{\beta\sharp}
    &\defeq
      \bigl\{\,
        \langle e, \sigma^\sharp \rangle
          \in \Exp \times \Mem^\sharp
      \bigm|
        \exists \sT \in \sType \st \beta \vdash_I e : \sT
      \,\bigr\}, \\
  \NTd^{\beta\sharp}
    &\defeq
      \bigl\{\,
        \langle d, \sigma^\sharp \rangle
          \in \Decl \times \Mem^\sharp
      \bigm|
         \exists \delta \in \TEnv \st \beta \vdash_I d : \delta
      \,\bigr\}, \\
  \NTg^{\beta\sharp}
    &\defeq
      \bigl\{\,
        \langle g, \sigma^\sharp \rangle
          \in \Glob \times \Mem^\sharp
      \bigm|
         \exists \delta \in \TEnv \st \beta \vdash_I g : \delta
      \,\bigr\}, \\
  \NTs^{\beta\sharp}
    &\defeq
      \bigl\{\,
        \langle s, \sigma^\sharp \rangle
          \in \Stmt \times \Mem^\sharp
      \bigm|
        \beta \vdash_I s
      \,\bigr\}, \\
  \NTb^{\beta\sharp}
    &\defeq
      \bigl\{\,
        \langle \funbody, \sigma^\sharp \rangle
          \in \funBody \times \Mem^\sharp
      \bigm|
        \exists \sT \in \sType \st \beta \vdash_I \funbody : \sT
      \,\bigr\}, \\
  \NTk^{\beta\sharp}
    &\defeq
      \bigl\{\,
        \langle k, \exceptstate^\sharp \rangle
          \in \Catch \times \ExceptState^\sharp
      \bigm|
        \beta \vdash_I k
      \,\bigr\}.
\end{align*}
We write $N^\sharp$ to denote a non-terminal abstract configuration.

The approximation relation
between concrete and abstract non-terminal configurations
is defined as follows.
For each
$q \in \{ \mathrm{e}, \mathrm{d}, \mathrm{g}, \mathrm{s}, \mathrm{b} \}$,
\(
  N = \langle q_1, \sigma \rangle \in \NTq^{\beta}
\)
and
\(
  N^\sharp = \langle q_2, \sigma^\sharp \rangle \in \NTq^{\beta\sharp}
\),
\begin{align}
\label{def:propto-nonterminal:expr-decl-glob-statement}
  N \propto N^\sharp
    &\iff
      (q_1 = q_2 \land \sigma \propto \sigma^\sharp).
\intertext{%
For each
\(
  N = \langle k_1, \exceptstate \rangle
    \in \NTk^{\beta}
\)
and
\(
  N^\sharp
    = \langle k_2, \exceptstate^\sharp \rangle
    \in \NTk^{\beta\sharp}
\),
}
\label{def:propto-nonterminal:catch}
  N \propto N^\sharp
    &\iff
      (k_1 = k_2 \land \exceptstate \propto \exceptstate^\sharp).
\end{align}
\end{definition}

\begin{definition} \summary{(Terminal abstract configurations.)}
\label{def:terminal-abstract-configuration}
The sets of \emph{terminal abstract configurations} for expressions,
local and global declarations, statements, function bodies and catch
clauses are given, respectively, by
\begin{align*}
  \Te^\sharp
    &\defeq
      \ValState^\sharp \uplus^\sharp \ExceptState^\sharp, \\
  \Td^\sharp
    &\defeq
  \Tg^\sharp
    \defeq
      (\Env \times \Mem^\sharp)
        \uplus^\sharp \ExceptState^\sharp, \\
  \Ts^\sharp
    &\defeq
  \Tb^\sharp
    \defeq
      \Mem^\sharp
        \uplus^\sharp \ExceptState^\sharp, \\
  \Tk^\sharp
    &\defeq
      \Ts^\sharp
        \uplus^\sharp
          \ExceptState^\sharp.
\end{align*}
We write $\eta^\sharp$ to denote a terminal abstract configuration.

The approximation relation $\eta \propto \eta^\sharp$
between concrete and abstract terminal configurations
is defined as follows.
For expressions,
\begin{align}
\label{def:propto-terminal:expr}
  \eta
    \propto
      \langle
        \valstate^\sharp,
        \exceptstate^\sharp
      \rangle
  &\iff
    \begin{cases}
      \valstate \propto \valstate^\sharp,
        &\text{if $\eta = \valstate$;} \\
      \exceptstate \propto \exceptstate^\sharp,
        &\text{if $\eta = \exceptstate$.}
    \end{cases} \\
\intertext{%
For local and global declarations,
}
\label{def:propto-terminal:decl}
  \eta
    \propto
      \bigl\langle
        (\rho_2, \sigma^\sharp),
        \exceptstate^\sharp
      \bigr\rangle
  &\iff
    \begin{cases}
      (\rho_1 = \rho_2
         \land
       \sigma \propto \sigma^\sharp),
          &\text{if $\eta = \langle \rho_1, \sigma \rangle$}; \\
      \exceptstate \propto \exceptstate^\sharp,
          &\text{if $\eta = \exceptstate$}.
    \end{cases} \\
\intertext{%%
For statements and function bodies,
}
\label{def:propto-terminal:statement}
  \eta
    \propto
      \langle \sigma^\sharp, \exceptstate^\sharp \rangle
  &\iff
    \begin{cases}
      \sigma \propto \sigma^\sharp,
          &\text{if $\eta = \sigma$}; \\
      \exceptstate \propto \exceptstate^\sharp,
          &\text{if $\eta = \exceptstate$}.
    \end{cases} \\
\intertext{%%
For catch sequences,
}
\label{def:propto-terminal:catch}
  \eta
    \propto
      \langle \eta^\sharp_\mathrm{s}, \exceptstate^\sharp \rangle
  &\iff
    \begin{cases}
      \eta_\mathrm{s} \propto \eta^\sharp_\mathrm{s},
          &\text{if $\eta = \langle \caught, \eta_\mathrm{s} \rangle$}; \\
      \exceptstate \propto \exceptstate^\sharp,
          &\text{if $\eta = \langle \uncaught, \exceptstate \rangle$}.
    \end{cases}
\end{align}
\end{definition}

The approximation relation for sequents is trivially obtained from the
approximation relations defined above for configurations.

\begin{definition}\summary{(`$\propto$' on sequents.)}
\label{def:propto-sequents}
The approximation relation
between concrete (positive and negative) sequents and
abstract sequents is defined,
for each $\beta \in \TEnv_I$,
for each $\rho_0, \rho_1 \in \Env_J$ such that
$\rho_0 : \beta \restrict{J}$ and
$\rho_1 : \beta \restrict{J}$,
for each
\(
   q \in
     \{
        \mathrm{e}, \mathrm{d}, \mathrm{g},
        \mathrm{s}, \mathrm{b}, \mathrm{k}
     \}
\),
$N \in \NTq^\beta$,
$\eta \in \Tq$,
$N^\sharp \in \NTq^{\beta\sharp}$ and
$\eta^\sharp \in \Tq^\sharp$, by
\begin{align}
\label{def:propto-sequents:term}
  (\rho_0 \vdash_\beta N \rightarrow \eta)
    \propto
      (\rho_1 \vdash_\beta N^\sharp \rightarrow \eta^\sharp)
  &\iff
    (\rho_0 = \rho_1
       \land N \propto N^\sharp
         \land \eta \propto \eta^\sharp); \\
\label{def:propto-sequents:diverge}
  (\rho_0 \vdash_\beta N \diverges)
    \propto
      (\rho_1 \vdash_\beta N^\sharp \rightarrow \eta^\sharp)
  &\iff
    (\rho_0 = \rho_1
       \land N \propto N^\sharp).
\end{align}
\end{definition}

\subsection{Supported Expressions, Declarations and Statements}

Each abstract domain has to provide a relation saying
which (abstract configuration for) expressions,
declarations and statements it directly supports,
as well as an abstract evaluation function providing safe approximations
of any supported expressions, declarations and statements.

\begin{definition} \summary{($\asupported$, $\aeval$.)}
\label{def:supports-eval}
For each
$q \in \{ \mathrm{e}, \mathrm{d}, \mathrm{g}, \mathrm{s} \}$,
we assume there exists a computable relation
and a partial and computable operation,
\[
  \reld{\asupported}%
       {\Env}%
       {\NTq^{\beta\sharp}}
  \qquad\text{and}\qquad
  \pard{\aeval}%
       {(\Env \times \NTq^{\beta\sharp})}%
       {\Tq^\sharp},
\]
such that whenever
$\rho : \beta$ and
$\asupported(\rho, N^\sharp)$ holds,
$\aeval(\rho, N^\sharp)$ is defined and has value
$\eta^\sharp \in \Tq^\sharp$
and, for each
$N \in \NTq^\beta$ and each $\eta \in \Tq$
such that $N \propto N^\sharp$ and
$\rho \vdash_\beta N \rightarrow \eta$,
we have $\eta \propto \eta^\sharp$.
\end{definition}

An appropriate use of `$\asupported$' and  `$\aeval$' allows
the design of the domain of abstract memory structures to be decoupled from
the design of the analyzer.  In particular, it enables
the use of relational as well as non-relational domains.
For example, using the domain of convex polyhedra the proper way,
one can easily implement a safe evaluation function for
(the non-terminal abstract configuration of) any affine expression $e$.
As a consequence, one can specify the support relation so that
$\asupported\bigl(\rho, \langle e, \sigma^\sharp\rangle \bigr)$ holds.
Similarly, one can specify
$\asupported\bigl(\rho, \langle \id := e, \sigma^\sharp \rangle \bigr)$ holds
for any affine assignment, i.e., an assignment where $e$
is an affine expression.
Other implementation choices are possible.
For instance, besides supporting affine expressions, the implementer
could specify that
$\asupported\bigl(\rho, \langle \id_1*\id_2, \sigma^\sharp \rangle \bigr)$ holds
provided $\rho : I$, $\id_1, \id_2 \in I$ and, for at least one
$i \in \{1, 2\}$,
$\gamma\bigl(\sigma^\sharp\bigl[\rho(\id_i)\bigr]\bigr) = \{ m \}$,
for some integer value $m$.
Similarly, the design can impose that
$\asupported\bigl(\rho, \langle \id*\id, \sigma^\sharp \rangle \bigr)$
always holds.

\subsection{Abstract Evaluation Relations}
\label{sec:abstract-evaluation-relations}

The abstract evaluation relations that provide the first part
of the specification of the abstract interpreter for CPM
are now defined.
These relations are of the form
\[
  \rho \vdash_\beta N^\sharp \rightarrow \eta^\sharp,
\]
where $\beta \in \TEnv$, $\rho : \beta$ and,
for some
\(
   q \in
     \{
        \mathrm{e}, \mathrm{d}, \mathrm{g},
        \mathrm{s}, \mathrm{b}, \mathrm{k}
     \}
\),
$N^\sharp \in  \NTq^{\beta\sharp}$
and $\eta^\sharp \in \Tq^\sharp$.
The definition is again by structural induction from a set of rule schemata.
In order to allow for the arbitrary weakening of the abstract descriptions
in the conclusion, without having to introduce precondition strengthening
and postcondition weakening rules, and to save typing at the same time,
we will use the notation
\begin{align*}
&\prooftree
  P_0 \cdots P_{\ell-1}
\justifies
  \rho \vdash_\beta N^\sharp \rightsquigarrow \eta^\sharp_0
\using\quad\text{(side condition)}
\endprooftree \\
\intertext{%
to denote
}
&\prooftree
  P_0 \cdots P_{\ell-1}
\justifies
  \rho \vdash_\beta N^\sharp \rightarrow \eta^\sharp
\using\quad\text{(side condition) and
$\eta^\sharp_0 \sqsubseteq \eta^\sharp$}
\endprooftree
\end{align*}
where `$\sqsubseteq$' is the natural ordering relation on the
appropriate abstract lattice (i.e., one of the $\Tq^\sharp$,
for
\(
   q \in
     \{
        \mathrm{e}, \mathrm{d}, \mathrm{g},
        \mathrm{s}, \mathrm{b}, \mathrm{k}
     \}
\).

Recalling the shorthand notation introduced
in Section~\ref{sec:approximation-disjoint-union},
when an abstract storable value $\sval^\sharp$ is expected and we write
an abstract integer $m^\sharp$ or an abstract Boolean $t^\sharp$,
then we are actually meaning the abstract storable value
$(m^\sharp, \bot)$ or $(\bot, t^\sharp)$, respectively;
similarly, when an abstract exception $\except^\sharp$ is expected
and we write an abstract RTS exception $\rtsexcept^\sharp$
or an abstract storable value $\sval^\sharp$,
then we are actually meaning the abstract exceptions
$(\rtsexcept^\sharp, \bot)$ or $(\bot, \sval^\sharp)$, respectively.

\subsubsection{Unsupported Expressions}

The following rules for the abstract evaluation of expressions
apply only if
$\asupported\bigl(\rho, \langle e, \sigma^\sharp \rangle \bigr)$
does not hold,
where $e$ is the expression being evaluated.
This side condition will be left implicit in order not to clutter
the presentation.

\begin{description}
\item[Constant]
\begin{gather}
\label{rule:abstr_constant}
\prooftree
  \nohyp
\justifies
  \rho \vdash_\beta \langle \con, \sigma^\sharp \rangle
    \rightsquigarrow
      \bigl\langle
        \spair{\alpha(\{\con\})}{\sigma^\sharp},
        \none^\sharp
      \bigr\rangle
\endprooftree
\end{gather}
\item[Identifier]
\begin{gather}
\label{rule:abstr_identifier}
\prooftree
  \nohyp
\justifies
  \rho \vdash_\beta \langle \id, \sigma^\sharp \rangle
    \rightsquigarrow
      \sigma^\sharp\bigl[\rho(\id)\bigr]
\endprooftree
\end{gather}
\item[Unary minus]
\begin{gather}
\label{rule:abstr_unary_minus}
\prooftree
  \rho \vdash_\beta \langle e, \sigma^\sharp \rangle
    \rightarrow
      \bigl\langle
        (m^\sharp, \sigma^\sharp_0),
        \exceptstate^\sharp
      \bigr\rangle
\justifies
  \rho \vdash_\beta \langle -e, \sigma^\sharp \rangle
    \rightsquigarrow
      \bigl\langle
        (\absuminus m^\sharp, \sigma^\sharp_0),
        \exceptstate^\sharp
      \bigr\rangle
\endprooftree
\end{gather}
\item[Binary arithmetic operations]
Let
\(
  \mathord{\boxcircle}
    \in
      \{ \mathord{+}, \mathord{-}, \mathord{*}, \divop, \modop \}
\)
be a syntactic operator and
$\mathord{\circledcirc} \in \{ \absadd, \abssub, \absmul, \absdiv, \absmod\}$
denote the corresponding abstract operation.
Then the abstract rules for addition, subtraction, multiplication,
division and remainder are given by the following schemata:
\begin{gather}
\label{rule:abstr_arith_bop}
\prooftree
  \rho \vdash_\beta \langle e_0, \sigma^\sharp \rangle
    \rightarrow
      \bigl\langle
        (m^\sharp_0, \sigma^\sharp_0),
        \exceptstate^\sharp_0
      \bigr\rangle
\quad
  \rho \vdash_\beta \langle e_1, \sigma^\sharp_0 \rangle
    \rightarrow
      \bigl\langle
        (m^\sharp_1, \sigma^\sharp_1),
        \exceptstate^\sharp_1
      \bigr\rangle
\justifies
  \rho \vdash_\beta \langle e_0 \boxcircle e_1, \sigma^\sharp \rangle
    \rightsquigarrow
      \bigl\langle
        (m^\sharp_0 \circledcirc m^\sharp_1, \sigma^\sharp_1),
	\exceptstate^\sharp_0 \sqcup \exceptstate^\sharp_1
      \bigr\rangle
\endprooftree
\end{gather}
if $\mathord{\boxcircle} \notin \{\divop, \modop \}$
or $0 \not\propto m^\sharp_1$.
\begin{gather}
\label{rule:abstr_div_mod_exc}
\prooftree
  \rho \vdash_\beta \langle e_0, \sigma^\sharp \rangle
    \rightarrow
      \bigl\langle
        (m^\sharp_0, \sigma^\sharp_0),
        \exceptstate^\sharp_0
      \bigr\rangle
\quad
  \rho \vdash_\beta \langle e_1, \sigma^\sharp_0 \rangle
    \rightarrow
      \bigl\langle
        (m^\sharp_1, \sigma^\sharp_1),
        \exceptstate^\sharp_1
      \bigr\rangle
\justifies
  \rho \vdash_\beta \langle e_0 \boxcircle e_1, \sigma^\sharp \rangle
    \rightsquigarrow
      \bigl\langle
        (m^\sharp_0 \circledcirc m^\sharp_1, \sigma^\sharp_1),
        \exceptstate^\sharp_0
          \sqcup \exceptstate^\sharp_1
            \sqcup \exceptstate^\sharp_2
      \bigr\rangle
\endprooftree
\end{gather}
if $\mathord{\boxcircle} \in \{\divop, \modop \}$,
$0 \propto m^\sharp_1$ and
\(
  \exceptstate^\sharp_2
    = \spair{\sigma^\sharp_1}{\alpha(\{\divbyzero\})}
\).
\item[Arithmetic tests]
Let
\(
  \mathord{\boxast}
    \in
      \{
        \mathord{=}, \mathord{\neq}, \mathord{<},
        \mathord{\leq}, \mathord{\geq}, \mathord{>}
      \}
\)
be an abstract syntax operator and let
$\fund{\mathord{\bowtie}}{(\Integer^\sharp \times \Integer^\sharp)}{\Bool^\sharp}$
denote the corresponding abstract test operation
in $\{ \abseq, \absneq, \abslt, \absleq, \absgeq, \absgt \}$.
Then the rules for the abstract arithmetic tests
are given by
\begin{gather}
\label{rule:abstr_arith_test}
\prooftree
  \rho \vdash_\beta \langle e_0, \sigma^\sharp \rangle
    \rightarrow
      \bigl\langle
        (m^\sharp_0, \sigma^\sharp_0),
        \exceptstate^\sharp_0
      \bigr\rangle
\quad
  \rho \vdash_\beta \langle e_1, \sigma^\sharp_0 \rangle
    \rightarrow
      \bigl\langle
        (m^\sharp_1, \sigma^\sharp_1),
        \exceptstate^\sharp_1
      \bigr\rangle
\justifies
  \rho \vdash_\beta \langle e_0 \boxast e_1, \sigma^\sharp \rangle
    \rightsquigarrow
      \bigl\langle
        (m^\sharp_0 \bowtie m^\sharp_1, \sigma^\sharp_1),
        \exceptstate^\sharp_0 \sqcup \exceptstate^\sharp_1
      \bigr\rangle
\endprooftree
\end{gather}
\item[Negation]
\begin{gather}
\label{rule:abstr_negation}
\prooftree
  \rho \vdash_\beta \langle b, \sigma^\sharp \rangle
    \rightarrow
      \bigl\langle
        (t^\sharp, \sigma^\sharp_0),
        \exceptstate^\sharp
      \bigr\rangle
\justifies
  \rho \vdash_\beta \langle \notop\ b, \sigma^\sharp \rangle
    \rightsquigarrow
      \bigl\langle
        (\absneg t^\sharp, \sigma^\sharp_0),
        \exceptstate^\sharp
      \bigr\rangle
\endprooftree
\end{gather}
\item[Conjunction]
\begin{gather}
\label{rule:abstr_conjunction}
\prooftree
  \rho \vdash_\beta \langle b_0, \sigma^\sharp \rangle
    \rightarrow
      \langle
        \valstate^\sharp_0,
        \exceptstate^\sharp_0
      \rangle
\quad
  \rho \vdash_\beta \langle b_1, \sigma^\sharp_\ttv \rangle
    \rightarrow
      \langle
        \valstate^\sharp_1,
        \exceptstate^\sharp_1
      \rangle
\justifies
  \rho \vdash_\beta \langle b_0 \andop b_1, \sigma^\sharp \rangle
    \rightsquigarrow
      \bigl\langle
        \valstate^\sharp_\ffv \sqcup \valstate^\sharp_1,
        \exceptstate^\sharp_0 \sqcup \exceptstate^\sharp_1
      \bigr\rangle
\endprooftree,
\end{gather}
if
\(
  \sigma^\sharp_\ttv = \phi(\rho, \sigma^\sharp, b_0)
\),
\(
  \sigma^\sharp_\ffv = \phi(\rho, \sigma^\sharp, \notop b_0)
\)
and
$\valstate^\sharp_\ffv = \spair{\alpha(\{\ffv\})}{\sigma^\sharp_\ffv}$.
\item[Disjunction]
\begin{gather}
\label{rule:abstr_disjunction}
\prooftree
  \rho \vdash_\beta \langle b_0, \sigma^\sharp \rangle
    \rightarrow
      \langle
        \valstate^\sharp_0,
        \exceptstate^\sharp_0
      \rangle
\quad
  \rho \vdash_\beta \langle b_1, \sigma^\sharp_\ffv \rangle
    \rightarrow
      \langle
        \valstate^\sharp_1,
        \exceptstate^\sharp_1
      \rangle
\justifies
  \rho \vdash_\beta \langle b_0 \orop b_1, \sigma^\sharp \rangle
    \rightsquigarrow
      \langle
        \valstate^\sharp_\ttv \sqcup \valstate^\sharp_1,
        \exceptstate^\sharp_0 \sqcup \exceptstate^\sharp_1
      \rangle
\endprooftree
\end{gather}
if
\(
  \sigma^\sharp_\ttv = \phi(\rho, \sigma^\sharp, b_0)
\),
\(
  \sigma^\sharp_\ffv = \phi(\rho, \sigma^\sharp, \notop b_0)
\)
and
$\valstate^\sharp_\ttv = \spair{\alpha(\{\ttv\})}{\sigma^\sharp_\ttv}$.
\end{description}

\subsubsection{Unsupported Declarations}

The following rules only apply if the condition
$\asupported\bigl(\rho, \langle q, \sigma^\sharp \rangle \bigr)$
does not hold,
where $q \in \Decl \uplus \Glob$ is the declaration being evaluated.
Again, this side condition is left implicit.

\begin{description}
\item[Nil]
\begin{gather}
\label{rule:abstr_decl_nil}
\prooftree
   \nohyp
\justifies
 \rho \vdash_\beta \langle \kw{nil}, \sigma^\sharp \rangle
   \rightsquigarrow
     \bigl\langle
       (\emptyset, \sigma^\sharp),
       \none^\sharp
     \bigr\rangle
\endprooftree
\end{gather}
\item[Environment]
\begin{gather}
\label{rule:abstr_decl_environment}
\prooftree
  \nohyp
\justifies
  \rho \vdash_\beta \langle \rho_0, \sigma^\sharp \rangle
    \rightsquigarrow
      \bigl\langle
        (\rho_0, \sigma^\sharp),
        \none^\sharp
      \bigr\rangle
\endprooftree
\end{gather}
\item[Recursive environment]
\begin{gather}
\label{rule:abstr_decl_rec_environment}
\prooftree
  \nohyp
\justifies
  \rho
    \vdash_\beta
      \langle \kw{rec} \rho_0, \sigma^\sharp \rangle
        \rightsquigarrow
      \bigl\langle
        (\rho_1, \sigma^\sharp),
        \none^\sharp
      \bigr\rangle
\endprooftree
\end{gather}
\begin{align*}
&\text{if }
  \rho_1
    = \bigl\{\,
         \id \mapsto \rho_0(\id)
       \bigm|
         \rho_0(\id) = \lambda \fps \st \kw{extern} : \sT
       \,\bigr\} \\
& \qquad
    \union
       \sset{
         \id \mapsto \abs_1
       }{
         \forall i \in \{0,1\}
           \itc
             \abs_i = \lambda \fps
                        \st \kw{let} d_i \,\kw{in} s \kw{result} e, \\
         \rho_0(\id) = \abs_0,
         d_1 = \kw{rec} \bigl( \rho_0 \setminus \DI(\fps) \bigr); d_0
       }.
\end{align*}
\item[Global variable declaration]
\begin{gather}
\label{rule:abstr_global_decl}
\prooftree
  \rho \vdash_\beta \langle e, \sigma^\sharp \rangle
    \rightarrow
      \langle
        \valstate^\sharp,
        \exceptstate^\sharp_0
      \rangle
\justifies
  \rho \vdash_\beta \langle \kw{gvar} \id : \sT = e, \sigma^\sharp \rangle
    \rightsquigarrow
      \bigl\langle
        (\rho_1, \sigma^\sharp_1),
        \datcleanup^\sharp(\exceptstate^\sharp_0 \sqcup \exceptstate^\sharp_1)
      \bigr\rangle
\endprooftree
\end{gather}
if
\(
  \datnew^\sharp(\valstate^\sharp)
    = \bigl( (\sigma^\sharp_1, l), \exceptstate^\sharp_1 \bigr)
\)
and
$\rho_1 = \bigl\{\id \mapsto (l, \sT)\bigr\}$.
\item[Local variable declaration]
\begin{gather}
\label{rule:abstr_local_decl}
\prooftree
  \rho \vdash_\beta \langle e, \sigma^\sharp \rangle
    \rightarrow
      \langle
        \valstate^\sharp,
        \exceptstate^\sharp_0
      \rangle
\justifies
  \rho \vdash_\beta \langle \kw{lvar} \id : \sT = e, \sigma^\sharp \rangle
    \rightsquigarrow
      \bigl\langle
        (\rho_1, \sigma^\sharp_1),
        \asunmark(\exceptstate^\sharp_0 \sqcup \exceptstate^\sharp_1)
      \bigr\rangle
\endprooftree
\end{gather}
if
\(
  \stknew^\sharp(\valstate^\sharp)
    = \bigl( (\sigma^\sharp_1, i), \exceptstate^\sharp_1 \bigr)
\)
and
$\rho_1 = \bigl\{ \id \mapsto (i, \sT) \bigr\}$.
\item[Function declaration]
\begin{gather}
\label{rule:abstr_decl_function}
\prooftree
  \nohyp
\justifies
  \rho
    \vdash_\beta
      \bigl\langle
        \kw{function} \id(\fps) = \funbody_0,
        \sigma^\sharp
      \bigr\rangle
  \rightsquigarrow
    \bigl\langle
      (\rho_0, \sigma^\sharp), \none^\sharp
    \bigr\rangle
\endprooftree
\end{gather}
if $\rho_0 = \{ \id \mapsto \lambdaop \fps \st \funbody_1 \}$ and
either $\funbody_0 = \funbody_1 = \kw{extern} : \sT$
or, for each $i \in \{0, 1\}$,
$\funbody_i = \kw{let} d_i \,\kw{in} s \kw{result} e$,
$I = \FI(\funbody_0) \setminus \DI(\fps)$ and
$d_1 = \rho \restrict{I}; d_0$.
\item[Recursive declaration]
\begin{gather}
\label{rule:abstr_recursive_decl}
\prooftree
  (\rho \setminus J) \vdash_{\beta[\beta_1]} \langle g, \sigma^\sharp \rangle
    \rightarrow
      \bigl\langle
        (\rho_0, \sigma^\sharp_0),
        \none^\sharp
      \bigr\rangle
\quad
  \rho \vdash_\beta \langle \kw{rec} \rho_0, \sigma^\sharp_0 \rangle
    \rightarrow
      \eta^\sharp
\justifies
  \rho \vdash_\beta \langle \kw{rec} g, \sigma^\sharp \rangle
    \rightsquigarrow
      \eta^\sharp
\endprooftree
\end{gather}
if $J = \FI(g) \inters \DI(g)$,
$\beta \vdash_{\FI(g)} g : \beta_0$ and
$\beta_1 = \beta_0 \restrict{J}$.
\item[Global sequential composition]
\begin{gather}
\label{rule:abstr_global_seq}
\prooftree
  \rho \vdash_\beta \langle g_0, \sigma^\sharp \rangle
    \rightarrow
      \bigl\langle
        (\rho_0, \sigma^\sharp_0),
        \exceptstate^\sharp_0
      \bigr\rangle
\quad
  \rho[\rho_0] \vdash_{\beta[\beta_0]} \langle g_1, \sigma^\sharp_0 \rangle
    \rightarrow
      \bigl\langle
        (\rho_1, \sigma^\sharp_1),
        \exceptstate^\sharp_1
      \bigr\rangle
\justifies
  \rho \vdash_\beta \langle g_0;g_1, \sigma^\sharp \rangle
    \rightsquigarrow
      \bigl\langle
        (\rho_0[\rho_1], \sigma^\sharp_1),
        \exceptstate^\sharp_0 \sqcup \exceptstate^\sharp_1
      \bigr\rangle
\endprooftree
\end{gather}
if $\beta \vdash_I g_0 : \beta_0$ and $\FI(g_0) \sseq I$.
\item[Local sequential composition]
\begin{gather}
\label{rule:abstr_local_seq}
\prooftree
  \rho \vdash_\beta \langle d_0, \sigma^\sharp \rangle
    \rightarrow
      \bigl\langle
        (\rho_0, \sigma^\sharp_0),
        \exceptstate^\sharp_0
      \bigr\rangle
\quad
  \rho[\rho_0] \vdash_{\beta[\beta_0]} \langle d_1, \sigma^\sharp_0 \rangle
    \rightarrow
      \bigl\langle
        (\rho_1, \sigma^\sharp_1),
        \exceptstate^\sharp_1
      \bigr\rangle
\justifies
  \rho \vdash_\beta \langle d_0;d_1, \sigma^\sharp \rangle
    \rightsquigarrow
      \bigl\langle
        (\rho_0[\rho_1], \sigma^\sharp_1),
        \exceptstate^\sharp_0 \sqcup \exceptstate^\sharp_1
      \bigr\rangle
\endprooftree
\end{gather}
if $\beta \vdash_I d_0 : \beta_0$ and $\FI(d_0) \sseq I$.
\end{description}

\subsubsection{Unsupported Statements}

The following rules only apply if the implicit side condition
$\asupported\bigl(\rho, \langle s, \sigma^\sharp \rangle \bigr)$ does not hold,
where $s$ is the statement being evaluated.

\begin{description}
\item[Nop]
\begin{gather}
\label{rule:abstr_nop}
\prooftree
  \nohyp
\justifies
  \rho \vdash_\beta \langle \kw{nop}, \sigma^\sharp \rangle
    \rightsquigarrow
      \sigma^\sharp
\endprooftree
\end{gather}
\item[Assignment]
\begin{gather}
\label{rule:abstr_assignment}
\prooftree
  \rho \vdash_\beta \langle e, \sigma^\sharp \rangle
    \rightarrow
      \bigl\langle
        (\sval^\sharp, \sigma^\sharp_0),
        \exceptstate^\sharp_0
      \bigr\rangle
\justifies
  \rho \vdash_\beta \langle \id := e, \sigma^\sharp \rangle
    \rightsquigarrow
      \langle
        \sigma^\sharp_1,
        \exceptstate^\sharp_0 \sqcup \exceptstate^\sharp_1
      \rangle
\using\quad\text{if
\(
  \sigma^\sharp_0\bigl[\rho(\id) :=^\sharp \sval^\sharp\bigr]
    = (\sigma^\sharp_1, \exceptstate^\sharp_1)
\)
}
\endprooftree
\end{gather}
\item[Statement sequence]
\begin{gather}
\label{rule:abstr_sequence}
\prooftree
  \rho \vdash_\beta \langle s_0, \sigma^\sharp \rangle
    \rightarrow
      \langle \sigma^\sharp_0, \exceptstate^\sharp_0 \rangle
\quad
  \rho \vdash_\beta \langle s_1, \sigma^\sharp_0 \rangle
    \rightarrow
      \langle \sigma^\sharp_1, \exceptstate^\sharp_1 \rangle
\justifies
  \rho \vdash_\beta \langle s_0 ; s_1, \sigma^\sharp \rangle
    \rightsquigarrow
      \bigl\langle
        \sigma^\sharp_1,
        \exceptstate^\sharp_0 \sqcup \exceptstate^\sharp_1
      \bigr\rangle
\endprooftree
\end{gather}
\item[Block]
\begin{gather}
\label{rule:abstr_block}
\prooftree
  \rho \vdash_\beta \bigl\langle d, \asmark(\sigma^\sharp) \bigr\rangle
    \rightarrow
      \bigl\langle
        (\rho_0, \sigma^\sharp_0), \exceptstate^\sharp_0
      \bigr\rangle
\quad
  \rho[\rho_0] \vdash_{\beta[\beta_0]} \langle s, \sigma^\sharp_0 \rangle
    \rightarrow
       \langle \sigma^\sharp_1, \exceptstate^\sharp_1 \rangle
\justifies
  \rho \vdash_\beta \langle d ; s, \sigma^\sharp \rangle
    \rightsquigarrow
       \bigl\langle
         \asunmark(\sigma^\sharp_1),
         \exceptstate^\sharp_0
           \sqcup
             \asunmark(\exceptstate^\sharp_1)
       \bigr\rangle
\endprooftree
\end{gather}
if $\beta \vdash_{\FI(d)} d : \beta_0$.
\item[Conditional]
\begin{gather}
\label{rule:abstr_conditional}
\prooftree
\begin{aligned}
  \rho \vdash_\beta \langle e, \sigma^\sharp \rangle
    \rightarrow
      \langle
        \valstate^\sharp_0,
        \exceptstate^\sharp_0
      \rangle
&\quad
  \rho \vdash_\beta \langle s_0, \sigma^\sharp_\ttv \rangle
    \rightarrow
      \langle \sigma^\sharp_1, \exceptstate^\sharp_1 \rangle \\
&\quad
  \rho \vdash_\beta \langle s_1, \sigma^\sharp_\ffv \rangle
    \rightarrow
      \langle \sigma^\sharp_2, \exceptstate^\sharp_2 \rangle
\end{aligned}
\justifies
  \rho
    \vdash_\beta
      \langle \kw{if} e \kw{then} s_0 \kw{else} s_1, \sigma^\sharp \rangle
    \rightsquigarrow
      \langle
        \sigma^\sharp_1 \sqcup \sigma^\sharp_2,
        \exceptstate^\sharp_0
          \sqcup \exceptstate^\sharp_1
            \sqcup \exceptstate^\sharp_2
      \rangle
\endprooftree
\end{gather}
if
\(
  \sigma^\sharp_\ttv = \phi(\rho, \sigma^\sharp, e)
\)
and
\(
  \sigma^\sharp_\ffv = \phi(\rho, \sigma^\sharp, \notop e)
\).
\item[While]
\begin{gather}
\label{rule:abstr_while}
\prooftree
\begin{aligned}
  \rho \vdash_\beta \langle e, \sigma^\sharp \rangle
    \rightarrow
      \langle
        \valstate^\sharp_0,
        \exceptstate^\sharp_0
      \rangle
&\quad
  \rho \vdash_\beta \langle s, \sigma^\sharp_\ttv \rangle
    \rightarrow
      \langle \sigma^\sharp_1, \exceptstate^\sharp_1 \rangle \\
&\quad
  \rho
    \vdash_\beta
      \langle \kw{while} e \kw{do} s, \sigma^\sharp_1 \rangle
    \rightarrow
      \langle \sigma^\sharp_2, \exceptstate^\sharp_2 \rangle
\end{aligned}
\justifies
  \rho
    \vdash_\beta
      \langle \kw{while} e \kw{do} s, \sigma^\sharp \rangle
    \rightsquigarrow
      \langle
        \sigma^\sharp_\ffv \sqcup \sigma^\sharp_2,
        \exceptstate^\sharp_0
          \sqcup \exceptstate^\sharp_1
            \sqcup \exceptstate^\sharp_2
      \rangle
\endprooftree
\end{gather}
if
\(
  \sigma^\sharp_\ttv = \phi(\rho, \sigma^\sharp, e)
\)
and
\(
  \sigma^\sharp_\ffv = \phi(\rho, \sigma^\sharp, \notop e)
\).
\item[Throw]
\begin{gather}
\label{rule:abstr_throw_except}
\prooftree
  \nohyp
\justifies
  \rho \vdash_\beta \langle \kw{throw} \rtsexcept, \sigma^\sharp \rangle
    \rightsquigarrow
      \langle \bot, \exceptstate^\sharp \rangle
\using\quad\text{if
\(
  \exceptstate^\sharp
    = \spair{\sigma^\sharp}{\alpha(\{\rtsexcept\})}
\)}
\endprooftree \\[1ex]
\label{rule:abstr_throw_expr}
\prooftree
  \rho \vdash_\beta \langle e, \sigma^\sharp \rangle
    \rightarrow
      \bigl\langle
        (\sval^\sharp, \sigma^\sharp_0),
        \exceptstate^\sharp_0
      \bigr\rangle
\justifies
  \rho \vdash_\beta \langle \kw{throw} e, \sigma^\sharp \rangle
    \rightsquigarrow
      \langle
        \bot,
        \exceptstate^\sharp_0 \sqcup \exceptstate^\sharp_1
      \rangle
\using\quad\text{if
$\exceptstate^\sharp_1 = \spair{\sigma^\sharp_0}{\sval^\sharp}$}
\endprooftree
\end{gather}
\item[Try blocks]
\begin{gather}
\label{rule:abstr_try_catch}
\prooftree
  \rho \vdash_\beta \langle s, \sigma^\sharp \rangle
    \rightarrow
      \langle \sigma^\sharp_0, \exceptstate^\sharp_0 \rangle
\quad
  \rho
    \vdash_\beta
      \langle k, \exceptstate^\sharp_0 \rangle
    \rightarrow
      \bigl\langle
        (\sigma^\sharp_1, \exceptstate^\sharp_1), \exceptstate^\sharp_2
      \bigr\rangle
\justifies
  \rho
    \vdash_\beta
      \langle \kw{try} s \kw{catch} k, \sigma^\sharp \rangle
    \rightsquigarrow
      \langle
        \sigma^\sharp_0 \sqcup \sigma^\sharp_1,
        \exceptstate^\sharp_1 \sqcup \exceptstate^\sharp_2
      \rangle
\endprooftree
\end{gather}

\begin{gather}
\label{rule:abstr_try_finally}
\prooftree
\begin{aligned}
  \rho \vdash_\beta \langle s_0, \sigma^\sharp \rangle
    \rightarrow
      \bigl\langle
        \sigma^\sharp_0,
        (\sigma^\sharp_1, \except^\sharp_1)
      \bigr\rangle
\quad
& \rho \vdash_\beta \langle s_1, \sigma^\sharp_0 \rangle
    \rightarrow
      \langle \sigma^\sharp_2, \exceptstate^\sharp_2 \rangle \\
& \rho \vdash_\beta \langle s_1, \sigma^\sharp_1 \rangle
    \rightarrow
      \langle \sigma^\sharp_3, \exceptstate^\sharp_3 \rangle
\end{aligned}
\justifies
  \rho
    \vdash_\beta
      \langle
        \kw{try} s_0 \kw{finally} s_1,
        \sigma^\sharp
      \rangle
    \rightsquigarrow
      \bigl\langle
        \sigma^\sharp_2,
        \exceptstate^\sharp_2
          \sqcup \exceptstate^\sharp_3
          \sqcup (\spair{\sigma^\sharp_3}{\except^\sharp_1})
      \bigr\rangle
\endprooftree
\end{gather}
\item[Function call]
With reference to conditions \eqref{eq:function-call-condition-beta-rho-d}
and \eqref{eq:function-call-condition-rho0-rho1} of the concrete rules for
function calls, the corresponding abstract rule schema is
\begin{gather}
\label{rule:abstr_function_call_eval}
\prooftree
\begin{aligned}
& \rho
    \vdash_\beta
      \bigl\langle
        d, \asmark(\sigma^\sharp)
      \bigr\rangle
    \rightarrow
      \bigl\langle
        (\rho_0, \sigma^\sharp_0),
        \exceptstate^\sharp_0
      \bigr\rangle \\
& \rho[\rho_1]
    \vdash_{\beta[\beta_1]}
      \bigl\langle \funbody, \aslink(\sigma^\sharp_0) \bigr\rangle
        \rightarrow
          \langle
            \sigma^\sharp_1,
            \exceptstate^\sharp_1
          \rangle \\
& \rho[\rho_0]
    \vdash_{\beta[\beta_0]}
      \bigl\langle
        \id_0 := \ridx_0,
        \asunlink(\sigma^\sharp_1)
      \bigr\rangle
        \rightarrow
          \langle
            \sigma^\sharp_2,
            \exceptstate^\sharp_2
          \rangle
\end{aligned}
\justifies
  \rho
    \vdash_\beta
      \bigl\langle
        \id_0 := \id(e_1, \ldots, e_n),
        \sigma^\sharp
      \bigr\rangle
    \rightsquigarrow
      \bigl\langle
        \asunmark(\sigma^\sharp_2),
        \exceptstate^\sharp
      \bigr\rangle
\endprooftree
\end{gather}
if \eqref{eq:function-call-condition-beta-rho-d}
and \eqref{eq:function-call-condition-rho0-rho1} hold and
\(
  \exceptstate^\sharp
    = \exceptstate^\sharp_0
        \sqcup
      \asunmark\bigl(\asunlink(\exceptstate^\sharp_1)\bigr)
        \sqcup
      \asunmark(\exceptstate^\sharp_2)
\).
\end{description}

\subsubsection{Function Bodies}
\begin{gather}
\label{rule:abstr_function_body}
\prooftree
\begin{aligned}
& \rho
    \vdash_\beta
      \bigl\langle
        d, \asmark(\sigma^\sharp)
      \bigr\rangle
    \rightarrow
      \bigl\langle
        (\rho_0, \sigma^\sharp_0),
        \exceptstate^\sharp_0
      \bigr\rangle \\
& \rho[\rho_0]
    \vdash_{\beta[\beta_0]}
      \langle s, \sigma^\sharp_0 \rangle
    \rightarrow
      \langle \sigma^\sharp_1, \exceptstate^\sharp_1 \rangle \\
& \rho[\rho_0]
    \vdash_{\beta[\beta_0]}
      \langle \ridx_0 := e, \sigma^\sharp_1 \rangle
    \rightarrow
      \langle
        \sigma^\sharp_2,
        \exceptstate^\sharp_2
      \rangle
\end{aligned}
\justifies
  \rho
    \vdash_\beta
      \langle
        \kw{let} d \,\kw{in} s \kw{result} e,
        \sigma^\sharp
      \rangle
    \rightsquigarrow
      \bigl\langle
        \asunmark(\sigma^\sharp_2),
        \exceptstate^\sharp_3
      \bigr\rangle
\endprooftree
\end{gather}
if
$\beta \vdash_{\FI(d)} d : \beta_0$,
\(
  \exceptstate^\sharp_3
    = \exceptstate^\sharp_0
        \sqcup
      \asunmark(\exceptstate^\sharp_1
                        \sqcup \exceptstate^\sharp_2)
\).

\begin{gather}
\label{rule:abstr_function_body_extern}
\prooftree
  \nohyp
\justifies
  \rho
    \vdash_\beta
      \langle
        \kw{extern} : \sT,
        \sigma^\sharp
      \rangle
    \rightsquigarrow
      \bigl\langle
        \sigma^\sharp_0,
        (\sigma_0^\sharp, \top)
      \bigr\rangle
\endprooftree
\end{gather}
if
\(
  \forall \sigma, \sigma_0 \in \Mem
    \itc
      \bigl(
        \sigma = (\mu, w)
          \land
        \sigma \propto \sigma^\sharp
          \land
        \sigma_0 = (\mu_0, w)
      \bigr)
        \implies
      \sigma_0 \propto \sigma^\sharp_0
\).

\subsubsection{Catch Clauses}

\begin{description}
\item[Catch]
\begin{gather}
\label{rule:abstr_catch_maybe_caught}
\prooftree
  \rho \vdash_\beta \bigl\langle s, \mem(\exceptstate^\sharp_0) \bigr\rangle
    \rightarrow
      \eta^\sharp_1
\justifies
  \rho
    \vdash_\beta
      \bigl\langle
        (p) \, s,
        \exceptstate^\sharp
      \bigr\rangle
    \rightsquigarrow
      \langle
        \eta^\sharp_1,
        \exceptstate^\sharp_1
      \rangle
\endprooftree \\[1ex]
\intertext{%
if $p = \kw{any}$ or $p = \rtsexcept$ or $p = \cT$,
$\exceptstate^\sharp_0 = \phi^+(p, \exceptstate^\sharp)$ and
$\exceptstate^\sharp_1 = \phi^-(p, \exceptstate^\sharp)$.
}
\label{rule:abstr_catch_expr_maybe_caught}
\prooftree
  \rho\bigl[ \{\id \mapsto (i,\sT) \} \bigr]
    \vdash_{\beta[ \{\id \mapsto \location{\sT}\} ]}
      \langle s, \sigma^\sharp_2 \rangle
    \rightarrow
      \langle \sigma^\sharp_3, \exceptstate^\sharp_3 \rangle
\justifies
  \rho
    \vdash_\beta
      \bigl\langle
        (\id : \sT) \, s,
        \exceptstate^\sharp
      \bigr\rangle
    \rightsquigarrow
      \bigl\langle
        (\sigma^\sharp_4, \exceptstate^\sharp_4),
        \exceptstate^\sharp_1
      \bigr\rangle
\endprooftree
\end{gather}
if $\exceptstate^\sharp_0 = \phi^+(\sT, \exceptstate^\sharp)$,
$\exceptstate^\sharp_1 = \phi^-(\sT, \exceptstate^\sharp)$,
\(
  \stknew^\sharp\Bigl(
                  \sT(\exceptstate^\sharp_0),
                  \asmark\bigl( \mem(\exceptstate^\sharp_0) \bigr)
                \Bigr)
    = \bigl( (\sigma^\sharp_2, i), \exceptstate^\sharp_2 \bigr)
\),
$\sigma^\sharp_4 = \asunmark(\sigma^\sharp_3)$
and
\(
  \exceptstate^\sharp_4
    = \asunmark(\exceptstate^\sharp_2)
        \sqcup \asunmark(\exceptstate^\sharp_3)
\).
\item[Catch sequence]
\begin{gather}
\label{rule:abstr_catch_seq}
\prooftree
  \rho
    \vdash_\beta
      \langle k_0, \exceptstate^\sharp \rangle
    \rightarrow
      \bigl\langle
        (\sigma^\sharp_0, \exceptstate^\sharp_0), \exceptstate^\sharp_1
      \bigr\rangle
\quad
  \rho
    \vdash_\beta
      \langle k_1, \exceptstate^\sharp_1 \rangle
    \rightarrow
      \bigl\langle
        (\sigma^\sharp_1, \exceptstate^\sharp_2), \exceptstate^\sharp_3
      \bigr\rangle
\justifies
  \rho
    \vdash_\beta
      \langle k_0 ; k_1, \exceptstate^\sharp \rangle
    \rightsquigarrow
      \bigr\langle
        (\sigma^\sharp_0 \sqcup \sigma^\sharp_1,
         \exceptstate^\sharp_0 \sqcup \exceptstate^\sharp_2),
        \exceptstate^\sharp_3
      \bigl\rangle
\endprooftree
\end{gather}
\end{description}

\subsubsection{Supported Expressions, Declarations and Statements}
\label{sec:supported-rules}

Let $q \in \{ \mathrm{e}, \mathrm{d}, \mathrm{g}, \mathrm{s} \}$
and $N^\sharp \in \NTq^{\beta\sharp}$.
Then, whenever $\asupported(\rho, N^\sharp)$ holds,
alternate versions of the rules above apply.
For each of the rules above,
\begin{align*}
&\prooftree
  P_0 \quad \cdots \quad P_{\ell-1}
\justifies
  \rho \vdash_\beta N^\sharp
    \rightsquigarrow
      \eta^\sharp
\using\quad\text{if (side condition) and
                 not $\asupported(\rho, N^\sharp)$}
\endprooftree \\
\intertext{%
we also have the rule
}
&\prooftree
  P_0 \quad \cdots \quad P_{\ell-1}
\justifies
  \rho \vdash_\beta N^\sharp
    \rightsquigarrow
      \aeval(\rho, N^\sharp)
\using\quad\text{if (side condition) and
                 $\asupported(\rho, N^\sharp)$}
\endprooftree
\end{align*}
Notice that even if $\aeval(\rho, N^\sharp)$
does not depend on the rule antecedents $P_0$, \dots,~$P_{\ell-1}$,
these cannot be omitted, as this would neglect the sub-computations
spawned by the unsupported evaluation of $N^\sharp$.

\subsection{Abstract Semantics Trees}
\label{sec:abstract-semantics-trees}

We now define possibly infinite abstract semantics trees along
the lines of what we did in Section~\ref{sec:concrete-semantics-trees}.
Notice that the need to consider infinite abstract trees goes
beyond the need to observe infinite concrete computations.
For instance, there is no finite abstract tree corresponding to a
program containing a $\kw{while}$ command,
because~\eqref{rule:abstr_while} is the only abstract
rule for $\kw{while}$ and it recursively introduces a new $\kw{while}$
node into the tree.

\begin{definition} \summary{(Abstract semantics rules.)}
\label{def:abstract-semantics-rules}
The set $\calR^\sharp$ of \emph{abstract semantics rules}
is the infinite set obtained by instantiating the rule schemata
of \textup{Section~\ref{sec:abstract-evaluation-relations}}
in all possible ways (respecting the side conditions).
\end{definition}

Let $\calS^\sharp$ be the (infinite) set of sequents occurring in the premises
and conclusions of the rules in $\calR^\sharp$.
Matching Definition~\ref{def:concrete-semantics-universe},
the \emph{abstract semantics universe}, denoted by $\calU^\sharp$, is the
set of finitely branching trees of at most $\omega$-depth with labels
in $\calS^\sharp$.

\begin{definition} \summary{(Abstract semantics trees.)}
\label{def:abstract-semantics-trees}
Let $\fund{\calF^\sharp}{\wp(\calU^\sharp)}{\wp(\calU^\sharp)}$ be given,
for each $U^\sharp \in \wp(\calU^\sharp)$, by
\[
  \calF^\sharp(U^\sharp)
    \defeq
      \biggl\{\,
        \prooftree
          \theta^\sharp_0 \; \cdots \; \theta^\sharp_{\ell-1}
        \justifies
          s
        \endprooftree
      \biggm|
        \{ \theta^\sharp_0, \ldots, \theta^\sharp_{\ell-1} \}
            \subseteq U^\sharp,
        \;
        \prooftree
          \theta^\sharp_0(\emptystring)
            \; \cdots \;
              \theta^\sharp_{\ell-1}(\emptystring)
        \justifies
          s
        \endprooftree
          \in \calR^\sharp
      \,\biggr\}.
\]
The set of \emph{abstract semantics trees} is
$\Theta^\sharp \defeq \gfp_{\mathord{\subseteq}}(\calF^\sharp)$.
\end{definition}

We now show that, for every non-terminal abstract configuration,
there exists an abstract tree with that in the root.
\begin{proposition}
\label{prop:abstract-tree-exists}
For each $\beta \in \TEnv$,
$\rho \in \Env$ such that $\rho : \beta$ and
$N^\sharp \in \NTq^{\beta\sharp}$, where
\(
  q
    \in
      \{
        \mathrm{e}, \mathrm{d}, \mathrm{g},
        \mathrm{s}, \mathrm{b}, \mathrm{k}
      \}
\),
there exists $\theta^\sharp \in \Theta^\sharp$ such that,
\[
  \theta^\sharp(\emptystring)
  \in \bigl\{\,
        (\rho \vdash_\beta N^\sharp \rightarrow \eta^\sharp)
      \bigm|
        \eta^\sharp \in \Tq^\sharp
      \,\bigr\}.
\]
\end{proposition}
\begin{proof}
For the proof, let%
\footnote{%%
For the definition of a well-typed sequent,
see the proof of Proposition~\ref{prop:concrete-tree-exists}.%
}
\begin{equation*}
  S^\sharp_+(\rho, \beta, N^\sharp)
    \defeq
      \bigl\{\,
        s^\sharp
      \bigm|
        s^\sharp = (\rho \vdash_\beta N^\sharp \rightarrow \eta^\sharp),
        (s \propto s^\sharp \implies
        \text{$s$ is well-typed})
      \,\bigr\}.
\end{equation*}

We now assume that $N^\sharp \in \NTq^{\beta\sharp}$ is a fixed but
arbitrary non-terminal abstract configuration.
Suppose that $\asupported(\rho, N^\sharp)$ does not hold.
By inspecting the abstract evaluation rules given in
Section~\ref{sec:abstract-evaluation-relations}, it can be
seen that there exists $\ell \geq 0$
and a nonempty set of rules $R_0 \in \calR^\sharp$
with $\ell$ premises and a conclusion in $S^\sharp_+(\rho, \beta, N^\sharp)$.
If, on the other hand,
$\asupported(\rho, N^\sharp)$ does hold,
then it follows from Section~\ref{sec:supported-rules} that,
by Definition~\ref{def:supports-eval},
$\aeval(\rho, N^\sharp)$ is defined and, for each rule in $R_0$,
there is a rule with the same set of premises
but where the conclusion
$\bigl(\rho \vdash_\beta N^\sharp \rightarrow \aeval(\rho, N^\sharp)\bigr)$
is also in $S^\sharp_+(\rho, \beta, N^\sharp)$.
Thus, in both cases, by definition of $\calU^\sharp$,
there exists a tree in $\calU^\sharp$ with root
in $S^\sharp_+(\rho, \beta, N^\sharp)$.

We prove that,
for any $n \in \Nset$,
there exists a tree $\theta^\sharp \in \calF^{\sharp n}(\calU^\sharp)$
such that
\(
   \theta^\sharp(\emptystring) \in S^\sharp_+(\rho, \beta, N^\sharp)
\).
To this end, we reason by induction on $n \geq 0$.
In the case $n=0$, $\calU = \calF^{\sharp n}(\calU^\sharp)$
so that the hypothesis holds.

We now suppose that $n > 0$.
Let $j \in \{0, \ldots, \ell\}$ be the maximal value for which
there exist trees
\(
   \theta^\sharp_0, \ldots, \theta^\sharp_{j-1}
      \in \calF^{\sharp (n-1)}(\calU^\sharp)
\)
where
\(
   P_0 = \theta^\sharp_0(\emptystring),
     \ldots, P_{j-1} = \theta^\sharp_{j-1}(\emptystring)
\)
are the first $j$ premises of a rule in $R_0$;
let $R_j \sseq R_0$ be the set of all rules in $R_0$ with
$P_0, \ldots, P_{j-1}$ as their first $j$ premises; then $R_j \neq \emptyset$.
We assume that $j < \ell$ and derive a contradiction.
By inspecting the rule schemata in
Section~\ref{sec:abstract-evaluation-relations},
it can be seen that, if there exists
$\frac{P_0 \; \cdots \; P_{j-1} \; P'_j \; \cdots}{\acute{s}^\sharp} \in R_j$
for some $P'_j \in S^\sharp_+(\rho_j, \beta_j, N^\sharp_j)$ and
$\acute{s}^\sharp \in S^\sharp_+(\rho, \beta, N^\sharp)$,
then
\begin{equation}
\label{prop:abstract-tree-exists:j_plus_1_premise}
  \forall P_j \in S^\sharp_+(\rho_j, \beta_j, N^\sharp_j)
    \itc
      \exists s^\sharp \in S^\sharp_+(\rho, \beta, N^\sharp)
        \st
          \frac{P_0 \; \cdots \; P_{j-1} \; P_j \; \cdots}{s^\sharp} \in R_j.
\end{equation}
By the inductive hypothesis, there exists
$\theta^\sharp_j \in \calF^{\sharp (n-1)}(\calU^\sharp)$
such that
\(
   P_j = \theta^\sharp_j(\emptystring)
     \in S^\sharp_+(\rho_j, \beta_j, N^\sharp_j)
\);
hence, by~\eqref{prop:abstract-tree-exists:j_plus_1_premise},
there must be a rule in $R_j$
whose $(j+1)$-th premise is $P_j$;
contradicting the assumption that $j < \ell$ is maximal.
Hence $j = \ell$.
Thus there exists a rule
\(
   \frac{P_0
           \; \cdots
              \; P_{\ell-1}}%
        {s^\sharp}
    \in
      R_0
\)
for some $s^\sharp \in S^\sharp_+(\rho, \beta, N^\sharp)$;
hence, by Definition~\ref{def:abstract-semantics-trees},
the tree
\(
\frac{\theta^\sharp_0
           \; \cdots
              \; \theta^\sharp_{\ell-1}}%
        {s^\sharp}
  \in \calF^{\sharp n}(\calU^\sharp)
\).
Therefore since, by Definition~\ref{def:abstract-semantics-trees},
$\Theta^\sharp = \gfp_{\mathord{\subseteq}}(\calF^\sharp)$,
there exists a tree $\theta^\sharp$ in $\Theta^\sharp$ such that
$\theta^\sharp(\emptystring) \in S^\sharp_+(\rho, \beta, N^\sharp)$.
\qed
\end{proof}

\section{Correctness of the Abstract Semantics}
\label{sec:abstract-semantics-correctness}

In Section~\ref{sec:abstract-dynamic-semantics},
we introduced the notion of \emph{sound approximation}
for configurations and sequents
in terms of the concretization function $\gamma$ defined
for each abstract domain.
We now proceed to define the notion of sound approximation for trees.

\begin{definition}\summary{(`$\propto$' for trees.)}
\label{def:propto-trees}
Let
\(
  \fund{\overline{\propto}}
       {\wp(\Theta\times\Theta^\sharp)}
       {\wp(\Theta\times\Theta^\sharp)}
\)
be given, for each $U \in \wp(\Theta\times\Theta^\sharp)$, by
\[
  \mathop{\overline{\propto}}(U)
    \defeq
      \sset{%
        (\theta, \theta^\sharp) \in \Theta\times\Theta^\sharp
      }{%
        \theta(\emptystring) \propto \theta^\sharp(\emptystring), \\
        \forall i \in \dom(\theta) \inters \Nset
          \itc \\
            \qquad
            \exists j \in \dom(\theta^\sharp) \inters \Nset
              \st
                \bigl(\theta_{[i]}, \theta^\sharp_{[j]}\bigr) \in U
      }.
\]
Then $\theta \propto \theta^\sharp$
if and only if
\(
  (\theta, \theta^\sharp)
    \in
      \gfp_{\mathord{\subseteq}}({\mathord{\overline{\propto}}})
\).
\end{definition}
In words, $\theta \propto \theta^\sharp$ means that
the root of $\theta$ is approximated by the root of $\theta^\sharp$ and
every immediate subtree of $\theta$ is approximated by
some immediate subtrees of $\theta^\sharp$.
Notice that one immediate subtree in $\theta^\sharp$ may be related by
`$\propto$' to none, one or more than one immediate subtree of $\theta$.

The following result states that, for each concrete tree, there is
always an abstract tree that is generated from a corresponding
non-terminal abstract configuration.
\begin{theorem}
\label{thm:abstract-tree-exists}
Let $\theta \in \Theta$ be a concrete tree such that
\(
  \theta(\emptystring)
    = (\rho \vdash_\beta N \rightarrow \eta)
\)
or
\(
  \theta(\emptystring)
    = (\rho \vdash_\beta N \diverges)
\).
Then there exists $\theta^\sharp \in \Theta^\sharp$ such that,
\(
  \theta^\sharp(\emptystring)
    = (\rho \vdash_\beta N^\sharp \rightarrow \eta^\sharp)
\)
and $N \propto N^\sharp$.
\end{theorem}

\begin{proof}
Suppose first that
$N = \langle q, \sigma \rangle$ where $q \in \{ e, d, g, s, b \}$.
By Definition~\ref{def:abstract-memory-structure},
we can always find $\sigma^\sharp \in \Mem^\sharp$ such that
$\sigma \propto \sigma^\sharp$.
Hence,
letting $N^\sharp = \langle q, \sigma^\sharp \rangle$,
by~\eqref{def:propto-nonterminal:expr-decl-glob-statement}
in Definition~\ref{def:nonterminal-abstract-configuration},
we obtain $N \propto N^\sharp$.
Next suppose $N = \langle k, \exceptstate \rangle$,
where $\exceptstate = (\sigma, \except)$.
As before, by Definition~\ref{def:abstract-memory-structure},
we can always find $\sigma^\sharp \in \Mem^\sharp$ such that
$\sigma \propto \sigma^\sharp$.
Moreover, by the definition of the approximation for exceptions,
we can always find $\except^\sharp \in \Except^\sharp$ such that
$\except \propto \except^\sharp$.
Hence, letting
$N^\sharp = \langle k, \spair{\sigma^\sharp}{\except^\sharp} \rangle$,
by~\eqref{def:propto-nonterminal:catch}
in Definition~\ref{def:nonterminal-abstract-configuration},
we again obtain $N \propto N^\sharp$.
In both cases, by Proposition~\ref{prop:abstract-tree-exists},
there exists an abstract tree $\theta^\sharp$
such that
\(
  \theta^\sharp(\emptystring)
    = (\rho \vdash_\beta N^\sharp \rightarrow \eta^\sharp)
\)
and $N \propto N^\sharp$.
\qed
\end{proof}

The next result states that our abstract rules only generate
abstract trees that are correct approximations of their concrete
counterparts (i.e., concrete trees rooted with the same statement, the same
environment and initial memory structure).
\begin{theorem}
\label{thm:abstract-tree-is-safe}
Let
$\theta \in \Theta$ and $\theta^\sharp \in \Theta^\sharp$
be such that
\(
  \theta(\emptystring)
    =
     \bigl(
       \rho \vdash_\beta N
         \rightarrow
           \eta
     \bigr)
\)
or
\(
  \theta(\emptystring)
    =
     \bigl(
       \rho \vdash_\beta N
         \diverges
     \bigr)
\)
and
\(
  \theta^\sharp(\emptystring)
    =
      \bigl(
        \rho \vdash_\beta N^\sharp
          \rightarrow
            \eta^\sharp
      \bigr)
\),
where $N \propto N^\sharp$.
Then $\theta \propto \theta^\sharp$.
\end{theorem}

Theorem~\ref{thm:abstract-tree-is-safe} is a trivial corollary
of the following
\begin{proposition}
\label{prop:abstract-tree-is-safe}
Let
\begin{align}
\label{eq:def_S}
  S
    &\defeq
      \sset{
	(\theta, \theta^\sharp) \in \Theta\times\Theta^\sharp
      }{
       \theta(\emptystring)
         \in \bigl\{
               \rho \vdash_\beta N
                 \rightarrow
                   \eta,\;
               \rho \vdash_\beta N
                 \diverges
             \bigr\}, \\
       \theta^\sharp(\emptystring) =
         \rho \vdash_\beta N^\sharp
           \rightarrow
             \eta^\sharp, \\
       N \propto N^\sharp
      }.
\end{align}
Then, for all $(\theta, \theta^\sharp) \in S$,
$\theta \propto \theta^\sharp$.
\end{proposition}

\begin{proof}
Let $\theta \in \Theta$ and $\theta^\sharp \in \Theta^\sharp$.
We define:
\begin{align*}
  r &\defeq
  \prooftree
    \theta_{[0]}(\emptystring) \; \cdots \; \theta_{[h-1]}(\emptystring)
  \justifies
    \theta(\emptystring)
  \endprooftree
\qquad & \qquad
  r^\sharp &\defeq
  \prooftree
    \theta^\sharp_{[0]}(\emptystring)
      \; \cdots \;
        \theta^\sharp_{[\ell-1]}(\emptystring)
  \justifies
    \theta^\sharp(\emptystring)
  \endprooftree
\end{align*}
where, for some $h, \ell \geq 0$,
$\{ 0, \ldots, h-1 \} \subseteq \dom(\theta)$,
$\{ 0, \ldots, \ell-1 \} \subseteq \dom(\theta^\sharp)$,
$h \notin \dom(\theta)$
and $\ell \notin \dom(\theta^\sharp)$.
By Definitions~\ref{def:concrete-semantics-trees}
and~\ref{def:abstract-semantics-trees},
$r \in \calR$ and $r^\sharp \in \calR^\sharp$.
Note that, to simplify the proof,
we will use the schematic concrete and abstract
rules given in Sections~\ref{sec:concrete-evaluation-relations}
and~\ref{sec:abstract-evaluation-relations} to denote the actual rule
instances $r$ and $r^\sharp$.

Letting $(\theta, \theta^\sharp) \in S$,
we need to show that $\theta \propto \theta^\sharp$;
by Definition~\ref{def:propto-trees}, this is equivalent to
showing that
\(
  (\theta, \theta^\sharp)
    \in
      \gfp_{\mathord{\subseteq}}({\mathord{\overline{\propto}}})
\).
To this end, by the principle of fixpoint coinduction,
we will show that
$(\theta, \theta^\sharp) \in \mathop{\overline{\propto}}(S)$.

By Definition~\ref{def:propto-trees}, we need to show that the following
properties hold:
\begin{enumerate}[(i)]
\item
\label{enum:abstract-tree-is-safe:root}
$\theta(\emptystring) \propto \theta^\sharp(\emptystring)$;
\item
\label{enum:abstract-tree-is-safe:immediate-subtrees}
for each $i = 0$, \dots,~$h-1$ there exists $j \in \{ 0, \ldots, \ell-1 \}$
such that $(\theta_{[i]}, \theta^\sharp_{[j]}) \in S$.
\end{enumerate}

The proof that properties~\eqref{enum:abstract-tree-is-safe:root}
and~\eqref{enum:abstract-tree-is-safe:immediate-subtrees} hold
is by (well-founded) induction on the structure of the concrete tree $\theta$.
Observe that the ``immediate subtree'' relation between trees
in $\Theta_+$ is a well-founded partial ordering because,
if $\theta \in \Theta_+$ then,
by Definition~\ref{def:concrete-semantics-trees},
there are no infinite descending chains.
We extend this ordering relation to the
\emph{immediate positive subtree} relation between trees in $\Theta$:
$\theta'$ is said to be an \emph{immediate positive subtree} of $\theta$
if and only if $\theta' \in \Theta_+$ and is an immediate subtree of $\theta$.
Clearly, by Definition~\ref{def:concrete-semantics-trees},
the immediate positive subtree ordering on trees in $\Theta$
is also well-founded.

We first note that it is not restrictive to only consider
unsupported expressions, declarations or statements: as noted in
Section~\ref{sec:abstract-evaluation-relations},
the tree for any supported expression (resp., declaration or statement)
has the same structure as the tree for the same expression
(resp., declaration or statement) as if it were unsupported. Hence,
once correctness of the approximation for unsupported expressions,
declarations or statements is proved, the correctness for their
supported counterparts will immediately follow
from Definition~\ref{def:supports-eval}.

Let
\begin{gather*}
  \theta(\emptystring)
    = \bigl(
        \rho \vdash_\beta N
          \rightarrow \eta
      \bigr)
\,\text{ or }\,
  \theta(\emptystring)
    = \bigl(
        \rho \vdash_\beta N
          \diverges
      \bigr),\\
  \theta^\sharp(\emptystring)
    = \bigl(
        \rho \vdash_\beta N^\sharp
          \rightsquigarrow \eta^\sharp
      \bigr).
\end{gather*}
By \eqref{eq:def_S}, $N \propto N^\sharp$.
Therefore, by condition~\eqref{def:propto-sequents:diverge}
of Definition~\ref{def:propto-sequents},
property~\eqref{enum:abstract-tree-is-safe:root} holds trivially
whenever $\theta \in \Theta_-$ (i.e., when $r$ is a negative concrete rule).
In addition,
to prove that property~\eqref{enum:abstract-tree-is-safe:root} holds for
each $\theta \in \Theta_+$ (i.e., when $r$ is a positive concrete rule),
by condition~\eqref{def:propto-sequents:term}
of Definition~\ref{def:propto-sequents},
we just need to show $\eta \propto \eta^\sharp$.

Consider next property~\eqref{enum:abstract-tree-is-safe:immediate-subtrees}.
The base cases are when the concrete rule $r$ has no premises
(i.e., $h = 0$);
and this property holds trivially in these cases.
For the inductive steps (i.e., $h > 0$)
suppose $i \in \{0, \ldots, h-1\}$ and $j \in \{0, \ldots, \ell-1\}$ are
such that $(\theta_{[i]}, \theta^\sharp_{[j]}) \in S$.
If $\theta \in \Theta_+$ then, by the inductive hypothesis, we can assume that
$(\theta_{[i]}, \theta^\sharp_{[j]}) \in \mathop{\overline{\propto}}(S)$;
similarly, if $\theta \in \Theta_-$ and $i \neq h-1$,
by the inductive hypothesis, we can assume that,
$(\theta_{[i]}, \theta^\sharp_{[j]}) \in \mathop{\overline{\propto}}(S)$.
Hence, in both cases, by Definition~\ref{def:propto-trees},
$\theta_{[i]}(\emptystring) \propto \theta^\sharp_{[j]}(\emptystring)$.
Also, if $\theta \in \Theta_-$,
by Definition~\ref{def:concrete-semantics-trees},
$\theta_{[h-1]}(\emptystring)$ is a divergent sequent
so that, by Definitions~\ref{def:propto-sequents}
and~\ref{def:propto-trees},
$\theta_{[h-1]}(\emptystring) \propto \theta^\sharp_{[j]}(\emptystring)$.
Thus, for all concrete trees $\theta \in \Theta$,
we can safely assume the following:
\begin{equation}
\label{enum:abstract-tree-is-safe:inductive_assume}
   \forall i \in \{ 0, \dots, h-1\},  j \in \{ 0, \dots, \ell-1\} \itc
          (\theta_{[i]}, \theta^\sharp_{[j]}) \in S
     \implies
       \theta_{[i]}(\emptystring) \propto \theta^\sharp_{[j]}(\emptystring).
\end{equation}
Moreover, we need only explicitly prove
property~\eqref{enum:abstract-tree-is-safe:immediate-subtrees}
for each of the positive rules
since, by the definition of the concrete divergence (negative) rules,
\eqref{enum:abstract-tree-is-safe:inductive_assume} and
Definition~\ref{def:nonterminal-abstract-configuration},
if property~\eqref{enum:abstract-tree-is-safe:immediate-subtrees}
holds for any positive rule
it also holds for the corresponding negative rules.
Thus in the detailed proofs of
properties~\eqref{enum:abstract-tree-is-safe:root}
and~\eqref{enum:abstract-tree-is-safe:immediate-subtrees}
for the inductive steps,
we only consider the positive rules.

To help the reader,
Tables~\ref{tab:abstract-tree-is-safe:expressions},
\ref{tab:abstract-tree-is-safe:declarations},
\ref{tab:abstract-tree-is-safe:statements},
\ref{tab:abstract-tree-is-safe:function-bodies}
and~\ref{tab:abstract-tree-is-safe:catch-clauses},
contain a summary of the conclusions of rules
$r$ and $r^\sharp$.
The first column $Q \in \{E, D, G, S, B, K\}$, gives
the syntactic forms in the first component
of the non-terminal configurations $N$ and $N^\sharp$ (which,
by Definition~\ref{def:nonterminal-abstract-configuration},
must be the same);
the second and third columns give a concrete rule $r$ and
abstract rule $r^\sharp$, respectively, that apply to $Q$.
Note that we do not pair concrete rules with abstract rules that
have mutually inconsistent side conditions.
Justification for the omission of any abstract rules
for a particular concrete rule $r$ is given in the detailed proof
for that case.
The column headed $\eta_q$,
where
\(
  q \in
    \{
       \mathrm{e}, \mathrm{d}, \mathrm{g},
       \mathrm{s}, \mathrm{b}, \mathrm{k}
    \}
\)
gives the concrete terminal configuration for $r$,
while the columns headed by $\eta^\sharp_q$ give the
components of the abstract terminal configuration for $r^\sharp$.
A blank entry in any table cell means that
the value is exactly the same as the value found
in the same column of the previous row.
To save space in Tables~\ref{tab:abstract-tree-is-safe:declarations},
\ref{tab:abstract-tree-is-safe:statements},
\ref{tab:abstract-tree-is-safe:function-bodies}
and~\ref{tab:abstract-tree-is-safe:catch-clauses},
we have denoted the operations
`$\mathord{\datcleanup}$', `$\mathord{\sunmark}$', `$\mathord{\sunlink}$',
`$\mathord{\asunmark}$' and `$\mathord{\asunlink}$'
by `$\mathord{\datcleanupshort}$', `$\mathord{\sunmarkshort}$',
`$\mathord{\sunlinkshort}$', `$\mathord{\sunmarkshort^\sharp}$' and
`$\mathord{\sunlinkshort^\sharp}$', respectively.
Note that the premises and the side conditions
for the rules are not provided in any of the tables;
reference must be made to the actual rules for this information.

\subsection{Expressions}

For this part of the proof,
we use Table~\ref{tab:abstract-tree-is-safe:expressions}.
By \eqref{eq:def_S}, $N \propto N^\sharp$.
Thus letting $N = \langle E, \sigma \rangle$
and $N^\sharp = \langle E, \sigma^\sharp \rangle$,
by Definition~\ref{def:nonterminal-abstract-configuration},
we have the implicit hypothesis $\sigma \propto \sigma^\sharp$.
We show using~\eqref{def:propto-terminal:expr}
in Definition~\ref{def:terminal-abstract-configuration},
that $\eta_\mathrm{e} \propto \eta^\sharp_\mathrm{e}$.

\begin{table}
\caption{Corresponding concrete and abstract rules and
         terminals for expressions}
\label{tab:abstract-tree-is-safe:expressions}
\centering
\footnotesize
\begin{tabular}
{|c|l|l|l|l|l|}
\hline
  \multicolumn{1}{|c|}{$E$}
      &\multicolumn{1}{c|}{$r$}
      &\multicolumn{1}{c|}{$r^\sharp$} &\multicolumn{1}{c|}{$\eta_\mathrm{e}$}
      &\multicolumn{2}{c|}{
        \(\eta^\sharp_\mathrm{e} =
           \bigl\langle
              (\sval^\sharp_a,\sigma^\sharp_a),
              \exceptstate^\sharp_a
           \bigr\rangle
         \) }\\
\cline{5-6}
  &&&
      & \multicolumn{1}{c|}{$(\sval^\sharp_a,\sigma^\sharp_a)$}
      & \multicolumn{1}{c|}{$\exceptstate^\sharp_a$}\\
\hline
\hline
  $\con$
     & \ref{rule:conc_constant}
     & \ref{rule:abstr_constant}
     & $\langle \con, \sigma \rangle$
     & $\spair{\alpha(\{\con\})}{\sigma^\sharp}$
     & $\none^\sharp$ \\
\hline
  $\id$
     & \ref{rule:conc_identifier}
     & \ref{rule:abstr_identifier}
     & $\sigma[\rho(\id)]$
     & \multicolumn{2}{c|}{$\sigma^\sharp[\rho(\id)]$} \\
\hline
  $-e$
     & \ref{rule:conc_uminus_error}
     & \ref{rule:abstr_unary_minus}
     & $\exceptstate$
     & $(\mathop{\absuminus} m^\sharp, \sigma^\sharp_0)$
     & $\exceptstate^\sharp$ \\
     & \ref{rule:conc_uminus_ok}
     &
     & $\langle -m, \sigma_0 \rangle$
     &
     & \\
\hline
  $e_0 \boxcircle e_1$
     & \ref{rule:conc_arith_bop_0}/\ref{rule:conc_arith_bop_1}
     & \ref{rule:abstr_arith_bop}
     & $\exceptstate$
     & $(m^\sharp_0 \circledcirc m^\sharp_1, \sigma^\sharp_1)$
     & $\exceptstate^\sharp_0 \sqcup \exceptstate^\sharp_1$ \\
     &
     & \ref{rule:abstr_div_mod_exc}
     &
     & $(m^\sharp_0 \circledcirc m^\sharp_1, \sigma^\sharp_1)$
     & \(
         \exceptstate^\sharp_0
           \sqcup \exceptstate^\sharp_1
             \sqcup \exceptstate^\sharp_2
       \) \\
\cline{2-6}
     & \ref{rule:conc_arith_bop_2}
     & \ref{rule:abstr_arith_bop}
     & $\langle m_0 \circ m_1, \sigma_1 \rangle$
     & $(m^\sharp_0 \circledcirc m^\sharp_1, \sigma^\sharp_1)$
     & $\exceptstate^\sharp_0 \sqcup \exceptstate^\sharp_1$ \\
     &
     & \ref{rule:abstr_div_mod_exc}
     &
     & $(m^\sharp_0 \circledcirc m^\sharp_1, \sigma^\sharp_1)$
     & \(
         \exceptstate^\sharp_0
           \sqcup \exceptstate^\sharp_1
             \sqcup \exceptstate^\sharp_2
       \) \\
\cline{2-6}
     & \ref{rule:conc_arith_bop_exc_0}
     & \ref{rule:abstr_div_mod_exc}
     & $\langle \sigma_1, \divbyzero \rangle$
     & $(m^\sharp_0 \circledcirc m^\sharp_1, \sigma^\sharp_1)$
     & \(
         \exceptstate^\sharp_0
           \sqcup \exceptstate^\sharp_1
             \sqcup \exceptstate^\sharp_2
       \) \\
\hline
  $m_0 \boxast m_1$
     & \ref{rule:conc_arith_test_error_0}/\ref{rule:conc_arith_test_error_1}
     & \ref{rule:abstr_arith_test}
     & $\exceptstate$
     & $(m_0^\sharp \bowtie m^\sharp_1, \sigma^\sharp_1)$
     & $\exceptstate^\sharp_0 \sqcup \exceptstate^\sharp_1$ \\
     & \ref{rule:conc_arith_test_ok}
     &
     & $\langle m_0 \lessgtr m_1, \sigma_1 \rangle$
     &
     & \\
\hline
  $\notop\ b$
     & \ref{rule:conc_negation_error}
     & \ref{rule:abstr_negation}
     & $\exceptstate$
     & $(\absneg t^\sharp, \sigma^\sharp_0)$
     & $\exceptstate^\sharp$ \\
     & \ref{rule:conc_negation_ok}
     &
     & $\langle \bneg t, \sigma_0 \rangle$
     &
     & \\
\hline
  $b_0 \andop\ b_1$
     & \ref{rule:conc_conjunction_0}
     & \ref{rule:abstr_conjunction}
     & $\exceptstate$
     & $\valstate^\sharp_\ffv \sqcup \valstate^\sharp_1$
     & $\exceptstate^\sharp_0 \sqcup \exceptstate^\sharp_1$ \\
     & \ref{rule:conc_conjunction_1}
     &
     & $\langle \ffv, \sigma_0 \rangle$
     &
     & \\
     & \ref{rule:conc_conjunction_2}
     &
     & $\eta$
     &
     & \\
\hline
  $b_0 \orop\ b_1$
     & \ref{rule:conc_disjunction_0}--\ref{rule:conc_disjunction_2}
     & \ref{rule:abstr_disjunction}
     &\multicolumn{3}{c|}{Similar to the rows for `$b_0 \andop\ b_1$'} \\
\hline
\end{tabular}
\end{table}

\paragraph*{Constant}

Suppose $r$ is an instance of~\eqref{rule:conc_constant}.
By definition of $\pard{\alpha}{\wp(\Integer)}{\Integer^\sharp}$
and $\pard{\alpha}{\wp(\Bool)}{\Bool^\sharp}$,
we have $\con \propto \alpha(\{\con\})$;
by hypothesis, $\sigma \propto \sigma^\sharp$
so that
$\langle \con, \sigma \rangle \propto \spair{\alpha(\{\con\})}{\sigma^\sharp}$.
Hence $\eta_\mathrm{e} \propto \eta^\sharp_\mathrm{e}$.

\paragraph*{Identifier}

Suppose $r$ is an instance of~\eqref{rule:conc_identifier}.
Since, by hypothesis, $\sigma \propto \sigma^\sharp$,
by Definition~\ref{def:abstract-memory-structure} we obtain
\(
  \sigma\bigl[ \rho(\id) \bigr]
    \propto \sigma^\sharp\bigl[ \rho(\id) \bigr]
\).
Hence, $\eta_\mathrm{e} \propto \eta^\sharp_\mathrm{e}$.

\paragraph*{Unary Minus}

Suppose $r$ is an instance of~\eqref{rule:conc_uminus_error}
or~\eqref{rule:conc_uminus_ok}.
Then, by hypothesis, $(\theta_{[0]}, \theta^\sharp_{[0]}) \in S$
and, hence, as $h=1$,
property~\eqref{enum:abstract-tree-is-safe:immediate-subtrees} holds.
By~\eqref{enum:abstract-tree-is-safe:inductive_assume},
$\theta_{[0]}(\emptystring) \propto \theta^\sharp_{[0]}(\emptystring)$.
Thus, if $r$ is an instance of~\eqref{rule:conc_uminus_error},
then $\exceptstate \propto \exceptstate^\sharp$;
if $r$ is an instance of~\eqref{rule:conc_uminus_ok},
then $m \propto m^\sharp$ and $\sigma_0 \propto \sigma^\sharp_0$.
In the latter case, by the soundness of `$\mathop{\absuminus}$',
$-m \propto \mathop{\absuminus} m^\sharp$.
Hence, in both cases,
$\eta_\mathrm{e} \propto \eta^\sharp_\mathrm{e}$.

\paragraph*{Binary Arithmetic Operations}

Suppose that  $r$ is an instance of one of
the rules \eqref{rule:conc_arith_bop_0}--\eqref{rule:conc_arith_bop_exc_0}.
Then, by hypothesis,
$(\theta_{[0]}, \theta^\sharp_{[0]}) \in S$.
By~\eqref{enum:abstract-tree-is-safe:inductive_assume},
$\theta_{[0]}(\emptystring) \propto \theta^\sharp_{[0]}(\emptystring)$.
Note that, in the condition for
abstract rule~\eqref{rule:abstr_div_mod_exc},
\(
  \exceptstate^\sharp_2
    = \spair{\sigma^\sharp_1}{\alpha(\{\divbyzero\})}
\).

If $r$ is an instance of~\eqref{rule:conc_arith_bop_0},
then $h=1$
so that property~\eqref{enum:abstract-tree-is-safe:immediate-subtrees} holds.
The property
$\theta_{[0]}(\emptystring) \propto \theta^\sharp_{[0]}(\emptystring)$
implies $\exceptstate \propto \exceptstate^\sharp_0$.
Therefore,%%
\footnote{Here and in the following, whenever
we need to prove $\iota \propto \iota^\sharp_0 \sqcup \iota^\sharp_1$,
we just prove either $\iota \propto \iota^\sharp_0$ or
$\iota \propto \iota^\sharp_1$
and implicitly use the monotonicity of $\gamma$.}
$\eta_\mathrm{e} \propto \eta^\sharp_\mathrm{e}$.

If $r$ is an instance of~\eqref{rule:conc_arith_bop_1},
\eqref{rule:conc_arith_bop_2}
or~\eqref{rule:conc_arith_bop_exc_0}, then $h=2$.
Property
$\theta_{[0]}(\emptystring) \propto \theta^\sharp_{[0]}(\emptystring)$
implies $\sigma_0 \propto \sigma^\sharp_0$ and $m_0 \propto m^\sharp_0$;
hence
$(\theta_{[1]}, \theta^\sharp_{[1]}) \in S$ and
property~\eqref{enum:abstract-tree-is-safe:immediate-subtrees} holds.
By~\eqref{enum:abstract-tree-is-safe:inductive_assume},
$\theta_{[1]}(\emptystring) \propto \theta^\sharp_{[1]}(\emptystring)$.

If $r$ is an instance of~\eqref{rule:conc_arith_bop_1},
then property
$\theta_{[1]}(\emptystring) \propto \theta^\sharp_{[1]}(\emptystring)$
implies $\exceptstate \propto \exceptstate^\sharp_1$;
thus $\eta_\mathrm{e} \propto \eta^\sharp_\mathrm{e}$.
If $r$ is an instance of~\eqref{rule:conc_arith_bop_2},
then property
$\theta_{[1]}(\emptystring) \propto \theta^\sharp_{[1]}(\emptystring)$
implies $\sigma_1 \propto \sigma^\sharp_1$
and $m_1 \propto m^\sharp_1$ so that,
by the soundness of `$\mathord{\circledcirc}$',
$(m_0 \boxcircle m_1) \propto (m^\sharp_0 \circledcirc m^\sharp_1)$;
and hence
$\eta_\mathrm{e} \propto \eta^\sharp_\mathrm{e}$.
If  $r$ is an instance of~\eqref{rule:conc_arith_bop_exc_0},
then the condition
$\theta_{[1]}(\emptystring) \propto \theta^\sharp_{[1]}(\emptystring)$
 implies $\sigma_1 \propto \sigma^\sharp_1$ and $0 \propto m^\sharp_1$.
Hence, by the side conditions,
$r^\sharp$ must be an instance of~\eqref{rule:abstr_div_mod_exc};
so that, as
\(
  \langle \sigma_1, \divbyzero \rangle
    \propto
      \spair{\sigma^\sharp_1}{\alpha(\{\divbyzero\})}
\), we have
$\eta_\mathrm{e} \propto \eta^\sharp_\mathrm{e}$.

\paragraph*{Test Operators}

Suppose $r$ is an instance of one of
rules~\eqref{rule:conc_arith_test_error_0}--\eqref{rule:conc_arith_test_ok}.
Then, by hypothesis,
$(\theta_{[0]}, \theta^\sharp_{[0]}) \in S$.
By~\eqref{enum:abstract-tree-is-safe:inductive_assume},
$\theta_{[0]}(\emptystring) \propto \theta^\sharp_{[0]}(\emptystring)$.

If $r$ is an instance of~\eqref{rule:conc_arith_test_error_0},
then $h=1$ and
property~\eqref{enum:abstract-tree-is-safe:immediate-subtrees} holds.
$\theta_{[0]}(\emptystring) \propto \theta^\sharp_{[0]}(\emptystring)$
implies $\exceptstate \propto \exceptstate^\sharp_0$.
Hence $\eta_\mathrm{e} \propto \eta^\sharp_\mathrm{e}$.

If r is an instance of~\eqref{rule:conc_arith_test_error_1}
or~\eqref{rule:conc_arith_test_ok}, then $h=2$.
$\theta_{[0]}(\emptystring) \propto \theta^\sharp_{[0]}(\emptystring)$
implies $\sigma_0 \propto \sigma^\sharp_0$ and $m_0 \propto m^\sharp_0$.
Thus
$(\theta_{[1]}, \theta^\sharp_{[1]}) \in S$ and
property~\eqref{enum:abstract-tree-is-safe:immediate-subtrees} holds.
By~\eqref{enum:abstract-tree-is-safe:inductive_assume},
$\theta_{[1]}(\emptystring) \propto \theta^\sharp_{[1]}(\emptystring)$.
If $r$ is an instance of~\eqref{rule:conc_arith_test_error_1},
then $\exceptstate \propto \exceptstate^\sharp_1$;
and if $r$ is an instance of~\eqref{rule:conc_arith_test_ok},
$\sigma_1 \propto \sigma^\sharp_1$ and $m_1 \propto m^\sharp_1$ so that,
by soundness of `$\bowtie$',
$(m_0 \lessgtr m_1) \propto (m^\sharp_0 \bowtie m^\sharp_1)$.
Hence, for both concrete rules,
$\eta_\mathrm{e} \propto \eta^\sharp_\mathrm{e}$.

\paragraph*{Negation}

The proof when $r$ is an instance
of~\eqref{rule:conc_negation_error} or~\eqref{rule:conc_negation_ok}
has the same structure of the proof for the unary minus case shown before.

\paragraph*{Conjunction}

Suppose $r$ is an instance of one of
rules~\eqref{rule:conc_conjunction_0}--\eqref{rule:conc_conjunction_2}.
By hypothesis,
$(\theta_{[0]}, \theta^\sharp_{[0]}) \in S$.

If $r$ is an instance of~\eqref{rule:conc_conjunction_0}
or~\eqref{rule:conc_conjunction_1},
then $h=1$ and
property~\eqref{enum:abstract-tree-is-safe:immediate-subtrees} holds.
If $r$ is an instance of~\eqref{rule:conc_conjunction_0},
by~\eqref{enum:abstract-tree-is-safe:inductive_assume},
we have $\theta_{[0]}(\emptystring) \propto \theta^\sharp_{[0]}(\emptystring)$,
which implies $\exceptstate \propto \exceptstate^\sharp_0$.
If $r$ is an instance of~\eqref{rule:conc_conjunction_1},
by Definition~\ref{def:memstruct-filter},
\(
  \sigma_0
    \propto
      \sigma^\sharp_{\ffv}
        = \phi(\rho, \sigma^\sharp, \kw{not} b_0)
\).
Thus, since $\ffv \propto \alpha(\{\ffv\})$ holds by definition,
we have $\langle \ffv, \sigma_0 \rangle \propto \valstate^\sharp_\ffv$.
Hence, for both concrete rules,
$\eta_\mathrm{e} \propto \eta^\sharp_\mathrm{e}$.

If $r$ is an instance of~\eqref{rule:conc_conjunction_2}, then $h = 2$.
By Definition~\ref{def:memstruct-filter},
$\sigma_0 \propto \sigma^\sharp_\ttv$, so that
$(\theta_{[1]},\theta^\sharp_{[1]}) \in S$ and
property~\eqref{enum:abstract-tree-is-safe:immediate-subtrees} holds.
By~\eqref{enum:abstract-tree-is-safe:inductive_assume},
$\theta_{[1]}(\emptystring) \propto \theta^\sharp_{[1]}(\emptystring)$
so that
\(
   \eta
     \propto
       \langle \valstate^\sharp_1, \exceptstate^\sharp_1 \rangle
\).
Hence, $\eta_\mathrm{e} \propto \eta^\sharp_\mathrm{e}$.

\paragraph*{Disjunction}
The proof when $r$ is an instance
of one of
rules~\eqref{rule:conc_disjunction_0}--\eqref{rule:conc_disjunction_2}
is similar to that for conjunction.

\subsection{Declarations}

In Table~\ref{tab:abstract-tree-is-safe:declarations},
$Q$ denotes a local declaration $D$ or a global declaration $G$.
Moreover, $\eta_\mathrm{q} \in \{\Td, \Tg\}$ and
$\eta^\sharp_\mathrm{q} \in \{\Td^\sharp, \Tg^\sharp\}$,
the actual domains for $\eta_\mathrm{q}$ and $\eta^\sharp_\mathrm{q}$
will depend on context.

By \eqref{eq:def_S} we have $N \propto N^\sharp$.
Thus letting $N = \langle Q, \sigma \rangle$
and $N^\sharp = \langle Q, \sigma^\sharp \rangle$
for any $Q \in \{D, G\}$,
by Definition~\ref{def:nonterminal-abstract-configuration},
we have the implicit hypothesis $\sigma \propto \sigma^\sharp$.
We show using~\eqref{def:propto-terminal:decl}
in Definition~\ref{def:terminal-abstract-configuration},
that $\eta_\mathrm{q} \propto \eta^\sharp_\mathrm{q}$.

\begin{table}
\caption{Corresponding concrete and abstract rules and
         terminals for declarations}
\label{tab:abstract-tree-is-safe:declarations}
\centering
\footnotesize
\begin{tabular}
{|c|l|l|l|l|l|}
\hline
  \multicolumn{1}{|c|}{$Q$}
      &\multicolumn{1}{c|}{$r$}
      &\multicolumn{1}{c|}{$r^\sharp$}
      &\multicolumn{1}{c|}{$\eta_\mathrm{q}$}
      &\multicolumn{2}{c|}{
        \(\eta^\sharp_\mathrm{q} =
           \bigl\langle
              (\rho^\sharp_a,\sigma^\sharp_a),
              \exceptstate^\sharp_a
           \bigr\rangle
         \) }\\
\cline{5-6}
  &&&
      & \multicolumn{1}{c|}{$(\rho^\sharp_a, \sigma^\sharp_a)$}
      & \multicolumn{1}{c|}{$\exceptstate^\sharp_a$}\\
\hline
\hline
  $\kw{nil}$
     & \ref{rule:conc_decl_nil}
     & \ref{rule:abstr_decl_nil}
     & $\langle \emptyset, \sigma \rangle$
     & $(\emptyset, \sigma^\sharp)$
     & $\none^\sharp$ \\
\hline
  $\rho_0$
     & \ref{rule:conc_decl_environment}
     & \ref{rule:abstr_decl_environment}
     & $\langle \rho_0, \sigma \rangle$
     & $(\rho_0, \sigma^\sharp)$
     & $\none^\sharp$ \\
\hline
  $\kw{rec} \rho_0$
     & \ref{rule:conc_decl_rec_environment}
     & \ref{rule:abstr_decl_rec_environment}
     & $\langle \rho_1, \sigma \rangle$
     & $(\rho_1, \sigma^\sharp)$
     & $\none^\sharp$ \\
\hline
  $\kw{gvar} \id : \sT = e$
     & \ref{rule:conc_gvar_decl_expr_err}/\ref{rule:conc_gvar_decl_dat_err}
     & \ref{rule:abstr_global_decl}
     & $\datcleanupshort(\exceptstate)$
     & $(\rho_1, \sigma^\sharp_1)$
     & $\datcleanupshort^\sharp(\exceptstate^\sharp_0 \sqcup \exceptstate^\sharp_1)$\\
     & \ref{rule:conc_gvar_decl_ok}
     &
     & \(
         \langle
           \rho_1, \sigma_1
         \rangle
       \)
     &
     & \\
\hline
  $\kw{lvar} \id : \sT = e$
     & \ref{rule:conc_lvar_decl_expr_err}/\ref{rule:conc_lvar_decl_stk_err}
     & \ref{rule:abstr_local_decl}
     & $\sunmarkshort(\exceptstate)$
     & $(\rho_1, \sigma^\sharp_1)$
     & $\sunmarkshort^\sharp(\exceptstate^\sharp_0 \sqcup \exceptstate^\sharp_1)$\\
     & \ref{rule:conc_lvar_decl_ok}
     &
     & \(
         \langle
           \rho_1, \sigma_1
         \rangle
       \)
     &
     & \\
\hline
  $\kw{function} \id(\fps) : \sT = e$
     & \ref{rule:conc_decl_function}
     & \ref{rule:abstr_decl_function}
     & $\langle \rho_0, \sigma \rangle$
     & $(\rho_0, \sigma^\sharp)$
     & $\none^\sharp$ \\
\hline
  $\kw{rec} g$
     & \ref{rule:conc_decl_rec}
     & \ref{rule:abstr_recursive_decl}
     & $\eta$
     & \multicolumn{2}{c|}{$\eta^\sharp$} \\
\hline
  $g_0;g_1$
     & \ref{rule:conc_glob_seq_err_0}/\ref{rule:conc_glob_seq_err_1}
     & \ref{rule:abstr_global_seq}
     & $\exceptstate$
     & $(\rho_0[\rho_1], \sigma^\sharp_1)$
     & $\exceptstate^\sharp_0 \sqcup \exceptstate^\sharp_1$\\
     & \ref{rule:conc_glob_seq_ok}
     &
     & $\bigl\langle \rho_0[\rho_1], \sigma_1 \bigr\rangle$
     &
     & \\
\hline
  $d_0;d_1$
     & \ref{rule:conc_decl_seq_err_0}--\ref{rule:conc_decl_seq_ok}
     & \ref{rule:abstr_local_seq}
     &\multicolumn{3}{c|}{Similar to the rows for `$g_0;g_1$'} \\
\hline
\end{tabular}
\end{table}

\paragraph*{Nil}

If $r$ is an instance of~\eqref{rule:conc_decl_nil}
then, by the hypothesis,
$\eta_\mathrm{q} \propto \eta^\sharp_\mathrm{q}$.

\paragraph*{(Recursive) Environment}

If $r$ is an instance of~\eqref{rule:conc_decl_environment}
or~\eqref{rule:conc_decl_rec_environment} then, by the hypothesis,
$\eta_\mathrm{q} \propto \eta^\sharp_\mathrm{q}$.

\paragraph*{Global Variable Declaration}

If $r$ is an instance of one of
rules~\eqref{rule:conc_gvar_decl_expr_err}--\eqref{rule:conc_gvar_decl_ok}
then, by the hypothesis
$(\theta_{[0]}, \theta^\sharp_{[0]}) \in S$
so that, as $h=1$,
property~\eqref{enum:abstract-tree-is-safe:immediate-subtrees} holds.
By~\eqref{enum:abstract-tree-is-safe:inductive_assume},
$\theta_{[0]}(\emptystring) \propto \theta^\sharp_{[0]}(\emptystring)$.
If $r$ is an instance of~\eqref{rule:conc_gvar_decl_expr_err},
then $\theta_{[0]}(\emptystring) \propto \theta^\sharp_{[0]}(\emptystring)$
implies $\exceptstate \propto \exceptstate^\sharp_0$;
by Definition~\ref{def:abstract-memory-structure} and monotonicity
of $\gamma$, we have
\(
  \datcleanup(\exceptstate)
    \propto
      \datcleanup^\sharp(\exceptstate^\sharp_0 \sqcup \exceptstate^\sharp_1)
\),
i.e., $\eta_\mathrm{q} \propto \eta^\sharp_\mathrm{q}$.
If $r$ is an instance
of~\eqref{rule:conc_gvar_decl_dat_err}
or~\eqref{rule:conc_gvar_decl_ok},
then $\theta_{[0]}(\emptystring) \propto \theta^\sharp_{[0]}(\emptystring)$
implies $\valstate \propto \valstate^\sharp$.
By Definition~\ref{def:abstract-memory-structure},
\(
  \datnew(\valstate)
    \propto
      \datnew^\sharp(\valstate^\sharp)
\).
By the side condition for abstract rule~\eqref{rule:abstr_global_decl},
\(
   \datnew^\sharp(\valstate)
     = \bigl((\sigma^\sharp_1, l), \exceptstate^\sharp_1\bigr)
\).
By the side conditions for~\eqref{rule:conc_gvar_decl_dat_err}
and~\eqref{rule:conc_gvar_decl_ok},
either $\datnew(\valstate) = \exceptstate \propto \exceptstate^\sharp_1$
---and hence
\(
  \datcleanup(\exceptstate)
    \propto
      \datcleanup^\sharp(\exceptstate^\sharp_0 \sqcup \exceptstate^\sharp_1)
\)
by Definition~\ref{def:abstract-memory-structure}---
or $\datnew(\valstate) = (\sigma_1, l) \propto (\sigma^\sharp_1, l)$.
Thus, in both cases, $\eta_\mathrm{q} \propto \eta^\sharp_\mathrm{q}$.

\paragraph*{Local Variable Declaration}

The proof for local variable declaration,
when $r$ is an instance of one of
rules~\eqref{rule:conc_lvar_decl_expr_err}--\eqref{rule:conc_lvar_decl_ok},
is the same as that for global variable declaration,
with the few necessary adjustments
(i.e., using $\sunmark$, $\asunmark$, $\stknew$,
$\stknew^\sharp$ and $i$
in place of $\datcleanup$, $\datcleanup^\sharp$, $\datnew$,
$\datnew^\sharp$ and $l$).

\paragraph*{Function Declaration}

If $r$ is an instance of~\eqref{rule:conc_decl_function}
then, by the hypothesis,
$\eta_\mathrm{q} \propto \eta^\sharp_\mathrm{q}$.

\paragraph*{Recursive Declaration}

If $r$ is an instance
of~\eqref{rule:conc_decl_rec},
then $h = 2$ and, by the hypothesis,
$(\theta_{[0]}, \theta^\sharp_{[0]}) \in S$.
By~\eqref{enum:abstract-tree-is-safe:inductive_assume},
$\theta_{[0]}(\emptystring) \propto \theta^\sharp_{[0]}(\emptystring)$,
which implies that $\rho_0$ denotes the same environment in both
$r$ and $r^\sharp$ and $\sigma_0 \propto \sigma^\sharp_0$.
Hence, $(\theta_{[1]}, \theta^\sharp_{[1]}) \in S$ and
property~\eqref{enum:abstract-tree-is-safe:immediate-subtrees} holds.
By~\eqref{enum:abstract-tree-is-safe:inductive_assume},
$\theta_{[1]}(\emptystring) \propto \theta^\sharp_{[1]}(\emptystring)$
which implies $\eta \propto \eta^\sharp$.
Hence, $\eta_\mathrm{q} \propto \eta^\sharp_\mathrm{q}$.

\paragraph*{Global Sequential Composition}

If $r$ is an instance of one of rules
\eqref{rule:conc_glob_seq_err_0}--\eqref{rule:conc_glob_seq_ok},
then $1 \leq h \leq 2$ and
$(\theta_{[0]}, \theta^\sharp_{[0]}) \in S$.
By~\eqref{enum:abstract-tree-is-safe:inductive_assume},
$\theta_{[0]}(\emptystring) \propto \theta^\sharp_{[0]}(\emptystring)$.

If $r$ is an instance of~\eqref{rule:conc_glob_seq_err_0},
then $h=1$ and
property~\eqref{enum:abstract-tree-is-safe:immediate-subtrees} holds.
Also, $\theta_{[0]}(\emptystring) \propto \theta^\sharp_{[0]}(\emptystring)$
implies $\exceptstate \propto \exceptstate^\sharp_0$
and hence $\eta_\mathrm{q} \propto \eta^\sharp_\mathrm{q}$.

If $r$ is an instance
of~\eqref{rule:conc_glob_seq_err_1} or
\eqref{rule:conc_glob_seq_ok},
then $h = 2$ and,
since $\sigma_0 \propto \sigma^\sharp_0$,
$(\theta_{[1]} \propto \theta^\sharp_{[1]}) \in S$, so that
property~\eqref{enum:abstract-tree-is-safe:immediate-subtrees} holds.
By~\eqref{enum:abstract-tree-is-safe:inductive_assume}, we have
$\theta_{[1]}(\emptystring) \propto \theta^\sharp_{[1]}(\emptystring)$.
If $r$ is an instance of~\eqref{rule:conc_glob_seq_err_1},
then $\theta_{[1]}(\emptystring) \propto \theta^\sharp_{[1]}(\emptystring)$
implies $\exceptstate \propto \exceptstate^\sharp_1$,
so that $\eta_\mathrm{q} \propto \eta^\sharp_\mathrm{q}$.
If $r$ is an instance of~\eqref{rule:conc_glob_seq_ok},
then $\theta_{[0]}(\emptystring) \propto \theta^\sharp_{[0]}(\emptystring)$
and $\theta_{[1]}(\emptystring) \propto \theta^\sharp_{[1]}(\emptystring)$
imply that $\sigma_1 \propto \sigma^\sharp_1$ and that the two
environments $\rho_0$ and $\rho_1$ are the same in both $r$ and $r^\sharp$.
Hence, their composition $\rho_0[\rho_1]$ is the same in both rules
$r$ and $r^\sharp$,
so that $\eta_\mathrm{q} \propto \eta^\sharp_\mathrm{q}$.

\paragraph*{Local Sequential Composition}

The proof when $r$ is an instance of one of
rules \eqref{rule:conc_decl_seq_err_0}--\eqref{rule:conc_decl_seq_ok}
is similar to that for global sequential composition.

\subsection{Statements}

For this part of the proof,
we use Table~\ref{tab:abstract-tree-is-safe:statements}.
By \eqref{eq:def_S}, $N \propto N^\sharp$.
Thus letting $N = \langle s, \sigma \rangle$
and $N^\sharp = \langle s, \sigma^\sharp \rangle$,
by Definition~\ref{def:nonterminal-abstract-configuration},
we have the implicit hypothesis $\sigma \propto \sigma^\sharp$.
We show using~\eqref{def:propto-terminal:statement}
in Definition~\ref{def:terminal-abstract-configuration},
that $\eta_\mathrm{s} \propto \eta^\sharp_\mathrm{s}$.

\begin{table}
%% \begin{sidewaystable}
\caption{Corresponding concrete and abstract rules and
         terminals for statements}
\label{tab:abstract-tree-is-safe:statements}
\centering
\footnotesize
\begin{tabular}
{|c|l|l|l|l|l|}
\hline
  \multicolumn{1}{|c|}{$S$}
      &\multicolumn{1}{c|}{$r$}
      &\multicolumn{1}{c|}{$r^\sharp$}
      &\multicolumn{1}{c|}{$\eta_\mathrm{s}$}
      &\multicolumn{2}{c|}{
        \(\eta^\sharp_\mathrm{s} =
           \langle
              \sigma^\sharp_a,
              \exceptstate^\sharp_a
           \rangle
         \) }\\
\cline{5-6}
  &&&
      & \multicolumn{1}{c|}{$\sigma^\sharp_a$}
      & \multicolumn{1}{c|}{$\exceptstate^\sharp_a$}\\
\hline
\hline
  $\kw{nop}$
     & \ref{rule:conc_nop}
     & \ref{rule:abstr_nop}
     & $\sigma$
     & $\sigma^\sharp$
     & $\none^\sharp$ \\
\hline
  $\id := e$
     & \ref{rule:conc_assignment_error}
     & \ref{rule:abstr_assignment}
     & $\exceptstate$
     & $\sigma^\sharp_1$
     & $\exceptstate^\sharp_0 \sqcup \exceptstate^\sharp_1$ \\
     & \ref{rule:conc_assignment_ok}
     &
     & $\sigma_0\bigl[ \rho(\id) := \sval \bigr]$
     &
     & \\
\hline
  $s_0 ; s_1$
     & \ref{rule:conc_sequence_0}
     & \ref{rule:abstr_sequence}
     & $\exceptstate$
     & $\sigma^\sharp_1$
     & $\exceptstate^\sharp_0 \sqcup \exceptstate^\sharp_1$ \\
     & \ref{rule:conc_sequence_1}
     &
     & $\eta$
     &
     & \\
\hline
  $d ; s$
     & \ref{rule:conc_block_0}
     & \ref{rule:abstr_block}
     & $\exceptstate$
     & $\sunmarkshort^\sharp(\sigma^\sharp_1)$
     & \(
          \exceptstate^\sharp_0
           \sqcup
             \sunmarkshort^\sharp(\exceptstate^\sharp_1)
       \) \\
     & \ref{rule:conc_block_1}
     &
     & $\sunmarkshort(\eta)$
     &
     & \\
\hline
  $\kw{if} e \kw{then} s_0 \kw{else}\, s_1$
     & \ref{rule:conc_conditional_error}
     & \ref{rule:abstr_conditional}
     & $\exceptstate$
     & $\sigma^\sharp_1 \sqcup \sigma^\sharp_2$
     & \(
         \exceptstate^\sharp_0
           \sqcup \exceptstate^\sharp_1
             \sqcup \exceptstate^\sharp_2
       \) \\
     & \ref{rule:conc_conditional_true}/\ref{rule:conc_conditional_false}
     &
     & $\eta$
     &
     & \\
\hline
  $\kw{while} e \kw{do} s_0$
     & \ref{rule:conc_while_guard_error}/\ref{rule:conc_while_true_body_error}
     & \ref{rule:abstr_while}
     & $\exceptstate$
     & $\sigma^\sharp_\ffv \sqcup \sigma^\sharp_2$
     & \(
         \exceptstate^\sharp_0
           \sqcup \exceptstate^\sharp_1
             \sqcup \exceptstate^\sharp_2
       \) \\
     & \ref{rule:conc_while_false_ok}
     &
     & $\sigma_0$
     &
     & \\
     & \ref{rule:conc_while_true_body_ok}
     &
     & $\eta$
     &
     & \\
\hline
  $\kw{throw} s$
     & \ref{rule:conc_throw_except}
     & \ref{rule:abstr_throw_except}
     & $\langle \sigma, \rtsexcept \rangle$
     & $\bot$
     & $\exceptstate^\sharp$ \\
     & \ref{rule:conc_throw_expr_error}
     & \ref{rule:abstr_throw_expr}
     & $\exceptstate$
     &
     & $\exceptstate^\sharp_0 \sqcup \exceptstate^\sharp_1$ \\
     & \ref{rule:conc_throw_expr_ok}
     &
     & $\langle \sigma_0, \sval \rangle$
     &
     & \\
\hline
  $\kw{try} s \kw{catch} k$
     & \ref{rule:conc_try_no_except}
     & \ref{rule:abstr_try_catch}
     & $\sigma_0$
     & $\sigma^\sharp_0 \sqcup \sigma^\sharp_1$
     & $\exceptstate^\sharp_1 \sqcup \exceptstate^\sharp_2$ \\
     & \ref{rule:conc_try_except}
     &
     & $\eta$
     &
     & \\
\hline
  $\kw{try} s_0 \kw{finally} s_1$
     & \ref{rule:conc_try_finally}
     & \ref{rule:abstr_try_finally}
     & $\eta$
     & $\sigma^\sharp_2$
     & \(
           \exceptstate^\sharp_2 \sqcup
           \exceptstate^\sharp_3 \sqcup
           (\spair{\sigma^\sharp_3}{\except^\sharp_1})
       \) \\
     & \ref{rule:conc_try_finally_exc_0}
     &
     & $\langle \sigma_1, \except_0 \rangle$
     &
     & \\
     & \ref{rule:conc_try_finally_exc_1}
     &
     & $\exceptstate$
     &
     & \\
\hline
  $\id := \id_0(e_1, \ldots, e_n)$
     & \ref{rule:conc_function_call_param_err}
     & \ref{rule:abstr_function_call_eval}
     & $\exceptstate$
     & $\sunmarkshort^\sharp(\sigma^\sharp_2)$
     & $\exceptstate^\sharp = \exceptstate^\sharp_0$ \\
     & \ref{rule:conc_function_call_eval_err}
     &
     & $\sunmarkshort\bigl(\sunlinkshort(\exceptstate)\bigr)$
     &
     & \(\qquad
          \sqcup
          \sunmarkshort^\sharp
            \bigl(
              \sunlinkshort^\sharp(\exceptstate^\sharp_1)
            \bigr)
       \) \\
     & \ref{rule:conc_function_call_eval_ok}
     &
     & $\sunmarkshort(\eta_2)$
     &
     & $\qquad \sqcup \sunmarkshort(\exceptstate^\sharp_2)$ \\
\hline
\end{tabular}
%% \end{sidewaystable}
\end{table}

\paragraph*{Nop}

If $r$ is an instance of~\eqref{rule:conc_nop} then,
by the hypothesis,
$\eta_\mathrm{e} \propto \eta^\sharp_\mathrm{e}$.

\paragraph*{Assignment}

Suppose $r$ is an instance of
\eqref{rule:conc_assignment_error}
or~\eqref{rule:conc_assignment_ok}.
Then $h = 1$
and, by the hypothesis, $(\theta_{[0]}, \theta^\sharp_{[0]}) \in S$
and hence
property~\eqref{enum:abstract-tree-is-safe:immediate-subtrees} holds.
By~\eqref{enum:abstract-tree-is-safe:inductive_assume} we have
$\theta_{[0]}(\emptystring) \propto \theta^\sharp_{[0]}(\emptystring)$.
If $r$ is an instance of~\eqref{rule:conc_assignment_error},
$\exceptstate \propto \exceptstate^\sharp_0$.
Moreover, if $r$ is an instance of~\eqref{rule:conc_assignment_ok},
\(
  \langle \sval, \sigma_0 \rangle
    \propto
      \langle \sval_0^\sharp, \sigma^\sharp_0 \rangle
\)
so that, by Definition~\ref{def:abstract-memory-structure},
\(
  \sigma_0\bigl[ \rho(\id) := \sval \bigr]
    \propto
      \sigma^\sharp_0\bigl[ \rho(\id) := \sval^\sharp \bigr]
\);
letting
\(
   \sigma^\sharp_0\bigl[ \rho(\id) := \sval^\sharp \bigr]
     = (\sigma^\sharp_1, \exceptstate^\sharp_1)
\),
this means that either we have
$\sigma_0\bigl[ \rho(\id) := \sval \bigr] \in \ExceptState$,
so that
$\sigma_0\bigl[ \rho(\id) := \sval \bigr] \propto \exceptstate^\sharp_1$,
or we have
$\sigma_0\bigl[ \rho(\id) := \sval \bigr] \in \Mem$,
so that
$\sigma_0\bigl[ \rho(\id) := \sval \bigr] \propto \sigma^\sharp_1$.
In all cases, $\eta_\mathrm{s} \propto \eta^\sharp_\mathrm{s}$.

\paragraph*{Statement Sequence}

Suppose $r$ is an instance of
\eqref{rule:conc_sequence_0}
or~\eqref{rule:conc_sequence_1}.
Then $1 \le h \le 2$
and, by the hypothesis, $(\theta_{[0]}, \theta^\sharp_{[0]}) \in S$.
By~\eqref{enum:abstract-tree-is-safe:inductive_assume},
$\theta_{[0]}(\emptystring) \propto \theta^\sharp_{[0]}(\emptystring)$.
If $r$ is an instance of rule~\eqref{rule:conc_sequence_0}, as $h = 1$,
property~\eqref{enum:abstract-tree-is-safe:immediate-subtrees} holds
and also $\exceptstate \propto \exceptstate^\sharp_0$.
If $r$ is an instance of \eqref{rule:conc_sequence_1},
then $\sigma_0 \propto \sigma_0^\sharp$
so that $(\theta_{[1]}, \theta^\sharp_{[1]}) \in S$;
also, as $h = 2$,
property~\eqref{enum:abstract-tree-is-safe:immediate-subtrees} holds;
by~\eqref{enum:abstract-tree-is-safe:inductive_assume},
$\theta_{[1]}(\emptystring) \propto \theta^\sharp_{[1]}(\emptystring)$
so that $\eta \propto \langle \sigma^\sharp_1, \exceptstate^\sharp_1 \rangle$.
Hence, in both cases, $\eta_\mathrm{s} \propto \eta^\sharp_\mathrm{s}$.

\paragraph*{Block}

Suppose $r$ is an instance of
\eqref{rule:conc_block_0}
or~\eqref{rule:conc_block_1}.
Then $1 \le h \le 2$
and, by the hypothesis and Definition~\ref{def:abstract-memory-structure},
$(\theta_{[0]}, \theta^\sharp_{[0]}) \in S$.
By~\eqref{enum:abstract-tree-is-safe:inductive_assume},
$\theta_{[0]}(\emptystring) \propto \theta^\sharp_{[0]}(\emptystring)$.
If $r$ is an instance of~\eqref{rule:conc_block_0}, as $h = 1$,
property~\eqref{enum:abstract-tree-is-safe:immediate-subtrees} holds
and also $\exceptstate \propto \exceptstate^\sharp_0$.
If $r$ is an instance of \eqref{rule:conc_block_1},
then $\sigma_0 \propto \sigma_0^\sharp$
so that $(\theta_{[1]}, \theta^\sharp_{[1]}) \in S$;
also, as $h = 2$,
property~\eqref{enum:abstract-tree-is-safe:immediate-subtrees} holds;
by~\eqref{enum:abstract-tree-is-safe:inductive_assume},
$\theta_{[1]}(\emptystring) \propto \theta^\sharp_{[1]}(\emptystring)$;
so that $\eta \propto \langle \sigma^\sharp_1, \exceptstate^\sharp_1 \rangle$
and therefore, by Definition~\ref{def:abstract-memory-structure},
\(
   \sunmark(\eta)
     \propto
       \bigl\langle
         \asunmark(\sigma^\sharp_1),
         \asunmark(\exceptstate^\sharp_1)
       \bigr\rangle
\).
Hence, in both cases $\eta_\mathrm{s} \propto \eta^\sharp_\mathrm{s}$.

\paragraph*{Conditional}

Suppose $r$ is an instance of one of rules
\eqref{rule:conc_conditional_error}--\eqref{rule:conc_conditional_false}.
Then $1 \le h \le 2$ and,
by the hypothesis, $(\theta_{[0]}, \theta^\sharp_{[0]}) \in S$.
By~\eqref{enum:abstract-tree-is-safe:inductive_assume},
$\theta_{[0]}(\emptystring) \propto \theta^\sharp_{[0]}(\emptystring)$.

If $r$ is an instance of~\eqref{rule:conc_conditional_error}, $h = 1$,
property~\eqref{enum:abstract-tree-is-safe:immediate-subtrees} holds
and, as $\exceptstate \propto \exceptstate^\sharp_0$,
$\eta_\mathrm{s} \propto \eta^\sharp_\mathrm{s}$.

If $r$ is an instance of~\eqref{rule:conc_conditional_true}
or~\eqref{rule:conc_conditional_false},
then $h = 2$ and $\sigma_0 \propto \sigma^\sharp_0$.
By the side conditions and Definition~\ref{def:memstruct-filter},
if $\ttv \propto t^\sharp$, then
$\langle \ttv, \sigma_0 \rangle \propto \langle t^\sharp, \sigma_\ttv \rangle$
and, if $\ffv \propto t^\sharp$, then
$\langle \ffv, \sigma_0 \rangle \propto \langle t^\sharp, \sigma_\ffv \rangle$.
Hence, if~\eqref{rule:conc_conditional_true} applies,
$\theta_{[1]}(\emptystring) \propto \theta^\sharp_{[1]}(\emptystring)$
so that $\eta \propto \langle \sigma^\sharp_1, \exceptstate^\sharp_1 \rangle$;
and, if~\eqref{rule:conc_conditional_false} applies,
$\theta_{[1]}(\emptystring) \propto \theta^\sharp_{[2]}(\emptystring)$
so that $\eta \propto \langle \sigma^\sharp_2, \exceptstate^\sharp_2 \rangle$.
Hence, in both cases,
$\eta_\mathrm{s} \propto \eta^\sharp_\mathrm{s}$.

\paragraph*{While}

Suppose $r$ is an instance of one of rules
\eqref{rule:conc_while_guard_error}--\eqref{rule:conc_while_true_body_ok}.
Then $1 \le h \le 3$ and, by hypothesis,
$(\theta_{[0]}, \theta^\sharp_{[0]}) \in S$.
By~\eqref{enum:abstract-tree-is-safe:inductive_assume},
$\theta_{[0]}(\emptystring) \propto \theta^\sharp_{[0]}(\emptystring)$.

If $r$ is an instance of~\eqref{rule:conc_while_guard_error},
$h = 1$,
property~\eqref{enum:abstract-tree-is-safe:immediate-subtrees} holds
and, as $\exceptstate \propto \exceptstate^\sharp_0$,
$\eta_\mathrm{s} \propto \eta^\sharp_\mathrm{s}$.

Suppose $r$ is an instance of
\eqref{rule:conc_while_false_ok},
\eqref{rule:conc_while_true_body_error} or
\eqref{rule:conc_while_true_body_ok}.
By the side conditions and Definition~\ref{def:memstruct-filter},
if $\ttv \propto t^\sharp$, then
$\langle \ttv, \sigma_0 \rangle \propto \langle t^\sharp, \sigma_\ttv \rangle$
and, if $\ffv \propto t^\sharp$, then
$\langle \ffv, \sigma_0 \rangle \propto \langle t^\sharp, \sigma_\ffv \rangle$.

If $r$ is an instance of~\eqref{rule:conc_while_false_ok},
then, as $h = 1$,
property~\eqref{enum:abstract-tree-is-safe:immediate-subtrees} holds
and hence $\eta_\mathrm{s} \propto \eta^\sharp_\mathrm{s}$.

If $r$ is an instance of~\eqref{rule:conc_while_true_body_error},
then $h = 2$.
Thus $(\theta_{[1]}, \theta^\sharp_{[1]}) \in S$ and
property~\eqref{enum:abstract-tree-is-safe:immediate-subtrees} holds.
By~\eqref{enum:abstract-tree-is-safe:inductive_assume},
$\theta_{[1]}(\emptystring) \propto \theta^\sharp_{[1]}(\emptystring)$
so that $\exceptstate \propto \exceptstate^\sharp_1$.
Hence $\eta_\mathrm{s} \propto \eta^\sharp_\mathrm{s}$.

If $r$ is an instance of~\eqref{rule:conc_while_true_body_ok},
then $h = 3$.
Thus  $(\theta_{[1]}, \theta^\sharp_{[1]}) \in S$.
By~\eqref{enum:abstract-tree-is-safe:inductive_assume},
$\theta_{[1]}(\emptystring) \propto \theta^\sharp_{[1]}(\emptystring)$
so that $\sigma_1 \propto \sigma^\sharp_1$.
Thus $(\theta_{[2]}, \theta^\sharp_{[2]}) \in S$ and
property~\eqref{enum:abstract-tree-is-safe:immediate-subtrees} holds.
By~\eqref{enum:abstract-tree-is-safe:inductive_assume},
$\theta_{[2]}(\emptystring) \propto \theta^\sharp_{[2]}(\emptystring)$
so that $\eta \propto \langle \sigma^\sharp_2, \exceptstate^\sharp_2 \rangle$.
Hence $\eta_\mathrm{s} \propto \eta^\sharp_\mathrm{s}$.

\paragraph*{Throw}

Suppose $r$ is an instance of~\eqref{rule:conc_throw_except}.
Then $s = \rtsexcept \in \RTSExcept$
(so that rule~\eqref{rule:abstr_throw_expr}
is not applicable).
By definition of
$\pard{\alpha}{\wp(\RTSExcept)}{\RTSExcept^\sharp}$,
$\rtsexcept \propto \alpha(\{\rtsexcept\})$.
Since, by hypothesis, $\sigma \propto \sigma^\sharp$,
\(
  \spair{\sigma^\sharp}{\alpha(\{\rtsexcept\})}
    = \bigl\langle \sigma^\sharp, \alpha(\{\rtsexcept\}) \bigr\rangle
\)
so that, by the side condition for~\eqref{rule:abstr_throw_except},
$(\sigma, \rtsexcept) \propto \exceptstate^\sharp$.
Hence $\eta_\mathrm{s} \propto \eta^\sharp_\mathrm{s}$.

Suppose $r$ is an instance of
\eqref{rule:conc_throw_expr_error}
or~\eqref{rule:conc_throw_expr_ok}.
Then $s = e \in \Exp$ (so that rule~\eqref{rule:abstr_throw_except}
is not applicable).
By hypothesis, $(\theta_{[0]}, \theta^\sharp_{[0]}) \in S$ and,
as $h = 1$,
property~\eqref{enum:abstract-tree-is-safe:immediate-subtrees} holds.
By~\eqref{enum:abstract-tree-is-safe:inductive_assume},
$\theta_{[0]}(\emptystring) \propto \theta^\sharp_{[0]}(\emptystring)$.
If $r$ is an instance of~\eqref{rule:conc_throw_expr_error},
then $\exceptstate \propto \exceptstate^\sharp_0$,
while, if $r$ is an instance of~\eqref{rule:conc_throw_expr_ok},
$\sval \propto \sval^\sharp$ and
$\sigma_0 \propto \sigma^\sharp_0$.
Hence, in both cases, $\eta_\mathrm{s} \propto \eta^\sharp_\mathrm{s}$.

\paragraph*{Try Blocks}

Suppose $r$ is an instance of
\eqref{rule:conc_try_no_except}--\eqref{rule:conc_try_finally_exc_1}.
By hypothesis, $(\theta_{[0]}, \theta^\sharp_{[0]}) \in S$.
By~\eqref{enum:abstract-tree-is-safe:inductive_assume},
$\theta_{[0]}(\emptystring) \propto \theta^\sharp_{[0]}(\emptystring)$.
Note that if $r$ is an instance of
\eqref{rule:conc_try_no_except} or \eqref{rule:conc_try_except},
only abstract rule \eqref{rule:abstr_try_catch} will be applicable while
if $r$ is an instance of
\eqref{rule:conc_try_finally}--\eqref{rule:conc_try_finally_exc_1},
only abstract rule \eqref{rule:abstr_try_finally} will be applicable.

If $r$ is an instance of \eqref{rule:conc_try_no_except},
$h = 1$,
property~\eqref{enum:abstract-tree-is-safe:immediate-subtrees} holds
and, as $\sigma_0 \propto \sigma^\sharp_0$,
$\eta_\mathrm{s} \propto \eta^\sharp_\mathrm{s}$.

If $r$ is an instance of \eqref{rule:conc_try_except},
then $\exceptstate_0 \propto \exceptstate^\sharp_0$
so that $(\theta_{[1]}, \theta^\sharp_{[1]}) \in S$.
Thus, as $h=2$,
property~\eqref{enum:abstract-tree-is-safe:immediate-subtrees} holds.
By~\eqref{enum:abstract-tree-is-safe:inductive_assume},
$\theta_{[1]}(\emptystring) \propto \theta^\sharp_{[1]}(\emptystring)$
so that
\(
   \langle u, \eta \rangle \propto
      \bigl\langle
        (\sigma^\sharp_1, \exceptstate^\sharp_1), \exceptstate^\sharp_2
      \bigr\rangle
\)
where $u \in \{ \caught, \uncaught \}$.
By Definition~\ref{def:terminal-abstract-configuration},
if $u = \caught$, then
$\eta  \propto \langle \sigma^\sharp_1, \exceptstate^\sharp_1 \rangle$
and, if $u = \uncaught$, then
\(
   \eta  \propto
     \exceptstate^\sharp_2
\).
Hence, in both cases, $\eta_\mathrm{s} \propto \eta^\sharp_\mathrm{s}$.

If $r$ is an instance of rule~\eqref{rule:conc_try_finally},
$\sigma_0 \propto \sigma_0^\sharp$;
hence $(\theta_{[1]}, \theta^\sharp_{[1]}) \in S$ and
property~\eqref{enum:abstract-tree-is-safe:immediate-subtrees} holds.
By~\eqref{enum:abstract-tree-is-safe:inductive_assume},
$\theta_{[1]}(\emptystring) \propto \theta^\sharp_{[1]}(\emptystring)$
so that
$\eta \propto \langle \sigma^\sharp_2, \exceptstate^\sharp_2 \rangle$.
Hence $\eta_\mathrm{s} \propto \eta^\sharp_\mathrm{s}$.

If $r$ is an instance of \eqref{rule:conc_try_finally_exc_0}
or~\eqref{rule:conc_try_finally_exc_1},
\(
   \langle \sigma_0, \except_0 \rangle \propto
     \langle \sigma^\sharp_1, \except^\sharp_1 \rangle
\);
hence $\sigma_0 \propto \sigma^\sharp_1$
and $\except_0 \propto \except^\sharp_1$
so that $(\theta_{[1]}, \theta^\sharp_{[2]}) \in S$ and
property~\eqref{enum:abstract-tree-is-safe:immediate-subtrees} holds.
By~\eqref{enum:abstract-tree-is-safe:inductive_assume},
$\theta_{[1]}(\emptystring) \propto \theta^\sharp_{[2]}(\emptystring)$.
Thus, if \eqref{rule:conc_try_finally_exc_0} applies,
$\sigma_1 \propto \langle \sigma^\sharp_3, \exceptstate^\sharp_3 \rangle$
so that
\(
    \langle \sigma_1, \except_0 \rangle \propto
      (\spair{\sigma^\sharp_3}{\except^\sharp_1})
\);
and, if \eqref{rule:conc_try_finally_exc_1} applies,
$\exceptstate \propto \langle \sigma^\sharp_3, \exceptstate^\sharp_3 \rangle$
so that $\exceptstate \propto \exceptstate^\sharp_3$.
Hence, in both cases, $\eta_\mathrm{s} \propto \eta^\sharp_\mathrm{s}$.

\paragraph*{Function call}

If $r$ is an instance of one of rules
\eqref{rule:conc_function_call_param_err}--\eqref{rule:conc_function_call_eval_ok},
then $1 \leq h \leq 3$ and $\ell = 3$.
Then the conditions  \eqref{eq:function-call-condition-beta-rho-d}
and \eqref{eq:function-call-condition-rho0-rho1}
are also conditions for abstract
rule~\eqref{rule:abstr_function_call_eval}.
By hypothesis and Definition~\ref{def:abstract-memory-structure},
$(\theta_{[0]}, \theta^\sharp_{[0]}) \in S$;
by~\eqref{enum:abstract-tree-is-safe:inductive_assume},
$\theta_{[0]}(\emptystring) \propto \theta^\sharp_{[0]}(\emptystring)$.

If $r$ is an instance
of~\eqref{rule:conc_function_call_param_err}, then
$\exceptstate \propto \exceptstate^\sharp_0$, $h = 1$ and
property~\eqref{enum:abstract-tree-is-safe:immediate-subtrees} holds.
Hence $\eta_\mathrm{s} \propto \eta^\sharp_\mathrm{s}$.

If $r$ is an instance of~\eqref{rule:conc_function_call_eval_err},
then $\sigma_0 \propto \sigma_0^\sharp$
so that, by Definition~\ref{def:abstract-memory-structure},
 $(\theta_{[1]}, \theta^\sharp_{[1]}) \in S$;
also, as $h = 2$,
property~\eqref{enum:abstract-tree-is-safe:immediate-subtrees} holds.
By~\eqref{enum:abstract-tree-is-safe:inductive_assume},
$\theta_{[1]}(\emptystring) \propto \theta^\sharp_{[1]}(\emptystring)$
and $\exceptstate \propto \exceptstate^\sharp_1$;
by Definition~\ref{def:abstract-memory-structure},
\(
   \sunmark\bigl(\sunlink(\exceptstate)\bigr)
     \propto
       \asunmark\bigl(\asunlink(\exceptstate^\sharp_1)\bigr)
\).
Hence $\eta_\mathrm{s} \propto \eta^\sharp_\mathrm{s}$.

If $r$ is an instance of~\eqref{rule:conc_function_call_eval_ok},
then $\sigma_1 \propto \sigma_1^\sharp$
so that, by Definition~\ref{def:abstract-memory-structure},
$(\theta_{[2]}, \theta^\sharp_{[2]}) \in S$;
also, as $h = 3$,
property~\eqref{enum:abstract-tree-is-safe:immediate-subtrees} holds.
By~\eqref{enum:abstract-tree-is-safe:inductive_assume},
$\theta_{[2]}(\emptystring) \propto \theta^\sharp_{[2]}(\emptystring)$
and
$\eta_2 \propto \langle \sigma^\sharp_2, \exceptstate^\sharp_2 \rangle$;
by Definition~\ref{def:abstract-memory-structure},
\(
   \sunmark(\eta_2)
     \propto
       \bigl\langle
         \asunmark(\sigma^\sharp_2),
         \asunmark(\exceptstate_2^\sharp)
       \bigr\rangle
\).
Hence $\eta_\mathrm{s} \propto \eta^\sharp_\mathrm{s}$.

\subsection{Function Bodies}

\begin{table}
\caption{Corresponding concrete and abstract rules and
         terminals for function bodies}
\label{tab:abstract-tree-is-safe:function-bodies}
\centering
\footnotesize
\begin{tabular}
{|c|l|l|l|l|l|}
\hline
  \multicolumn{1}{|c|}{$B$}
      &\multicolumn{1}{c|}{$r$}
      &\multicolumn{1}{c|}{$r^\sharp$} &\multicolumn{1}{c|}{$\eta_\mathrm{b}$}
      &\multicolumn{2}{c|}{
        \(\eta^\sharp_\mathrm{b} =
           \bigl\langle
              (\sval^\sharp_a,\sigma^\sharp_a),
              \exceptstate^\sharp_a
           \bigr\rangle
         \) }\\
\cline{5-6}
  &&&
      & \multicolumn{1}{c|}{$(\sval^\sharp_a,\sigma^\sharp_a)$}
      & \multicolumn{1}{c|}{$\exceptstate^\sharp_a$}\\
\hline
\hline
  $\kw{let} d \,\kw{in} s \kw{result} e$
     & \ref{rule:conc_function_body_0}
     & \ref{rule:abstr_function_body}
     & $\exceptstate$
     & $\sunmarkshort^\sharp(\sigma^\sharp_2)$
     & $\exceptstate^\sharp_3 = \exceptstate^\sharp_0$ \\
     & \ref{rule:conc_function_body_1}
     &
     & $\sunmarkshort(\exceptstate)$
     &
     & $\qquad\; \sqcup \sunmarkshort^\sharp(\exceptstate^\sharp_1 \sqcup \exceptstate^\sharp_2)$ \\
     & \ref{rule:conc_function_body_2}
     &
     & $\sunmarkshort(\eta_0)$
     &
     & \\
\hline
  $\kw{extern} : \sT$
     & \ref{rule:conc_function_body_extern}
     & \ref{rule:abstr_function_body_extern}
     & $\sigma_0 \vbar \langle\sigma_0, \except\rangle$
     & $\sigma^\sharp_0$
     & $(\sigma^\sharp_0, \top)$ \\
\hline
\end{tabular}
\end{table}

For this part of the proof,
we use Table~\ref{tab:abstract-tree-is-safe:function-bodies}.
By \eqref{eq:def_S}, $N \propto N^\sharp$.
Thus letting $N = \langle B, \sigma \rangle$
and $N^\sharp = \langle B, \sigma^\sharp \rangle$,
by Definition~\ref{def:nonterminal-abstract-configuration},
we have the implicit hypothesis $\sigma \propto \sigma^\sharp$.
We show using~\eqref{def:propto-terminal:statement}
in Definition~\ref{def:terminal-abstract-configuration},
that $\eta_\mathrm{b} \propto \eta^\sharp_\mathrm{b}$.

Suppose $r$ is an instance
of one of rules
\eqref{rule:conc_function_body_0}--\eqref{rule:conc_function_body_2}.
By hypothesis and Definition~\ref{def:abstract-memory-structure},
$\smark(\sigma) \propto \asmark(\sigma^\sharp)$, so that
$(\theta_{[0]}, \theta^\sharp_{[0]}) \in S$.
By~\eqref{enum:abstract-tree-is-safe:inductive_assume},
$\theta_{[0]}(\emptystring) \propto \theta^\sharp_{[0]}(\emptystring)$.

If $r$ is an instance of~\eqref{rule:conc_function_body_0},
$\exceptstate \propto \exceptstate^\sharp_0$,
$h=1$ and
property~\eqref{enum:abstract-tree-is-safe:immediate-subtrees} holds.
Hence $\eta_\mathrm{b} \propto \eta^\sharp_\mathrm{b}$.

If $r$ is an instance of~\eqref{rule:conc_function_body_1},
$\sigma_0 \propto \sigma^\sharp_0$;
hence
$(\theta_{[1]}, \theta^\sharp_{[1]}) \in S$
and, as $h = 2$,
property~\eqref{enum:abstract-tree-is-safe:immediate-subtrees} holds.
By~\eqref{enum:abstract-tree-is-safe:inductive_assume},
$\theta_{[1]}(\emptystring) \propto \theta^\sharp_{[1]}(\emptystring)$
so that $\exceptstate \propto \exceptstate^\sharp_1$.
By Definition~\ref{def:abstract-memory-structure},
$\sunmark(\exceptstate) \propto \asunmark(\exceptstate^\sharp_1)$;
hence $\eta_\mathrm{b} \propto \eta^\sharp_\mathrm{b}$.

If $r$ is an instance of~\eqref{rule:conc_function_body_2},
$\sigma_0 \propto \sigma^\sharp_0$;
hence $(\theta_{[1]}, \theta^\sharp_{[1]}) \in S$.
By~\eqref{enum:abstract-tree-is-safe:inductive_assume} we have
$\theta_{[1]}(\emptystring) \propto \theta^\sharp_{[1]}(\emptystring)$,
so that $\sigma_1 \propto \sigma^\sharp_1$; hence
$(\theta_{[2]}, \theta^\sharp_{[2]}) \in S$;
as $h = 3$,
property~\eqref{enum:abstract-tree-is-safe:immediate-subtrees} holds.
Again, by~\eqref{enum:abstract-tree-is-safe:inductive_assume},
$\theta_{[2]}(\emptystring) \propto \theta^\sharp_{[2]}(\emptystring)$;
hence,
$\eta_0 \propto \langle \sigma^\sharp_2, \exceptstate^\sharp_2 \rangle$.
By Definition~\ref{def:abstract-memory-structure},
\(
  \sunmark(\eta_0)
    \propto
      \bigl\langle
        \asunmark(\sigma^\sharp_2),
	\asunmark(\exceptstate^\sharp_2)
      \bigr\rangle
\);
hence $\eta_\mathrm{b} \propto \eta^\sharp_\mathrm{b}$.

Suppose $r$ is an instance of~\eqref{rule:conc_function_body_extern}.
Then $\sigma = (\mu, w)$ and $\sigma_0 = (\mu_0, w)$.
By the hypothesis, $\sigma \propto \sigma^\sharp$;
hence, by the side conditions, $\sigma_0 \propto \sigma^\sharp_0$;
also, $\except \propto \top$,
so that $\eta_\mathrm{b} \propto \eta^\sharp_\mathrm{b}$.

\subsection{Catch Clauses}

For this part of the proof,
we use Table~\ref{tab:abstract-tree-is-safe:catch-clauses}.
By \eqref{eq:def_S}, $N \propto N^\sharp$.
Thus, letting $N = \langle K, \exceptstate \rangle$
and $N^\sharp = \langle K, \exceptstate^\sharp \rangle$,
by Definition~\ref{def:nonterminal-abstract-configuration},
we have the implicit hypothesis $\exceptstate \propto \exceptstate^\sharp$.
We show using~\eqref{def:propto-terminal:catch}
in Definition~\ref{def:terminal-abstract-configuration},
that $\eta_\mathrm{k} \propto \eta^\sharp_\mathrm{k}$.

\begin{table}
\caption{Corresponding concrete and abstract rules and
         terminals for catch clauses}
\label{tab:abstract-tree-is-safe:catch-clauses}
\centering
\footnotesize
\begin{tabular}
{|c|l|l|l|l|l|}
\hline
  \multicolumn{1}{|c|}{$K$}
      &\multicolumn{1}{c|}{$r$}
      &\multicolumn{1}{c|}{$r^\sharp$}
      &\multicolumn{1}{c|}{$\eta_\mathrm{k}$}
      &\multicolumn{2}{c|}{
        \(\eta^\sharp_\mathrm{k} =
           \langle
              \eta^\sharp_a,
              \exceptstate^\sharp_a
           \rangle
         \) }\\
\cline{5-6}
  &&&
      & \multicolumn{1}{c|}{$\eta^\sharp_a$}
      & \multicolumn{1}{c|}{$\exceptstate^\sharp_a$}\\
\hline
\hline
  $(\kw{any}) \, s \vbar (\rtsexcept) \, s \vbar (\sT) \, s$
     & \ref{rule:conc_catch_caught}
     & \ref{rule:abstr_catch_maybe_caught}
     & $\langle \caught, \eta_0 \rangle$
     & $\eta^\sharp_1$
     & $\exceptstate^\sharp_1$ \\
\hline
  $(\id : \sT) \, s$
     & \ref{rule:conc_catch_expr_caught_stkerr}
     & \ref{rule:abstr_catch_expr_maybe_caught}
     & $\bigl\langle \caught, \sunmarkshort(\exceptstate_0) \bigr\rangle$
     & $(\sigma_4, \exceptstate_4) = \bigl(\sunmarkshort^\sharp(\sigma^\sharp_3)$,
     & $\exceptstate^\sharp_1$ \\
     & \ref{rule:conc_catch_expr_caught_noerr}
     &
     & $\langle \caught, \eta_0 \rangle$
     & $\qquad \sunmarkshort^\sharp(\exceptstate^\sharp_2) \sqcup \sunmarkshort^\sharp(\exceptstate^\sharp_3)\bigr)$
     & \\
\hline
  $(\rtsexcept) \, s \vbar (\sT) \, s$
     & \ref{rule:conc_catch_uncaught}
     & \ref{rule:abstr_catch_maybe_caught}
     & $\bigl\langle \uncaught, (\sigma, \except) \bigr\rangle$
     & $\eta^\sharp_1$
     & $\exceptstate^\sharp_1$ \\
\hline
  $(\id : \sT) \, s$
     & \ref{rule:conc_catch_uncaught}
     & \ref{rule:abstr_catch_expr_maybe_caught}
     & $\bigl\langle \uncaught, (\sigma, \except) \bigr\rangle$
     & $(\sigma^\sharp_3, \exceptstate^\sharp_2 \sqcup \exceptstate^\sharp_3)$
     & $\exceptstate^\sharp_1$ \\
\hline
  $k_0;k_1$
     & \ref{rule:conc_catch_seq_caught}
     & \ref{rule:abstr_catch_seq}
     & $\langle \caught, \eta_0 \rangle$
     & \(
         (\sigma^\sharp_0 \sqcup \sigma^\sharp_1,
          \exceptstate^\sharp_0 \sqcup \exceptstate^\sharp_2)
       \)
     & $\exceptstate^\sharp_3$ \\
     & \ref{rule:conc_catch_seq_uncaught}
     &
     & $\eta$
     &
     & \\
\hline
\end{tabular}
\end{table}

\paragraph*{Catch}

Let $K$ have the form $(p) \, s$ for some exception declaration $p$.

Suppose $r$ is an instance of one of rules
\eqref{rule:conc_catch_caught}--\eqref{rule:conc_catch_expr_caught_noerr}.
Then, by the hypothesis and Definition~\ref{def:exceptstate-filter},
$\exceptstate \propto \phi^+(p, \exceptstate^\sharp)$;
by the side conditions for the abstract rules,
$\exceptstate \propto \exceptstate^\sharp_0$.

If $r$ is an instance of~\eqref{rule:conc_catch_caught}
then $\exceptstate = (\sigma, \except)$;
by Definition~\ref{def:abstract-exception-state},
$\sigma \propto \mem(\exceptstate^\sharp_0)$;
Hence $(\theta_{[0]}, \theta^\sharp_{[0]}) \in S$ and, as $h = 1$,
property~\eqref{enum:abstract-tree-is-safe:immediate-subtrees} holds.
By~\eqref{enum:abstract-tree-is-safe:inductive_assume},
$\theta_{[0]}(\emptystring) \propto \theta^\sharp_{[0]}(\emptystring)$,
which implies $\eta_0 \propto \eta^\sharp_1$ so that
$\eta_\mathrm{k} \propto \eta^\sharp_\mathrm{k}$.

If $r$ is an instance of~\eqref{rule:conc_catch_expr_caught_stkerr}
or~\eqref{rule:conc_catch_expr_caught_noerr},
then $\exceptstate = (\sigma, \sval)$ and $\type(\sval) = \sT$;
by Definition~\ref{def:abstract-exception-state},
$\sigma \propto \mem(\exceptstate^\sharp_0)$
and $\sval \propto \sT(\exceptstate^\sharp_0)$.
Hence,
by Definition~\ref{def:abstract-memory-structure},
\begin{equation}
\label{eq:abstract-tree-exists:catch-property}
  \stknew\bigl(\sval, \smark(\sigma) \bigr)
    \propto
      \stknew^\sharp\Bigl(
        \sT(\exceptstate^\sharp_0),
        \asmark\bigl(\mem(\exceptstate^\sharp_0)\bigr)
                    \Bigr)
          = \bigl( (\sigma^\sharp_2, i), \exceptstate^\sharp_2 \bigr).
\end{equation}
If~\eqref{rule:conc_catch_expr_caught_stkerr} applies,
then $h=0$, so that
property~\eqref{enum:abstract-tree-is-safe:immediate-subtrees} holds trivially,
and, by the side condition,
$\exceptstate_0 = \stknew\bigl(\sval, \smark(\sigma) \bigr)$
so that by \eqref{eq:abstract-tree-exists:catch-property},
$\exceptstate_0 \propto \exceptstate^\sharp_2$;
by Definition~\ref{def:abstract-memory-structure},
$\sunmark(\exceptstate_0) \propto \asunmark(\exceptstate^\sharp_2)$.
If~\eqref{rule:conc_catch_expr_caught_noerr} applies,
then, by the side condition,
$(\sigma_0, i)  = \stknew\bigl(\sval, \smark(\sigma) \bigr)$
so that by \eqref{eq:abstract-tree-exists:catch-property},
$\sigma_0 \propto \sigma^\sharp_2$.
Hence,
$(\theta_{[0]}, \theta^\sharp_{[0]}) \in S$ and, as $h = 1$,
property~\eqref{enum:abstract-tree-is-safe:immediate-subtrees} holds.
By~\eqref{enum:abstract-tree-is-safe:inductive_assume},
$\theta_{[0]}(\emptystring) \propto \theta^\sharp_{[0]}(\emptystring)$,
which implies
$\eta_0 \propto \langle \sigma^\sharp_3, \exceptstate^\sharp_3 \rangle$.
Thus, by Definition~\ref{def:abstract-memory-structure},
\(
  \sunmark(\eta_0)
    \propto
      \bigl(
        \asunmark(\sigma^\sharp_3),
        \asunmark(\exceptstate^\sharp_2)
      \bigr).
\)
Hence, in both cases,
$\eta_\mathrm{k} \propto \eta^\sharp_\mathrm{k}$.

If $r$ is an instance of~\eqref{rule:conc_catch_uncaught},
then $h=0$, so that
property~\eqref{enum:abstract-tree-is-safe:immediate-subtrees} holds trivially.
We have $\exceptstate = (\sigma, \except)$ and,
by the side condition,
$p \notin \{ \except, \cT, \kw{any} \}$, where $\cT = \type(\except)$.
If $p \in \{\rtsexcept, \sT\}$
then abstract rule~\eqref{rule:abstr_catch_maybe_caught} applies
so that, by the hypothesis, the side conditions
and Definition~\ref{def:exceptstate-filter},
\(
  (\sigma, \except)
    \propto \phi^-(p, \exceptstate^\sharp)
      = \exceptstate^\sharp_1
\).
Similarly, if $p = \id : \sT$ and
abstract rule~\eqref{rule:abstr_catch_expr_maybe_caught} applies,
\(
  (\sigma, \except)
    \propto \phi^-(\sT, \exceptstate^\sharp)
      = \exceptstate^\sharp_1
\).
Hence, in both cases,
$\eta_\mathrm{k} \propto \eta^\sharp_\mathrm{k}$.

\paragraph*{Catch Sequence}

If $r$ is an instance of~\eqref{rule:conc_catch_seq_caught},
then as $h = 1$ and $(\theta_{[0]}, \theta^\sharp_{[0]}) \in S$,
property~\eqref{enum:abstract-tree-is-safe:immediate-subtrees} holds.
By~\eqref{enum:abstract-tree-is-safe:inductive_assume},
$\theta_{[0]}(\emptystring) \propto \theta^\sharp_{[0]}(\emptystring)$,
so that
\(
  \langle \caught, \eta_0 \rangle
    \propto
      \bigl\langle
        (\sigma^\sharp_0, \exceptstate^\sharp_0), \exceptstate^\sharp_1
      \bigr\rangle
\).
By~\eqref{def:propto-terminal:catch}
in Definition~\ref{def:terminal-abstract-configuration},
\(
  \eta_0 \propto (\sigma^\sharp_0, \exceptstate^\sharp_0)
\),
which implies $\eta_\mathrm{k} \propto \eta^\sharp_\mathrm{k}$.

If $r$ is an instance of~\eqref{rule:conc_catch_seq_uncaught},
then $(\theta_{[0]}, \theta^\sharp_{[0]}) \in S$ and,
by~\eqref{enum:abstract-tree-is-safe:inductive_assume},
$\theta_{[0]}(\emptystring) \propto \theta^\sharp_{[0]}(\emptystring)$.
Thus,
\(
  \langle \uncaught, \exceptstate_0 \rangle
    \propto
      \bigl\langle
        (\sigma^\sharp_0, \exceptstate^\sharp_0), \exceptstate^\sharp_1
      \bigr\rangle
\),
so that, by~\eqref{def:propto-terminal:catch}
in Definition~\ref{def:terminal-abstract-configuration},
 $\exceptstate_0 \propto \exceptstate^\sharp_1$.
Hence $(\theta_{[1]}, \theta^\sharp_{[1]}) \in S$ and,
as $h = 2$,
property~\eqref{enum:abstract-tree-is-safe:immediate-subtrees} holds.
By~\eqref{enum:abstract-tree-is-safe:inductive_assume},
$\theta_{[1]}(\emptystring) \propto \theta^\sharp_{[1]}(\emptystring)$,
so that
\(
  \eta
    \propto
      \bigl\langle
        (\sigma^\sharp_1, \exceptstate^\sharp_2), \exceptstate^\sharp_3
      \bigr\rangle
\),
which implies
$\eta_\mathrm{k} \propto \eta^\sharp_\mathrm{k}$.
\qed
\end{proof}

A few observations regarding the precision of the proposed
approximations are in order.
Consider an abstract tree $\theta^\sharp \in \Theta^\sharp$ such that
\(
  \theta^\sharp(\emptystring)
    = (\rho \vdash_\beta N^\sharp \rightarrow \eta^\sharp)
\),
where $N^\sharp \in \NTs^{\beta\sharp}$ and $\eta^\sharp \in \Ts^\sharp$.
If the concretization functions relating the concrete and abstract domains
are strict, then the abstract tree above will encode the following
\emph{definite information}:
\begin{itemize}
\item
non-terminating computations (i.e., unreachable code),
if $\eta^\sharp = \bot$;
\item
non-exceptional computations, if
$\eta^\sharp = \langle \sigma^\sharp, \none^\sharp \rangle$ and
$\sigma^\sharp \neq \bot$;
\item
exceptional computations, if
$\eta^\sharp = \langle \bot, \exceptstate^\sharp \rangle$ and
$\exceptstate^\sharp \neq \none^\sharp$.
\end{itemize}
Obviously, a precise propagation of this definite information
requires that all of the abstract domain operators are strict too.
Hence, if
\(
  \theta^\sharp(\emptystring)
    = \bigl(
        \rho \vdash_\beta \langle s, \bot \rangle \rightarrow \eta^\sharp
      \bigr)
\),
we will also have $\eta^\sharp = \bot$.
Similar properties hold when considering
expressions, declarations and catch clauses.

\section{Computing Abstract Trees}
\label{sec:computable-approximations}

The results of the previous section
(Theorems~\ref{thm:abstract-tree-exists} and \ref{thm:abstract-tree-is-safe})
guarantee that each concrete tree can be safely approximated by an
abstract tree, provided the non-terminal configurations in the roots
satisfy the approximation relation.

For expository purposes,
suppose we are interested in a whole-program analysis.
For each (concrete and abstract) pair of initial memories
satisfying $\sigma_\mathrm{i} \propto \sigma^\sharp_\mathrm{i}$
and each
\(
  g_0 = (g; \textup{$\kw{gvar} \ridx : \tinteger = 0$})
\),
where $g$ is a valid program,
we obtain that any abstract tree
$\theta^\sharp_0 \in \Theta^\sharp$
such that
\(
  \theta^\sharp_0(\emptystring)
    =
      \bigl(
        \emptyset
          \vdash_\emptyset
            \langle g_0, \sigma^\sharp_\mathrm{i} \rangle
              \rightarrow
                \eta^\sharp_0
      \bigr)
\)
correctly approximates each concrete tree $\theta_0 \in \Theta$
such that
\(
  \theta_0(\emptystring)
    =
      \bigl(
        \emptyset
          \vdash_\emptyset
            \langle g_0, \sigma_\mathrm{i} \rangle
              \rightarrow
                \eta_0
      \bigr)
\).
Notice that $\theta^\sharp_0$ is a finite tree.
Letting
\(
  \eta^\sharp_0
    = \bigl\langle
        (\rho_0, \sigma^\sharp_0),
        \exceptstate^\sharp_0
      \bigr\rangle
\)
and assuming $\eta_0 \notin \ExceptState$,
we obtain $\eta_0 = \langle \rho_0, \sigma_0 \rangle$,
where $\sigma_0 \propto \sigma^\sharp_0$.
Hence,
letting $s_0 = \bigl(\ridx := \main(\emptysequence)\bigr)$
and $\rho_0 : \beta$,
any abstract tree $\theta^\sharp_1 \in \Theta^\sharp$
such that
\(
  \theta^\sharp_1(\emptystring)
    =
      \bigl(
        \rho_0
          \vdash_\beta
            \langle s_0, \sigma^\sharp_0 \bigr\rangle
              \rightarrow
                \eta^\sharp_1
      \bigr)
\)
correctly approximates each concrete tree $\theta_1 \in \Theta$
such that either
\(
  \theta_1(\emptystring)
    =
      \bigl(
        \rho_0
          \vdash_\beta
            \langle s_0, \sigma_0 \bigr\rangle
              \rightarrow
                \eta_1
      \bigr)
\)
or
\(
  \theta_1(\emptystring)
    =
      \bigl(
        \rho_0
          \vdash_\beta
            \langle s_0, \sigma_0 \bigr\rangle
              \diverges
      \bigr)
\).
We are thus left with the problem of computing (any) one of these
abstract trees, which are usually infinite.
In particular, we are interested in choosing $\theta^\sharp_1$
in a subclass of trees admitting finite representations
and, within this class, in maintaining a level of accuracy that is
compatible with the complexity/precision trade-off dictated
by the application.

A classical choice is to restrict attention to rational trees,
that is, trees with only finitely many subtrees:
the algorithm sketched in \cite{Schmidt95,Schmidt97,Schmidt98},
which assumes that the abstract domain is \emph{Noetherian}
(i.e., all of its ascending chains are finite),
guides the analysis toward the computation of a rational tree
by forcing each infinite path to contain a repetition node.
Here below we describe a variation, also working for
abstract domains that admit infinite ascending chains, that exploits
\emph{widening operators} \cite{CousotC76,CousotC77,CousotC92plilp}.

\begin{definition}\summary{(Widening operators.)}
\label{def:widening-operator}
Let $(D^\sharp, \sqsubseteq, \bot, \sqcup)$ be an abstract domain.
The partial operator $\pard{\widen}{D^\sharp \times D^\sharp}{D^\sharp}$
is a \emph{widening} if:
\begin{itemize}
\item
for all $x^\sharp, y^\sharp \in D^\sharp$,
$y^\sharp \sqsubseteq x^\sharp$ implies that
$y^\sharp \widen x^\sharp$ is defined and
$x^\sharp \sqsubseteq y^\sharp \widen x^\sharp$;
\item
for all increasing chains
$x^\sharp_0 \sqsubseteq x^\sharp_1 \sqsubseteq \cdots$,
the increasing chain defined by
$y^\sharp_0 \defeq x^\sharp_0$ and
\(
  y^\sharp_{i+1}
    \defeq
      y^\sharp_i \widen (y^\sharp_i \sqcup x^\sharp_{i+1})
\),
for $i \in \Nset$, is not strictly increasing.
\end{itemize}
\end{definition}

The algorithm works by recursively constructing a finite approximation
for the abstract subtree rooted in the current node
(initially, the root of the whole tree).
Let
\(
  n = \bigl(
        \rho
          \vdash_\beta
            \langle q, y^\sharp_n \rangle
              \rightarrow
                r_n
      \bigr)
\)
be the current node,
where $q$ is a uniquely labeled program phrase,%
\footnote{Unique labels (e.g., given by the address of the root node
for $q$ in the program parse tree) ensure that different occurrences
of the same syntax are not confused~\cite{Schmidt95}; this also means
that, in each node $n$, the type and execution environments $\rho$
and $\beta$ are uniquely determined by $q$.}
$y^\sharp \in D^\sharp$ is either
an abstract memory $\sigma^\sharp \in \Mem^\sharp$ or
an abstract exception state $\exceptstate^\sharp \in \ExceptState^\sharp$,
and $r_n$ is a placeholder for the ``yet to be computed'' conclusion.
The node $n$ is processed according to the following alternatives.
\begin{enumerate}[(i)]
\item
\label{enum-case:expand}
If no ancestor of $n$ is labeled by the program phrase $q$,
the node has to be expanded using an applicable abstract rule instance.
Namely, descendants of the premises of the rule are (recursively)
processed, one at a time and from left to right.
When the expansion of all the premises has been completed, including
the case when the rule has no premise at all, the marker $r_n$
is replaced by an abstract value computed according to the conclusion
of the rule.
\item
\label{enum-case:subsumed}
If there exists an ancestor node
\(
  m = \bigl(
        \rho
          \vdash_\beta
            \langle q, y^\sharp_m \rangle
              \rightarrow
                r_m
      \bigr)
\)
of $n$ labeled by the same program phrase $q$ and such that
$y^\sharp_n \sqsubseteq y^\sharp_m$,
i.e., if node $n$ is subsumed by node $m$,
then the node is not expanded further and
the placeholder $r_n$ is replaced by the least fixpoint of the equation
$r_n = f_m(r_n)$, where $f_m$ is the expression corresponding to
the conclusion of the abstract rule that was used for the expansion
of node $m$.%
\footnote{As explained in \cite{Schmidt95,Schmidt97,Schmidt98},
the computation of such a \emph{least} fixpoint
(in the context of a coinductive interpretation of the abstract rules)
is justified by the fact that here we only need to approximate
the conclusions produced by the terminating concrete computations,
i.e., by the concrete rules that are interpreted inductively.
Also note that the divergence rules have no conclusion at all.}
Intuitively, an infinite subtree rooted in node $m$ has been identified
and the ``repetition node'' $n$ is transformed to a back edge to
the root $m$ of this subtree.
\item
\label{enum-case:widen}
Otherwise, there must be an ancestor node
\(
  m = \bigl(
        \rho
          \vdash_\beta
            \langle q, y^\sharp_m \rangle
              \rightarrow
                r_m
      \bigr)
\)
of $n$ labeled by the same program phrase $q$,
but the subsumption condition
$y^\sharp_n \sqsubseteq y^\sharp_m$ does not hold.
Then, to ensure convergence, the abstract element $y^\sharp_n$ in node $n$
is further approximated by $y^\sharp_m \widen (y^\sharp_m \sqcup y^\sharp_n)$
and we proceed as in case~\eqref{enum-case:expand}.
\end{enumerate}

Termination of the algorithm can be proved thanks to the following
observations:
an infinite abstract tree necessarily has infinite paths
(since the tree is finitely branching);
each infinite path necessarily has an infinite number of nodes labeled by
the same program phrase (since the set of program phrases is finite);
the application of case~\eqref{enum-case:widen} leads to the computation,
along each infinite path, of increasing chains of abstract elements and,
by Definition~\ref{def:widening-operator},
these chains are necessarily finite;
hence, case~\eqref{enum-case:subsumed} is eventually applied
to all infinite paths, leading to a finite representation of the
rational tree where all the infinite paths are expressed by using back edges.

It should be stressed that, as far as efficiency is concerned,
the algorithm outlined above can be improved by the adoption of
well studied memoization techniques;
as noted in \cite{Schmidt97},
by clearly separating design concerns from implementation concerns,
the adopted methodology produces simpler proofs of correctness.
Also note that the choice of the widening operator has a deep impact
on the precision of the results obtained and, moreover, even a precise
widening can lead to inaccurate results if applied too eagerly.
However, precision problems can be mitigated by the application
of suitable ``widening delay'' techniques
\cite{CousotC92plilp,HalbwachsPR97,BagnaraHRZ05SCP}.

\section{Extensions}
\label{sec:extensions}

In this section we outline how the techniques presented in the first part
of the paper can be extended so as to encompass the C language and
all the imperative aspects of \Cplusplus{} (including, of course, exceptions):
Section~\ref{sec:arithemetic-types} shows how the set of primitive types
can be extended by discussing the introduction of bounded integer
and floating-point types;
Section~\ref{sec:c-like-pointers-and-arrays} provides a sketch of how
C-like pointers, arrays and records can be dealt with;
dynamic memory allocation and deallocation is treated in
Section~\ref{sec:heap-memory-management};
and Section~\ref{sec:non-structured-control-flow-mechanisms} illustrates how
all the non-structured control flow mechanisms of C and \Cplusplus{}
can be accounted for.

Once an ABI (\emph{Application Binary Interface}) has been fixed and
its characteristics have been reflected into concrete and abstract
memory structures, C struct and union compound types can be
accommodated, even in presence of pointer casts and unrestricted
pointer arithmetics, by compiling down all their uses to memory reads
and writes performed through pointer dereferencing \cite{Mine06a}.

While we have not yet tried to incorporate object-oriented features
(like classes, inheritance, method calls with dynamic binding
and so forth) we do not see what, in the current design,
would prevent such an extension.

\subsection{Additional Arithmetic Types}
\label{sec:arithemetic-types}

The addition of more arithmetic types such as (signed and unsigned)
finite integer and floating-point types is fairly straightforward.
It is assumed that a preprocessor will add, as needed, a value cast
operator that, for a given numeric type and constant expression,
ensures that either the returned value is in the domain of that type
or an appropriate exception is thrown.
With this assumption, all the operations need only to be specified
for operands of the very same type.

\subsubsection{Syntax}

For floating-point numbers, we add a new basic type $\tfloat$ that represents
a fixed and finite subset of the reals together with a set
of special values denoting infinities, NaN (\emph{Not a Number}) value
and so forth.
The exact format and range of a floating-point literal is unspecified.
The addition of other floating-point types to represent double and extended
precision numbers can be done the same way.
To exemplify the inclusion of signed and unsigned bounded integer types,
we also add the $\tschar$ and $\tuchar$ basic types.
\begin{description}
\item[Integer types]
$\iT \in \iType \defeq \{ \tinteger, \tschar, \tuchar, \ldots \}$;
\item[Numeric types]
$\nT \in \nType \defeq \iType \union \{ \tfloat, \ldots \}$;
\item[Basic types]
$T \in \Type \defeq \nType \union \{ \tboolean \}$;
\item[Floating-point literals]
$\flcon \in \Float$;
\item[Signed char literals]
$\sccon \in \sChar$;
\item[Unsigned char literals]
$\uccon \in \uChar$.
\item[Expressions and constants]
Expressions are extended with floating-point constants,
bounded integer constants, and $\kw{vcast}$,
a \emph{value cast} operator for converting values from
one basic type to another, when possible, or yielding an appropriate
exception:
\begin{align*}
  \Exp \ni
  e &::= \ldots \vbar \flcon \vbar \sccon \vbar \uccon
                \vbar \kw{vcast}(\nT, e) \\
  \Con \ni
  \con &::= \ldots \vbar \flcon \vbar \sccon \vbar \uccon.
\end{align*}
\end{description}

The functions
$\fund{\dom}{\cType}{\{\Integer, \Bool, \RTSExcept, \Float, \sChar, \uChar\}}$
and $\pard{\type}{\sVal}{\sType}$
are easily extended:
\begin{align*}
  \dom(\tfloat)
    &\defeq
        \Float,
&
  \type(\flcon)
    &\defeq
        \tfloat, \\
  \dom(\tschar)
    &\defeq
        \sChar,
&
  \type(\sccon)
    &\defeq
        \tschar, \\
  \dom(\tuchar)
    &\defeq
        \uChar,
&
  \type(\uccon)
    &\defeq
        \tuchar.
\end{align*}

\subsubsection{Static Semantics}

The required adjustments to functions $\FI$ and $\DI$ are straightforward
and thus omitted.
Then, we add the following static semantic rules, where
\(
  \mathord{\boxcircle}
    \in
      \{ \mathord{+}, \mathord{-}, \mathord{*}, \divop, \modop \}
\)
and
\(
  \mathord{\boxast}
    \in
      \{
        \mathord{=}, \mathord{\neq}, \mathord{<},
        \mathord{\leq}, \mathord{\geq}, \mathord{>}
      \}
\):
\begin{description}
\item[Expressions]
\begin{gather*}
\begin{aligned}
&
\prooftree
  \nohyp
\justifies
  \beta \vdash_I \flcon : \tfloat
\endprooftree
&\qquad
&\prooftree
  \nohyp
\justifies
  \beta \vdash_I \sccon : \tschar
\endprooftree \\[1ex]
&
\prooftree
 \beta \vdash_I e : \nT
\justifies
 \beta \vdash_I -e : \nT
\endprooftree
&
&\prooftree
  \nohyp
\justifies
  \beta \vdash_I \uccon : \tuchar
\endprooftree \\[1ex]
&
\prooftree
 \beta \vdash_I e_0 : \nT
\quad
 \beta \vdash_I e_1 : \nT
\justifies
  \beta \vdash_I e_0 \boxcircle e_1 : \nT
\endprooftree
&
&\prooftree
 \beta \vdash_I e_0 : \nT
\quad
 \beta \vdash_I e_1 : \nT
\justifies
  \beta \vdash_I e_0 \boxast e_1 : \tboolean
\endprooftree \\[1ex]
&
\prooftree
 \beta \vdash_I e : T_0
\justifies
  \beta \vdash_I \kw{vcast}(T_1, e) : T_1
\using\quad\text{\hbox to0pt{if casting $T_0$ to $T_1$ is legal.}}
\endprooftree
\end{aligned}
\end{gather*}

\end{description}

\subsubsection{Concrete Dynamic Semantics}
The added numeric types and the operations upon them
bring in a considerable degree of complexity.  Consider the C language,
for example: unsigned bounded integers employ modular arithmetic;
for signed bounded integers, overflow yields undefined behavior;
the results of floating-point operations depend on the rounding mode
in effect and on the settings that cause floating-point exceptions to be
trapped or ignored;
relational operators may or may not raise a floating-point exception
when one or both arguments are NaN.
In order to factor out these details and delegate them to the memory
structure, we resort to a device like the one used to model supported
and unsupported language elements in the abstract semantics.
We thus postulate the existence of the partial functions
\begin{align*}
  \pard{\ceval{\mathrm{vc}}}{&(\nType \times \Con \times \Mem)}%
                            {\ValState \uplus \ExceptState}, \\
  \pard{\ceval{-_1}}{&(\Con \times \Mem)}%
                    {\ValState \uplus \ExceptState}, \\
  \pard{\ceval{\boxcircle}}{&(\Con \times \Con \times \Mem)}%
                           {\ValState \uplus \ExceptState}, \\
  \pard{\ceval{\boxast}}{&(\Con \times \Con \times \Mem)}%
                        {\ValState \uplus \ExceptState},
\end{align*}
that model the cast operator, unary minus, binary operators
\(
  \mathord{\boxcircle}
    \in
      \{ \mathord{+}, \mathord{-}, \mathord{*}, \divop, \modop \}
\)
and relational operators
\(
  \mathord{\boxast}
    \in
      \{
        \mathord{=}, \mathord{\neq}, \mathord{<},
        \mathord{\leq}, \mathord{\geq}, \mathord{>}
      \}
\),
respectively.
Such functions need not be always defined:
for example, there is no need to define $\ceval{+}(\con_0, \con_1, \sigma)$
for the case $\type(\con_0) \neq \type(\con_1)$.
\begin{description}
\item[Value casts]
The following concrete rule schemata use the corresponding evaluation
function to specify the execution of the $\kw{vcast}$ operator.
\begin{align*}
&\prooftree
  \rho \vdash_\beta \langle e, \sigma \rangle
    \rightarrow
      \exceptstate
\justifies
  \rho \vdash_\beta \langle \kw{vcast}(\nT, e), \sigma \rangle
    \rightarrow
      \exceptstate
\endprooftree
&
&\prooftree
  \rho \vdash_\beta \langle e, \sigma \rangle
    \rightarrow
      \langle \con, \sigma_0 \rangle
\justifies
  \rho \vdash_\beta \langle \kw{vcast}(\nT, e), \sigma \rangle
    \rightarrow
      \ceval{\mathrm{vc}}(\nT, \con, \sigma_0)
\endprooftree
\end{align*}
\item[Arithmetic evaluation]
By using the evaluation functions, we can substitute rules
\eqref{rule:conc_uminus_ok}, \eqref{rule:conc_arith_bop_2}
and \eqref{rule:conc_arith_bop_exc_0}
with the following (note that they also capture the case when
a divide-by-zero exception is thrown):
\begin{align*}
&\prooftree
  \rho \vdash_\beta \langle e, \sigma \rangle
    \rightarrow
      \langle \con, \sigma_0 \rangle
\justifies
  \rho \vdash_\beta \langle -e, \sigma \rangle
    \rightarrow
      \ceval{-_1}(\nT, \con, \sigma_0)
\endprooftree \\[1ex]
&\prooftree
  \rho \vdash_\beta \langle e_0, \sigma \rangle
    \rightarrow
      \langle \con_0, \sigma_0 \rangle
\quad
  \rho \vdash_\beta \langle e_1, \sigma_0 \rangle
    \rightarrow
      \langle \con_1, \sigma_1 \rangle
\justifies
  \rho \vdash_\beta \langle e_0 \boxcircle e_1, \sigma \rangle
    \rightarrow
      \ceval{\boxcircle}(\con_0, \con_1, \sigma_1)
\endprooftree
\end{align*}
\item[Arithmetic tests]
Similarly, rule \eqref{rule:conc_arith_test_ok}
is replaced by the more general rule
\[
\prooftree
  \rho \vdash_\beta \langle e_0, \sigma \rangle
    \rightarrow
      \langle \con_0, \sigma_0 \rangle
\quad
  \rho \vdash_\beta \langle e_1, \sigma_0 \rangle
    \rightarrow
      \langle \con_1, \sigma_1 \rangle
\justifies
  \rho \vdash_\beta \langle e_0 \boxast e_1, \sigma \rangle
    \rightarrow
      \ceval{\boxast}(\con_0, \con_1, \sigma_1)
\endprooftree
\]
\end{description}

\subsection{C-like Pointers, Arrays and Records}
\label{sec:c-like-pointers-and-arrays}

\subsubsection{Syntax}

Recall that in Sections~\ref{sec:syntax} and~\ref{sec:static-semantics}
we defined the set of storable types,
whose values can be read from and written to memory,
and the set of denotable types, that can occur in declarations.
The introduction of pointer, array and record types requires the adoption
of a finer classification.
The set of all \emph{memory types} is partitioned into \emph{object types}
and \emph{function types}: the latter differ in that we cannot read
or update the ``value'' of a function; rather, we execute it.
Object types are further partitioned into \emph{elementary types}
(also called scalar types, including basic types and pointer types)
and \emph{aggregate types} (arrays and records).
All the elementary types are storable, meaning that their values
can be read directly from or written directly to memory, as well as
passed to and returned from functions.
Regarding aggregate types, the C language prescribes that
record types are storable, whereas array types are not.
Pointer, array and record type derivations can be applied
repeatedly to obtain, e.g., multi-dimensional arrays.
\begin{description}
\item[Types]
\begin{align*}
  \eType \ni
  \eT &::= T \vbar \pT
& \oType \ni
  \oT &::= \sT \vbar \aT \\
  \pType \ni
  \pT &::= \pointer{\mT}
& \fType \ni
  \fT &::= \fps \to \sT \\
  \sType \ni
  \sT &::= \eT \vbar \rT
& \mType \ni
  \mT &::= \oT \vbar \fT \\
  \aType \ni
  \aT &::= \arraytype{m}{\oT}
& \dType \ni
  \dT &::= \location{\mT} \\
  \rType \ni
  \rT &::= \recordtype{\id}{\id_1 : \oT_1, \ldots, \id_j : \oT_j}\hspace{-1cm}
\end{align*}
\end{description}
We assume, without loss of generality, that the field names
of record types are unique across the entire program
(for example, $\id_1$, \dots,~$\id_j$ could contain $\id$ as some
kind of special prefix).

Identifiers are no longer the only way to denote
a memory structure location. This can also be referred to by
combining a pointer with the indirection operator `$\mathord{\ast}$',
an array with the indexing operator,
or a record with the field selection operator.
Hence, we introduce the concept of \emph{lvalue},
which can be read as ``location-valued expression.''

\begin{description}
\item[Offsets and lvalues]
\begin{align*}
& \Offset \ni
  o ::= \nooffset \vbar
    \indexoffset{e} \cdot o \vbar
    \fieldoffset{\id} \cdot o \\
& \lValue \ni
  \lvalue ::= \id \cdot o
    \vbar (\indirection{e}) \cdot o
\end{align*}
\end{description}
Consequently, the syntactic production for expressions generating
identifiers, as well as the productions for statements generating
assignments and function calls, are replaced by more general versions
using lvalues;
expressions and declarations are also extended
with the address-of operator, null pointers and array variables.
\begin{description}
\item[Expressions, declarations and statements]
\begin{align*}
& \Exp \ni
  e ::= \ldots
    \vbar \kw{val} \lvalue
    \vbar \maddress{\lvalue}
    \vbar (\pT)\,0
&\quad
& \Glob \ni
  g ::= \ldots
      \vbar \kw{gvar} \id : \aT = e \\
& \Stmt \ni
  s ::= \ldots
    \vbar \lvalue := e
    \vbar \lvalue := e(\es)
&\quad
& \Decl \ni
  d ::= \ldots
      \vbar \kw{lvar} \id : \aT = e
\end{align*}
\end{description}

\subsubsection{Static Semantics}

The required adjustments to functions $\FI$ and $\DI$ are straightforward
and thus omitted.
The well-typedness of offsets and lvalues
is encoded by the following predicates:
\begin{align*}
\beta, \dT_0 &\vdash_I o : \dT_1,
  &\text{$o$ is compatible with $\dT_0$ and has type $\dT_1$ in $\beta$;} \\
\beta &\vdash_I \lvalue : \dT,
  &\text{$\lvalue$ is well-formed and has type $\dT$ in $\beta$.}
\end{align*}
The static semantics is thus extended by the following rules.%
\footnote{The previous rules for identifier, assignment and function call
are no longer used.}
Note that the evaluation of an lvalue as an expression
---$\kw{val} \lvalue$---
causes a suitable type conversion,
sometimes referred to as ``type decay.''
Pointer arithmetics can only be applied to object types.
In function calls, the callee is specified via an expression
having function pointer type (typically resulting from a type decay).

\begin{description}
\item[Offset]
\begin{align*}
&
\begin{aligned}
%% No offset.
\prooftree
  \nohyp
\justifies
  \beta, \dT \vdash_I \nooffset : \dT
\endprooftree
&\quad&
%% Indexing.
\prooftree
  \beta \vdash_I e : \tinteger
\quad
  \beta, \location{\oT} \vdash_I o : \dT
\justifies
  \beta, \location{(\arraytype{m}{\oT})}
    \vdash_I
      \indexoffset{e} \cdot o : \dT
\endprooftree
\end{aligned} \\[1ex]
&
%% Field offset.
\prooftree
  \beta, \location{\oT_i} \vdash_I o : \dT
\justifies
  \beta, \location{(\recordtype{\id}{\id_1 : \oT_1; \ldots; \id_j : \oT_j})}
    \vdash_I
      \fieldoffset{\id_i} \cdot o : \dT
\using\quad\text{if $i \in \{1, \ldots, j\}$}
\endprooftree
\end{align*}
\item[Lvalue]
\begin{align*}
&
%% Identifier.
\prooftree
  \beta, \dT_0 \vdash_I o : \dT_1
\justifies
  \beta \vdash_I \id \cdot o : \dT_1
\using\quad\text{if $\beta(\id) = \dT_0$}
\endprooftree
&\quad&
%% Indirection.
\prooftree
  \beta \vdash_I e : \pointer{\mT}
\quad
  \beta, \location{\mT} \vdash_I o : \dT
\justifies
  \beta \vdash_I (\indirection{e}) \cdot o : \dT
\endprooftree
\end{align*}
\item[Null pointer and address-of operator]
\begin{gather*}
%% Null pointer constant.
\prooftree
  \nohyp
\justifies
  \beta \vdash_I (\pT)\,0 : \pT
\endprooftree
\qquad
%% Address operator.
\prooftree
  \beta \vdash_I \lvalue : \location{\mT}
\justifies
  \beta \vdash_I \maddress{\lvalue} : \pointer{\mT}
\endprooftree
\end{gather*}
\item[Type decay]
\begin{gather*}
%% lvalue-to-value: storable type.
\prooftree
  \beta \vdash_I \lvalue : \location{\sT}
\justifies
  \beta \vdash_I \kw{val} \lvalue : \sT
\endprooftree
\qquad
%% id-to-value: array to pointer decay.
\prooftree
  \beta \vdash_I \lvalue : \location{(\arraytype{m}{\oT})}
\justifies
  \beta \vdash_I \kw{val} \lvalue : \pointer{\oT}
\endprooftree
\qquad
%% lvalue-to-value: function to pointer decay.
\prooftree
  \beta \vdash_I \lvalue : \location{\fT}
\justifies
  \beta \vdash_I \kw{val} \lvalue : \pointer{\fT}
\endprooftree
\end{gather*}
\item[Pointer arithmetics]
\begin{gather*}
\begin{aligned}
&
%% Pointer plus integer.
\prooftree
  \beta \vdash_I e_0 : \pointer{\oT}
\quad
  \beta \vdash_I e_1 : \tinteger
\justifies
  \beta \vdash_I e_0 + e_1 : \pointer{\oT}
\endprooftree
&\quad&
%% Integer plus pointer.
\prooftree
  \beta \vdash_I e_0 : \tinteger
\quad
  \beta \vdash_I e_1 : \pointer{\oT}
\justifies
  \beta \vdash_I e_0 + e_1 : \pointer{\oT}
\endprooftree \\[1ex]
&
%% Pointer minus integer.
\prooftree
  \beta \vdash_I e_0 : \pointer{\oT}
\quad
  \beta \vdash_I e_1 : \tinteger
\justifies
  \beta \vdash_I e_0 - e_1 : \pointer{\oT}
\endprooftree
&\quad&
%% Pointer difference.
\prooftree
  \beta \vdash_I e_0 : \pointer{\oT}
\quad
  \beta \vdash_I e_1 : \pointer{\oT}
\justifies
  \beta \vdash_I e_0 - e_1 : \tinteger
\endprooftree
\end{aligned}
\end{gather*}
\item[Pointer comparison]
\[
%% Pointer comparison.
\prooftree
  \beta \vdash_I e_0 : \pT
\quad
  \beta \vdash_I e_1 : \pT
\justifies
  \beta \vdash_I e_0 \boxast e_1 : \tboolean
\using\quad\text{where
\(
  \mathord{\boxast}
    \in
      \{
        \mathord{=}, \mathord{\neq}, \mathord{<},
        \mathord{\leq}, \mathord{\geq}, \mathord{>}
      \}
\).}
\endprooftree
\]
\item[Assignment and function call]
\begin{gather*}
\begin{aligned}
&
%% Assignment.
\prooftree
  \beta \vdash_I \lvalue : \location{\sT}
\quad
  \beta \vdash_I e : \sT
\justifies
  \beta \vdash_I \lvalue := e
\endprooftree \\[1ex]
&
%% Function call.
\prooftree
  \beta \vdash_I \lvalue : \location{\sT}
    \quad
  \beta \vdash_I e : \pointer{(\fps \rightarrow \sT)}
\quad
  \beta, \fps \vdash_I \es
\justifies
  \beta \vdash_I \lvalue := e(\es)
\endprooftree
\end{aligned}
\end{gather*}
\item[(Multi-dimensional) Global array declaration]
\begin{align*}
\prooftree
  \beta \vdash_I \kw{gvar} \id : \oT = e
    : \{ \id \mapsto \location{\oT} \}
\justifies
  \beta
    \vdash_I
      \kw{gvar} \id : \arraytype{m}{\oT} = e
       : \bigl\{
            \id \mapsto \location{(\arraytype{m}{\oT})}
         \bigr\}
\using\quad\text{if $m > 0$}
\endprooftree
\end{align*}
The static semantics rule for a local array declaration is similar.
\end{description}

\subsubsection{Concrete Dynamic Semantics}

Concrete execution environments now map function identifiers to
(properly typed) locations, rather than function abstracts:
hence, we redefine
\(
  \dVal
    \defeq
      \Addr \times \mType
\).

A proper handling of aggregate and function types in memory structures
requires a few semantic adjustments and extensions.
New memory functions allow the allocation of function abstracts
in the text segment, as well as the contiguous allocation
of a number of memory cells, so as to model (multi-dimensional) arrays:
\begin{align*}
  \fund{\txtnew}%
       {&(\Abstract \times \Mem)}%
       {\bigl( (\Mem \times \Loc) \uplus \ExceptState \bigr)}, \\
  \fund{\arraydatnew}%
       {&(\Integer \times \ValState)}%
       {\bigl( (\Mem \times \Loc) \uplus \ExceptState \bigr)}, \\
  \fund{\arraystknew}%
       {&(\Integer \times \ValState)}%
       {\bigl( (\Mem \times \Ind) \uplus \ExceptState \bigr)}.
\end{align*}
It can be observed that the properties stated
in Definition~\ref{def:concrete-memory-structure}
still hold as long as we consider locations having non-aggregate type
and properly extend the domain and codomain of the absolute memory map:
\[
  \Map
    \defeq
      \left(\Loc \times (\eType \uplus \fType)\right)
        \rightarrowtail
          (\Con \uplus \Loc \uplus \Abstract).
\]
These ``elementary'' memory maps need to be extended to read or
update record values. To this end, we assume the existence
of a couple of helper functions working on locations having
aggregate type:
\begin{align*}
  \pard{\locfield}%
       {&(\Id \times \Loc \times \rType)}%
       {(\Loc \times \oType)}, \\
  \pard{\locindex}%
       {&(\Integer \times \Loc \times \aType)}%
       {(\Loc \times \oType)}.
\end{align*}
Intuitively, when defined, these functions map
a record (resp., array) typed location to the typed location
of one of its record fields (resp., array elements).
Hence, for each $\mu \in \Map$, the extension
\(
  \pard{\mu}{(\Loc \times \sType)}{\sVal}
\)
can be recursively obtained, for each $l \in \Loc$ and
$\rT = \recordtype{\id}{\id_1 : \oT_1; \ldots; \id_j : \oT_j}$,
as follows and under the following conditions:
\begin{gather*}
  \mu(l, \rT)
    \defeq
      \Bigl\langle
        \mu\bigl(\locfield(\id_1, l, \rT)\bigr),
        \ldots,
        \mu\bigl(\locfield(\id_j, l, \rT)\bigr)
      \Bigr\rangle, \\
\intertext{%
where, for each $l \in \Loc$ and $\aT = \arraytype{m}{\oT} \in \aType$,
}
  \mu(l, \aT)
    \defeq
      \Bigl[
        \mu\bigl(\locindex(0, l, \aT)\bigr),
        \ldots,
        \mu\bigl(\locindex(m-1, l, \aT)\bigr)
      \Bigr].
\end{gather*}
A similar extension is required for the memory update operator.
%% for instance, under the conditions above, we can recursively define
%% \begin{gather*}
%%   \mu\bigl[(l, \rT) := \langle \sval_1, \ldots, \sval_j \rangle \bigr]
%%     \defeq
%%       \mu_j, \\
%% \intertext{%
%% where $\mu_0 = \mu$ and, for each $i = 1$, \ldots, $j$,
%% }
%%   \mu_i
%%     \defeq
%%       \mu_{i-1}\bigl[ \field(\id_i, l, \rT) := \sval_i \bigr].
%% \end{gather*}
%%
Note that we will still use $\valstate$ as a syntactic meta-variable for
$\ValState = \sVal \times \Mem$, but now its first component can be
either a constant, or an absolute location, or a record value.

Pointer and array indexing errors are modeled via RTS exceptions.
%% For simplicity of exposition, the RTS exception $\memerror$ will be thrown;
%% a refined specification can (and should) provide a better distinction
%% between the various kinds of pointer errors.
%%
It is assumed there exists a special location $\locnull \in \Loc$
(the \emph{null pointer} value) such that
$(\locnull, \mT) \notin \dom(\sigma)$
for all $\sigma \in \Mem$ and $\mT \in \mType$;
this also implies that $\locnull$ cannot be returned by
the memory allocation operators.
Hence, any attempt to read from or write to memory
through this location will result in an exception state.
Suitable operators on memory structures are required to check
the constraints regarding pointer arithmetics
(e.g., out-of-bounds array accesses),
pointer comparisons (where $\mathord{\boxast}$ ranges over
\(
  \{
    \mathord{=}, \mathord{\neq}, \mathord{<},
    \mathord{\leq}, \mathord{\geq}, \mathord{>}
  \}
\)%
)
and to perform ``array-to-pointer decay'' conversions
or record field selections:
\begin{align*}
  \fund{\ptrmove}%
       {&(\Integer \times \Loc \times \Mem)}%
       {\ValState \uplus \ExceptState}, \\
  \fund{\ptrdiff}%
       {&(\Loc \times \Loc \times \Mem)}%
       {\ValState \uplus \ExceptState}, \\
  \fund{\ptrcmp_\boxast}%
       {&(\Loc \times \Loc \times \Mem)}%
       {\ValState \uplus \ExceptState}, \\
  \fund{\firstof}%
       {&(\Loc \times \Mem)}%
       {\ValState \uplus \ExceptState}, \\
  \fund{\field}%
       {&(\Id \times \Loc \times \Mem)}%
       {\ValState \uplus \ExceptState}.
\end{align*}
Note that array indexing is semantically equivalent to
a suitable combination of type decay, pointer arithmetics
and pointer indirection.
Nonetheless, for the sake of clarity and also to simplify the application
of pointer and array dependence analyses~\cite{EmamiGH94},
we keep the distinction of the two constructs and,
to simplify notation, we define%
\footnote{Functions `$\field$' and `$\arrayindex$' are similar to
`$\locfield$' and `$\locindex$', but they are also meant to check
their arguments against the memory structure, possibly returning
an RTS exception.}
\[
  \pard{\arrayindex}%
       {(\Loc \times \ValState)}%
       {\ValState \uplus \ExceptState}
\]
as follows:
\[
  \arrayindex\bigl(l, (m, \sigma)\bigr)
    \defeq
      \begin{cases}
        \exceptstate,
          &\text{if $\firstof(l, \sigma) = \exceptstate$;} \\
        \ptrmove(m, l_0, \sigma_0),
          &\text{if $\firstof(l, \sigma) = (l_0, \sigma_0)$.}
      \end{cases}
\]

Non-terminal and terminal configurations are extended so as to
allow for the syntactic categories of offsets and lvalues,
whose non-exceptional evaluation leads to a location:
\begin{align*}
  \NTo^\beta
    &\defeq
      \sset{%
        \langle o, l, \sigma \rangle
          \in \Offset \times \Loc \times \Mem
      }{%
        \exists \dT_0, \dT_1 \in \dType \st \\
          \beta, \dT_0 \vdash_I o : \dT_1
      }, \\
  \NTl^\beta
    &\defeq
      \bigl\{\,
        \langle \lvalue, \sigma \rangle
          \in \lValue \times \Mem
      \bigm|
        \exists \dT \in \dType \st \beta \vdash_I \lvalue : \dT
      \,\bigr\}, \\
  \To
    &\defeq
  \Tl
    \defeq
      (\Loc \times \Mem) \uplus \ExceptState,
\end{align*}
The dynamic concrete semantics is extended with the following rule schemata.

\begin{description}
\item[Offset]
\begin{align*}
%% No offset.
&
\prooftree
  \nohyp
\justifies
  \rho \vdash_\beta \langle \nooffset, l, \sigma \rangle
    \rightarrow
      \langle l, \sigma \rangle
\endprooftree \\[1ex]
%% Indexing offset.
&
\begin{aligned}
\prooftree
  \rho \vdash_\beta
    \langle e, \sigma \rangle
      \rightarrow
        \exceptstate
\justifies
  \rho \vdash_\beta
    \bigl\langle
      \indexoffset{e} \cdot o, l, \sigma
    \bigr\rangle
      \rightarrow
        \exceptstate
\endprooftree
&\quad&
\prooftree
  \rho \vdash_\beta
    \langle e, \sigma \rangle
      \rightarrow
        \valstate
\justifies
  \rho \vdash_\beta
    \bigl\langle
      \indexoffset{e} \cdot o, l, \sigma
    \bigr\rangle
      \rightarrow
        \exceptstate
\using\quad\text{if $\arrayindex(l, \valstate) = \exceptstate$}
\endprooftree
\end{aligned} \\[1ex]
&
\prooftree
  \rho \vdash_\beta
    \langle e, \sigma \rangle
      \rightarrow
        \valstate
\quad
  \rho \vdash_\beta
    \langle o, l_0, \sigma_0 \rangle
        \rightarrow
          \eta
\justifies
  \rho \vdash_\beta
    \bigl\langle
      \indexoffset{e} \cdot o, l, \sigma
    \bigr\rangle
      \rightarrow
        \eta
\using\quad\text{if $\arrayindex(l, \valstate) = (l_0, \sigma_0)$}
\endprooftree \\[1ex]
%% Field offset.
&
\prooftree
  \nohyp
\justifies
  \rho \vdash_\beta
    \langle \fieldoffset{\id_i} \cdot o, l, \sigma \rangle
      \rightarrow
        \exceptstate
\using\quad\text{if $\field(\id_i, l, \sigma) = \exceptstate$}
\endprooftree \\[1ex]
&
\prooftree
  \rho
    \vdash_\beta
      \langle o, l_0, \sigma_0 \rangle
        \rightarrow
          \eta
\justifies
  \rho \vdash_\beta
    \langle \fieldoffset{\id_i} \cdot o, l, \sigma \rangle
      \rightarrow
        \eta
\using\quad\text{if $\field(\id_i, l, \sigma) = (l_0, \sigma_0)$}
\endprooftree
\end{align*}
\item[Lvalue]
\begin{align*}
&
\prooftree
  \rho \vdash_\beta
    \langle o, \sigma \mathbin{@} a, \sigma \rangle
      \rightarrow
        \eta
\justifies
  \rho \vdash_\beta \langle \id \cdot o, \sigma \rangle
    \rightarrow
      \eta
\using\quad\text{if $\rho(\id) = (a, \mT)$}
\endprooftree \\[1ex]
&
\begin{aligned}
\prooftree
  \rho \vdash_\beta
    \langle e, \sigma \rangle
      \rightarrow
        \exceptstate
\justifies
  \rho \vdash_\beta
    \bigl\langle
      (\indirection{e}) \cdot o,
      \sigma
    \bigr\rangle
      \rightarrow
        \exceptstate
\endprooftree
&\quad&
\prooftree
  \rho \vdash_\beta
    \langle e, \sigma \rangle
      \rightarrow
        \langle l_0, \sigma_0 \rangle
\quad
  \rho \vdash_\beta \langle o, l_0, \sigma_0 \rangle \rightarrow \eta
\justifies
  \rho \vdash_\beta
    \bigl\langle
      (\indirection{e}) \cdot o,
      \sigma
    \bigr\rangle
      \rightarrow
        \eta
\endprooftree
\end{aligned}
\end{align*}
\item[Null pointer and address-of operator]
\begin{align*}
&
\prooftree
  \nohyp
\justifies
  \rho \vdash_\beta \bigl\langle (\pT)\,0, \sigma \bigr\rangle
    \rightarrow
      \langle \locnull, \sigma \rangle
\endprooftree
&\quad&
%% Address operator.
\prooftree
  \rho \vdash_\beta
    \langle \lvalue, \sigma \rangle
      \rightarrow
        \eta
\justifies
  \rho \vdash_\beta
    \langle \maddress{\lvalue}, \sigma \rangle
      \rightarrow
        \eta
\endprooftree
\end{align*}
\item[Type decay]
\begin{align*}
&
\begin{aligned}
\prooftree
  \rho \vdash_\beta \langle \lvalue, \sigma \rangle
    \rightarrow \exceptstate
\justifies
  \rho \vdash_\beta
    \langle \kw{val} \lvalue, \sigma \rangle
      \rightarrow
        \exceptstate
\endprooftree
\end{aligned} \\[1ex]
&
%% Lvalue-to-value: storable type.
\prooftree
  \rho \vdash_\beta \langle \lvalue, \sigma \rangle
    \rightarrow
      \langle l, \sigma_0 \rangle
\justifies
  \rho \vdash_\beta
    \langle \kw{val} \lvalue, \sigma \rangle
      \rightarrow
        \sigma_0[l, \sT]
\using\quad\text{if
$\beta \vdash_{\FI(\lvalue)} \lvalue : \location{\sT}$}
\endprooftree \\[1ex]
&
%% Lvalue-to-value: array to pointer decay.
\prooftree
  \rho \vdash_\beta \langle \lvalue, \sigma \rangle
    \rightarrow
      \valstate
\justifies
  \rho \vdash_\beta
    \langle \kw{val} \lvalue, \sigma \rangle
      \rightarrow
        \firstof(\valstate)
\using\quad\text{if
$\beta \vdash_{\FI(\lvalue)} \lvalue : \location{\aT}$}
\endprooftree \\[1ex]
&
%% Lvalue-to-value: function to pointer decay.
\prooftree
  \rho \vdash_\beta \langle \lvalue, \sigma \rangle
    \rightarrow
      \valstate
\justifies
  \rho \vdash_\beta
    \langle \kw{val} \lvalue, \sigma \rangle
      \rightarrow
        \valstate
\using\quad\text{if
$\beta \vdash_{\FI(\lvalue)} \lvalue : \location{\fT}$}
\endprooftree
\end{align*}
\item[Pointer arithmetics]
Let $\mathord{\boxcircle}$ denote a binary abstract syntax operator
in $\{ \mathord{+}, \mathord{-} \}$, as well as the corresponding
unary operation on integers. Then, the following are added to
rule schemata~\eqref{rule:conc_arith_bop_0}--\eqref{rule:conc_arith_bop_exc_0}.

\begin{gather*}
\prooftree
  \rho \vdash_\beta \langle e_0, \sigma \rangle
    \rightarrow
      \langle l, \sigma_0 \rangle
\quad
  \rho \vdash_\beta \langle e_1, \sigma_0 \rangle
    \rightarrow
      \exceptstate
\justifies
  \rho \vdash_\beta \langle e_0 \boxcircle e_1, \sigma \rangle
    \rightarrow
      \exceptstate
\endprooftree \\[1ex]
\prooftree
  \rho \vdash_\beta \langle e_0, \sigma \rangle
    \rightarrow
      \langle l, \sigma_0 \rangle
\quad
  \rho \vdash_\beta \langle e_1, \sigma_0 \rangle
    \rightarrow
      \langle m, \sigma_1 \rangle
\justifies
  \rho \vdash_\beta \langle e_0 \boxcircle e_1, \sigma \rangle
    \rightarrow
      \ptrmove(m_0, l, \sigma_1)
\using\quad\text{if $m_0 = \mathop{\boxcircle} m$}
\endprooftree \\[1ex]
\prooftree
  \rho \vdash_\beta \langle e_0, \sigma \rangle
    \rightarrow
      \langle m, \sigma_0 \rangle
\quad
  \rho \vdash_\beta \langle e_1, \sigma_0 \rangle
    \rightarrow
      \langle l, \sigma_1 \rangle
\justifies
  \rho \vdash_\beta \langle e_0 + e_1, \sigma \rangle
    \rightarrow
      \ptrmove(m, l, \sigma_1)
\endprooftree \\[1ex]
\prooftree
  \rho \vdash_\beta \langle e_0, \sigma \rangle
    \rightarrow
      \langle l_0, \sigma_0 \rangle
\quad
  \rho \vdash_\beta \langle e_1, \sigma_0 \rangle
    \rightarrow
      \langle l_1, \sigma_1 \rangle
\justifies
  \rho \vdash_\beta \langle e_0 - e_1, \sigma \rangle
    \rightarrow
      \ptrdiff(l_0, l_1, \sigma_1)
\endprooftree
\end{gather*}
\item[Pointer comparison]
Let $\mathord{\boxast}$ denote a binary abstract syntax operator
in the set
\(
  \{ \mathord{=}, \mathord{\neq}, \mathord{<},
    \mathord{\leq}, \mathord{\geq}, \mathord{>} \}
\).
Then, the following are added to rule
schemata~\eqref{rule:conc_arith_test_error_0}--\eqref{rule:conc_arith_test_ok}.

\begin{gather*}
\prooftree
  \rho \vdash_\beta \langle e_0, \sigma \rangle
    \rightarrow
      \langle l, \sigma_0 \rangle
\quad
  \rho \vdash_\beta \langle e_1, \sigma_0 \rangle
    \rightarrow
      \exceptstate
\justifies
  \rho \vdash_\beta \langle e_0 \boxast e_1, \sigma \rangle
    \rightarrow
      \exceptstate
\endprooftree \\[1ex]
\prooftree
  \rho \vdash_\beta \langle e_0, \sigma \rangle
    \rightarrow
      \langle l_0, \sigma_0 \rangle
\quad
  \rho \vdash_\beta \langle e_1, \sigma_0 \rangle
    \rightarrow
      \langle l_1, \sigma_1 \rangle
\justifies
  \rho \vdash_\beta \langle e_0 \boxast e_1, \sigma \rangle
    \rightarrow
      \ptrcmp_\boxast(l_0, l_1, \sigma_1)
\endprooftree
\end{gather*}
\item[Assignment]
\begin{align*}
&
\begin{aligned}
\prooftree
  \rho \vdash_\beta \langle \lvalue, \sigma \rangle
    \rightarrow
      \exceptstate
\justifies
  \rho \vdash_\beta \langle \lvalue := e, \sigma \rangle
    \rightarrow
      \exceptstate
\endprooftree
&\quad&
\prooftree
  \rho \vdash_\beta \langle \lvalue, \sigma \rangle
    \rightarrow
      (l, \sigma_0)
\quad
  \rho \vdash_\beta \langle e, \sigma_0 \rangle
    \rightarrow
      \exceptstate
\justifies
  \rho \vdash_\beta \langle \lvalue := e, \sigma \rangle
    \rightarrow
      \exceptstate
\endprooftree
\end{aligned} \\[1ex]
&
\prooftree
  \rho \vdash_\beta \langle \lvalue, \sigma \rangle
    \rightarrow
      (l, \sigma_0)
\quad
  \rho \vdash_\beta \langle e, \sigma_0 \rangle
    \rightarrow
      \langle \sval, \sigma_1 \rangle
\justifies
  \rho \vdash_\beta \langle \lvalue := e, \sigma \rangle
    \rightarrow
      \sigma_1\bigl[ (l, \sT) := \sval \bigr]
\using\quad\text{if $\beta \vdash_{\FI(e)} e : \sT$}
\endprooftree
\end{align*}
Similar changes are required for the case of a function call.
First, the lvalue is evaluated so as to obtain the target location
where the result of the function call will be stored;
then, the function designator (an expression) is evaluated to obtain
a location having function type; this location is fed to the memory
structure so as to obtain the function abstract.
All the other computation steps, including parameter passing,
are performed as before. On exit from the function call,
the return value is stored at the location computed in the first step.
Exceptions are eventually detected and propagated as usual.
Also note that, thanks to the rules for type decay,
arrays and functions can be passed to and returned from function calls.
\item[(Multi-dimensional) Global array declaration]
In the following rule schemata, let $n > 0$,
$\aT = \arraytype{m_1}{( \dots (\arraytype{m_n}{\sT}) \dots )}$
and $m = m_1 \times \ldots \times m_n$.
\begin{gather*}
\prooftree
  \rho \vdash_\beta \langle e, \sigma \rangle
    \rightarrow
      \eta
\justifies
  \rho
    \vdash_\beta
      \langle
        \kw{gvar} \id : \aT = e,
        \sigma
      \rangle
    \rightarrow
      \datcleanup(\exceptstate)
\endprooftree \\
\intertext{%
if either $\eta = \exceptstate$,
or $\eta = \valstate$ and $\arraydatnew(m, \valstate) = \exceptstate$;
}
\prooftree
  \rho \vdash_\beta \langle e, \sigma \rangle
    \rightarrow
      \valstate
\justifies
  \rho
    \vdash_\beta
      \langle
        \kw{gvar} \id : \aT = e,
        \sigma
      \rangle
    \rightarrow
      \langle \rho_0, \sigma_0 \rangle
\endprooftree
\end{gather*}
if $\arraydatnew(m, \valstate) = (\sigma_0, l)$ and
$\rho_0 = \bigl\{ \id \mapsto (l, \aT) \bigr\}$.
\end{description}
The rules for local array declaration are similar.
Since function abstracts are now stored in memory structures,
a few minor adaptations, omitted for space reasons, are also required
for the rule of function declarations (which uses $\txtnew$) and
the rules for recursive environments and declarations.

\subsection{Heap Memory Management}
\label{sec:heap-memory-management}

By adding a \emph{heap segment} to memory structures, as well as
suitable helper functions ($\heapnew$, $\heapdel$ and the
corresponding array versions), it is possible to further extend
the language to embrace dynamic memory allocation and deallocation.

\subsubsection{Syntax}

We add an allocation expression and a deallocation statement:
\begin{align*}
  \Exp &\ni
  e ::= \ldots
      \vbar \kw{new} \sT = e \\
  \Stmt &\ni
  s ::= \ldots
      \vbar \kw{delete} e
\end{align*}

\subsubsection{Static Semantics}

\begin{align*}
%% New storable.
\prooftree
  \beta \vdash_I e : \sT
\justifies
  \beta \vdash_I \kw{new} \sT = e : \pointer{\sT}
\endprooftree
&\quad&
%% Delete storable.
\prooftree
  \beta \vdash_I e : \pointer{\sT}
\justifies
  \beta \vdash_I \kw{delete} e
\endprooftree
\end{align*}

\subsubsection{Concrete Dynamic Semantics}
This is extended with the schemata:
\begin{description}
\item[New expression]
\begin{align*}
&
\begin{aligned}
\prooftree
  \rho \vdash_\beta \langle e, \sigma \rangle
    \rightarrow
      \exceptstate
\justifies
  \rho \vdash_\beta \langle \kw{new} \sT = e, \sigma \rangle
    \rightarrow
      \exceptstate
\endprooftree
&\quad&
\prooftree
  \rho \vdash_\beta \langle e, \sigma \rangle
    \rightarrow
      \valstate
\justifies
  \rho \vdash_\beta \langle \kw{new} \sT = e, \sigma \rangle
    \rightarrow
      \exceptstate
\using\quad\text{if $\heapnew(\valstate) = \exceptstate$}
\endprooftree
\end{aligned} \\[1ex]
&
\prooftree
  \rho \vdash_\beta \langle e, \sigma \rangle
    \rightarrow
      \valstate
\justifies
  \rho \vdash_\beta \langle \kw{new} \sT = e, \sigma \rangle
    \rightarrow
      \langle l, \sigma_0 \rangle
\using\quad\text{if $\heapnew(\valstate) = (\sigma_0, l)$}
\endprooftree
\end{align*}
\item[Delete operator]
\begin{align*}
\prooftree
  \rho \vdash_\beta \langle e, \sigma \rangle
    \rightarrow
      \exceptstate
\justifies
  \rho \vdash_\beta \langle \kw{delete} e, \sigma \rangle
    \rightarrow
      \exceptstate
\endprooftree
&\quad&
\prooftree
  \rho \vdash_\beta \langle e, \sigma \rangle
    \rightarrow
      \valstate
\justifies
  \rho \vdash_\beta \langle \kw{delete} e, \sigma \rangle
    \rightarrow
      \heapdel(\valstate)
\endprooftree
\end{align*}
\end{description}
Similar rules allow for allocation and deallocation of an array on the heap:
note that, contrary to the previous cases,
the dimensions of the array can be specified as expressions
that will be evaluated dynamically.

Regarding the abstract semantics, the extensions concerning C-like
pointers and arrays as well as heap memory management can be obtained
along the lines followed in
Section~\ref{sec:abstract-dynamic-semantics}.
In particular, the new memory structure operators described above are
provided with safe approximations and a new abstract domain $\Loc^\sharp$
for location-valued expressions has to be defined.
By generalizing the abstract memory read and update operators so as
to take as input an abstract location, we realize the so-called
\emph{weak read} and \emph{weak update} operators, so as to correctly
deal with, e.g., assignments or function calls whose target is not
statically known.
In practice, no fundamentally new issue has to be solved as far as
the specification of the abstract interpreter is concerned.
This is not to say that these extensions are trivial; rather,
the real issues (e.g., the efficient and accurate tracking of
aliasing information for pointers~\cite{Emami93th,EmamiGH94}
or the appropriate summarization
techniques for large arrays~\cite{GopanRS05} and
heap-allocated data~\cite{GopanDMDRS04,SagivRW02})
are orthogonal to the current approach and should be addressed elsewhere.

\subsection{Non-Structured Control Flow Mechanisms}
\label{sec:non-structured-control-flow-mechanisms}

It turns out that the approach we have chosen to model exceptional
behavior of programs can be easily generalized so as to capture all the
non-structured control flow mechanisms of languages such as
C and \Cplusplus{}.
To exemplify such a generalization, the abstract syntax of commands is
extended with branching and labeled statements:
\begin{align*}
  \Label \ni
  l &::= \id
     \vbar m
     \vbar \kw{default} \\
  \Stmt \ni
  s &::= \ldots
     \vbar \kw{goto} \id
     \vbar \kw{switch} e \kw{in} s
     \vbar \kw{break}
     \vbar \kw{continue}
     \vbar \kw{return} e
     \vbar l : s
\end{align*}
We assume that the static semantics ensures the labels used in a function
body are all distinct (if the language supports local labels, then a trivial
renaming will be required) and that every goto has access to a corresponding
labeled statement, respecting the constraints imposed by the language
(concerning, for instance, jumping into and outside blocks).

The state of a computation is captured, besides the current
program point, by a \emph{control mode} and a memory structure,
which together constitute what we call a \emph{control state}.
A control state is classified by the corresponding control mode
in either a plain execution state or an \emph{exception state};
a plain execution state can be further distinguished in either
a \emph{normal execution state},
or a \emph{branching state},
or a \emph{value state} (for computations yielding a proper value),
or an \emph{environment state} (for computations yielding an
execution environment).
\begin{definition}
\summary{($\GotoMode$, $\SwitchMode$, $\ValMode$,
          $\EnvMode$, $\ExceptMode$, $\CtrlMode$,
          $\CtrlState$.)}
\label{def:concrete-control-mode}
\label{def:concrete-control-state}
The sets of \emph{goto}, \emph{switch},
\emph{value},
\emph{environment}, \emph{exception} and all \emph{control modes}
are given, respectively, by
\begin{align*}
  \GotoMode
    &\defeq
      \bigl\{\, \cmgoto(\id) \bigm| \id \in \Id \,\bigr\}, \\
  \SwitchMode
    &\defeq
           \bigl\{\, \cmswitch(\sval) \bigm| \sval \in \sVal \,\bigr\},  \\
  \ValMode
    &\defeq
      \bigl\{\, \cmvalue(\sval) \bigm| \sval \in \sVal \,\bigr\}, \\
  \EnvMode
    &\defeq
      \bigl\{\, \cmenv(\rho) \bigm| \rho \in \Env \,\bigr\}, \\
  \ExceptMode
    &\defeq
      \bigl\{\, \cmexcept(\except) \bigm| \except \in \Except \,\bigr\}, \\
  \CtrlMode
    &\defeq
      \GotoMode \uplus \SwitchMode \uplus \ValMode
        \uplus \EnvMode \\
    & \qquad  \uplus \ExceptMode
                \uplus \{ \cmcontinue, \cmbreak, \cmreturn, \cmexec \},
\end{align*}
where $\cmcontinue$, $\cmbreak$ and $\cmreturn$
are the \emph{exit modes}
and $\cmexec$ is the \emph{plain execution mode}.
Control modes are denoted by $\cm$, $\cm_0$, $\cm_1$ and so forth.

A \emph{control state} is an element of
\(
  \CtrlState
    \defeq
      \CtrlMode \times \Mem
\).
Control states are denoted by
$\cs$, $\cs_0$, $\cs_1$ and so forth.
\end{definition}

The concrete semantics of the goto statement can now be expressed by
\[
\prooftree
  \nohyp
\justifies
  \rho \vdash_\beta
         \bigl\langle
           \kw{goto} \id, (\cm, \sigma)
         \bigr\rangle
    \rightarrow
      \langle \cm_0, \sigma \rangle
\endprooftree
\]
if $\cm = \cmexec$ and $\cm_0 = \cmgoto(\id)$ or
$\cm \neq \cmexec$ and $\cm_0 = \cm$.

The semantics of labeled statements is given by
\[
\prooftree
  \rho \vdash_\beta \bigl\langle s, (\cm_0, \sigma) \bigr\rangle
    \rightarrow
      \eta
\justifies
  \rho \vdash_\beta \bigl\langle l : s, (\cm, \sigma) \rangle
    \rightarrow
      \eta
\endprooftree
\]
where $\cm_0 = \cmexec$
if $\cm = \cmexec$,
or $\cm = \cmgoto(\id)$ and $l = \id$,
or $\cm = \cmswitch(\sval)$ and $l \in \{\kw{default}, \sval\}$;
otherwise $\cm_0 = \cm$.

Of course, the semantics of all statements must be suitably modified.
For instance, the assignment should behave like a nop unless the control
mode is the normal execution one.
Statements with non trivial control flow need more work.
For example, the semantics of the conditional statement can be captured
by\footnote{Recall that, in C, it is perfectly legal to jump into
the ``else branch'' from the ``then branch.''}
\begin{gather}
\notag
\prooftree
  \rho \vdash_\beta \bigl\langle e, (\cmexec, \sigma) \bigr\rangle
    \rightarrow
      \langle \cm_0, \sigma_0 \rangle
\justifies
  \rho \vdash_\beta
         \bigl\langle
           \kw{if} e \kw{then} s_0 \kw{else} s_1, (\cmexec, \sigma)
         \bigr\rangle
    \rightarrow
      \langle \cm_0, \sigma_0 \rangle
\using\quad\text{if $\cm_0 \in \ExceptMode$}
\endprooftree \\[1ex]
\label{rule:conc_conditional_then_jumps_into_else}
\prooftree
\begin{aligned}
  \rho \vdash_\beta \bigl\langle e, (\cmexec, \sigma) \bigr\rangle
    \rightarrow
      \bigl\langle \cmvalue(\ttv), \sigma_0 \bigr\rangle
\quad
  &\rho \vdash_\beta \bigl\langle s_0, (\cmexec, \sigma_0) \bigr\rangle
    \rightarrow
      \langle \cm_1, \sigma_1 \rangle \\
  &\rho \vdash_\beta
      \bigl\langle s_1, (\cm_1, \sigma_1) \bigr\rangle
    \rightarrow
      \eta
\end{aligned}
\justifies
  \rho \vdash_\beta
         \bigl\langle \kw{if} e \kw{then} s_0 \kw{else} s_1, (\cmexec, \sigma)
         \bigr\rangle
    \rightarrow
      \eta
\endprooftree
\end{gather}
if $\cm_1 \in \GotoMode$;
\begin{gather}
\notag
\prooftree
  \rho \vdash_\beta \bigl\langle e, (\cmexec, \sigma) \bigr\rangle
    \rightarrow
      \bigl\langle \cmvalue(\ttv), \sigma_0 \bigr\rangle
\quad
  \rho \vdash_\beta \bigl\langle s_0, (\cmexec, \sigma_0) \bigr\rangle
    \rightarrow
      \langle \cm_1, \sigma_1 \rangle
\justifies
  \rho \vdash_\beta
         \bigl\langle \kw{if} e \kw{then} s_0 \kw{else} s_1, (\cmexec, \sigma)
         \bigr\rangle
    \rightarrow
      \langle \cm_1, \sigma_1 \rangle
\endprooftree
\end{gather}
if $\cm_1 \notin \GotoMode$;
\begin{gather*}
\prooftree
  \rho \vdash_\beta \bigl\langle e, (\cmexec, \sigma) \bigr\rangle
    \rightarrow
      \bigl\langle \cmvalue(\ffv), \sigma_0 \bigr\rangle
\quad
  \rho \vdash_\beta \bigl\langle s_1, (\cmexec, \sigma_0) \bigr\rangle
    \rightarrow
      \eta
\justifies
  \rho \vdash_\beta
         \bigl\langle
           \kw{if} e \kw{then} s_0 \kw{else} s_1, (\cmexec, \sigma)
         \bigr\rangle
    \rightarrow
      \eta
\endprooftree \\[1ex]
\prooftree
  \rho \vdash_\beta \bigl\langle s_0, (\cm, \sigma) \bigr\rangle
    \rightarrow
      \langle \cm_0, \sigma_0 \rangle
\justifies
  \rho \vdash_\beta
         \bigl\langle
           \kw{if} e \kw{then} s_0 \kw{else} s_1, (\cm, \sigma)
         \bigr\rangle
    \rightarrow
      \langle \cm_0, \sigma_0 \rangle
\endprooftree
\end{gather*}
if $\cm \in \GotoMode \uplus \SwitchMode$
and $\cm_0 \notin \GotoMode \uplus \SwitchMode$;
\begin{gather*}
\prooftree
  \rho \vdash_\beta \bigl\langle s_0, (\cm, \sigma) \bigr\rangle
    \rightarrow
      \langle \cm_0, \sigma_0 \rangle
\quad
  \rho \vdash_\beta \bigl\langle s_1, (\cm_0, \sigma_0) \bigr\rangle
    \rightarrow
      \eta
\justifies
  \rho \vdash_\beta
         \bigl\langle \kw{if} e \kw{then} s_0 \kw{else} s_1, (\cm, \sigma)
         \bigr\rangle
    \rightarrow
      \eta
\endprooftree
\end{gather*}
if $\cm \in \GotoMode \uplus \SwitchMode$
and $\cm_0 \in \GotoMode \uplus \SwitchMode$;
\begin{gather*}
\prooftree
  \nohyp
\justifies
  \rho \vdash_\beta
         \bigl\langle
           \kw{if} e \kw{then} s_0 \kw{else} s_1, (\cm, \sigma)
         \bigr\rangle
    \rightarrow
      \langle \cm, \sigma \rangle
\endprooftree
\end{gather*}
if $\cm \notin \GotoMode \uplus \SwitchMode \uplus \{ \cmexec \}$.

Likewise, the semantics of the $\kw{switch}$ statement can be captured by:
\begin{gather}
\notag
\prooftree
  \rho \vdash_\beta \bigl\langle e, (\cmexec, \sigma) \bigr\rangle
    \rightarrow
      \langle \cm_0, \sigma_0 \rangle
\justifies
  \rho \vdash_\beta
         \bigl\langle
           \kw{switch} e \kw{in} s, (\cmexec, \sigma)
         \bigr\rangle
    \rightarrow
      \langle \cm_0, \sigma_0 \rangle
\using\quad\text{if $\cm_0 \in \ExceptMode$}
\endprooftree \\[1ex]
\notag
\prooftree
  \begin{aligned}
  &\rho \vdash_\beta \bigl\langle e, (\cmexec, \sigma) \bigr\rangle
    \rightarrow
      \bigl\langle \cmvalue(\sval_0), \sigma_0 \bigr\rangle \\
  &\qquad
  \rho \vdash_\beta \bigl\langle s, (\cmswitch(\sval_0), \sigma_0) \bigr\rangle
    \rightarrow
      \langle \cm_1, \sigma_1 \rangle
  \end{aligned}
\justifies
  \rho \vdash_\beta
         \bigl\langle
           \kw{switch} e \kw{in} s, (\cmexec, \sigma)
         \bigr\rangle
    \rightarrow
      \langle \cm_2, \sigma_1 \rangle
\endprooftree
\end{gather}
if
\(
  \cm_2 =
    \begin{cases}
      \cmexec,
        &\text{if $\cm_1 \in \SwitchMode \uplus \{ \cmbreak \}$,} \\
      \cm_1,
        &\text{otherwise;}
    \end{cases}
\)
\begin{gather}
\notag
\prooftree
  \rho \vdash_\beta \bigl\langle s, (\cmgoto(\id), \sigma) \bigr\rangle
    \rightarrow
      \langle \cm_0, \sigma_0 \rangle
\justifies
  \rho \vdash_\beta
         \bigl\langle
           \kw{switch} e \kw{in} s, (\cmgoto(\id), \sigma)
         \bigr\rangle
    \rightarrow
      \langle \cm_1, \sigma_0 \rangle
\endprooftree
\end{gather}
if
\(
  \cm_1 =
    \begin{cases}
      \cmexec,
        &\text{if $\cm_0 = \cmbreak$,} \\
      \cm_0,
        &\text{otherwise;}
    \end{cases}
\)
\begin{gather}
\notag
\prooftree
  \nohyp
\justifies
  \rho \vdash_\beta
         \bigl\langle
           \kw{switch} e \kw{in} s, (\cm, \sigma)
         \bigr\rangle
    \rightarrow
            \langle \cm, \sigma \rangle
\using\quad\text{if $\cm \notin \GotoMode \uplus \{ \cmexec \}$.}
\endprooftree
\end{gather}

While such a semantic treatment captures all forward jumps,
for backward jumps something more is required.
One simple possibility (which is not the only one) is to explicitly
introduce a looping construct that is (only) available in the abstract
syntax.  That is, we extend $\Stmt$ once again as
\[
  \Stmt \ni
  s ::= \ldots
     \vbar \kw{loop} s
\]
and assume that a set of such loops has been inserted so that all backward
jumps are enclosed in at least one loop (notice that at most one such
loop per function body suffices, but more can be used as a matter of
optimization).
For $s \in \Stmt$, let $\SL(s)$ denote the set of statement labels in $s$.
The concrete semantics of this looping construct is now given by
\begin{gather*}
\prooftree
  \rho \vdash_\beta \langle s, \cs \rangle
    \rightarrow
      \langle \cm, \sigma \rangle
\justifies
  \rho \vdash_\beta \langle \kw{loop} s, \cs \rangle
    \rightarrow
      \langle \cm, \sigma \rangle
\using\quad\text{if $\cm \neq \cmgoto(\id)$ for each $\id \in \SL(s)$}
\endprooftree \\[1ex]
\prooftree
  \rho \vdash_\beta \langle s, \cs \rangle
    \rightarrow
      \bigl\langle \cmgoto(\id), \sigma \bigr\rangle
\quad
  \rho \vdash_\beta
      \bigl\langle \kw{loop} s, \bigl(\cmgoto(\id), \sigma\bigr) \bigr\rangle
    \rightarrow
      \eta
\justifies
  \rho \vdash_\beta \langle \kw{loop} s, \cs \rangle
    \rightarrow
      \eta
\using\quad\text{if $\id \in \SL(s)$}
\endprooftree
\end{gather*}
Observe that the systematic use of the looping construct can make
rule schema~\eqref{rule:conc_conditional_then_jumps_into_else}
redundant.

Other rules are omitted for space reasons.  However, there are
no additional difficulties besides the ones just addressed:
the rules for $\kw{break}$ and $\kw{continue}$ are straightforward;
$\kw{return} e$ can be modeled as the assignment to the reserved
identifier $\ridx_0$
(see concrete rule~\eqref{rule:conc_function_body_2}),
followed by the setting of the control mode;
the rules for the $\kw{while}$ loop are a bit involved as they must
support the `$\cmbreak$' and `$\cmcontinue$' control modes in addition
to `$\cmgoto$' and `$\cmswitch$'.

The proposed approach handles non-structured control flow mechanisms
essentially by adding a sort of control register to the rule-based
interpreter of the language.
As far as the abstract semantics is concerned, a first choice to be made
concerns the approximation of the values that the control register can take.
As usual, there is a complexity/precision trade-off to be faced:
the simple solution is to approximate $\wp(\CtrlMode)$ by some (simple)
abstract domain $\CtrlMode^\sharp$ and then approximate
$\CtrlState = \CtrlMode \times \Mem$
by $\CtrlMode^\sharp \stimes \Mem^\sharp$;
a more precise solution is to approximate $\wp(\CtrlState)$
by an abstract domain $\CtrlState^\sharp$ that captures
relational information connecting the control modes to the
memory structures they can be coupled with.
The abstract rules schemata must of course be modified to match the
concrete world.
For instance, the abstract rule for the conditional statement becomes:
\begin{equation*}
\label{rule:abstr_conditional_extended}
\prooftree
  \rho \vdash_\beta \langle e, \cs^\sharp_\mathrm{cond} \rangle
    \rightarrow
      \cs^\sharp_0
\quad
  \rho \vdash_\beta \langle s_0, \cs^\sharp_\mathrm{then} \rangle
    \rightarrow
      \cs^\sharp_1
\quad
  \rho \vdash_\beta \langle s_1, \cs^\sharp_\mathrm{else} \rangle
    \rightarrow
      \cs^\sharp_2
\justifies
  \rho
    \vdash_\beta
      \langle \kw{if} e \kw{then} s_0 \kw{else} s_1, \cs^\sharp \rangle
    \rightsquigarrow
      \cs^\sharp_3
\endprooftree
\end{equation*}
where
\begin{align*}
  \cs^\sharp_\mathrm{cond}
     &=
       \Phi_\mathrm{e}(\rho, \cs^\sharp, \ttv), \\
  \cs^\sharp_\mathrm{then}
     &=
       \Phi_\mathrm{e}(\rho, \cs^\sharp, e)
     \sqcup
       \Phi_\mathrm{m}(\cs^\sharp, \GotoMode \uplus \SwitchMode), \\
  \cs^\sharp_\mathrm{else}
     &=
       \Phi_\mathrm{e}(\rho, \cs^\sharp, \kw{not} e)
     \sqcup
       \Phi_\mathrm{m}(\cs^\sharp_1, \GotoMode)
     \sqcup \cs^\sharp_\mathrm{jump}, \\
  \cs^\sharp_\mathrm{jump}
     &=
     \begin{cases}
       \bot,
         &\text{if
             $\Phi_\mathrm{m}(\cs^\sharp, \GotoMode \uplus \SwitchMode) = \bot$,} \\
       \Phi_\mathrm{m}(\cs^\sharp_1, C_\mathrm{jump}),
         &\text{otherwise,}
     \end{cases} \\
  C_\mathrm{jump}
     &=
       \GotoMode \union \bigl\{\,
                          \cm \in \CtrlMode
                        \bigm|
                          \exists \sigma \in \Mem
                            \st \gamma(\cs^\sharp) = (\cm, \sigma)
                        \,\bigl\}, \\
  \cs^\sharp_3
     &=
       \Phi_\mathrm{m}
         \Bigl(
           \cs^\sharp,
           \CtrlMode
             \setminus
               \bigl(\{\cmexec\} \uplus \GotoMode \uplus \SwitchMode\bigr)
         \Bigr) \\
     &\qquad
     \sqcup
       \Phi_\mathrm{m}
         \Bigl(
           \cs^\sharp_0,
           \CtrlMode
             \setminus
               \ValMode
         \Bigr)
     \sqcup
        \cs^\sharp_1
     \sqcup
        \cs^\sharp_2,
\end{align*}
and the two computable filter functions
\(
  \fund{\Phi_\mathrm{e}}%
       {(\Env \times \CtrlState^\sharp \times \Exp)}%
       {\CtrlState^\sharp}
\)
and
\(
  \fund{\Phi_\mathrm{m}}%
       {\bigl(\CtrlState^\sharp \times \wp(\CtrlMode)\bigr)}%
       {\CtrlState^\sharp}
\)
are defined as follows,
for each $\rho \in \Env$, $\cs^\sharp \in \CtrlState^\sharp$,
$e \in \Exp$ and $C \sseq \CtrlMode$ such that,
for some $\beta \in \TEnv$, $\beta : I$ with $\FI(e) \sseq I$
and $\beta \vdash_I e : \tboolean$:
\begin{align*}
  \gamma\bigl(\Phi_\mathrm{e}(\rho, \cs^\sharp, e)\bigr)
    &\supseteq
      \sset{%
        \cs \in \gamma(\cs^\sharp)
      }{%
        \exists \sigma \in \Mem \st \cs = (\cmexec, \sigma), \\
        \exists \sigma' \in \Mem \\
          \qquad
          \st
            \bigl(
              \rho
                \vdash_\beta
                  \langle e, \cs \rangle
                    \rightarrow
                      \bigl\langle
                        \cmvalue(\ttv), \sigma'
                      \bigr\rangle
            \bigr)
      }, \\
  \gamma\bigl(\Phi_\mathrm{m}(\cs^\sharp, C)\bigr)
    &\supseteq
      \bigl\{\,
        \cs \in \gamma(\cs^\sharp)
      \bigm|
        \exists \sigma \in \Mem \st \cs = (\cm, \sigma),
        \cm \in C
      \,\bigr\}.
\end{align*}

\section{Conclusion}
\label{sec:conclusion}

In this paper, we have confronted the problem of defining an analysis
framework for the specification and realization of precise static
analyzers for mainstream imperative programming languages, tools in
very short supply that, however, ought to become part of the current
programming practice.
A proposal put forward by Schmidt twelve years ago \cite{Schmidt95}
held, in our eyes, considerable promise, despite the fact it had not
been fully developed and applied in realistic contexts.
It was therefore natural to question whether the promise
could be fulfilled.
To investigate Schmidt's approach, which is based on structured
operational semantics and abstract interpretation, we have defined an
imperative language, CPM, that embodies all the ``problematic features'' of
single-threaded imperative languages now in widespread use.
We have presented a concrete semantics of CPM that is suitable for
abstraction while retaining all the nice features of SOS descriptions.
For a subset of the language we have formally defined an abstract semantics
that can fully exploit the precision offered by relational abstract
domains, and proved its soundness with respect to the concrete one.
We have also shown how approximations of the abstract semantics
can be effectively computed.
In order to provide an experimental evaluation of the ideas presented
in this paper, both the concrete and the abstract semantics
---instantiated over sophisticated numeric domains and together with
a suitable fixpoint computation engine--- have been incorporated
into the ECLAIR system.  This work allows us to conclude that
the proposal of Schmidt can play a crucial role in the development
of reliable and precise analyzers.
The key features of this approach are:
\begin{itemize}
\item
a fairly concise concrete semantics that experts can easily read
(and modify as needed) and that everyone can execute on non-trivial
examples in order to check its agreement with the applicable language
standards;
\item
a fairly concise abstract semantics that is fully parametric with
respect to the abstract domain, that is not difficult to prove correct
with respect to the concrete one (to the point that automatizing the
proof seems to be a reasonable goal), and that directly leads to
the implementation of static analyzers.
\end{itemize}
Of course, the story does not end here.  For instance, our analysis
framework is parametric on abstract memory structures.  While the
literature seems to provide all that is necessary to realize very
sophisticated ones, it is not difficult to predict that, among all the
code out there waiting to be analyzed, some will greatly exacerbate
the complexity/precision trade-off.  However, these are research
problems for the future --- now that we have, as given here, a formal
design on which analyzers can be built, our next goal is to complete
the build and make the technology described here truly available and
deployable.

\ifthenelse{\boolean{TOPLAS}}{
\begin{acks}
}{
\bigskip
\noindent
{\bfseries Acknowledgments.}
}
Anna Dolma Alonso, Irene Bacchi, Danilo Bonardi, Andrea Cimino,
Enrico Franchi, Davide Masi and Alessandro Vincenzi
(all students of the course on ``Analysis and Verification of Software''
taught by Roberto Bagnara at the University of Parma)
and Vajirapan Panumong (University of Leeds) collaborated on previous,
much more restricted versions of this work.
We are also grateful to David Merchat (formerly at the University of Parma)
and Katy Dobson (University of Leeds)
for the discussions we have had on the subject of this paper.
\ifthenelse{\boolean{TOPLAS}}{
\end{acks}
}{
}

\ifthenelse{\boolean{TOPLAS}}{
\bibliographystyle{amsalpha}
}{
\bibliographystyle{plain}
}
%\bibliography{ppl,ppl_citations,mybib}
\newcommand{\etalchar}[1]{$^{#1}$}
\newcommand{\noopsort}[1]{}\hyphenation{ Ba-gna-ra Bie-li-ko-va Bruy-noo-ghe
  Common-Loops DeMich-iel Dober-kat Di-par-ti-men-to Er-vier Fa-la-schi
  Fell-eisen Gam-ma Gem-Stone Glan-ville Gold-in Goos-sens Graph-Trace
  Grim-shaw Her-men-e-gil-do Hoeks-ma Hor-o-witz Kam-i-ko Kenn-e-dy Kess-ler
  Lisp-edit Lu-ba-chev-sky Ma-te-ma-ti-ca Nich-o-las Obern-dorf Ohsen-doth
  Par-log Para-sight Pega-Sys Pren-tice Pu-ru-sho-tha-man Ra-guid-eau Rich-ard
  Roe-ver Ros-en-krantz Ru-dolph SIG-OA SIG-PLAN SIG-SOFT SMALL-TALK Schee-vel
  Schlotz-hauer Schwartz-bach Sieg-fried Small-talk Spring-er Stroh-meier
  Thing-Lab Zhong-xiu Zac-ca-gni-ni Zaf-fa-nel-la Zo-lo }
\providecommand{\bysame}{\leavevmode\hbox to3em{\hrulefill}\thinspace}
\providecommand{\MR}{\relax\ifhmode\unskip\space\fi MR }
\providecommand{\MRhref}[2]{%
  \href{http://www.ams.org/mathscinet-getitem?mr=#1}{#2}
}
\providecommand{\href}[2]{#2}

\received
\endreceived

\end{document}